\documentclass[]{aa}
\usepackage[varg]{txfonts}
\usepackage{graphics}
\def\cc{\,{\rm cm^{-3}}} 
\def\cm2{\,{\rm cm^{-2}}}

\def\kms{\,{\rm {km\,s^{-1}}}} 
\def\kkms{\,{\rm {K\,km\,s^{-1}}}}

\def\co{\,{\rm ^{12}CO}} 
\def\thirco{\,{\rm ^{13}CO}} 
\def\h2{\,{\rm H_2}} 
\def\C34S{\,{\rm C^{34}S}} 
\def\hi{\,{\rm H I}}
\def\ci{\,{\rm [C I]}}
\def\hii{\,{\rm H II]}}
\def\cii{\,{\rm [C II]}}

\def\etal{et\,al.}
\def\to{\rightarrow} 
\def\aua{{\it A\&A} } 
\def\auas{{\it A\&A Suppl} } 
\def\apj{{\it ApJ} } 
\def\aj{{\it AJ} } 
\def\apjs{{\it ApJS} } 
\def\apjl{{\it ApJL} } 
\def\araa{{\it ARAA} } 
\def\mnras{{\it MNRAS} } 
 
\def\assl{{\it ASSL} } 
\def\pasj{{\it PASJ} } 
\def\pasp{{\it PASP} } 

\begin{document}

\title{Central molecular zones in galaxies: $\co$-to-$\thirco$ ratios, carbon budget, and $X$ factors}

\author{F.P. Israel 
        \inst{1} 
           } 

\offprints{F.P. Israel} 

\titlerunning{Central molecular zones in galaxies}

\institute{Sterrewacht Leiden, P.O. Box 9513, 2300 RA Leiden, the Netherlands}
 
\date{Accepted December 19. 2019} 
 
\abstract{We present ground-based measurements of 126 nearby galaxy
  centers in $\co$ and 92 in $\thirco$ in various low-$J$
  transitions. More than 60 galaxies were measured in at least four
  lines.  The average relative intensities of the first four $\co$ $J$
  transitions are 1.00 : 0.92 : 0.70 : 0.57. In the first three $J$
  transitions, the average $\co$-to-$\thirco$ intensity ratios are
  13.0, 11.6, and 12.8, with individual values in any transition
  ranging from 5 to 25.  The sizes of central CO concentrations are
  well defined in maps, but poorly determined by multi-aperture
  photometry.  On average, the $J$=1-0 $\co$ fluxes increase linearly
  with the size of the observing beam, but CO emission covers only a
  quarter of the $\hi$ galaxy disks.  Using radiative transfer models
  ({\it RADEX}), we derived model gas parameters.  The assumed carbon
  elemental abundances and carbon gas depletion onto dust are the main
  causes of uncertainty.  The new CO data and published $\ci$ and
  $\cii$ data imply that CO, C$^{\circ}$, and C$^{+}$ each represent
  about one-third of the gas-phase carbon in the molecular
  interstellar medium.  The mean beam-averaged molecular hydrogen
  column density is $N(\h2)\,=\,(1.5\pm0.2)\times10^{21}\,\cm2$.
  Galaxy center CO-to-$\h2$ conversion factors are typically ten times
  lower than the `standard' Milky Way $X_{\circ}$ disk value, with a
  mean $X$(CO) = $(1.9\pm0.2)\times10^{19}\,\cm2/\kkms$ and a
  dispersion $1.7\times10^{19}\,\cm2/\kkms$. The corresponding
  $\ci$-$\h2$ factor is five times higher than $X$(CO), with $X$[CI] =
  $(9\pm2)\times10^{19}\,\cm2/\kkms$. No unique conversion factor can
  be determined for $\cii$.  The low molecular gas content of galaxy
  centers relative to their CO intensities is explained in roughly
  equal parts by high central gas-phase carbon abundances, elevated
  gas temperatures, and large gas velocity dispersions relative to the
  corresponding values in galaxy disks.}


\maketitle

\section{Introduction}

The aim of this paper is to determine the carbon budget and the amount
of molecular hydrogen in the centers of nearby galaxies as accurately
as possible, based on extensive new observations and current chemical
and radiative transfer models. The bright inner disks of late-type
galaxies contain massive concentrations of circumnuclear molecular
hydrogen gas. These reservoirs feed central black holes, outflows, and
bursts of star formation.  Before their crucial role in inner galaxy
evolution can be understood and evaluated, the physical
characteristics of the gas must be determined.  Cool and quiescent
molecular hydrogen ($\h2$) gas is difficult to detect, and studies of
the molecular interstellar medium (ISM) in galaxies rely on the
observation of tracers such as continuum emission from thermal dust or
line emission from the CO molecule.  CO is one of the most common
molecules in the ISM after $\h2$, even though its relative abundance
is only about $10^{-5}$.  It has become the instrument of choice in
the investigation of the molecular ISM because it is comparatively
easy to detect and traces molecular gas already at low densities and
temperatures.

%Table 1 Sample
\begin{table*}
\begin{center}
{\small %
\caption[]{Galaxy sample} 
\begin{tabular}{lrrrc||lrrrc||lrrrc}
\noalign{\smallskip}     
\hline
\noalign{\smallskip} 
NGC & Dist.& lgFIR   &lg$L_{FIR}$ &$D_{25}$  & NGC & Dist.& lgFIR &lg$L_{FIR}$ & Size & NGC & Dist.& lgFIR &lg$L_{FIR}$ & Size \\
IC  & Mpc  & Wm$^{-2}$&$L_{\odot}$ & '       & IC  & Mpc  & Wm$^{-2}$&$L_{\odot}$& ' & IC & Mpc  & Wm$^{-2}$&$L_{\odot}$ & '\\
(1) & (2)  &     (3) & (4)       & (5)     &(1) & (2)  & (3)     & (4)      & (5) & (1) & (2)  &  (3)   &  (4) & (5)\\
\noalign{\smallskip}      
\hline
\noalign{\smallskip} 
N 134    & 21.5 & -11.88 & 10.26 & 8.1x2.6  & N2993    & 35.9 & -12.28 & 10.31 & 1.3x0.9  & N4666    & 27.5 & -11.74 & 10.62 & 4.6x1.3 \\
N 253    &  3.4 & -10.42 & 10.13 &  25x7.4  & N3034    &  5.9 & -10.28 & 10.74 &  11x4.3  & N4736    &  4.8 & -11.50 &  9.35 &  11x9.1  \\
N 275    & 23.6 & -12.56 &  9.67 & 1.5x1.1  & N3044    & 20.4 & -12.25 &  9.86 & 5.7x0.6  & N4826$^*$&  3.8 & -11.66 &  9.34 &  10x5.4  \\
N 278    & 11.3 & -11.87 &  9.76 & 2.2x2.1  & N3079$^*$& 20.7 & -11.60 & 10.51 & 7.9x1.4  & N4835    & 23.9 & -12.00 & 10.24 & 4.0x0.9  \\
N 300    &  1.9 & -11.77 &  8.26 &  22x16   & I2554    & 16.4 & -12.04 &  9.87 & 3.2x1.5  & N4945$^*$&  4.4 & -10.67 & 10.10 &  20x3.8  \\
N 470    & 31.7 & -12.44 & 10.04 & 2.8x1.7  & N3175    & 13.6 & -12.10 &  9.65 & 5.0x1.3  & N5033$^*$& 17.2 & -12.00 &  9.95 &  11x5.0  \\ 
N 520    & 30.5 & -11.79 & 10.66 & 4.5x1.8  & N3227$^*$& 20.3 & -12.32 &  9.77 & 5.4x3.6  & N5055    &  8.3 & -11.66 &  9.66 &  13x7.2  \\
N 613    & 19.7 & -11.90 & 10.17 & 5.5x4.2  & N3256    & 37.0 & -11.71 & 10.91 & 3.8x2.1  & N5135$^*$& 57.7 & -12.04 & 10.96 & 2.6x1.8  \\
N 628    &  9.9 & -12.60 &  8.87 &  11x9.5  & N3281$^*$& 44.7 & -12.50 & 10.28 & 3.3x1.8  & N5194$^*$&  9.1 & -11.59 &  9.81 &  11x6.9  \\
N 660    & 12.2 & -11.46 & 10.20 & 9.1      & N3310    & 19.2 & -11.79 & 10.26 & 3.1x2.4  & N5218    & 46.5 & -12.38 & 10.43 & 1.9x1.0  \\
N 695    &  130 & -12.36 & 11.35 & 0.8x0.7  & N3351    &  9.0 & -12.00 &  9.39 & 3.1x2.9  & N5236    &  4.0 & -11.22 &  9.46 &  13x12   \\
N 891    &  9.4 & -11.53 &  9.90 &  14x2.5  & N3504    & 27.8 & -11.98 & 10.39 & 2.7x2.2  & Circ$^*$ &  2.9 & -10.92 &  9.48 & 6.9x3.0  \\
N 908    & 19.9 & -12.00 & 10.08 & 6.0x2.6  & N3556    & 14.2 & -11.81 &  9.97 & 8.7x2.2  & N5433    & 67.8 & -12.44 & 10.70 & 1.6x0.4  \\
N 972    & 21.4 & -11.75 & 10.39 & 3.4x1.7  & N3593    &  5.6 & -11.97 &  9.01 & 5.75     & I4444    & 23.1 & -11.95 & 10.26 & 1.7x1.4  \\
Maff2    &  3.1 & -11.23 &  9.23 & 5.8x1.6  & N3620    & 20.4 & -11.61 & 10.49 & 2.8x1.1  & N5643$^*$& 14.4 & -11.93 &  9.87 & 4.6x4.0  \\
N1055    & 13.4 & -11.84 &  9.89 & 7.6x2.7  & N3621    &  6.5 & -11.96 &  9.15 &  12x7.1  & N5713    & 31.3 & -11.95 & 10.52 & 2.8x2.5  \\
N1068$^*$& 15.2 & -11.04 & 10.80 & 7.1x6.0  & N3627    &  6.5 & -11.61 &  9.50 & 9.1x4.2  & N5775    & 28.9 & -11.97 & 10.43 & 4.2x1.0  \\ 
N1084    & 18.6 & -12.33 &  9.69 & 3.3x1.2  & N3628    &  8.5 & -11.54 &  9.80 &  15x3.0  & N6000    & 31.0 & -11.71 & 10.75 & 1.9x1.6  \\ 
N1097$^*$& 16.5 & -11.81 & 10.10 & 9.3x6.6  & N3690    & 48.5 & -11.32 & 11.53 & 2.9x2.1  & N6090    &  126 & -12.47 & 11.21 & 1.7x0.7  \\ 
N1317    & 25.8 & -12.62 &  9.68 & 2.8x2.4  & N3783    & 36.1 & -12.76 &  9.83 & 1.9x1.7  & N6215    & 20.2 & -11.83 & 10.26 & 2.1x1.8  \\
N1365$^*$& 21.5 & -11.36 & 10.78 &  11x6.2  & N3982$^*$& 21.8 & -12.37 &  9.79 & 1.7x1.5  & N6221$^*$& 19.3 & -11.65 & 10.40 & 3.5x2.5  \\
I342     &  3.8 & -11.36 &  9.28 &  21x21   & N4030    & 26.4 & -11.95 & 10.37 & 4.3      & N6240    &  109 & -11.96 & 11.59 & 2.1x1.1  \\ 
N1433$^*$& 13.3 & -12.55 &  9.18 & 6.5x5.9  & N4038    & 23.3 & -11.65 & 10.56 & 5.2x3.1  & N6300$^*$& 14.0 & -12.02 &  9.75 & 4.5x3.0  \\ 
N1448    & 14.7 & -12.22 &  9.59 & 7.6x1.7  & N4039    & 23.3 & -11.65 & 10.56 & 3.1x1.6  & N6744    & 10.7 & -12.55 &  8.99 &  20x13   \\ 
N1482    & 25.4 & -11.79 & 10.50 & 2.5x1.4  & N4051$^*$& 12.9 & -12.27 &  9.43 & 5.2x3.9  & N6764    & 38.5 & -12.44 & 10.21 & 2.3x1.3  \\ 
N1559    & 16.3 & -11.83 & 10.07 & 3.5x2.0  & N4102    & 17.3 & -11.62 & 10.34 & 2.8x1.2  & N6810    & 28.8 & -11.99 & 10.41 & 3.2x0.9  \\ 
N1566$^*$& 19.4 & -12.02 & 10.04 & 8.3x6.6  & N4254    & 39.8 & -11.78 & 10.90 & 5.4x4.7  & N6946    &  5.5 & -11.46 &  9.50 &  12x9.8  \\ 
N1614    & 64.2 & -11.82 & 11.28 & 1.3x1.1  & N4258$^*$&  8.0 &   ...  &  ...  &  19x7.2  & N6951$^*$& 24.3 & -12.04 & 10.21 & 3.9x3.2  \\ 
N1667$^*$& 61.2 & -12.43 & 10.62 & 1.8x1.4  & N4293    & 14.1 & -12.55 &  9.23 & 5.6x2.6  & I5063$^*$& 49.4 & -12.60 & 10.27 & 2.1x1.4  \\
N1672    & 16.7 & -11.70 & 10.23 & 6.5x5.5  & N4303$^*$& 13.6 & -11.81 &  9.94 & 6.5x5.5  & I5179    & 48.8 & -11.96 & 10.90 & 2.3x1.1  \\ 
N1792    & 15.4 & -11.75 & 10.11 & 5.2x2.6  & N4321    & 14.1 & -11.88 &  9.90 & 7.4x6.3  & N7331    & 14.4 & -11.78 & 10.02 &  11x3.7  \\ 
N1808    & 12.3 & -11.31 & 10.35 & 6.5x3.9  & N4385    & 34.5 & -12.64 &  9.92 & 2.0x1.0  & N7469$^*$& 67.0 & -11.88 & 11.25 & 1.5x1.1  \\ 
N2146    & 16.7 & -11.16 & 10.77 & 6.0x3.4  & N4388$^*$& 41.4 & -12.24 & 10.47 & 4.8x0.9  & N7541    & 37.5 & -11.96 & 10.67 & 3.5x1.2  \\ 
N2273$^*$& 28.5 & -12.48 &  9.91 & 3.2x2.5  & N4414    &  9.0 & -11.77 &  9.62 & 3.6x2.0  & N7552    & 22.5 & -11.44 & 10.74 & 3.4x2.7  \\ 
N2369    & 45.2 & -11.94 & 10.85 & 3.5x1.1  & N4418    & 34.7 & -11.73 & 10.47 & 1.6x0.7  & N7582$^*$& 22.0 & -11.61 & 10.55 & 5.0x2.1  \\ 
N2397    & 16.6 & -12.29 &  9.63 & 2.5x1.2  & N4457    & 13.6 & -12.56 &  9.19 & 2.7x2.3  & N7590$^*$& 22.0 & -12.33 &  9.83 & 2.7x1.0  \\ 
N2415    & 54.3 & -12.35 & 10.60 & 0.9x0.9  & N4527    & 13.5 & -11.79 &  9.95 & 6.2x2.1  & N7599    & 23.1 & -12.37 &  9.84 & 4.4x1.3  \\ 
N2559    & 21.4 & -11.78 & 10.36 & 4.1x2.1  & N4536    & 30.8 & -11.81 & 10.65 & 7.6x3.2  & N7632    & 21.3 & -12.63 &  9.51 & 2.2x1.1  \\ 
N2623    & 79.4 & -11.94 & 11.34 & 2.4x0.7  & N4565    & 27.2 & -12.29 & 10.06 &  16x1.9  & N7674    &  117 & -12.54 & 11.08 & 1.1x1.0  \\ 
N2798    & 28.6 & -11.96 & 10.43 & 2.6x1.0  & N4593$^*$& 41.3 & -12.78 &  9.93 & 3.0x2.9  & N7714    & 38.5 & -12.30 & 10.35 & 1.9x1.4  \\ 
N2903    &  7.3 & -11.65 &  9.56 &  13x6.0  & N4631    &  7.6 & -11.50 &  9.74 &  16x2.7  & N7771    & 58.0 & -11.97 & 11.04 & 2.5x1.0  \\ 
N2992$^*$& 34.1 & -12.39 & 10.16 & 3.5x1.4  & N4647    & 13.9 & -12.44 &  9.33 &  2.9x2.3  & N7793    &  3.3 & -12.22 &  8.30 & 9.3x6.3  \\ 
\noalign{\smallskip}                                                                              
\hline                                                                                             
\end{tabular}                                                           
} %
\end{center}                                                         
\label{sample}
Notes: Column 1: NGC/IC; Col. 2: corrected
distances from the  NASA/IPAC Extragalactic Database (NED) 
(Virgo+Great Attractor+Shapley Super-cluster case, assuming 
$H_{0}\,=\,73.0\,\kms$); Col. 3: {\it IRAS}  $FIR$; Col. 4: $FIR$ luminosity following from Cols. 2 and 3;
Col. 5: optical size $D_{25}$ taken from the Second Reference Catalog 
of Bright Galaxies (2RCBG, de Vaucouleurs $\etal$, 1976). Seyfert galaxies 
(Huchra \& Burg 1992; Maiolino \& Rieke, 1995) are marked by an asterisk.
\end{table*}

Following the first detections in the mid-1970s, numerous galaxies
have been observed in various transitions of CO and its isotopologue
$\thirco$. Substantial surveys have been conducted in the $J$=1-0
transition of $\co$ (e.g., Stark $\etal$ 1987; Braine $\etal$ 1993a;
Sage 1993; Young $\etal$ 1995; Elfhag $\etal$ 1996; Nishiyama $\&$
Nakai 2001; Sauty $\etal$ 2003; Albrecht $\etal$ 2007; Kuno $\etal$
2007). These surveys sample the nucleus and sometimes also a limited
number of disk positions. Extensive surveys in higher $\co$
transitions are fewer in number and usually only sample the nucleus
($J$=2-1; Braine $\etal$ 1993a; Albrecht $\etal$ 2007; $J$=3-2:
Mauersberger $\etal$ (1999); Yao $\etal$ 2003; Mao $\etal$ 2010). The
survey by Dumke $\etal$ (2001) and especially the James Clerk Maxwell
Telescope (JCMT) legacy survey of nearby galaxies (NGLS: Wilson
$\etal$ 2012; Mok $\etal$ 2016) are exceptional because they provide
maps of almost 100 galaxies in the $J$=3-2 transition, many of them in
the Virgo cluster. Specific surveys of Virgo cluster galaxies have
also been published by Stark $\etal$ (1986, $J$=1-0), Kenney and Young
(1988, $J$=1-0), and Hafok $\&$ Stutzki (2003, $J$=2-1 and $J$=3-2).

The $\co$ lines in the survey are optically thick and cannot be used to
measure molecular gas column densities or masses.  Even the analysis
of a whole ladder of multiple $\co$ transitions either fails to break
the degeneracy between $\h2$ density, kinetic temperature, and CO
column density and leaves the mass an undetermined quantity, or samples only
a small fraction of the total gas content in the higher $J$
transitions. Consequently, most molecular gas masses quoted in the
literature are critically dependent on an assumed value for the
relation between velocity-integrated CO line intensity and $\h2$ column
density, $X_{\rm CO}\,=\,N(\h2)/I$(CO).  Unfortunately, this so-called
$X$-factor does not follow from basic physical considerations.
Instead, its empirically estimated value is rather sensitive to
assumptions made in the process, and it varies depending on author and
method. The most reliable method uses gamma-ray observations to trace
hydrogen nuclei, and a useful overview of $X$ values thus obtained can
be found in Table E.1 of Remy $\etal$ (2017). The empirically
determined $X$ values implicitly include both $\h2$ gas mixed with CO
and $\h2$ gas that contains no or very little CO (`CO-dark
gas'). There is some confusion in the literature as different $X$
values have been referred to as the `standard' CO-to-$\h2$ conversion
factor. In this paper, we define
$X_{\circ}({\rm CO})\,=\,2\times10^{20}\,\cm2/\kkms$ (corresponding to
4.3 M$_{\odot}$ pc$^{-2}$ when it also includes a helium contribution)
as the standard factor to convert CO intensity into $\h2$ column
density.

In one form or another, the `standard' factor is frequently applied to
other galaxies, often without caveats of any sort. These are
essential, however, as the effects of metallicity, irradiation, and excitation
may cause $X$ to vary by large factors in different environments such
as are found in low-metallicity dwarf galaxies, galaxy centers, luminous
star-forming galaxies, molecular outflows, and high-redshift galaxies,
as was already explained in the pioneering papers by Maloney $\&$ Black
(1988) and Maloney (1990). Even the $X$-factor of our own Galactic
center region has been known to be very different since Blitz $\etal$
(1985) discussed the remarkably low ratio of gamma-ray to CO
intensities in the central few hundred parsecs and suggested that it
is caused by $\h2$/$\co$ abundances that are an order of magnitude below those
in the rest of the disk. These low $X$ values were since confirmed,
for instance, by Sodroski $\etal$ (1995; $X\,=\,0.22\,X_{\circ}$),
Dahmen $\etal$ (1998; $X\,=\,(0.06-0.33)\,X_{\circ}$), and Oka $\etal$
(1998; $X\,=\,0.12\,X_{\circ}$).

Conversion factors much lower than the standard Milky Way disk factor
have also been ascribed to the central regions of other galaxies.
Stacey $\etal$ (1991) used a comparison of $\cii$ and $\co$
intensities to suggest a factor of three or more below $X_{\circ}$.
Solomon $\etal$ (1997) and Downes $\&$ Solomon (1998) argued that in
ultra-luminous galaxies the $X$-factor had to be well below standard
for the gas mass to avoid exceeding the dynamical mass, and adopted a
value five times lower based on dust mass considerations.

Dust emission is relatively easy to measure but not so easy to
interpret.  Because the nature of the emitting dust grains is poorly
known, uncertainties in interstellar dust composition, dielectric
properties, size distributions, and dust-to-gas ratios cannot be
avoided, and each of these properties may also change with
environment. It is not entirely obvious how the measured intensity of
infrared continuum emission should be
translated into dust column density, let alone gas column
density. These uncertainties allowed authors to err on the side of
caution and estimate only moderately low values
$X\,\sim\,0.5\,X_{\circ}$ (M~82, Smith $\etal$ 1991; M~51, Nakai \&
Kuno, 1995; NGC~7469, Davies $\etal$ 2004), although substantially
lower values $X\,\sim\,0.1-0.2\,X_{\circ}$ (NGC~3079, Braine $\etal$
1997; NGC~7469, Papadopoulos $\&$ Allen 2000; NGC~4258, Ogle $\etal$
2014) have also been suggested. Such rather low values were also
inferred from the local thermal equilibrium (LTE) analysis of
optically thin but weak C$^{18}$O isotopologue emission (NGC~1068,
Papadopoulos, 1999; NGC~6000, Mart\'in $\etal$ 2010).

The potentially problematical use of dust continuum emission for
determining the properties of molecular gas is thus not preferred when
actual molecular line measurements are available. Both observations
and models have increasingly allowed the detailed analysis of CO line
intensities using the more sophisticated non-LTE
large velocity gradient (LVG) radiative transfer codes.  An essential
step toward reliable molecular gas mass determinations consists of
reducing or breaking the crippling temperature-density degeneracies
that plague the analysis of $\co$ measurements. This is accomplished
by including measurements of CO isotopologues with lower optical
depth. However, even the strongest of these ($\thirco$) is a
relatively weak emitter. Consequently, only the brightest galaxies
have been analyzed in this way (M82, Weisz $\etal$, 2001; NGC~4945 and
the Circinus galaxy, Curran $\etal$ 2001; Hitschfeld $\etal$ 2008,
Zhang $\etal$ 2014; VV~114, Sliwa $\etal$ 2013). These all yield
values of $X\,=\,0.1-0.2\,X_{\circ}$.

Extensive $\thirco$ surveys of external galaxies have so far been
lacking in any transition.  The survey presented in this paper is
therefore a significant addition to the existing CO database on nearby
galaxies. The newly determined multi-transition $\co$-to-$\thirco$
isotopologue ratios allow us to determine more accurately the CO gas
column densities and their relation to the much more abundant $\h2$
gas, including the values of $X$ in a large number of galaxy central
regions.  The analysis of the $\co$ and $\thirco$ spectral lines is of
particular importance in the interpretation of the Herschel Space
Observatory (2009-2013) observations of galaxies in the two
submillimeter $\ci$ lines and the far-infrared $\cii$ line (Israel
$\etal$ 2015; Kamenetzky $\etal$ 2016; Fern\'andez-Ontiveros $\etal$
2016; Lu $\etal$ 2017; Croxall $\etal$ 2017; D\'iaz-Santos $\etal$
2017; and Herrera-Camus $\etal$ 2018). With it, we will place
significant constraints on the relation between molecular and atomic
carbon and determine the carbon budget in the observed galaxy
centers.

% Table 2 J=1-0 line Intensities
\begin{table*}
\caption[]{\label{sestdat}Galaxy center $J$=1-0 line intensities}
\begin{center} 
{\small % 
\begin{tabular}{l|rr|rr||l|rr|rr||l|rr|rr} 
\noalign{\smallskip}     
\hline
  \noalign{\smallskip}
 \multicolumn{15}{c}{$\int T_{\rm mb}$d$V$ ($\kkms$)}\\
\noalign{\smallskip}     
\hline
\noalign{\smallskip} 
NGC&\multicolumn{2}{c}{$\co$}&\multicolumn{2}{c}{$\thirco$}&NGC&\multicolumn{2}{c}{$\co$}&\multicolumn{2}{c}{$\thirco$}&NGC&\multicolumn{2}{c}{$\co$}&\multicolumn{2}{c}{$\thirco$}\\
IC      & S$45"$ & I$22"$ &S$47"$&I$23"$ & IC     & S$45"$ & I$22"$ &S$47"$&I$23"$ &  IC    & S$45"$ & I$22"$ &S$47"$&I$23"$ \\
(1)     & (2)    & (3)    & (4)  & (5    & (1)    & (2)    & (3)    & (4)  & (5    &  (1)   & (2)    & (3)    & (4)  & (5   \\
\noalign{\smallskip}     
\hline
\noalign{\smallskip} 
 134 &  17.1 &   ...  &  ... &   ... & 3044 &    4.6 &   11.4 & 1.33 &  0.84 & 4736 &   ...  &   42.2 & ...  &  4.27 \\
 253 & 321   &  1030  & 27.3 &  76.5 & 3079 & 93.5$^b$&   235 & ...  & 14.1  & 4826 &    ... &   90.5 & ...  &  10.9 \\
 278 &   ... &   20.5 &  ... &  2.29 & 2554 &   11.7 &    ... & ...  &  ...  & 4835 &   11.1 &    ... & ...  &  ...  \\
 300 &   3.0 &   ...  &  ... &   ... & 3175 &   18.9 &   42.8 & 1.69 &  4.26 & 4945 &   523  &    ... & 37.0 &  ...  \\
 470 &  ...  &   28.2 &  ... &  1.88 & 3227 &    ... &   61.7 & ...  &  3.46 & 5033 &    ... &   52.7 & ...  &  5.77 \\
 520 &  14.7 &    113 &  ... &  7.96 & 3256 &   68.6 &    ... & 2.8  &  ...  & 5055 &    ... &   70.4 & ...  &  9.37 \\
 613 &  25.0 &   69.7 & 2.72 &  5.04 & 3281 &    1.4 &    ... & ...  &  ...  & 5135 &   18.0 &   61.8 &  1.4 &  2.70 \\
 628 &   ... &    7.0 &  ... &  1.10 & 3310 &    ... &    7.8 & ...  &  0.65 & 5194 &    ... &   47.6 & ...  &  7.06 \\
 660 &  38.6 &    154 & 3.12 &  9.96 & 3351 &    ... & 17$^c$ & ...  &  ...  & 5218 &    ... &    ... & ...  &   ... \\
 891 &   ... &    137 &  ... & 17.5  & 3504 & 20.2$^d$&  56.4 & ...  &  4.30 & 5236 &   78.6 &    195 & ...  &  14.3 \\
 908 &  23.9 &   29.8 & 1.22 &  4.54 & 3556 &    ... &   53.9 & ...  &  4.31 & Circ  &   155  &    ... & 8.45 &   ... \\
 972 &   ... &   66.7 &  ... &  5.75 & 3593 &   24.5 &   63.1 & ...  &  5.08 & 4444 &    9.4 &    ... & 1.18 &   ... \\
Maf2 &   ... &    220 & 20.6 & 27.5  & 3620 &   47.3 &    ... & 3.37 &  ...  & 5643 &   12.7 &    ... & ...  &   ... \\
1055 &  28.8 &   76.7 & 3.92 & 10.7  & 3621 &   11.6 &    ... & 0.73 &  ...  & 5713 &   16.8 &   45.4 & 0.99 &  2.87 \\
1068 &   ... &    168 &  ... & 14.2  & 3627 &   27.3 &   74.4 & ...  &  5.59 & 5775 &    ... &   47.9 & ...  &  5.28 \\
1084 &  19.8 &   30.4 & 1.52 &  2.29 & 3628 &   74.9 &    203 & 7.07 & 15.2  & 6000 &   22.5 &   74.7 & 1.83 &  4.88 \\
1097 &  68.7 & 136    &  ... &  12.9 & 3690 &    ... &   68.8 & 3.50 &  2.97 & 6215 &   10.9 &    ... & ...  &   ... \\
1317 &   2.4 &   ...  &  ... &   ... & 3783 &    3.4 &   ...  &  ... &   ... & 6221 &   30.8 &    ... & 2.7  &   ... \\
1365 & 102   &    260 & 9.42 & 22.8  & 4030 &   23.1 &   42.1 & ...  &  6.40 & 6240 &   17.5 &   70.1 & ...  &  2.44 \\
342  &   ... &    161 & ...  & 15.8  & 4038 &  30$^e$&   46.8 & 1.9  &  3.50 & 6300 &   28.1 &    ... & 1.40 &  ...  \\
1433 &  14.6 &    ... & 2.18 &  ...  & 4039 &  31$^e$&   45.5 & ...  &  2.07 & 6744 &   10.3 &    ... & ...  &  ...  \\
1448 &  14.1 &    ... & 1.09 &  ...  & 4051 &    ... &   37.8 & ...  &  2.08 & 6764 &    ... &   30.3 & ...  &  1.64 \\
1482 &  15.5 &32.1$^a$& 1.12 &  ...  & 4102 &    ... &   74.7 & ...  &  6.00 & 6810 &   29.4 &    ... & ...  &   ... \\
1559 &   5.0 &    ... & 0.86 &  ...  & 4254 &   31.3 &   42.7 & ...  &  4.79 & 6946 &    ... &    228 & ...  &  16.7 \\
1566 &  23.2 &    ... & 1.45 &  ...  & 4258 &    ... &   75.8 &  ... &  ...  & 5063 &    5.4 &    ... & ...  &  ...  \\
1614 &  14.3 &   43.2 & ...  &  1.44 & 4293 &    ... &   36.0 & ...  &  3.03 & 5179 &   20.9 &    ... & 1.70 &  ...  \\
1672 &  23.5 &    ... & 2.21 &  ...  & 4303 &    ... &   55.2 & ...  &  2.96 & 6951 & ...    &   50.1 & ...  &  4.77 \\
1792 &  23.2 &   27.7 & 4.64 &  3.07 & 4321 &   23.7 &   81.5 & ...  &  8.14 & 7469 &   10.7 &   54.6 & ...  &  3.22 \\
1808 &  92.0 &    135 & 3.49 &  7.57 & 4388 &    8.6 &    ... & ...  &  ...  & 7541 &   21.2 &   28.4 & ...  &  2.90 \\
2146 &   ... &    187 & ...  &  11.8 & 4414 &    ... &   51.4 & ...  &  6.90 & 7552 &   38.8 &    ... & 3.59 &  ...  \\
2273 &   ... &   16.5 & ...  &  1.57 & 4414 &    ... &   54.8 & ...  &  ...  & 7582 &   32.5 &    ... & ...  &  ...  \\
2369 &  26.1 &    ... & 1.75 &  ...  & 4457 &    ... &   29.5 & ...  &  1.94 & 7590 &    7.7 &    ... & 0.37 &  ...  \\
2397 &  16.6 &    ... & 1.41 &  ...  & 4527 &   34.7 &   88.0 & 2.77 &  6.35 & 7599 &    2.4 &    ... & ...  &  ...  \\
2559 &  32.1 &   78.4 & 3.39 &  5.81 & 4536 &   14.8 &   61.6 & ...  &  3.27 & 7632 &    7.8 &    ... & ...  &  ...  \\
2623 &   ... &   18.2 & ...  &  2.6  & 4565 &    ... & 12$^c$ & ...  &  ...  & 7714 &    1.0 &    3.5 & ...  &  0.64 \\
2903 &   ... &   79.8 & ...  &  7.10 & 4593 &    1.7 &    7.5 & ...  &  ...  & 7771 &    ... &   99.5 & ...  &  7.18 \\
2992 &   8.2 &    ... & ...  &  ...  & 4631 &    ... &   43.9 & ...  &  2.91 & 7793 &    2.7 &    1.8 & ...  &  ...  \\ 
3034 &   ... &    680 & ...  &    37 & 4666 &   30.7 &   73.6 & ...  &  7.58 &       &        &        &      &       \\
\noalign{\smallskip}     
\hline
\end{tabular}
}%    
\end{center} 
Notes:
$^a$ IRAM, Albrecht $\etal$ (2007
$^b$ SEST, Elfhag et al. (1996);
$^c$ IRAM, Braine $\etal$ (1993a)
$^d$ SEST, Chini $\etal$ (1996)
$^e$ SEST, Aalto et al. (1995);
\end{table*}

%Figure 1  CO Line Profiles SEST 
\begin{figure*}
\hspace{-0.5cm}
  \resizebox{20cm}{!}{\rotatebox{270}{\includegraphics*{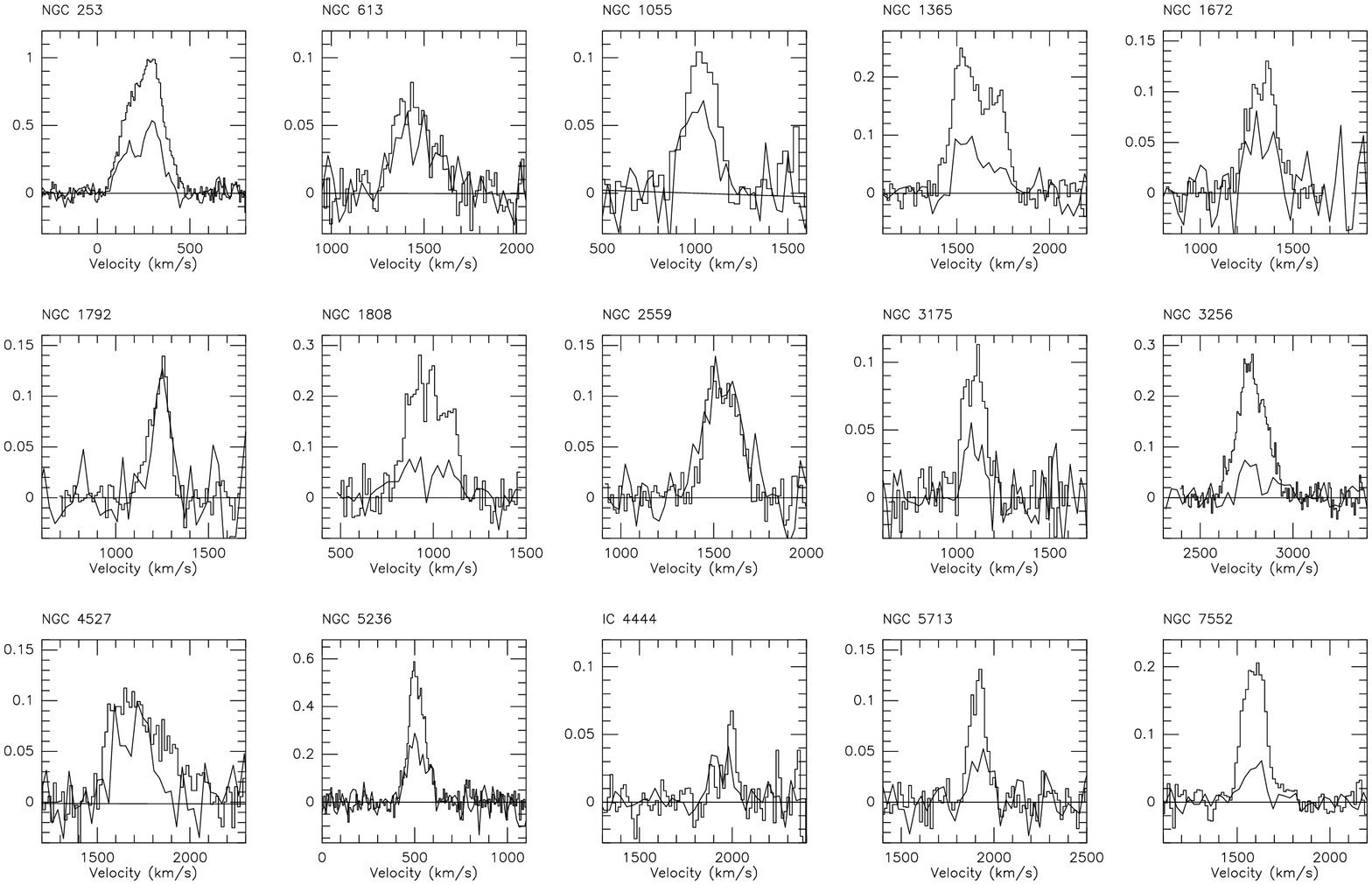}}}
\vspace{-1.53cm}
\caption[] {Sample of SEST $J$=1-0 CO observations of galaxy centers, showing  
     $^{12}$CO (histogram) and superposed $^{13}$CO (continuous lines)
     profiles; the intensities of the latter have been multiplied by
     a factor 5. Intensities are in $T_{\rm A}^{*}$ (K). Velocities are
     V(LSR) in $\kms$. Galaxies are identified at the top.  
}
\label{sestprofiles}
\end{figure*}

%Table 3 Line Intensities
\begin{table*}
\caption[]{\label{iramdat}Galaxy center $J$=2-1 line intensities}  
\begin{center} 
{\small % 
\begin{tabular}{l|rrr|rr||l|rrr|rr||l|rrr|rr} 
\noalign{\smallskip}     
\hline
  \noalign{\smallskip}
  \multicolumn{18}{c}{$\int T_{\rm mb}$d$V$ ($\kkms$)}\\
NGC&\multicolumn{3}{c}{$\co$}&\multicolumn{2}{c}{$\thirco$}&NGC&\multicolumn{3}{c}{$\co$} &\multicolumn{2}{c}{$\thirco$}&NGC&\multicolumn{3}{c}{$\co$}&\multicolumn{2}{c}{$\thirco$} \\
IC    &J$45"$&J$22"$&I$11"$ &J$23"$ & I$11"$&IC    &J$45"$&J$22"$&I$11"$ &J$23"$ & I$11"$&IC    &J$45"$&J$22"$&I$11"$ &J$23"$ & I$11"$ \\
(1)   &  (2) & (3)  & (4)   & (5)   & (6)   &(1)   &  (2) & (3)  & (4)   & (5)   & (6)   &(1)   &  (2) & (3)  & (4)   & (5)   & (6)    \\
\noalign{\smallskip}     
\hline
\noalign{\smallskip} 
 253 &  572 & 1360 &  1588 &   120 &   149 & 2903 &  ...  &   59.5 &   125 &  8.34 &  15.3 & 4647 &   9.7 &  15.9 &  ... &  ...  & ...   \\
 275 &  3.4 &  6.3 &   ... &   ... &   ... & 2992 &  ...  &   25.1 &   ... &  ...  &  ...  & 4666 &  26.2 &  53.1 & 75.5 &  5.91 & 6.35  \\ 
 278 & 12.6 & 19.6 &  23.5 &  2.93 &  2.34 & 3034 &  589  &    657 &   928 &  66.5 &    81 & 4736 &  15.4 &  42.7 & 65.7 &  4.33 & 6.96  \\
 470 &  ... & 23.2 &  62.7 &  1.65 &  4.34 & 3044 &  6.1  &    9.8 &  14.8 &  1.79 &$<$1.0 & 4826 &  49.7 &   102 & 75.3 &  14.3 & 10.5  \\
 520 &  ... & 94.9 &   221 &  4.75 &  15.5 & 3079 & 89.3  &  188.2 &   445 &  13.4 &  27.7 & 4945 &640$^b$&   931 &  ... &  81.2 & ...   \\
 613 &  ... & 74.2 &  73.7 &  4.65 &  6.99 & 3175 & 16.4  &   34.3 &  54.0 &  3.20 &  4.96 & 5033 &   ... &  42.5 & 50.5 &  7.19 & 7.31  \\
 628 &  4.9 &  4.2 &   6.4 &  1.45 &  0.49 & 3227 &  ...  &   48.3 &   104 &  5.55 &  9.37 & 5055 &   ... &  54.9 & 88.3 &  8.68 & 9.80  \\
 660 & 62.0 &  149 &   329 &  11.2 &  16.6 & 3256 &  ...  & 314$^b$&   ... &   ... &  ...  & 5135 &   ... &  39.1 & 69.7 &  4.24 & 6.58  \\
 695 & 22.2 & 38.3 &   ... &  ...  &  ...  & 3310 &  ...  &    8.7 &  12.8 &  0.74 &  1.04 & 5194 &  46.9 &  54.2 & 69.8 &  5.82 & 6.84  \\
 891 &  ... & 61.6 &   165 &  5.59 &  18.3 & 3504 &  ...  &   51.9 &   129 &  3.84 &  12.6 & 5218 &  17.8 &  52.6 &  ... &   ... & ...   \\
 908 &  ... & 18.4 &  19.2 &  1.59 &  2.36 & 3593 &  ...  &   41.7 &  73.5 &  5.46 &  8.48 & 5236 &   118 &   251 &  271 &  28.5 & 32.6  \\
 972 & 30.5 & 70.5 &  90.8 &  4.87 &  7.90 & 3620 &  ...  &76.4$^a$&   ... &  ...  &  ...  & Circ  &   ... &234$^c$&  ... &   ... & ...   \\
Maf2 &  104 &  247 &   239 & 26.7  &  34.3 & 3621 &  ...  & 8.5$^a$&   ... &  ...  &  ...  & 5433 &   ... &  21.3 &  ... &$<$0.6 & ...   \\
1055 &  ... & 54.4 &  90.1 &  6.25 &  10.1 & 3627 &  33.0 &   74.3 &  89.2 &  5.43 &  9.01 & 4444 &   ... &9.9$^d$&  ... &   ... & ...   \\
1068 &  103 &  239 &   236 &  21.2 &  16.4 & 3628 &  72.4 &    162 &   262 &  15.5 &  19.6 & 5713 &  37.0 &  56.6 & 65.9 &  6.14 & 4.77  \\
1084 &  ... & 30.7 &  28.4 &  3.59 &  3.15 & 3690 &  30.0 &   64.4 &  63.9 &  3.2  &  ...  & 5775 &   ... &  34.5 & 55.9 &  3.03 & 6.16  \\
1097 &  ... &  119 &   166 &  5.93 &  16.2 & 3982 &   ... &   13.9 &   ... &  ...  &  ...  & 6000 &   ... &  76.3 &  132 &  7.70 & 12.8  \\
1365 & 97.9 &  248 &   333 &  21.9 &  28.5 & 4030 &   ... &   35.1 &  37.5 &  3.57 &  3.81 & 6215 &   ... & 17$^c$&  ... &   ... & ...   \\
342  &  106 &  173 &   205 &  27.2 &  24.3 & 4038 &  40.6 &   63.6 &  52.9 &  3.78 &  3.80 & 6221 &   ... &30.4$^a$& ... &   ... & ...   \\
1433 &  ... &14.2$^a$& ... &  ...  &  ...  & 4039 &   ... &   35   &  41.7 &  2.84 &  3.09 & 6240 &  38.3 &  70.2 &  ... &  1.84 & ...   \\
1448 &  ... & 8.7$^a$& ... &  ...  &  ...  & 4051 &   ... &   22.2 &  59.3 &  1.06 &  2.88 & 6300 &   ... &20.4$^a$& ... &  ...  & ...   \\
1482 &  ... &18.3$^a$& ... &  ...  &  ...  & 4102 &   ... &   90.4 &  92.6 &  6.17 &  6.81 & 6764 &   ... &  22.6 & 73.1 &  1.00 & 2.98  \\
1559 &  ... & 4.4$^a$& ... &  ...  &  ...  & 4254 &   ... &   40.6 &  47.4 &  3.75 &  6.00 & 6946 &   113 &   240 &  361 &  17.7 & 24.1  \\
1566 &  ... &10.9$^a$& ... &  ...  &  ...  & 4258 &  22.9 &   44.3 &   118 &  ...  &  ...  & 6951 &   ... &  39.3 & 97.7 &  5.15 & 13.6  \\
1614 &  ... &  32.1 &  ... &  ...  &  ...  & 4293 &   ... &   26.9 &  52.3 &  5.44 &  6.45 & 7331 &  13.0 &  15.5 &  ... &  2.50 & ...   \\
1667 &  ... &  13.6 &  ... &  ...  &  ...  & 4303 &  23.8 &   42.6 &  80.0 &  3.27 &  7.09 & 7469 &  16.7 &  52.0 &  117 &  3.51 & 6.17  \\
1792 & 19.8 &  29.8 & 33.6 &  3.98 &  4.47 & 4321 &  31.2 &   55.6 &   121 &  4.87 &  12.2 & 7541 &   ... &  59.7 & 30.2 &  3.80 & 4.09  \\
1808 & 1030 &   165 &  185 &  13.5 &  14.3 & 4388 &   ... &   22.1 &   ... &  ...  &   ... & 7552 &   ... &123$^c$& ...&  ...  & ...   \\
2146 & 55.8 &   164 &  203 &  24.0 &  18.6 & 4414 &   ... &   37.1 &  35.4 &  4.64 &  3.97 & 7582 &   ... &116$^c$&...&  ...  & ...   \\
2273 &  ... & 16.9  & 31.6 &  1.28 &  2.85 & 4457 &   ... &   27.5 &  32.6 &  1.45 &  1.48 & 7590 &   ... & 7.3$^a$&  ...&  ...  & ...   \\
2369 &  ... &43.6$^a$& ... &  ...  &  ...  & 4527 &   ... &    103 &  89.2 &  5.26 &  5.75 & 7674 &   ... &  11.5 &  ... &  ...  & ...   \\
2397 &  ... &14.7$^a$& ... &  ...  &  ...  & 4536 &   ... &   63.5 &   101 &  5.04 &  10.8 & 7714 &   ... &   9.5 &  2.3 &$<$1.0 & ...   \\
2415 &  ... &  13.5 &  ... &  0.9  &  ...  & 4565 &  6.61 &   10.1 &   ... &  ...  &   ... & 7793 &   ... &   ... &  2.6 &  ...  & ...   \\
2559 & 32.6 &  69.7 &  121 &  8.24 &  11.6 & 4593 &   ... &    2.2 &   6.0 &  ...  &   ... &       &       &       &      &       &       \\
2623 &  ... &  25.9 & 22.1 &  $<$3 &  4.75 & 4631 &  27.1 &   34.3 &  41.1 &  1.96 &  3.14 &       &       &       &      &       &       \\
\noalign{\smallskip}     
\hline
\end{tabular}
}%    
\end{center} 
Notes:
$^a$ SEST, This Paper; 
$^b$ SEST, Ott et al. (2001
$^c$ SEST, Aalto et al. (1995);
$^d$ SEST, Chini et al. (1996);
\end{table*}

% Figure 2  CO Line Profiles IRAM 
\begin{figure*}
\resizebox{18cm}{!}{\rotatebox{0}{\includegraphics*{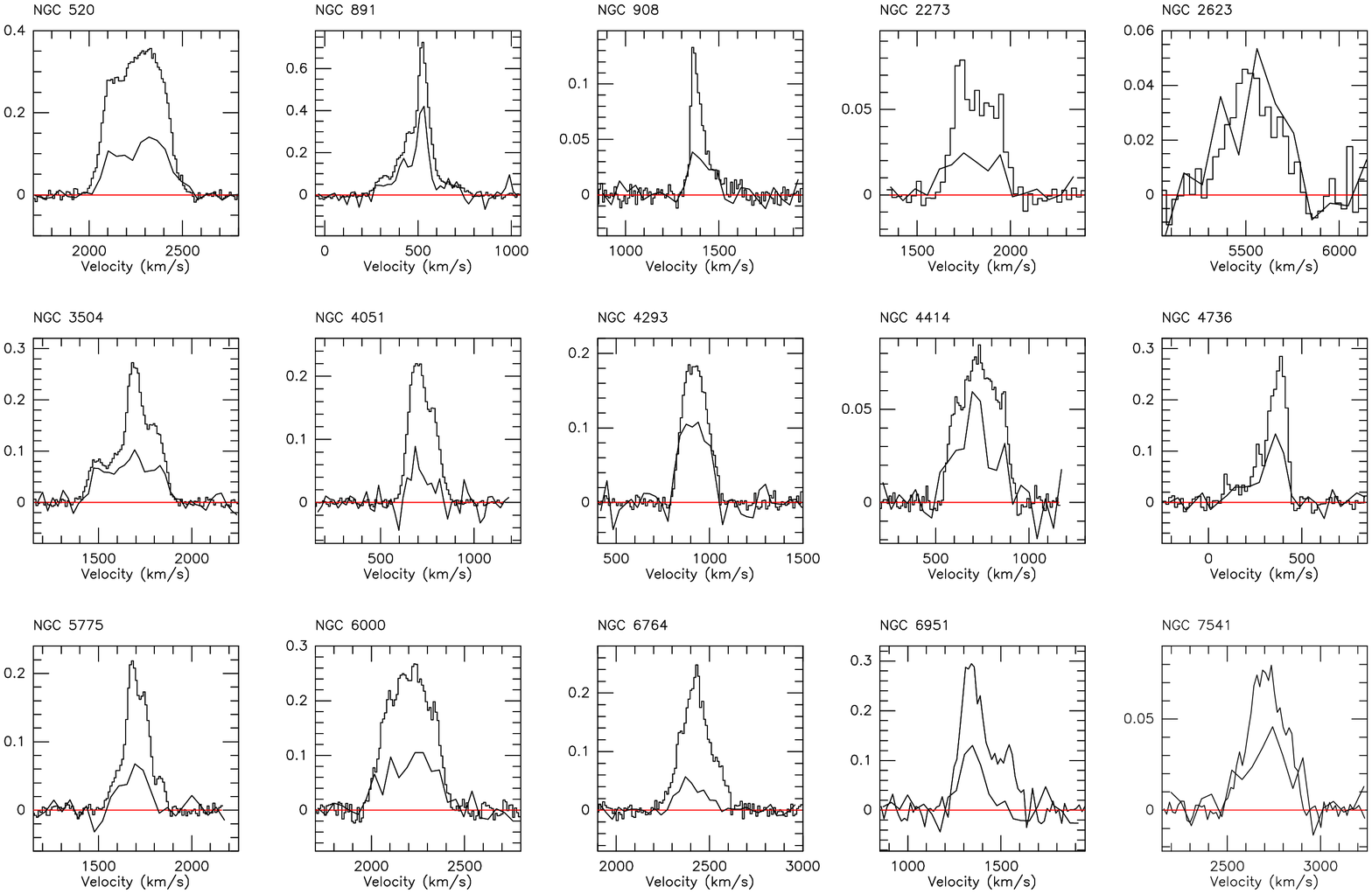}}}
%\vspace{-1.53cm}
\caption[] {Sample of IRAM $J$=2-1 CO observations of galaxy centers,
  showing $^{12}$CO (histogram) and superposed $^{13}$CO (continuous
  lines) profiles; the intensities of the latter have been multiplied
  by a factor 5. Intensities are in $T_{\rm A}^{*}$ (K). Velocities
  are V(LSR) in $\kms$. Galaxies are identified at the top. IRAM
  $J$=1-0 profiles (not shown) are similar, with better S/N.
}
\label{iramprofiles}
\end{figure*}

% Table 4 Line Intensities
\begin{table*}
\caption[]{\label{jcmtdat3}Galaxy center $J$=3-2 line intensities}
\begin{center} 
{\small % 
\begin{tabular}{l|rr|r||l|rr|r||l|rr|r||l|rr|r} 
\noalign{\smallskip}     
\hline
\noalign{\smallskip} 
 \multicolumn{16}{c}{$\int T_{\rm mb}$d$V$ ($\kkms$)}\\
NGC&\multicolumn{2}{c}{$\co$}&$\thirco$ &NGC&\multicolumn{2}{c}{$\co$}&$\thirco$ &NGC&\multicolumn{2}{c}{$\co$}&$\thirco$ &NGC&\multicolumn{2}{c}{$\co$}&$\thirco$ \\
IC    &J$22"$&J$14"$&J$14"$ &IC    &J$22"$&J$14"$&J$14"$ & IC    &J$22"$&J$14"$&J$14"$ &IC    &J$22"$&J$14"$&J$14"$ \\
(1)   & (2)  & (3)  & (4)   & (1)  & (2)  & (3)  & (4)   &(1)    & (2)  & (3)  & (4)   &(1)   & (2)  & (3)  & (4)   \\
\noalign{\smallskip}     
\hline
\noalign{\smallskip} 
 253 &  850 & 1449 &  124  & 1808 &  151 &  270 & 14.5& 3690 & 40.3 & 81.4 &  ... & 4736 & 27.4 &  44.0 &  3.05  \\ 
 275 &  ... &  5.3 &   ... & 2146 &  137 &  236 & 17.3& 3982 & 12.7 & 18.0 &  ... & 4826 & 49.9 &  91.9 &  10.9  \\  
 278 & 14.0 & 17.9 &  1.83 & 2273 & 11.4 & 21.4 & ... & 4030 &  ... & 19.1 & 1.92 & 4945 &  ... &   871 &  87.9  \\ 
 470 & 18.3 &35.0  &   ... & 2415 &  ... & 13.6 & 0.7 & 4038 & 52.0 & 87.6 & 6.63 & 5033 & 24.1 &  27.7 &  2.36  \\ 
 520 &  ... & 56.1 &  5.9  & 2559 & 40.3 & 65.8 & 2.94& 4039 & 23.8 & 59.5 & 4.08 & 5055 &  ... &  28.6 &  3.41  \\ 
 613 &  ... & 55.1 &  4.6  & 2623 & 18.9 & 40.3 & $<$4& 4051 & 29.9 & 63.4 &  ... & 5135 & 52.4 &  87.3 &  3.9   \\ 
 628 &  2.7 &  2.8 &   ... & 2798 & 43.6 & 64.8 & ... & 4102 &  ... & 68.0 & 5.30 & 5194 & 31.7 &  37.3 &  4.37  \\ 
 660 & 94.0 &  128 &  10.1 & 2903 & 53.8 & 78.8 & 6.32& 4254 &  ... & 24.3 & 4.87 & 5236 &  137 &   233 &  23.3  \\ 
 695 & 15.8 & 32.7 &   ... & 2992 & 18.7 & 25.8 & ... & 4258 & 32.7 & 50.3 & ...  & 5433 &  ... &  29.0 & $<$1.6 \\ 
 891 &  ... & 38.1 &  3.50 & 2993 & 17.4 & 22.7 & ... & 4293 &  ... & 31.7 & 2.14 & 5713 & 27.5 &  42.8 &  2.62  \\ 
 908 &  ... &  7.2 &  1.20 & 3034 &  548 &  649 & 72  & 4303 & 23.0 & 43.8 & 2.66 & 5775 & 20.0 &  30.0 & $<$0.8 \\ 
 972 & 39.0 & 61.4 &  3.46 & 3044 &  4.9 &  8.7 & ... & 4321 & 35.3 & 62.3 & 5.40 & 6000 &  ... &  58.5 &  ...   \\ 
Maf2 &  179 &  321 & 33.4  & 3079 &  ... &  154 & 26  & 4385 & 6.11 & 7.95 & ...  & 6090 & 17.6 &  45.0 &  ...   \\ 
1055 &  ... & 30.3 &  2.65 & 3175 & 23.2 & 32.6 & 2.28& 4388 & 10.9 & 16.5 & ...  & 6240 & 80.4 &   127 &  4.80  \\ 
1068 &  101 &  153 & 10.7  & 3227 & 41.7 & 92.6 & 5.32& 4414 & 18.7 & 20.6 & 3    & 6764 & 20.6 &  31.4 &  ...   \\ 
1084 &  ... & 17.9 &  1.8  & 3256 &  117 &  173 &  ...& 4418 &  ... & 47.4 & ...  & 6946 &  117 &   201 &  18.9  \\ 
1097 &  107 &  142 &  ...  & 3310 & 11.7 & 27.9 & 3.12& 4457 & 19.0 & 29.7 & 1.24 & 6951 & 46.8 &  56.5 &  3.72  \\
1365 &  170 &  284 &  23.3 & 3351 & 29.9 & 45.9 & ... & 4527 &  ... & 33.3 & 3.3  & 7331 &  7.1 &  10.1 &  1.76  \\ 
342  &  121 &  186 &  17.3 & 3504 & 34.5 & 57.0 & 6.26& 4536 &  ... & 77.3 & 4.79 & 7469 & 40.7 &  89.6 &  3.95  \\ 
1614 &  ... & 32.7 &  ...  & 3593 &  ... & 39.4 & 2.86& 4631 & 23.2 & 29.6 & 4.29 & 7541 &  ... &  35.6 &  3.46  \\ 
1667 &  7.1 & 13.9 &  ...  & 3627 & 48.7 & 84.1 & 6.93& 4647 &  8.0 & 11.7 & ...  & 7674 & 9.60 &  18.3 &  ...   \\ 
1792 &  ... & 16.4 &  2.13 & 3628 &  114 &  210 & 23.1& 4666 & 35.3 & 51.5 & 3.56 & 7714 &  ... &   8.5 &  ...   \\ 
\noalign{\smallskip}     
\hline
\end{tabular}
}%    
\end{center} 
\end{table*}   
         
% Figure 3  CO Line Profiles JCMT 
\begin{figure*}
\resizebox{18cm}{!}{\rotatebox{0}{\includegraphics*{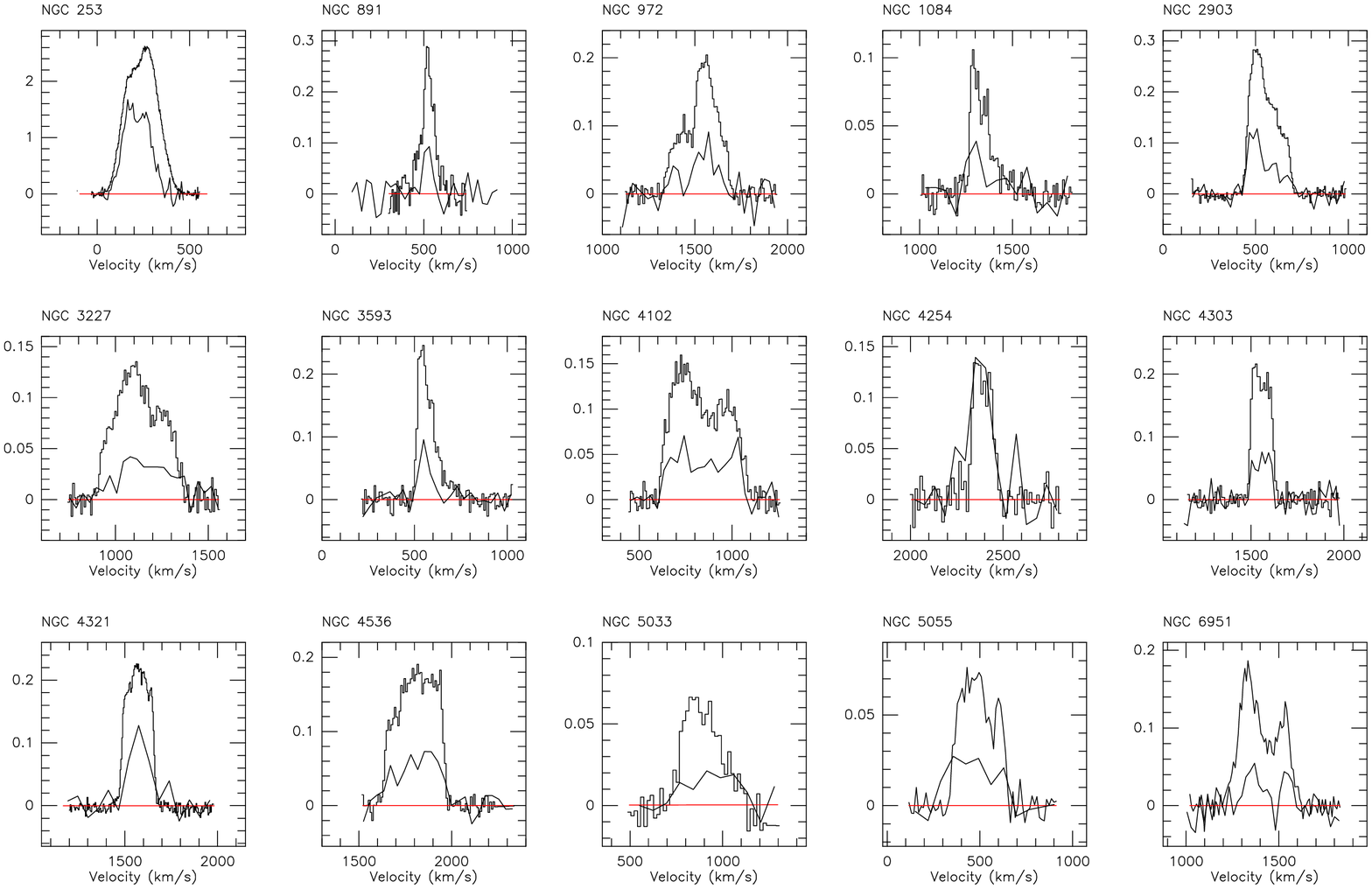}}}
\caption[] {Sample of JCMT $J$=3-2 CO observations of galaxy centers, showing  
     $^{12}$CO (histogram) and superposed $^{13}$CO (continuous lines)
     profiles; the intensities of the latter have been multiplied by
     a factor 5. Intensities are in $T_{\rm A}^{*}$ (K). Velocities are
     V(LSR) in $\kms$. Galaxies are identified at the top. JCMT $J$=2-1
     profiles (not shown) are similar, with better S/N.
}
\label{jcmtprofiles}
\end{figure*}

%Table 5 Line Intensities
\begin{table*}
\caption[]{\label{jcmtdat4}Galaxy center $J$=4-3 line intensities} 
\begin{center} 
{\small % 
\begin{tabular}{l|rrr||l|rrr||l|rrr||l|rrr} 
\noalign{\smallskip}     
\hline
\noalign{\smallskip} 
 \multicolumn{16}{c}{$\int T_{\rm mb}$d$V$ ($\kkms$)} \\
NGC&\multicolumn{3}{c}{$\co$ } & NGC&\multicolumn{3}{c}{$\co$ } &NGC&\multicolumn{3}{c}{$\co$ } &NGC&\multicolumn{3}{c}{$\co$ } \\
IC    &J$22"$&J$14"$&J$11"$&IC     &J$22"$&J$14"$&J$11"$& IC     &J$22"$&J$14"$&J$11"$   & IC    &J$22"$&J$14"$&J$11"$\\
(1)   & (2)  & (3)  & (4)  & (1)   & (2)  & (3)  & (4)  &  (1)   & (2)  & (3)  & (4)     & (1)   & (2)  & (3)  & (4) \\
\noalign{\smallskip}     
\hline
\noalign{\smallskip} 
 253 &  927 & 1484 & 2046 & 2146 &  ... &  122 &   139 & 3628 &  89.2 &   146 &   206 & 4826 &   66 &   114 &   121  \\
 278 &  8.4 &  9.1 & 11.5 & 2273 &  ... &  ... &   6.1 & 3690 &  47.2 &  51.9 &  59.4 & 5033 &  ... &   ... &   4.7  \\
 660 &   92 &  122 &  161 & 2623 &  ... &  ... &  18.7 & 4051 &   ... &   ... &  19.5 & 5194 & 25.0 &  28.3 &  30.6  \\
Maf2 &  155 &  284 &  407 & 3034 &  396 &  481 &   547 & 4321 &   ... &   ... &  12.2 & 5236 &  111 &   192 &   257  \\
1068 &  156 &  219 &  291 & 3079 &  131 &  208 &   243 & 4631 &   ... &  14.2 &  17.0 & 6946 &  110 &   183 &   245  \\
342  &  111 &  190 &  223 & 3175 &  ... &  ... &  28.8 & 4666 &  15.5 &  22.8 &  34.2 & 7469 &   47 &    68 &    92  \\
\noalign{\smallskip}     
\hline
\end{tabular}
}%    
\end{center} 
\end{table*}   
         
\section{Observations and data handling}

\subsection{SEST 15m observations}
 
With the 15 m Swedish-ESO Submillimetre Telescope (SEST) at La Silla
(Chile) \footnote{The Swedish-ESO Submm Telescope (SEST) was operated
  jointly by the European Southern Observatory (ESO) and the Swedish
  Science Research Council (NFR) from 1987 until 2003.} we conducted
seven observing runs between May 1988 and January 1992, and another
three runs between 1999 and 2003. Observations in the first period
were mostly in the $J$=1-0 $\co$ transition, with some $J$=1-0
$\thirco$ observations of the brightest galaxies. In the second period
we obtained additional $J$=2-1 $\co$ and $J$=1-0 $\thirco$
observations simultaneously. The SEST full width at half-maximum
(FWHM) beam sizes were 45$\arcsec$ at 115 GHz ($J$=1-0 $\co$) and
23$\arcsec$ at 230 GHz ($J$=2-1 $\co$). All observations were made in
a double beam-switching mode with a throw of 12$\arcmin$.  Using the
CLASS package, we binned the spectra to resolutions of 10-30 $\kms$
after which third-order baselines were subtracted if the spectral
coverage allowed it; otherwise, only a linear baseline was fit. A
sample of the SEST observations is shown in Fig.\,\ref{sestprofiles}.
Line parameters were determined by fitting with one or two Gaussians
as required by the shape of the profile. In the $\thirco$ profiles, we
set the fitting range to be the same as determined in the $\co$
profiles with higher signal-to-noise ratios (S/N). Intensities were
reduced to main-beam brightness temperatures $T_{\rm mb}$ =
$T_{A}^{*}/\eta_{\rm mb}$, using main-beam efficiencies at 115 GHz
$\eta_{\rm mb}(115)$ = 0.66 until October 1988, 0.74 until June 1990,
0.75 until October 1990, and 0.70 thereafter (L.E.B. Johansson,
private communication), and $\eta_{\rm mb}(230)$ = 0.50 for the whole
period. The resulting velocity-integrated line intensities are listed
in Tables\,\ref{sestdat} and \ref{iramdat}.

\subsection{IRAM 30m observations}

Using the IRAM 30 m telescope on Pico de Veleta (Granada, Spain)
\footnote{IRAM is supported by INSU/CNRS (France), MPG (Germany), and
  IGN (Spain). The IRAM observations in this paper have benefited from
  research funding by the European Community Sixth Framework Programme
  under RadioNet R113CT 2003 5058187.} , we conducted four observing
runs between December 2004 and July 2006, simultaneously observing the
$J$=1-0 and $J$=2-1 transitions of both $\co$ and $\thirco$ with the
facility 3mm and 1.3mm SIS receivers coupled to 4 MHz backends. All
observations were made in beam-switching mode with a throw of
4$\arcmin$. The FWHM beam sizes were 22$\arcsec$ at 110/115 GHz and
11$\arcsec$ at 220/230 GHz. The diameter of the IRAM telescope is
twice that of the JCMT (and the SEST) so that $J$=1-0 (IRAM) and
$J$=2-1 (JCMT) observations are beam-matched, as are the $J$=2-1
(IRAM) and $J$=4-3 (JCMT) observations. A sample of the IRAM
observations in the $J$=2-1 transition is shown in
Fig.\,\ref{iramprofiles}. The profile analysis was similar to that
described for the SEST.  Intensities were converted into main-beam
brightness temperatures using main-beam efficiencies $\eta_{\rm mb}$
of 0.79/0.80 at 110/115 GHz and 0.59/0.57 at 220/230 GHz. The
resulting velocity-integrated line intensities are listed in
Tables\,\ref{sestdat} and \ref{iramdat}.

\subsection{JCMT 15 m observations}

The observations with the 15 m JCMT on Mauna Kea (Hawaii)
\footnote{Between 1987 and 2015, the 
  (JCMT was operated by the Joint Astronomy Centre on behalf of the
  Particle Physics and Astronomy Research Council of the United
  Kingdom, the Netherlands Organization for Scientific Research (until
  2013), and the National Research Council of Canada.}  were obtained
at various periods between 1988 and 2005. When the JCMT changed from
PI-scheduling to queue-scheduling in the late 1990s, most of the
survey measurements were made in back-up service mode. In both the
$J$=2-1 and the $J$=3-2 transitions, $\co$ and $\thirco$ observations
were made closely together in time. The JCMT FWHM beam-sizes were
22$\arcsec$ at 220/230 GHz and 14$\arcsec$ at 330/345 GHz.  All
observations were made in a beam-switching mode with a throw of
3$\arcmin$. We have discarded almost all early observations,
preferring to use those obtained after 1992 with more sensitive
receivers and the more sophisticated Dutch Autocorrelator System (DAS)
back-end.  We included data extracted from the CADC/JCMT archives on
galaxies relevant to our purpose that had been observed by other
observers (e.g., Devereux $\etal$ 1994; Papadopoulos $\&$ Allen, 2000;
Zhu $\etal$ 2003; Petitpas $\&$ Wilson 2003).

We reduced the JCMT observations using the SPECX package, and
subtracted baselines up to order three, depending on source
line-width. We determined integrated intensities by summing channel
intensities over the full range of emission. In the $\thirco$
profiles, we set this range to be the same as determined in higher S/N
$\co$ profiles. Antenna temperatures were converted into main-beam
brightness temperatures with efficiencies $\eta_{\rm mb}(230)$ = 0.70
and $\eta_{\rm mb}(345)$ = 0.63. The velocity-integrated line
intensities are listed in Tables\,\ref{iramdat} and
\ref{jcmtdat3}. Available $J$=4-3 $\co$ observations, many of which
were discussed in earlier papers (Israel $\etal$ 2009b, and references
therein) were re-reduced and the results are listed in
Table\,\ref{jcmtdat4}.

\begin{figure*}
\vspace{-0.8cm}  
\resizebox{18.7cm}{!}{\rotatebox{0}{\includegraphics*{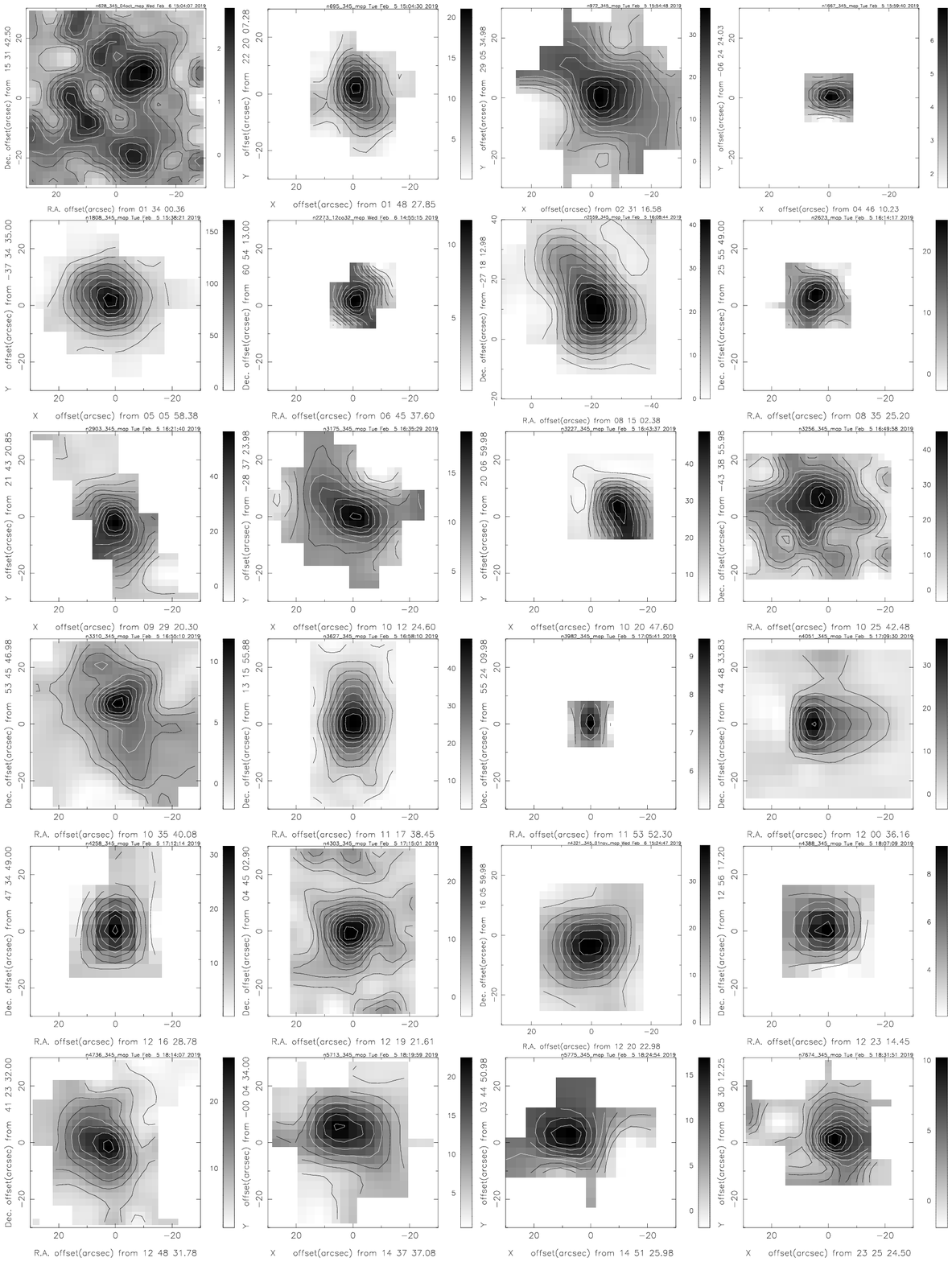}}}
\vspace{-1.0cm}
\caption{
  JCMT $\co$(3-2) $1'\times1'$ galaxy center maps. Linear
  contours $\int T_{mb}$d$V$ ($\kkms$) are superposed on
  grayscales $\int T_{A}^{*}$d$V$ ($\kkms$). Galaxy names, the values of the
  lowest white contour, and the contour step are as follows: 
  Row 1: NGC 628 (2, 0.5), NGC 695 (20, 4), NGC 972 (24, 6), NGC 1667 (7.5, 1.5);
  Row 2: NGC 1808 (150, 30), NGC 2273 (12, 2), NGC 2559 (36,6), NGC 2623 (24,4);
  Row 3: NGC 2903 (48, 8), NGC 3175 (15, 5), NGC 3227 (48, 8), NGC 3256 (48, 8);
  Row 4: NGC 3310 (12, 3), NGC 3627 (40, 8), NGC 3982 (14, 2), NGC 4051 (32, 8);
  Row 5: NGC 4258 (30, 6), NGC 4303 (20, 4), NGC 4321 (32, 8), NGC 4388 (9, 1.5);
  Row 6: NGC 4735 (25, 5), NGC 5713 (20, 5), NGC 5775 (12, 3), NGC 7674 (7.5, 1.5).  
}
\label{galmap}
\end{figure*}

For almost half of the sample, small maps of the $J$=3-2 and $J$=2-1
$\co$ emission from the central region were obtained in addition to
the central profiles. Maps and profiles of more than 16 galaxies
have already been published (Israel, 2009a, b, and references
therein). A sample of the new JCMT $J$=3-2 profiles is shown in
Fig.\,\ref{jcmtprofiles}. All JCMT $J$=3-2 $\co$ maps not included in
our previous papers are shown in Fig.\,\ref{galmap}.

% Figure 4 Galaxy J=3-2 CO maps JCMT

\subsection{Observational error}

We usually integrated until the peak signal-to-noise ratio in
individual 10-20 km s$^{-1}$ channels exceeded a value of
5-10. Especially for $\thirco$ line measurements, this required long
integration times, sometimes up to several hours. The JCMT B-band
receiver system had a relatively high system temperature spike around
330 GHz, resulting in a decreased sensitivity for the $J$=3-2
$\thirco$ line. The higher profile noise level and the limited
bandwidth of 920 MHz (800 km/s) caused additional uncertainties in the
line parameters that could only partly be alleviated using longer
integration times.  In the SEST 1999-2003 and all IRAM observing runs,
the need to obtain a good detection of the $\thirco$ line
automatically provided very high S/N for
simultaneously observed stronger $\co$ lines.

From repeated observations, and from comparison with published
measurements by others (summarized in Appendix A), we find the
uncertainty in individual intensities obtained with the SEST in
1988-1992 to be about $30\%$, and those obtained in 1999-2003 to be
about $20\%$. Depending on profile width, galaxies with intensities
above 40-70 $\kkms$ have somewhat lower uncertainties, whereas
galaxies with intensities below 10 $\kkms$ have larger uncertainties
of up to $50\%$.  The IRAM profiles in particular were obtained with
wide velocity coverage and well-defined baselines, which is especially
important for observations of galaxy center profiles with large
velocity widths. They have relatively high S/N and are generally
superior to those obtained in earlier measurements as well as to our
own SEST and JCMT data.  The uncertainty in the intensities observed
with IRAM is $\sim 10\%$ for $\co$, and $20-25\%$ for $\thirco$.
Again from repeated observations, individual intensities measured with
the JCMT have an uncertainty of $15-20\%$, except for those of $J$=3-2
$\thirco,$ where uncertainties range from $20\%$ for bright narrow
lines to $50\%$ for weak broad lines.  However, because the $\co$ and
$\thirco$ intensities were measured (almost) simultaneously, the
uncertainty in their ratio is lower, typically $10-20\%$ for the
$J$=1-0 and $J$=2-1 transitions and $15-25\%$ for the $J$=3-2
transition. A comparison of the $\co$-to-$\thirco$ ratios determined
in this paper and published in the literature may be found in Appendix
B.

%Figure 5 Transition ratios
\begin{figure}
\begin{minipage}[]{9cm} 
%  \begin{minipage}[]{2.6cm} 
    \resizebox{2.9cm}{!}{\rotatebox{0}{\includegraphics*{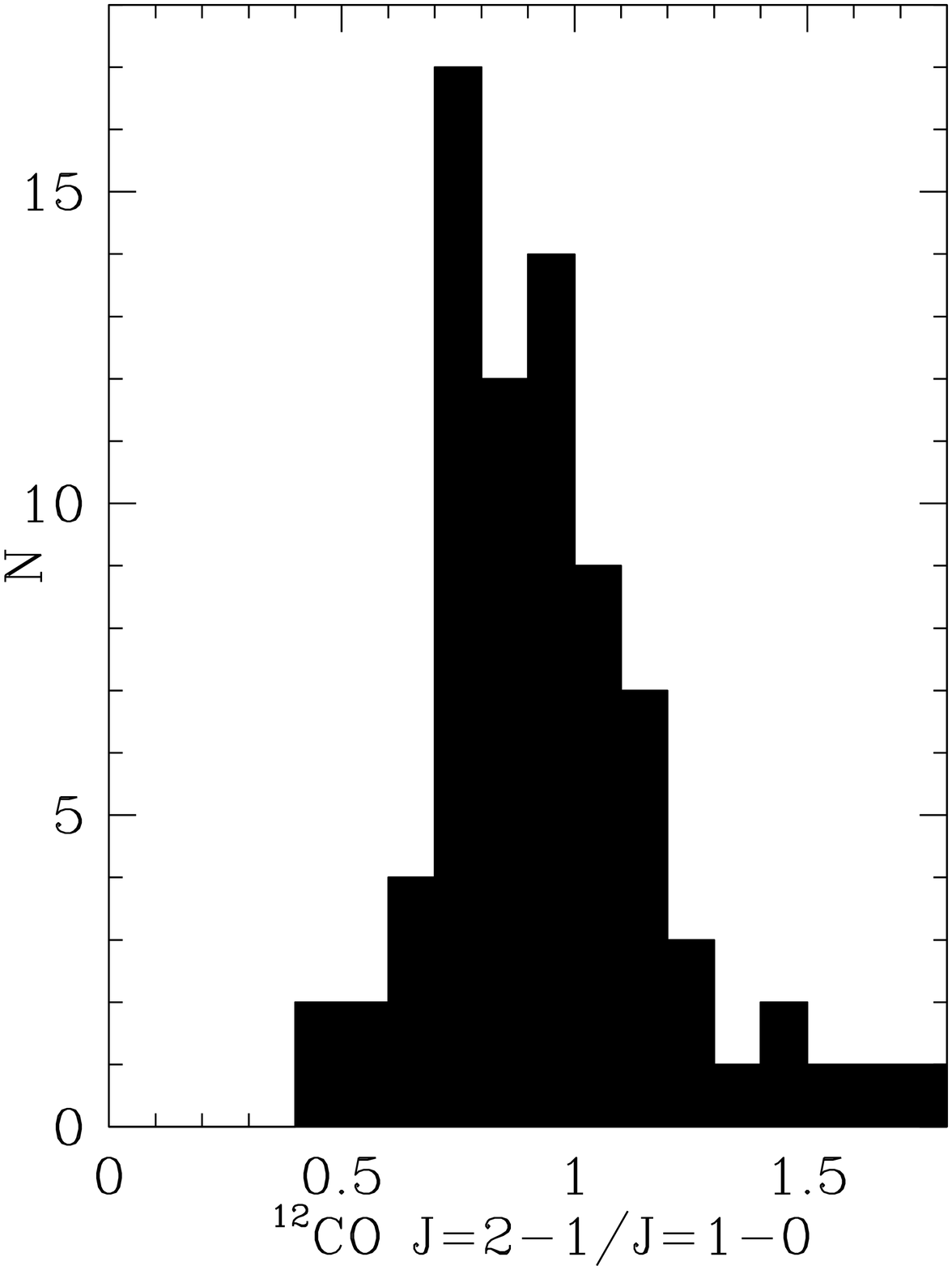}}}
%  \end{minipage}
%  \begin{minipage}[]{2.6cm} 
    \resizebox{2.9cm}{!}{\rotatebox{0}{\includegraphics*{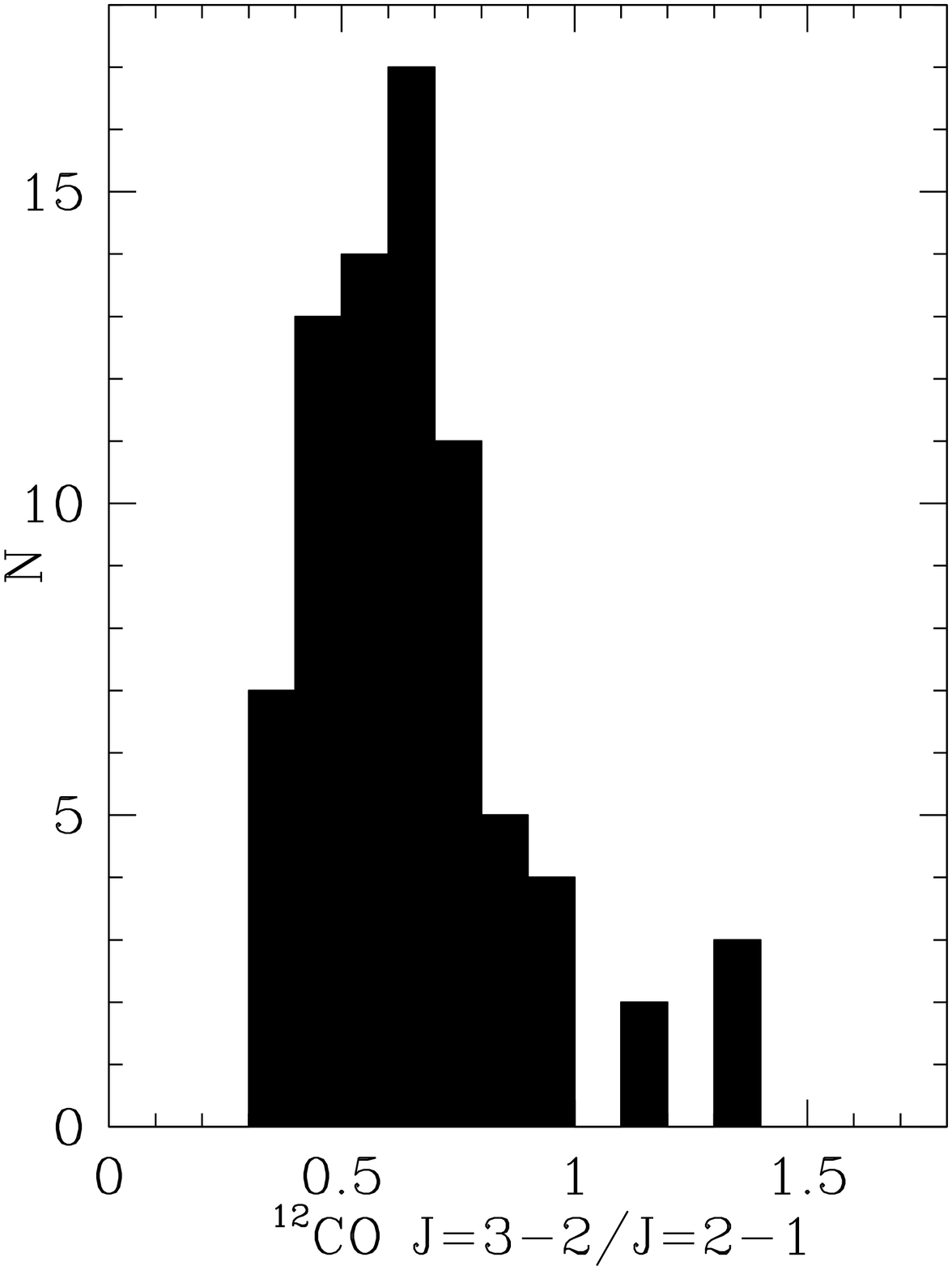}}}
%  \end{minipage}
%  \begin{minipage}[]{2.6cm} 
    \resizebox{2.9cm}{!}{\rotatebox{0}{\includegraphics*{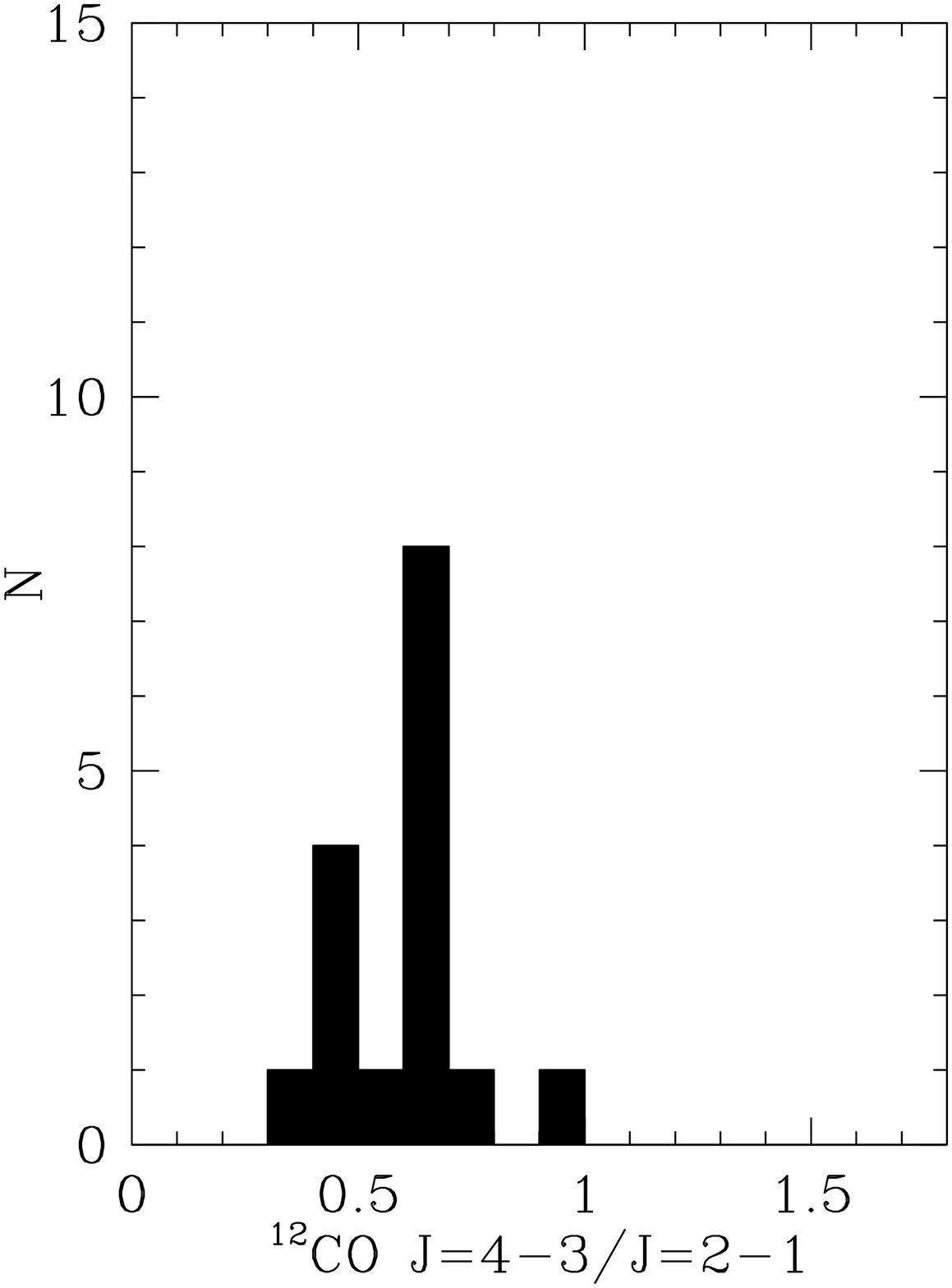}}}
%  \end{minipage}
\end{minipage}
\hfill
\begin{minipage}[]{9cm}
%  \resizebox{4.4cm}{!}{\rotatebox{270}{\includegraphics*{1021LFIR.eps}}}
\begin{center}
  \resizebox{5.9cm}{!}{\rotatebox{270}{\includegraphics*{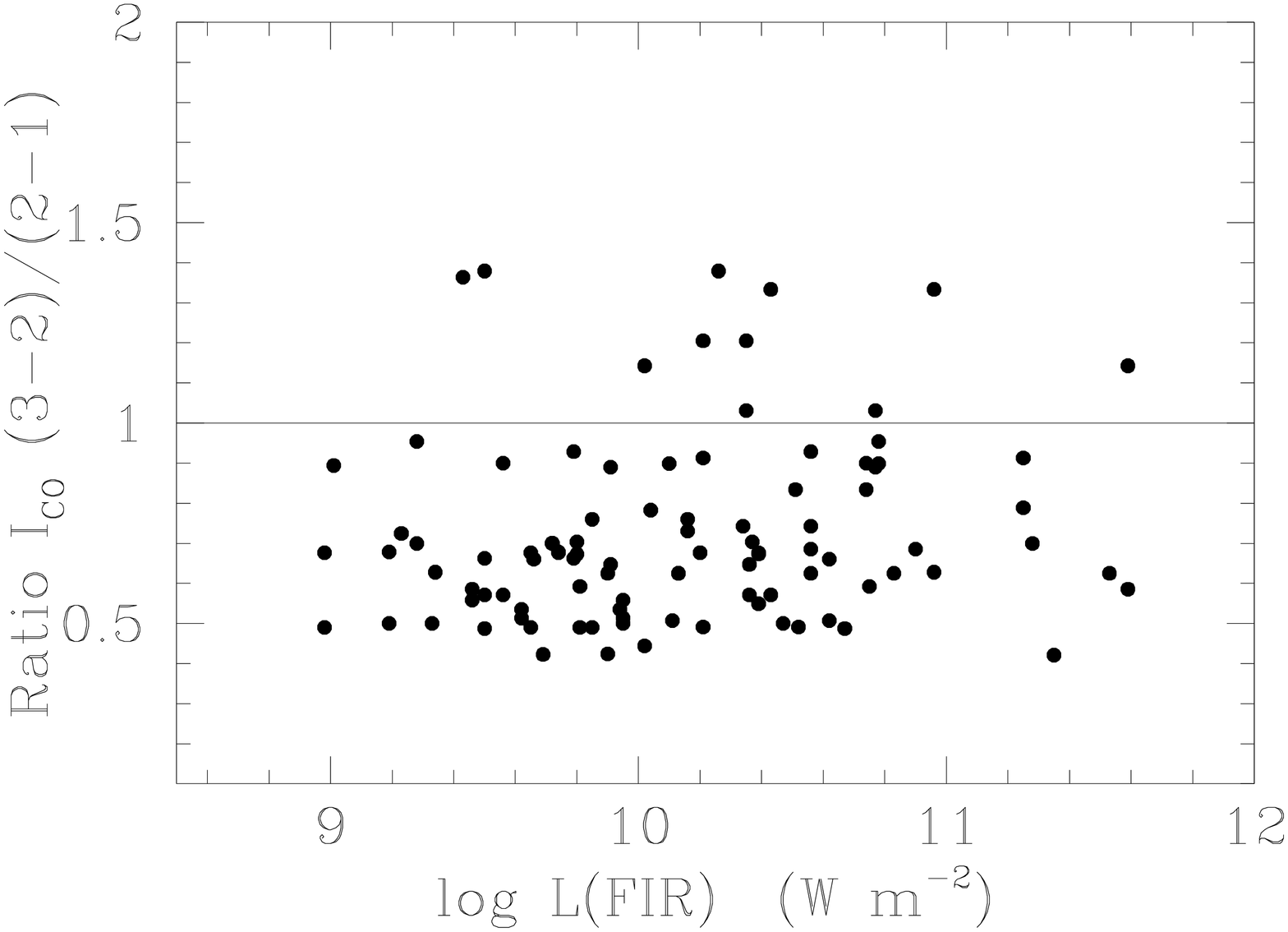}}}
\end{center}
\end{minipage}
\caption[] {Distribution of the
  $J$=2-1/$J$=1-0, the $J$=3-2/$J$=2-1, and the $J$=4-3/$J$=2-1 $\co$
  intensities.
  Bottom: $J$=3-2 intensities relative to the $J$=2-1 $\co$
  intensity as a function of galaxy total FIR luminosity.
}
\label{transrat}
\end{figure}

%Figure 6 Isotopologue ratios
\begin{figure}
\begin{minipage}[]{9cm} 
%  \begin{minipage}[]{2.6cm} 
    \resizebox{2.8cm}{!}{\rotatebox{0}{\includegraphics*{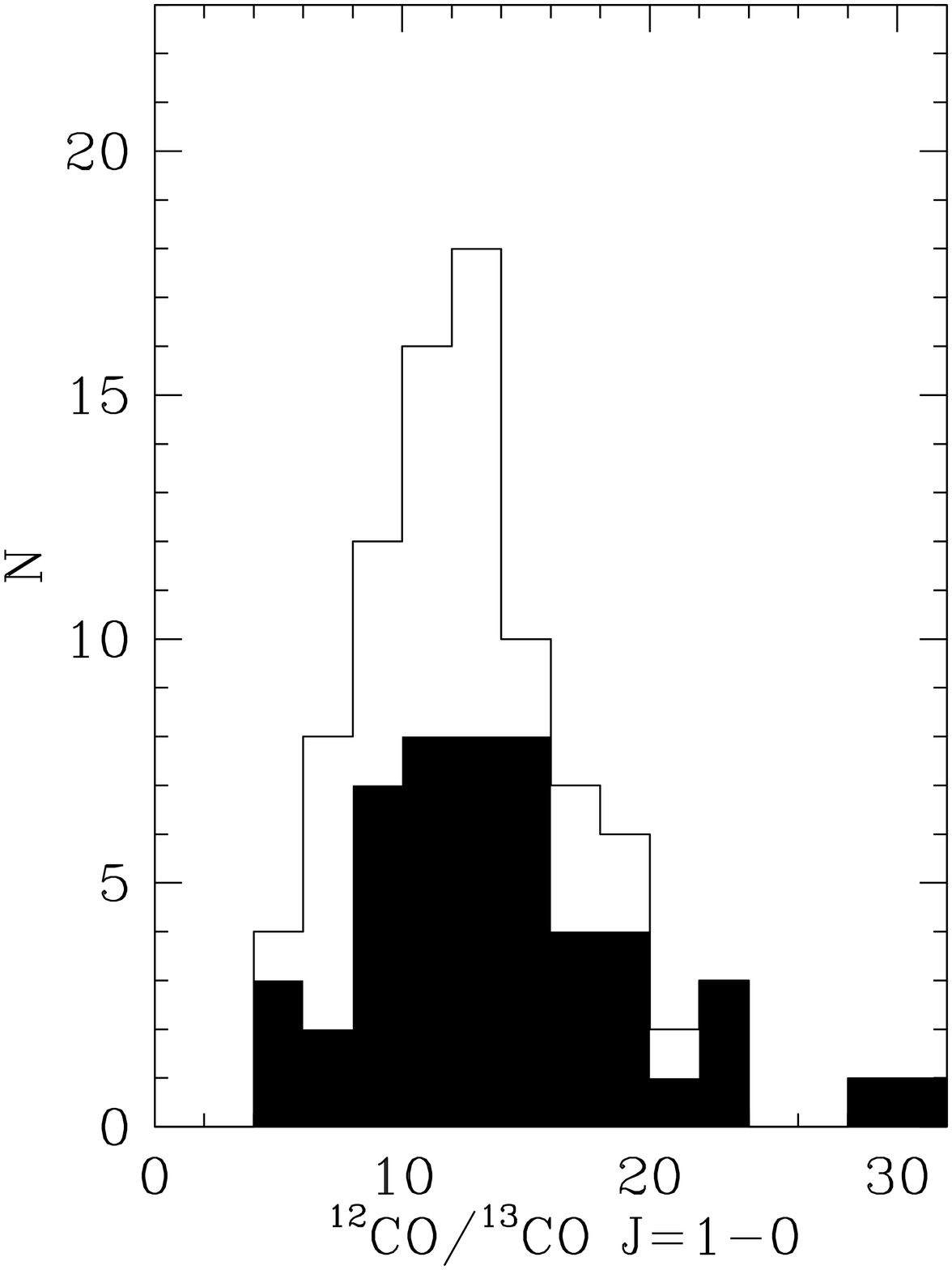}}}
%  \end{minipage}
%  \begin{minipage}[]{2.6cm} 
    \resizebox{2.8cm}{!}{\rotatebox{0}{\includegraphics*{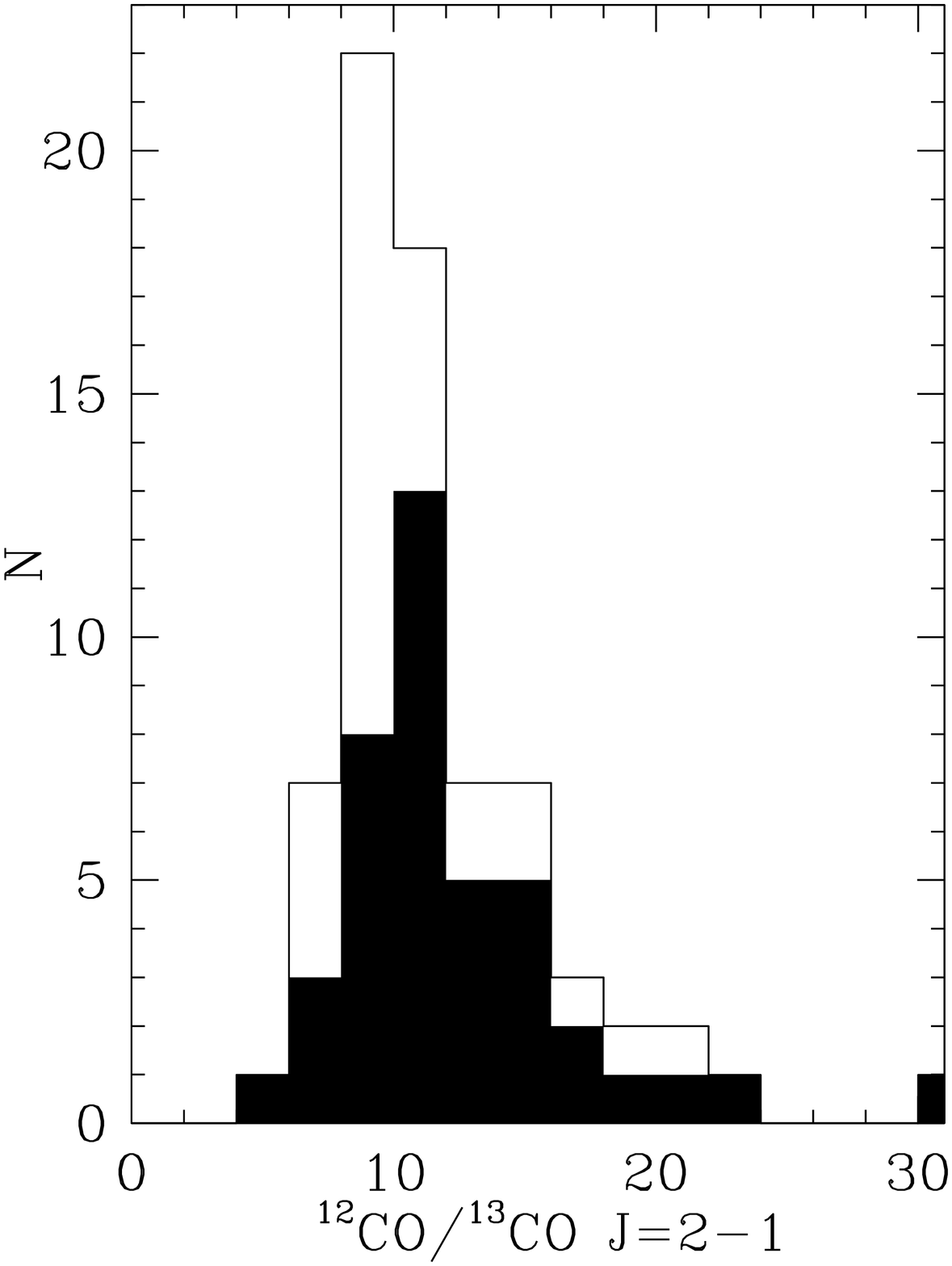}}}
%  \end{minipage}
%  \begin{minipage}[]{2.6cm} 
    \resizebox{2.8cm}{!}{\rotatebox{0}{\includegraphics*{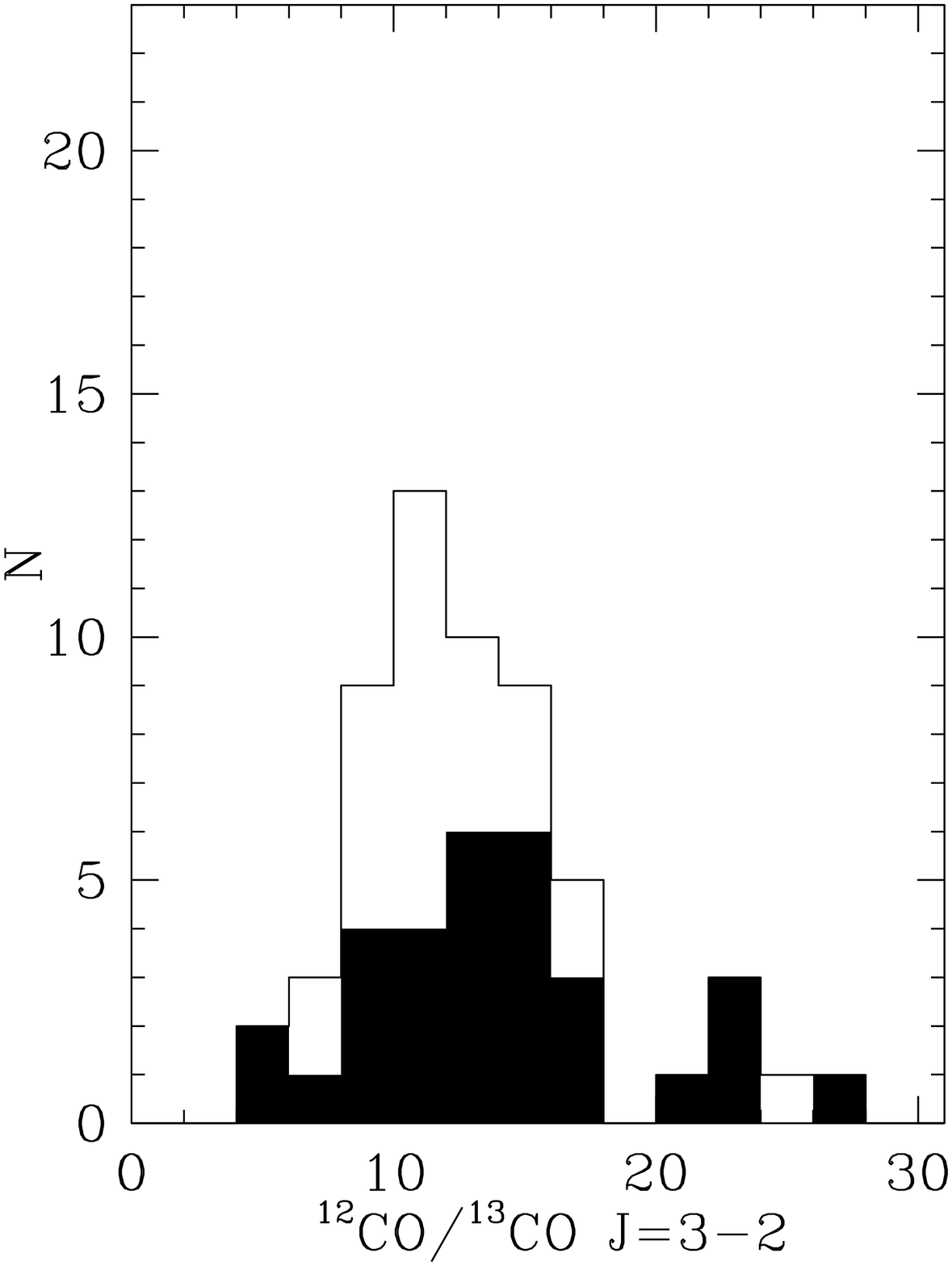}}}
  \end{minipage}
\begin{minipage}[]{9cm}
    \resizebox{4.3cm}{!}{\rotatebox{0}{\includegraphics*{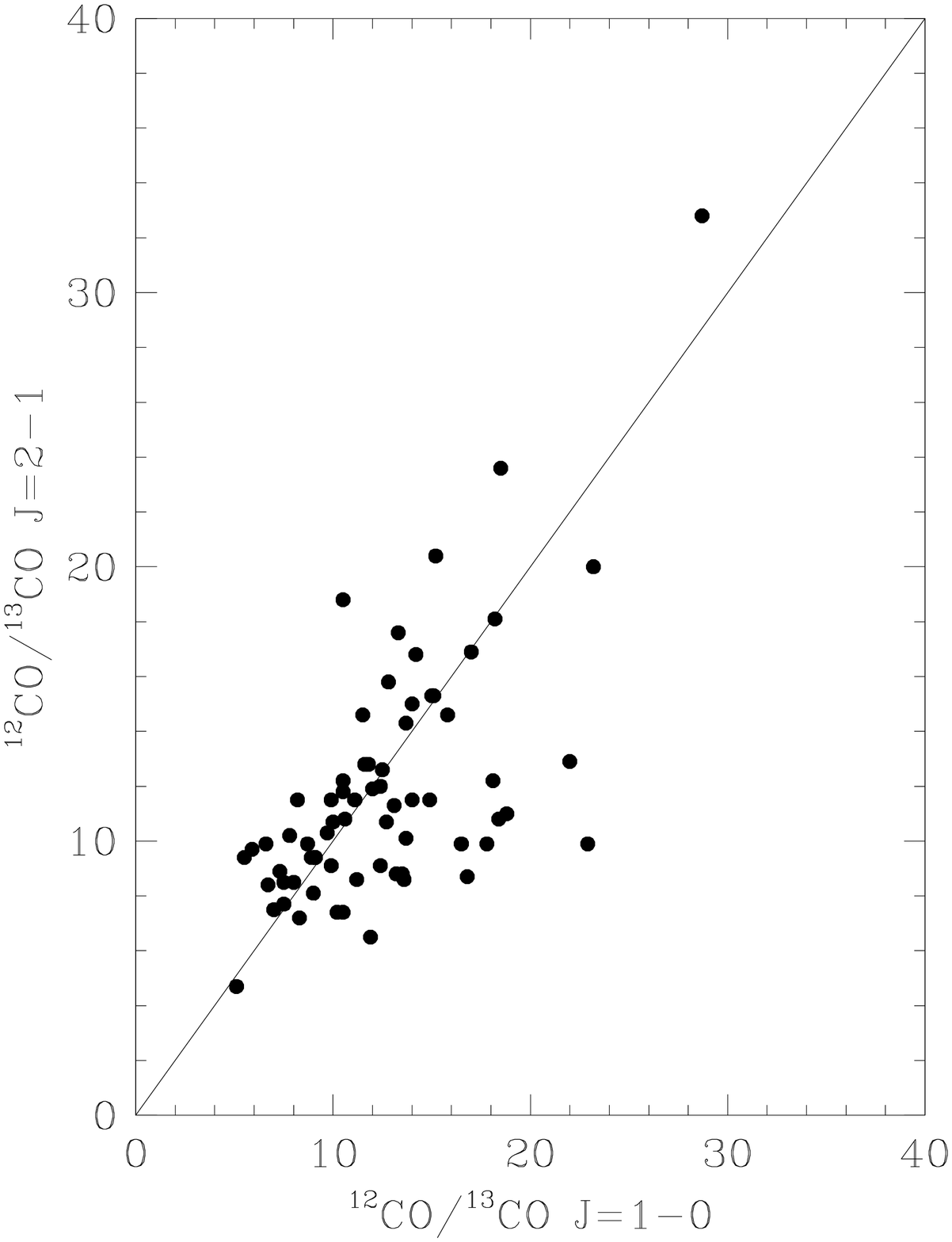}}}
    \resizebox{4.3cm}{!}{\rotatebox{0}{\includegraphics*{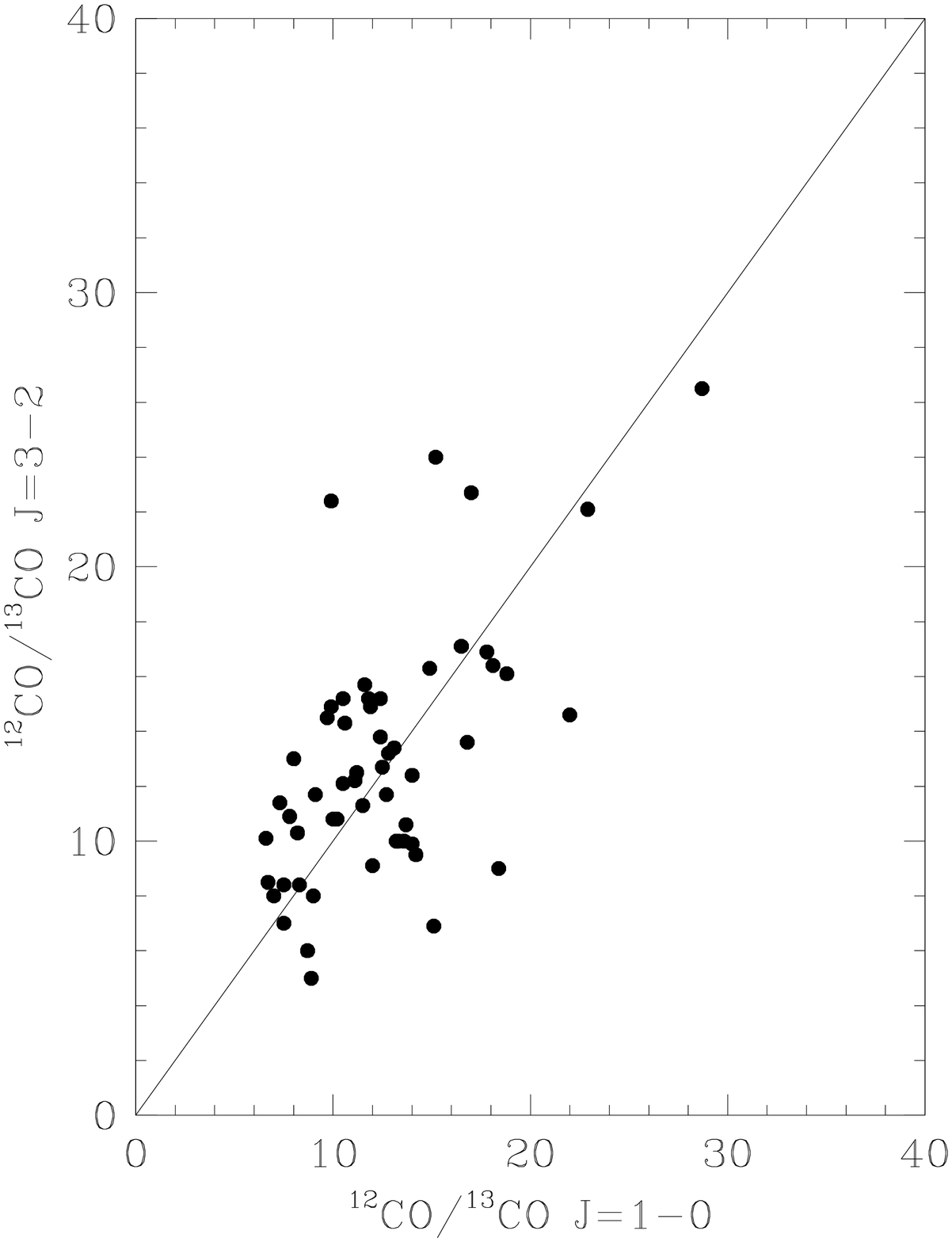}}}
  \end{minipage}
%\begin{minipage}[]{9cm}
%  \resizebox{4.3cm}{!}{\rotatebox{270}{\includegraphics*{iso10lum10.eps}}}
%  \resizebox{4.3cm}{!}{\rotatebox{270}{\includegraphics*{iso21lum21.eps}}}
%\end{minipage}
%\end{minipage}
\hfill
\caption[] {Top: Distribution of the $J$=1-0, the $J$=2-1, and the
  $J$=3-2 isotopologue ratios. The histogram fraction representing
  luminous galaxies (log $L_{FIR}$/L$_{\odot}\geq10$) is filled. The
  remainder represent the normal galaxies (log  L$_{FIR}$/L$_{\odot}<10$)
  in the sample.  Bottom:  $J$=2-1 and $J$=3-2 isotopologue ratio as
  a function of the $J$=1-0 ratio.
%  Bottom: the $J$=1-0 and $J$=2-1
%  isotopologue ratio as a function of the $\co$ luminosity in the same
%  transition.
}
\label{isorat}
\end{figure}

\section{Results}

Tables\,\ref{sestdat} through \ref{jcmtdat4} list all directly
observed $\co$ intensities measured with the SEST (S), the JCMT (J),
and the IRAM 30m (I) telescope, with the resolution in arcseconds
indicated in the headers. For comparison purposes, we also listed
additional intensities at lower resolutions determined by the
convolution of JCMT $\co$ maps such as those shown in
Fig.\,\ref{galmap}. In a few cases we have included published
measurements obtained by others with the same telescopes; these are
identified in the footnotes.

With Tables\,\ref{sestdat} throughj \ref{jcmtdat3} we have constructed
transition line ratios in matched beams. Individual ratios have
typical errors of $25\%$ to $30\%$. The histograms in
Fig.\,\ref{transrat} show the distributions of the transition line
ratios. These are clearly peaked, and their width reflects in roughly
equal parts the measurement error and the intrinsic variation. The
average (1-0):(2-1):(3-2):(4-3) $\co$ line intensities relate to one
another as ($1.09\pm0.04$):(1.00):($0.76\pm0.05$):($0.62\pm0.05$). As
a practical application, the quantities $1.1\times I_{CO}$(2-1) or
$1.4\times I_{CO}$(3-2) can thus be used to estimate the central
$I_{CO}$(1-0) intensities in gas-rich spiral galaxies when these are
needed but not measured. Oka $\etal$ (2012) found the identical
(3-2):(1-0) ratio for the central region of the Milky Way. The central
(2-1):(1-0) ratio of 0.9 exceeds the value 0.7 used by Sandstrom
$\etal$ (2013) for galaxy disks. The bottom diagram in
Fig.\,\ref{transrat} shows (3-2):(2-1) ratios as a function of the
parent galaxy FIR luminosity, ranging from log$L(FIR)$ = 9 for normal
galaxies over log$L(FIR)$ = 10 for star-burst galaxies to log$L(FIR)$
= 11 for luminous infrared galaxies (LIRGs). It does not reveal a
clear dependence on galaxy class, nor does any of the other transition
line ratios.

An essential part of this survey is the measurement of $\co$-to-$\thirco$
isotopologue ratios in the $J$=1-0, $J$=2-1, and $J$=3-2 transitions;
high atmospheric opacities render the $J$=4-3 $\thirco$ line
practically unobservable from the ground. In galaxy disks and centers,
the observed $\co$ lines are optically thick ($\tau>1$), but in the
observed three lowest transitions, the $\thirco$ lines have optical
depths (well) below unity. This is important because including lines
with low optical depth reduces the degeneracy that severely limits the
analysis of the optically thick $\co$ lines.  The measured $\thirco$
fluxes are listed in Tables\,\ref{sestdat}-\ref{jcmtdat3}, and
Figs. \ref{sestprofiles} through \ref{jcmtprofiles} show that the
$\thirco$ and $\co$ line profiles are very similar in width and
shape. The single but frequently occurring difference is a dip in the
central $\thirco$ profile at the systemic velocity where the $\co$
profile shows a flat top. This dip suggests an optical depth decrease
in the nuclear line of sight that is consistent with a lack of material (an
unresolved `hole') in the very galaxy center.

Taking into account the errors, the isotopologue ratios in the lower
two transitions do not depend on the aperture size. We therefore
averaged whenever possible the isotopologue ratios in the $45"$ and
$22"$ and the $22"$ and $11"$ apertures. The resulting distributions
in the lower three transitions are shown in Fig.\,\ref{isorat}. The
$\co$-to-$\thirco$ ratios peak around $R$=10 in the $J$=2-1 transition
and well above that in the other two transitions. The isotopologue
ratios in the three transitions are clearly related to one another.
In all three transitions, most isotopologue ratios occur between $R$=8
and $R$=16. Only a few galaxies have $R<8,$ which is characteristic of
the relatively high optical depths of dense star-forming molecular
clouds in the spiral arm disk of the Milky Way.

%Figure 7 Multi-aperture CO photometry
\begin{figure*}
  \resizebox{2.9cm}{!}{\rotatebox{0}{\includegraphics*{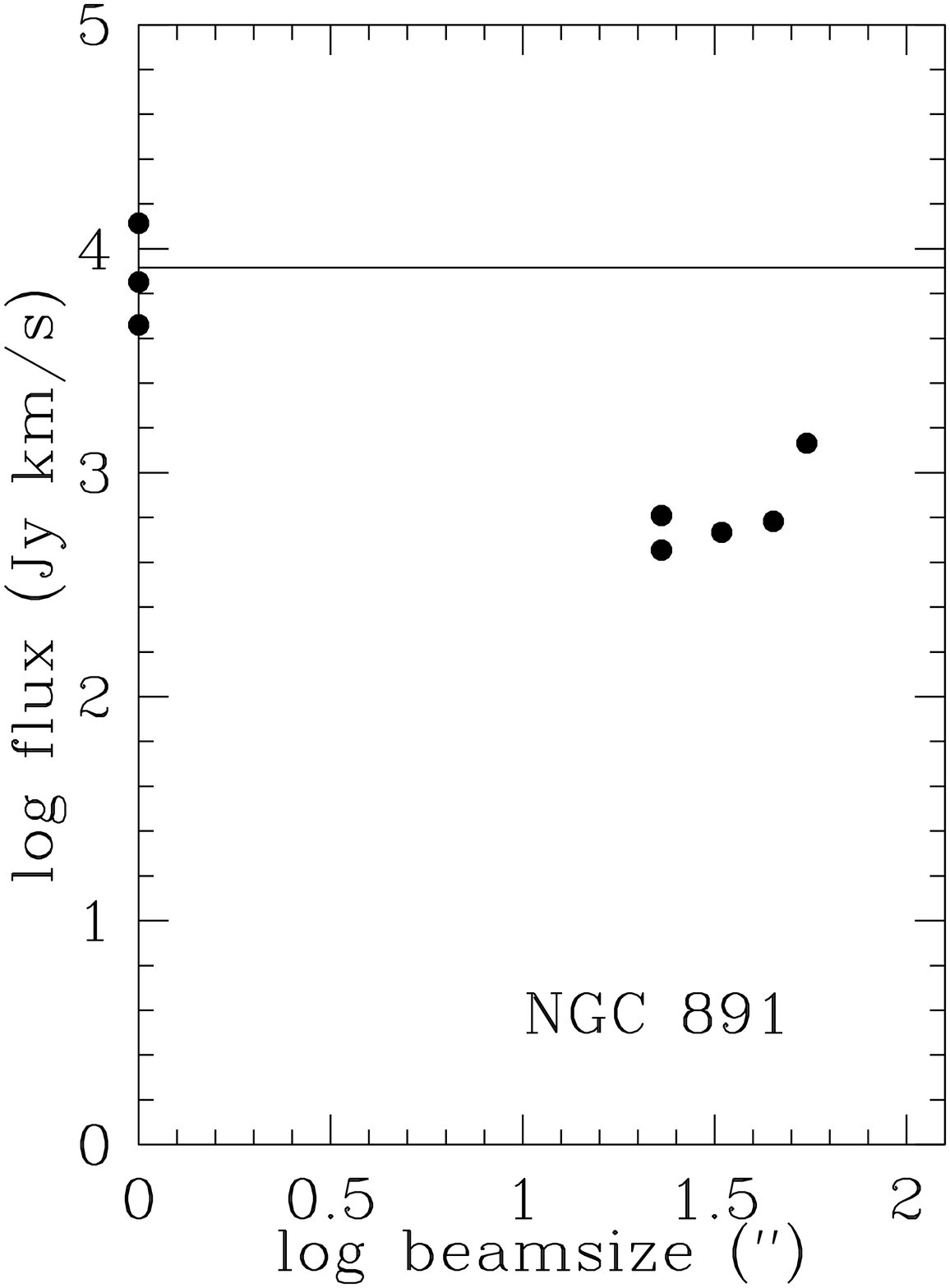}}}
   \hfill
  \resizebox{2.9cm}{!}{\rotatebox{0}{\includegraphics*{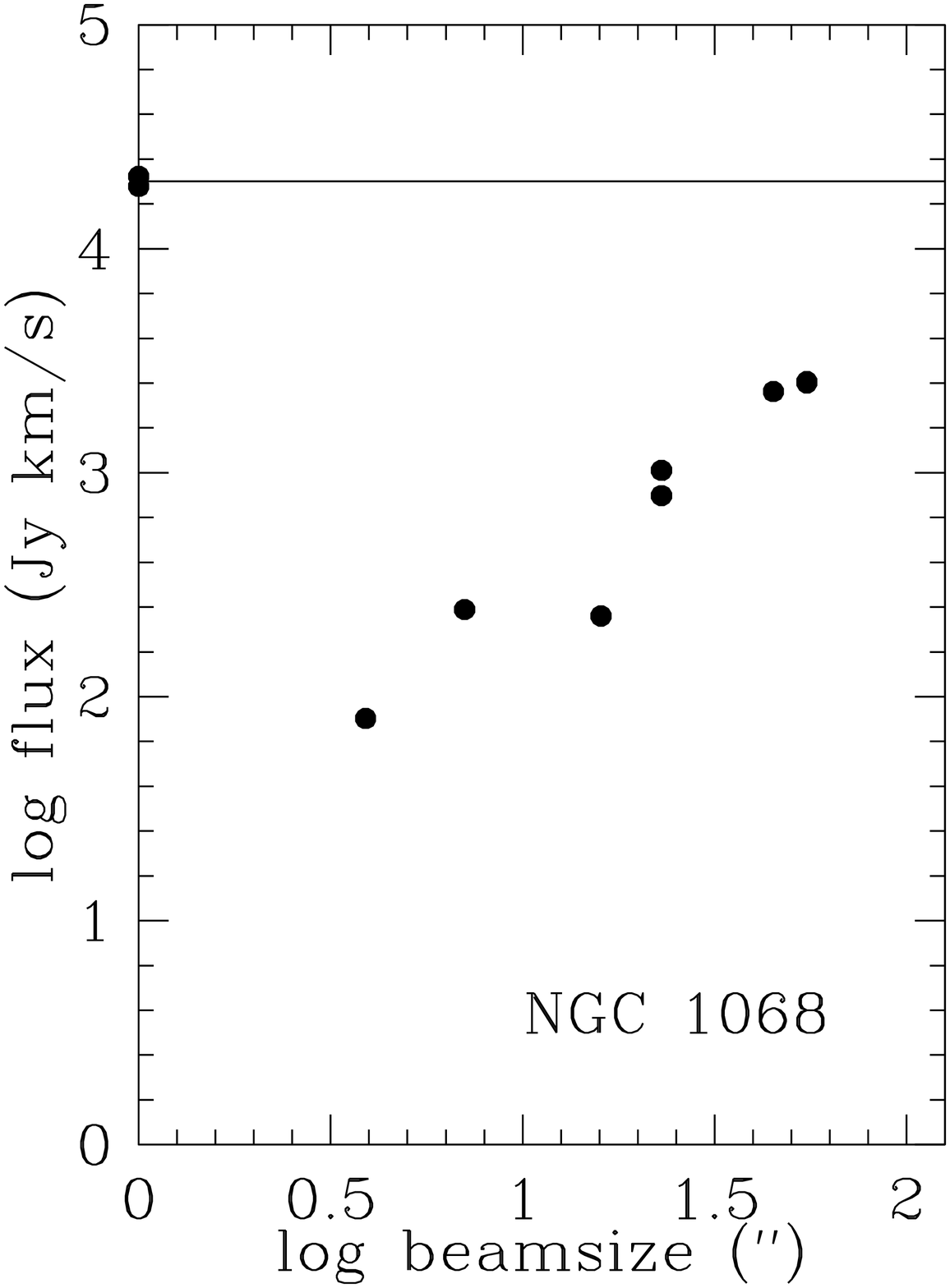}}}
   \hfill
  \resizebox{2.9cm}{!}{\rotatebox{0}{\includegraphics*{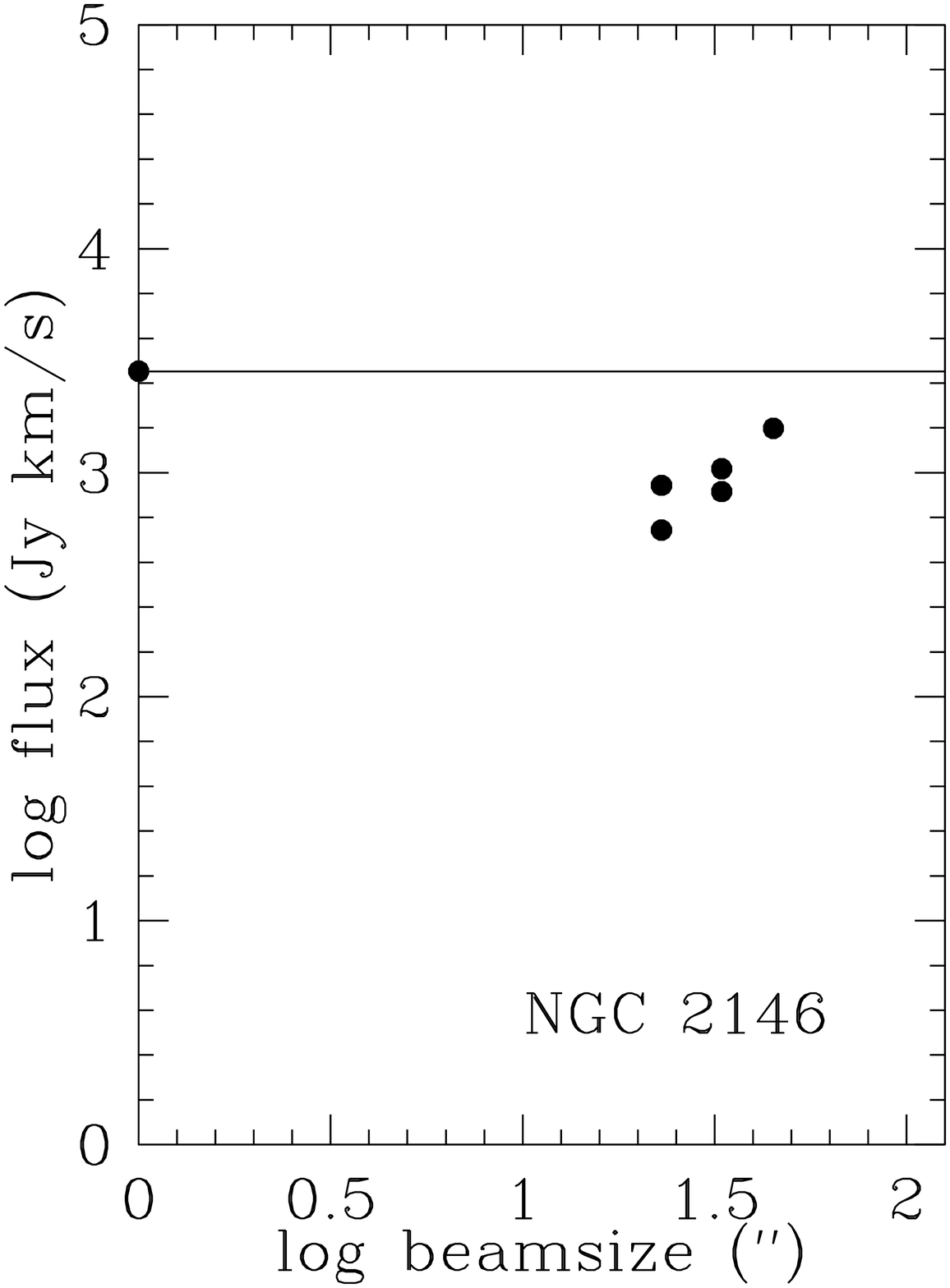}}}
   \hfill
  \resizebox{2.9cm}{!}{\rotatebox{0}{\includegraphics*{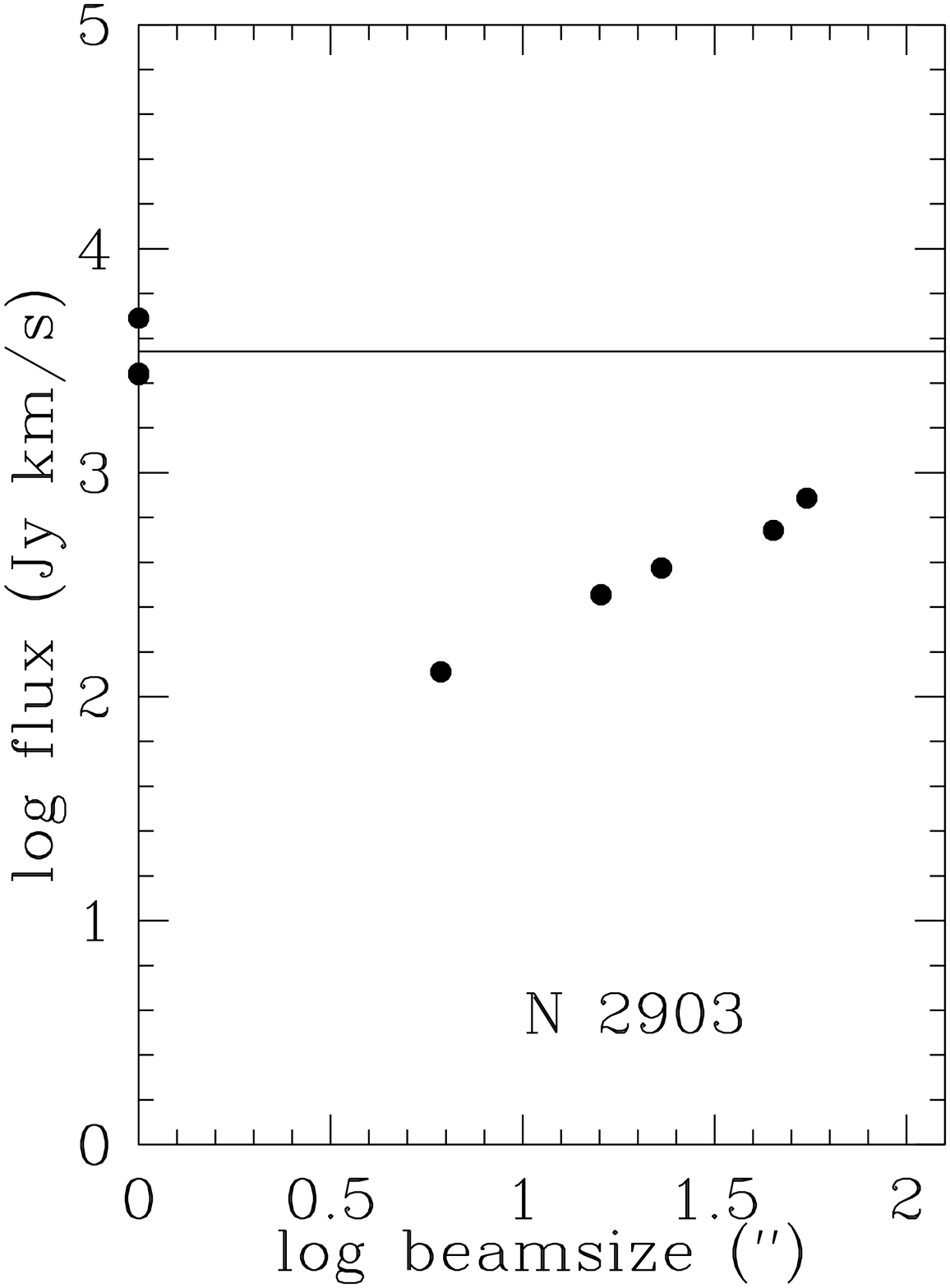}}}
   \hfill
  \resizebox{2.9cm}{!}{\rotatebox{0}{\includegraphics*{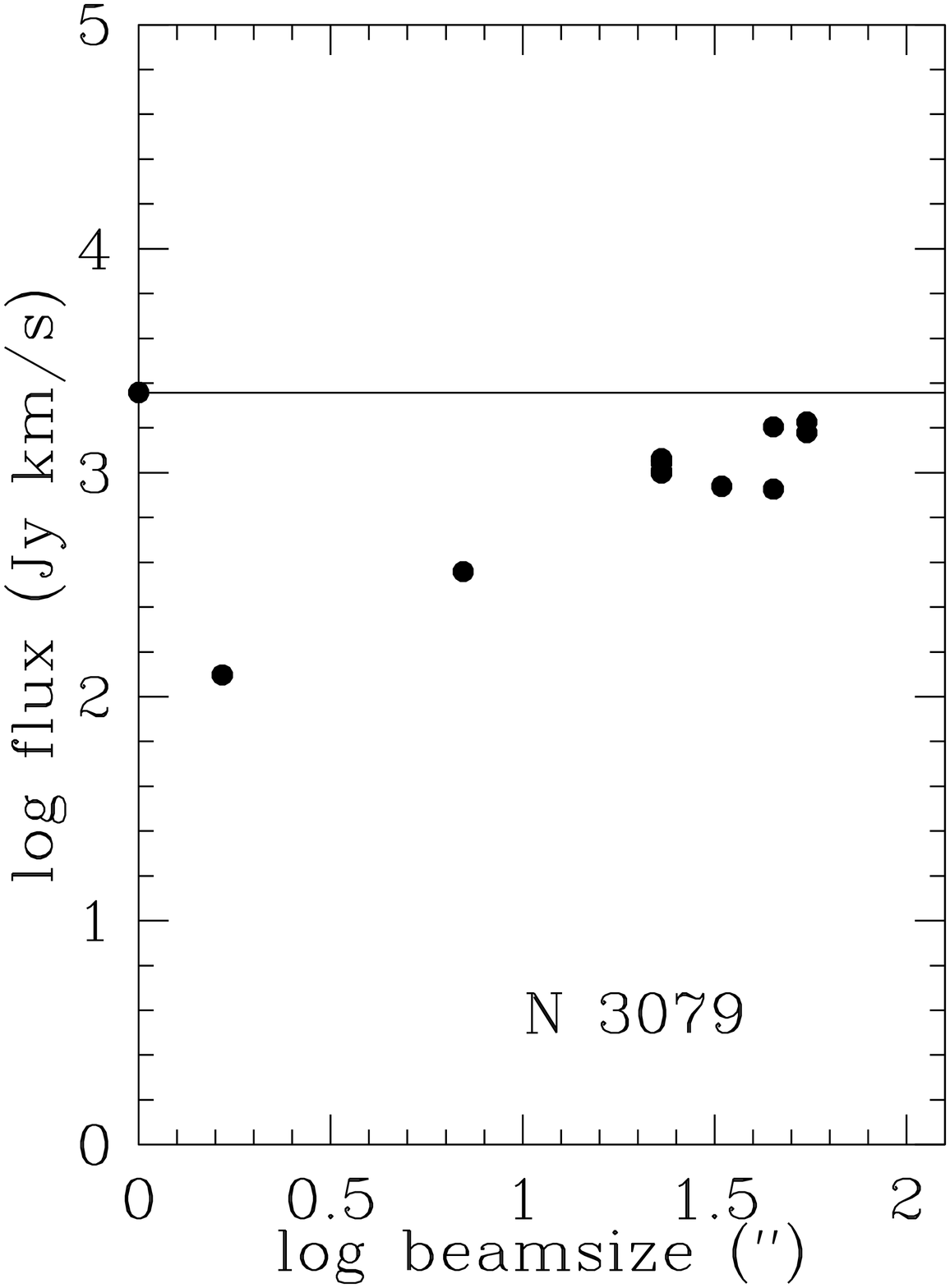}}}
   \hfill
  \resizebox{2.9cm}{!}{\rotatebox{0}{\includegraphics*{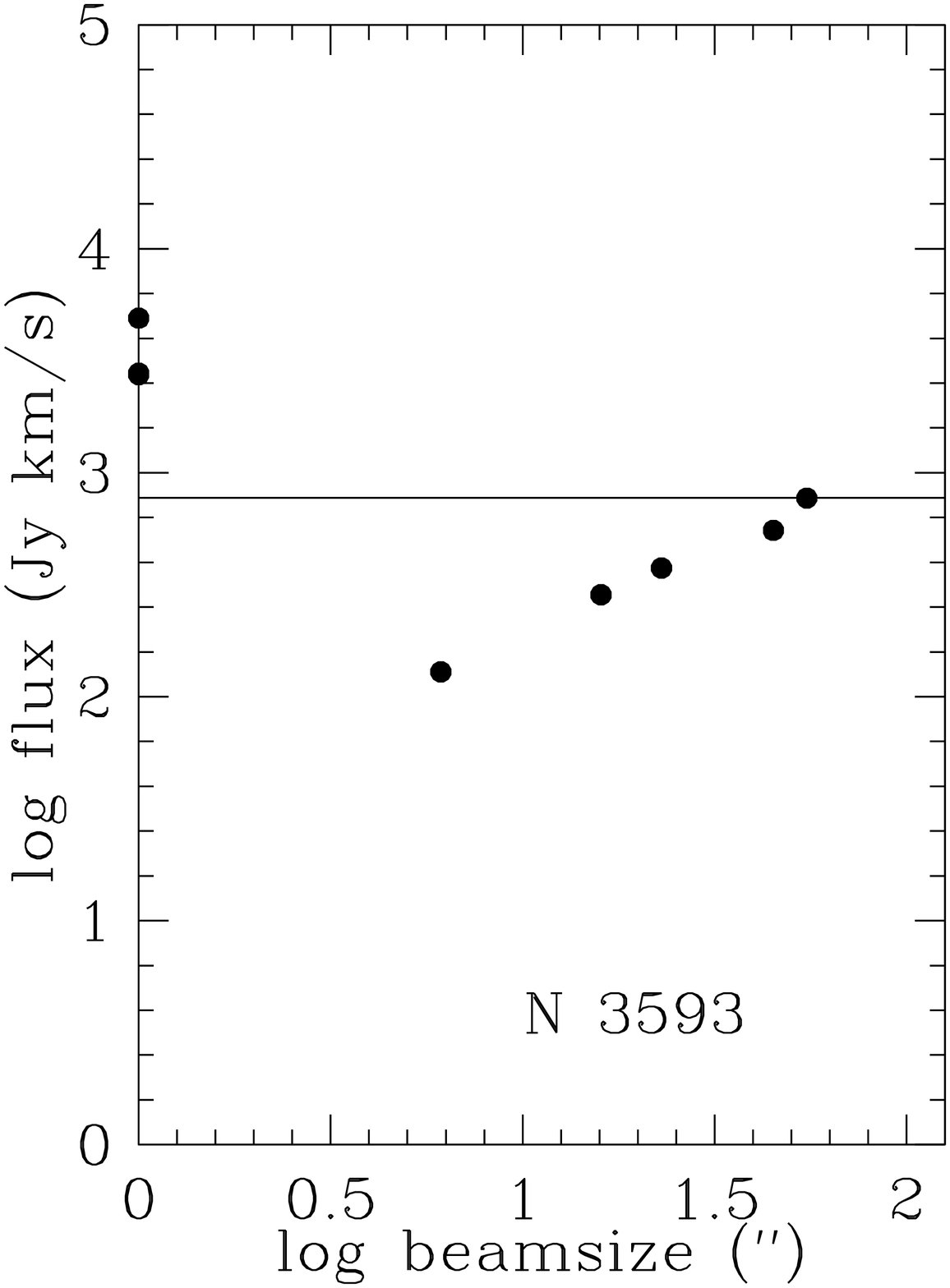}}}
  \hfill
  \resizebox{2.9cm}{!}{\rotatebox{0}{\includegraphics*{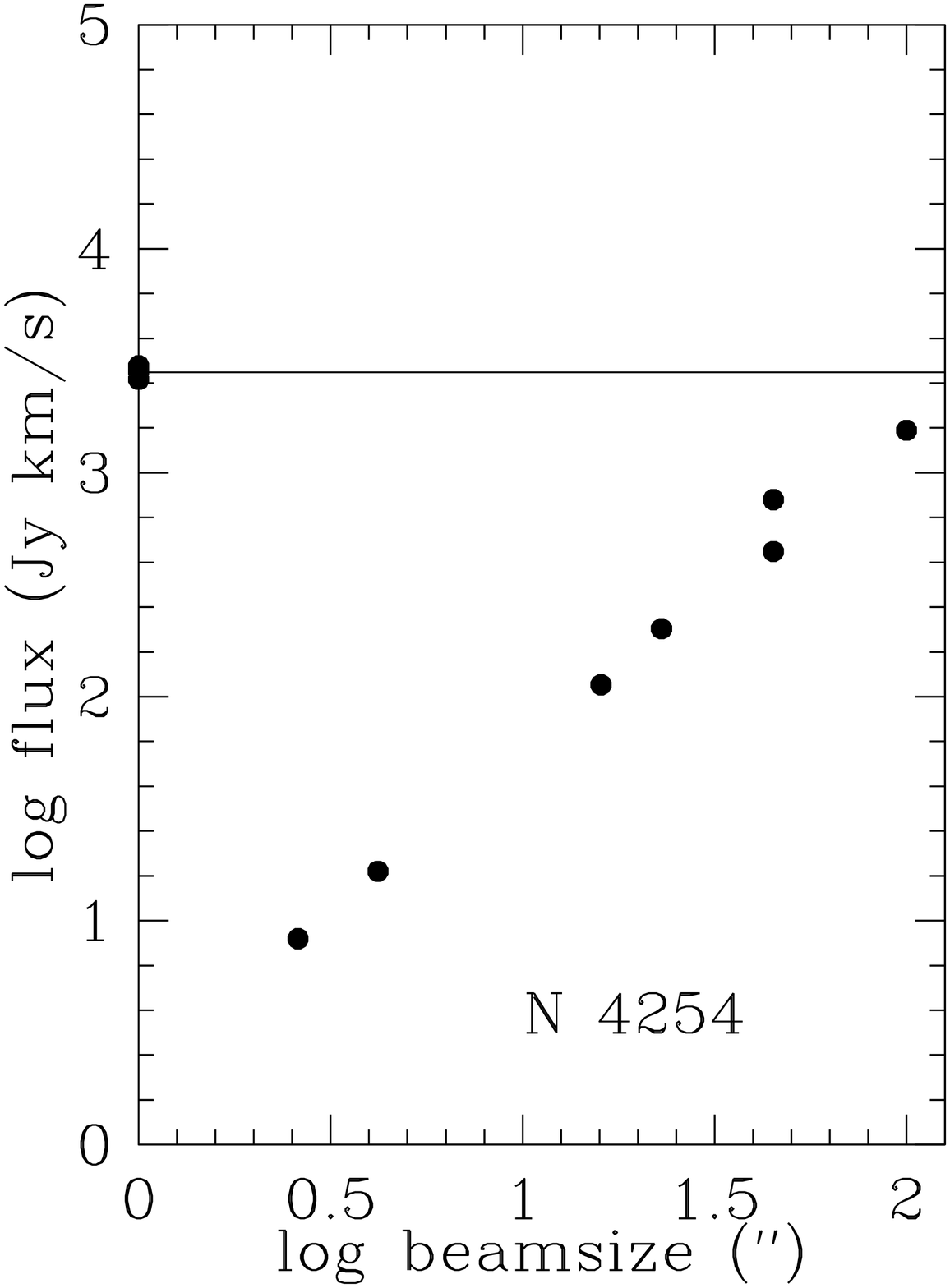}}}
   \hfill
  \resizebox{2.9cm}{!}{\rotatebox{0}{\includegraphics*{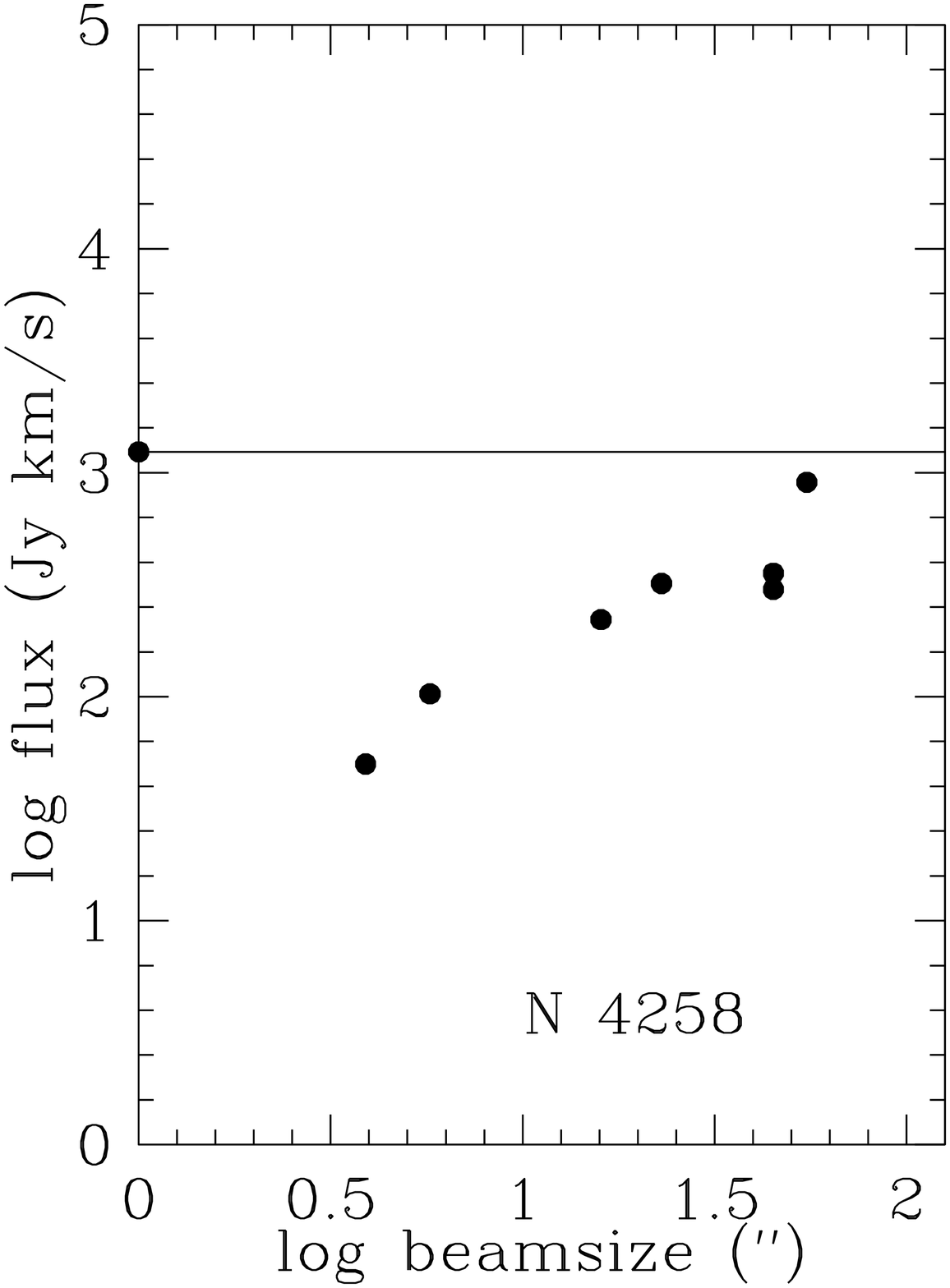}}}
   \hfill
 \resizebox{2.9cm}{!}{\rotatebox{0}{\includegraphics*{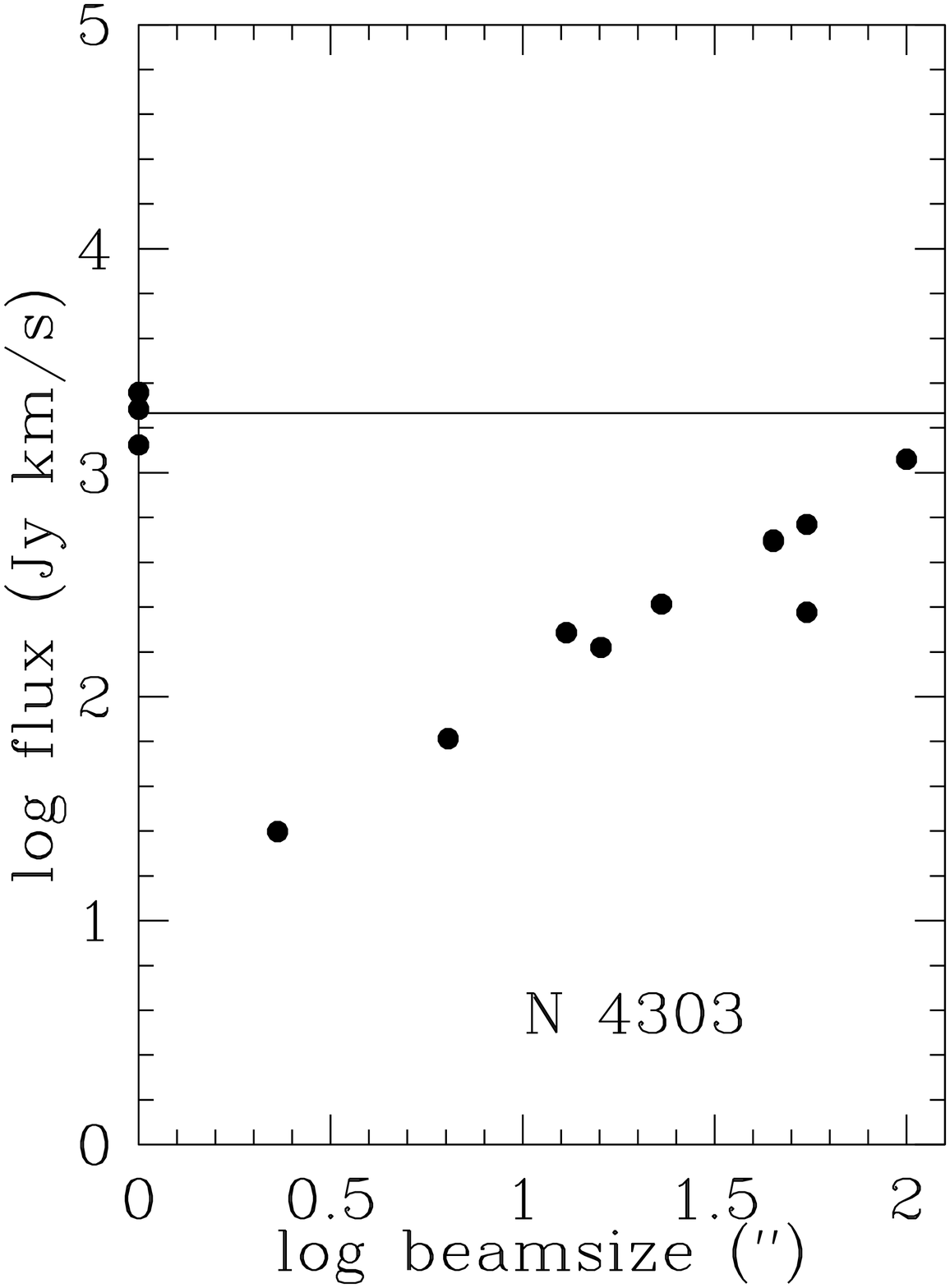}}}
   \hfill
 \resizebox{2.9cm}{!}{\rotatebox{0}{\includegraphics*{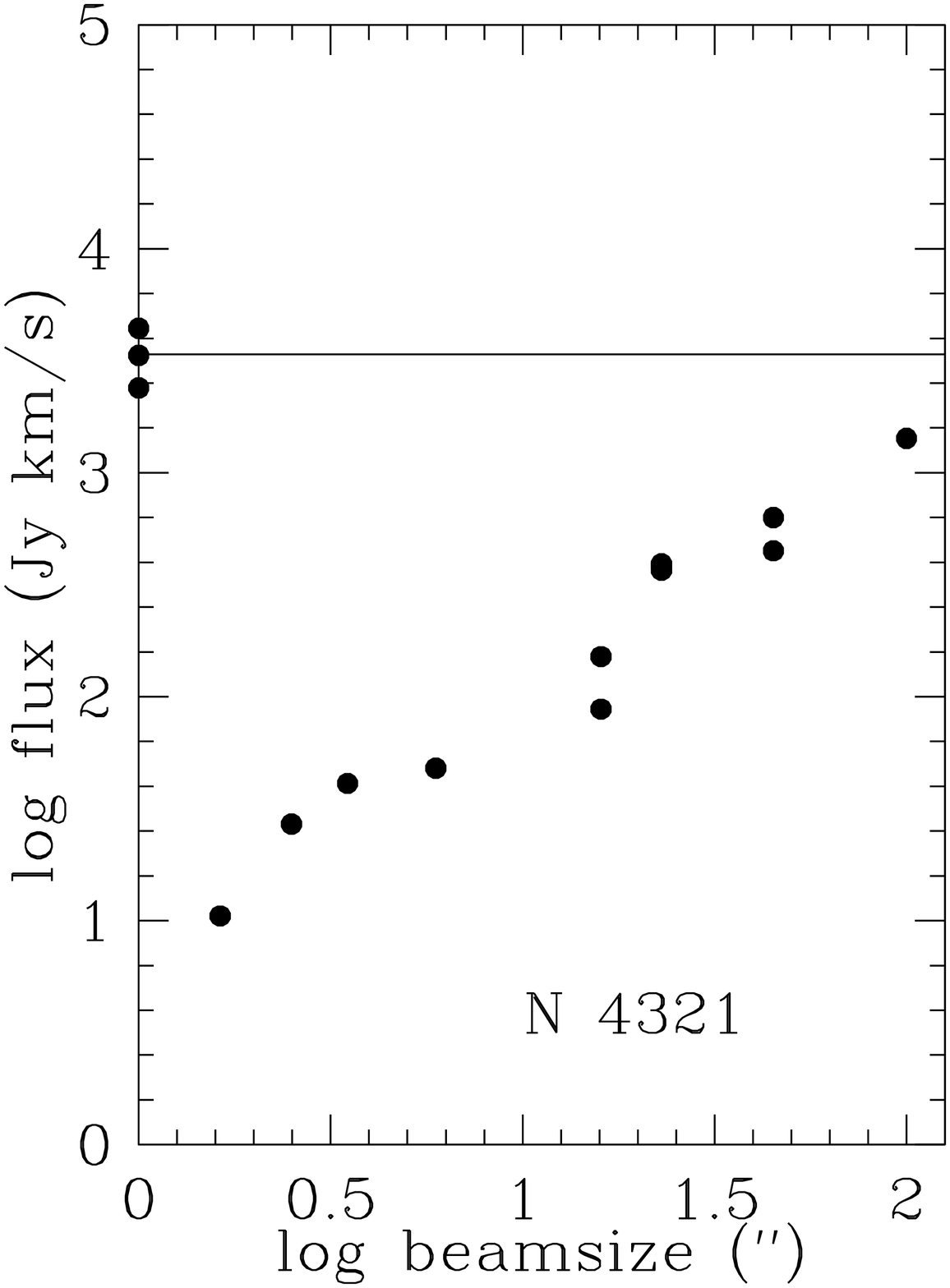}}}
   \hfill
 \resizebox{2.9cm}{!}{\rotatebox{0}{\includegraphics*{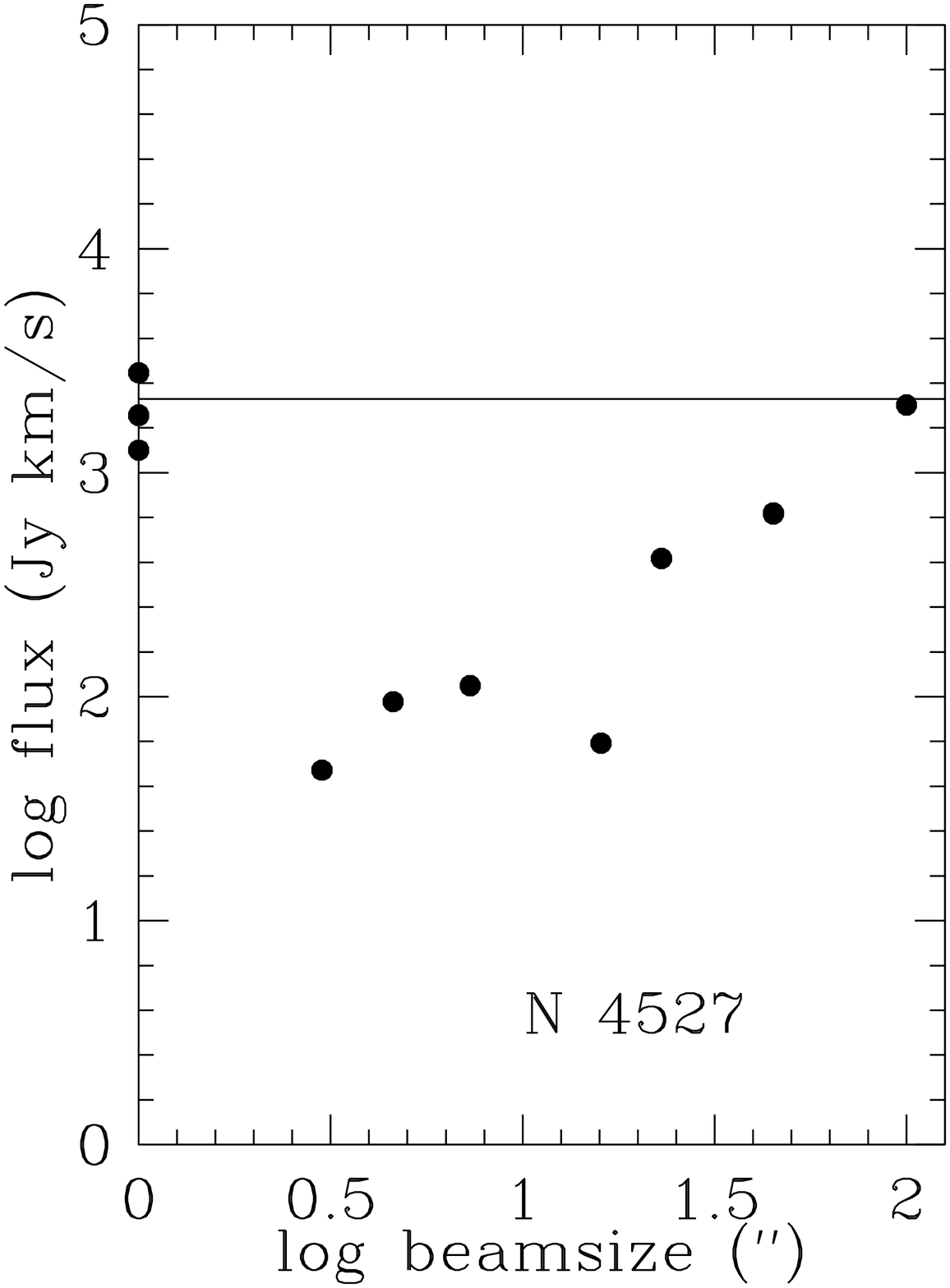}}}
   \hfill
 \resizebox{2.9cm}{!}{\rotatebox{0}{\includegraphics*{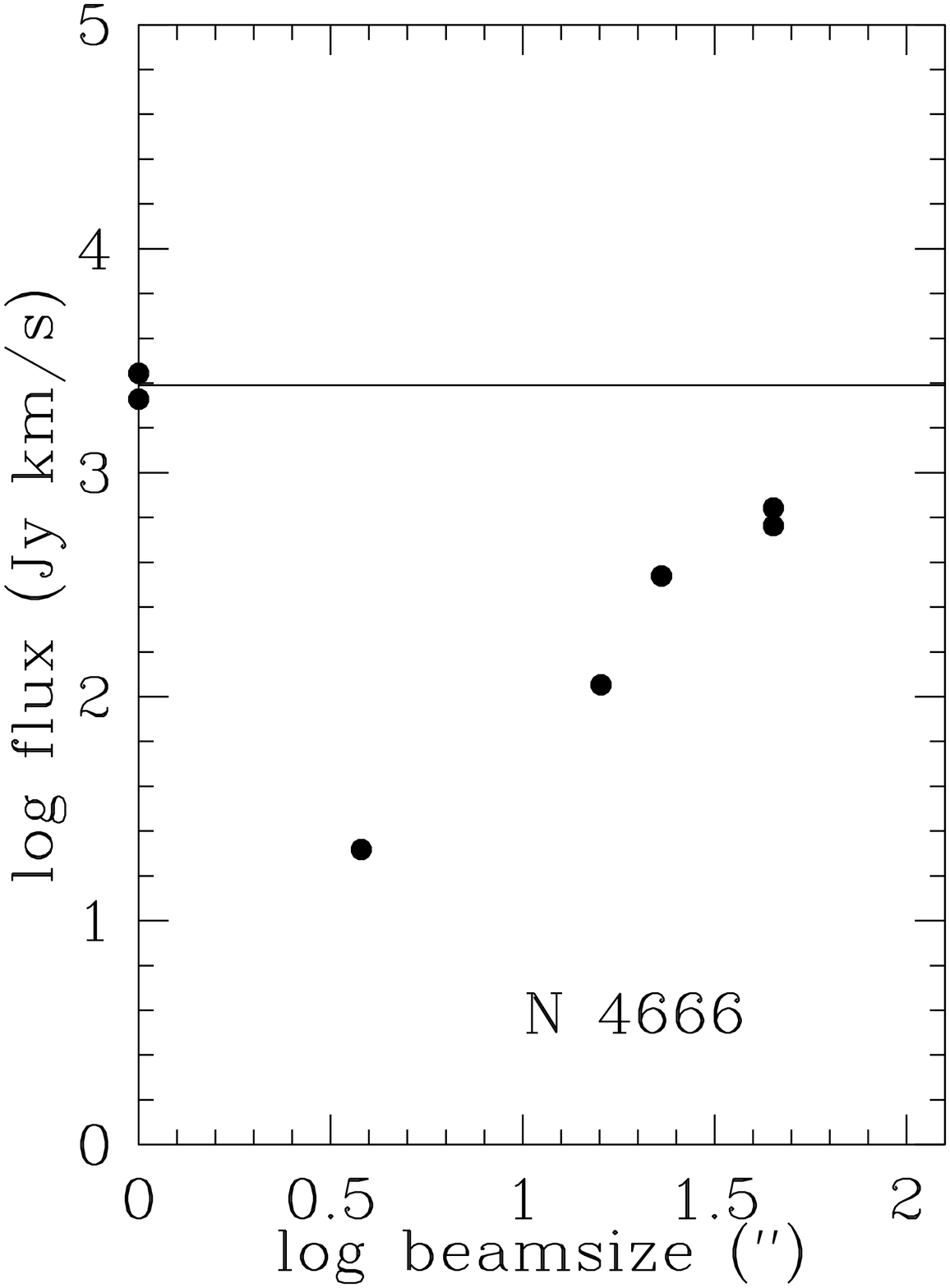}}}
   \hfill
 \resizebox{2.9cm}{!}{\rotatebox{0}{\includegraphics*{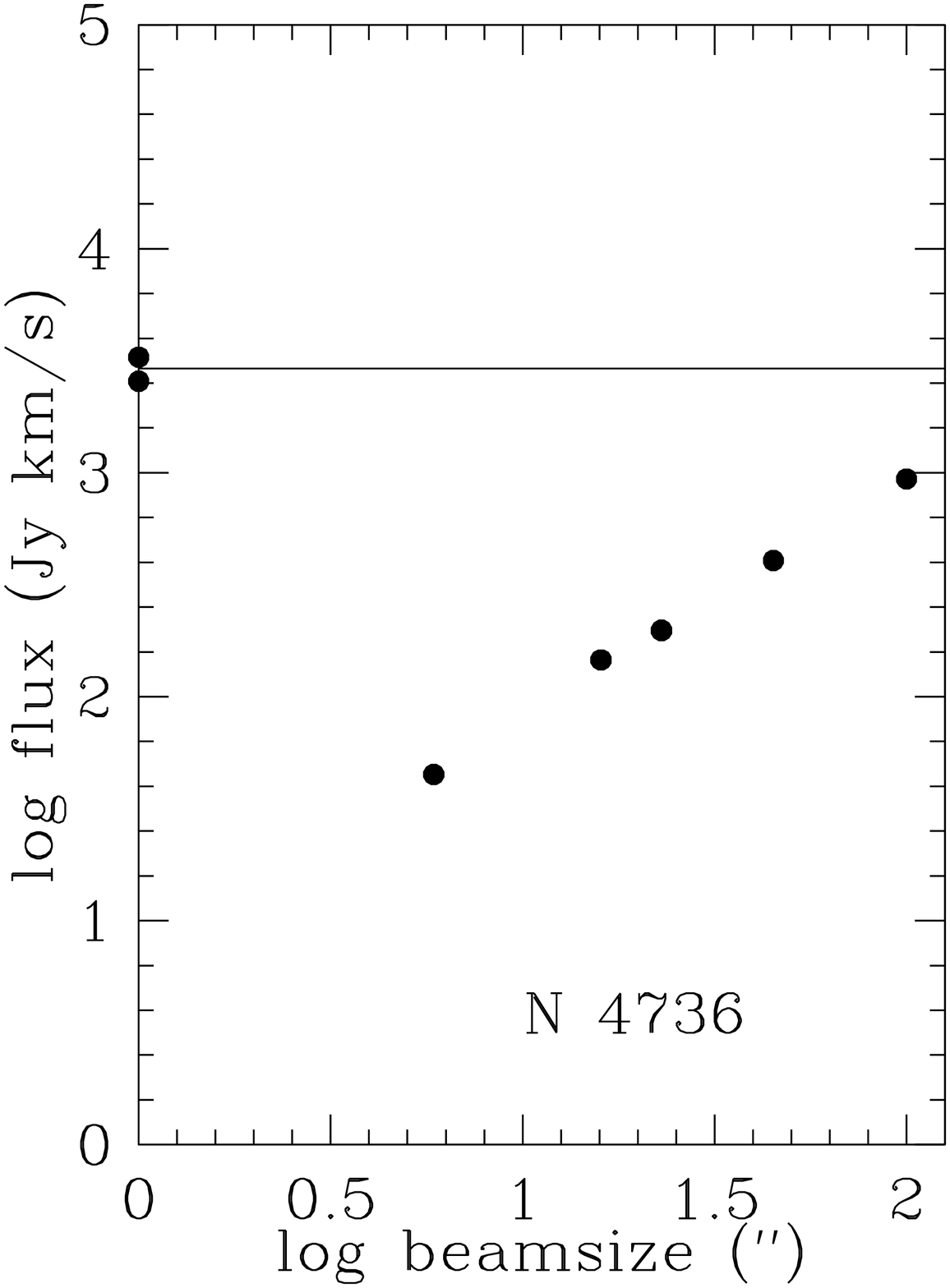}}}
   \hfill
 \resizebox{2.9cm}{!}{\rotatebox{0}{\includegraphics*{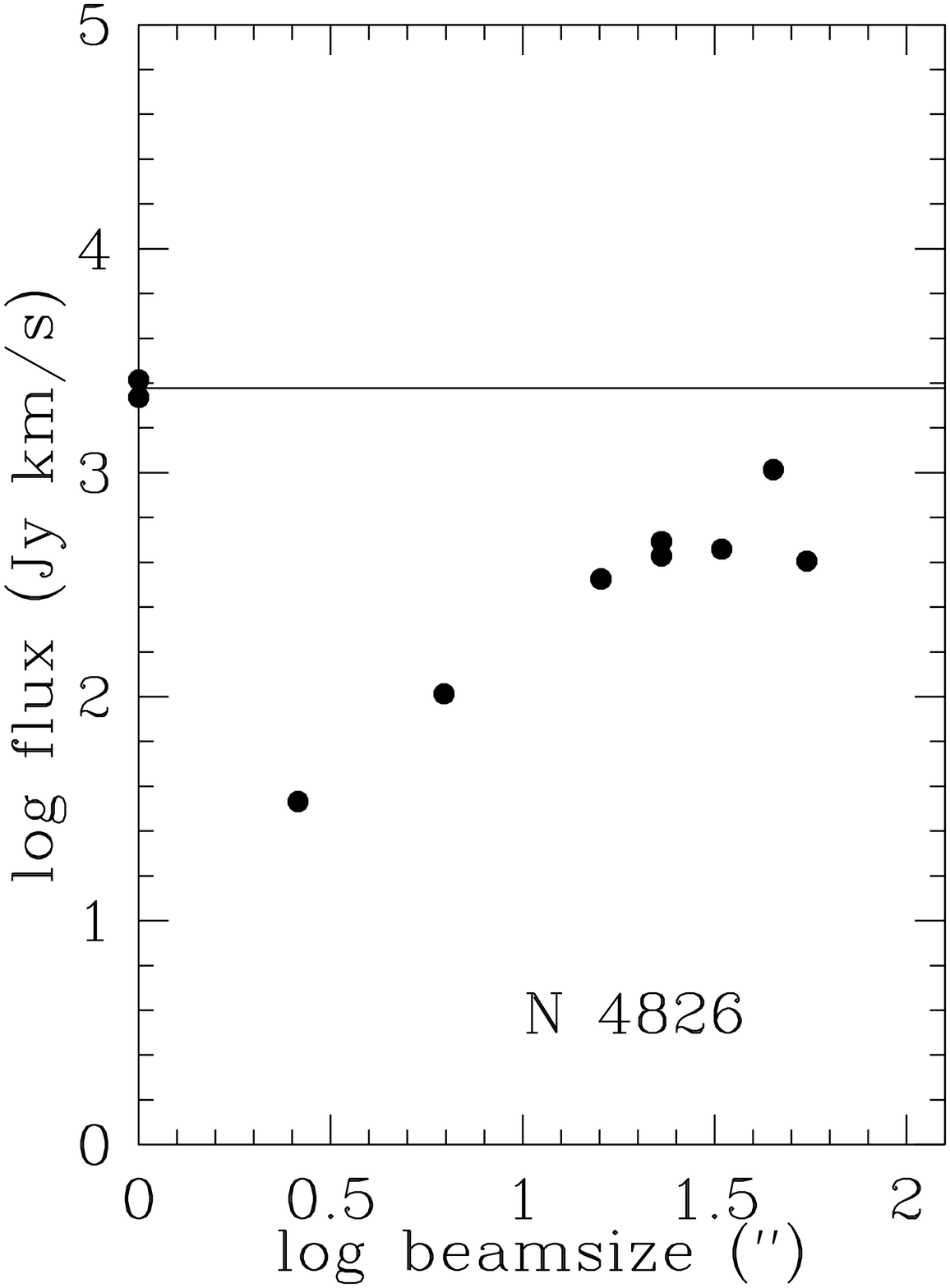}}}
   \hfill
 \resizebox{2.9cm}{!}{\rotatebox{0}{\includegraphics*{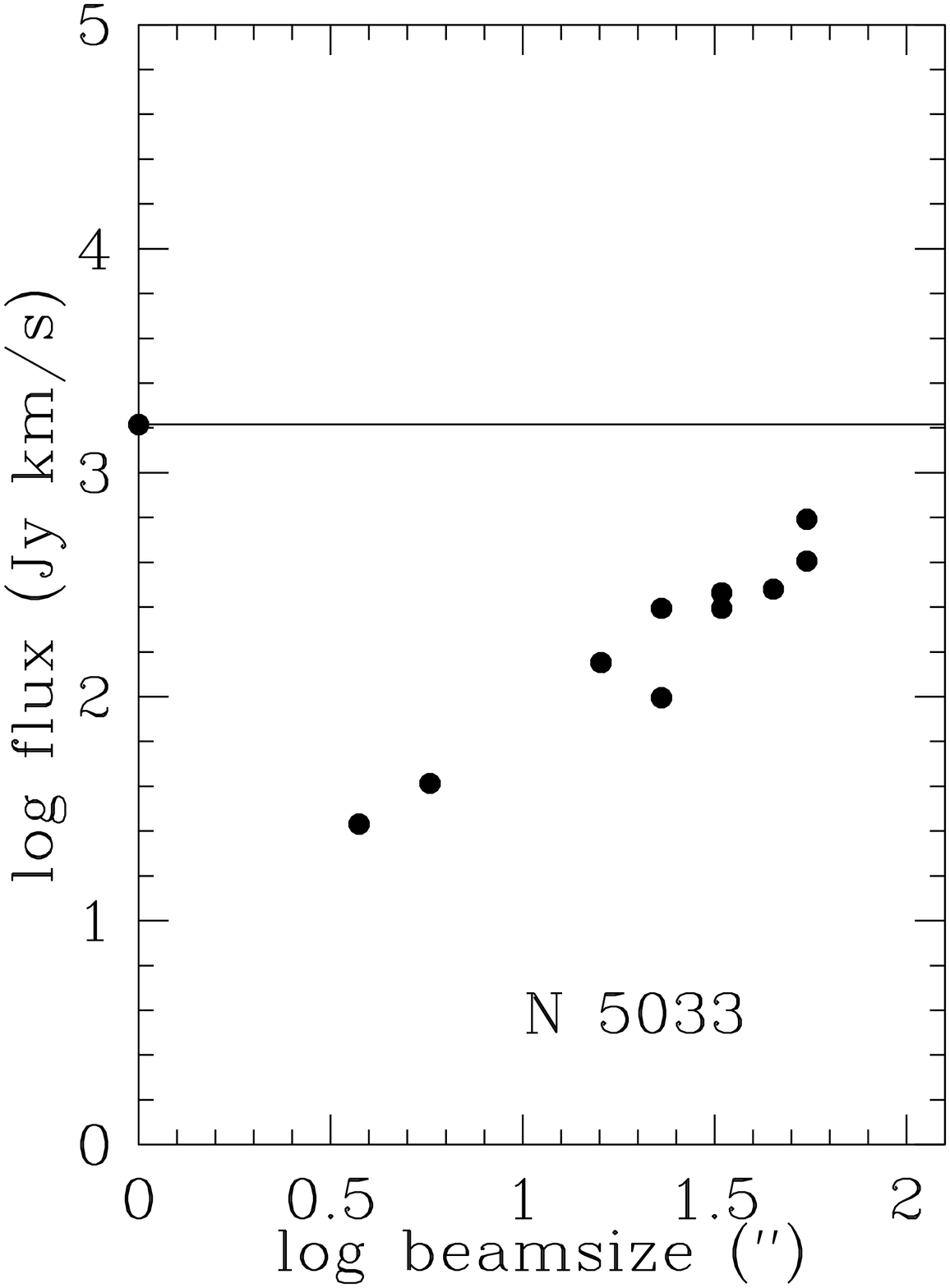}}}
   \hfill
 \resizebox{2.9cm}{!}{\rotatebox{0}{\includegraphics*{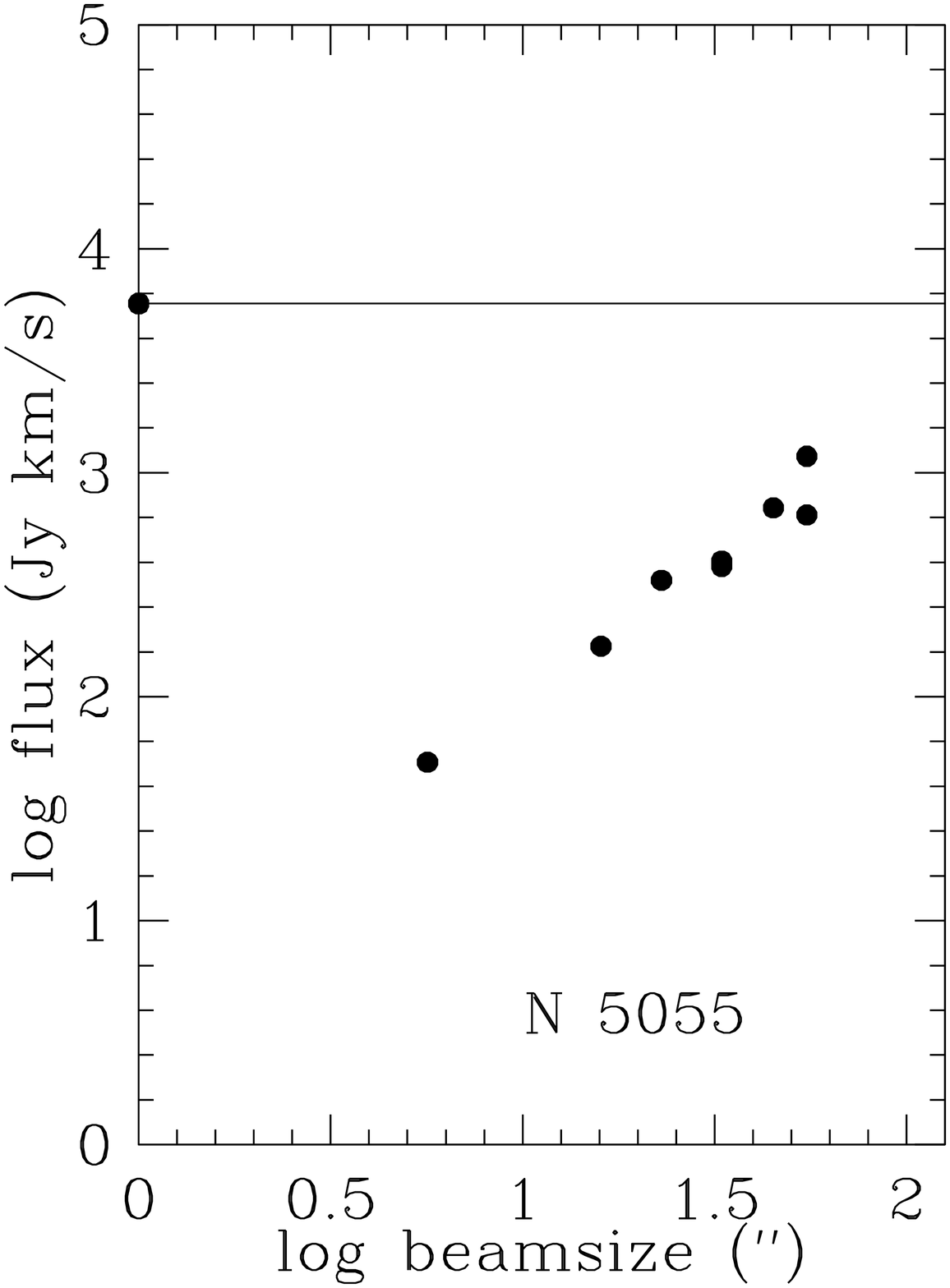}}}
   \hfill
 \resizebox{2.9cm}{!}{\rotatebox{0}{\includegraphics*{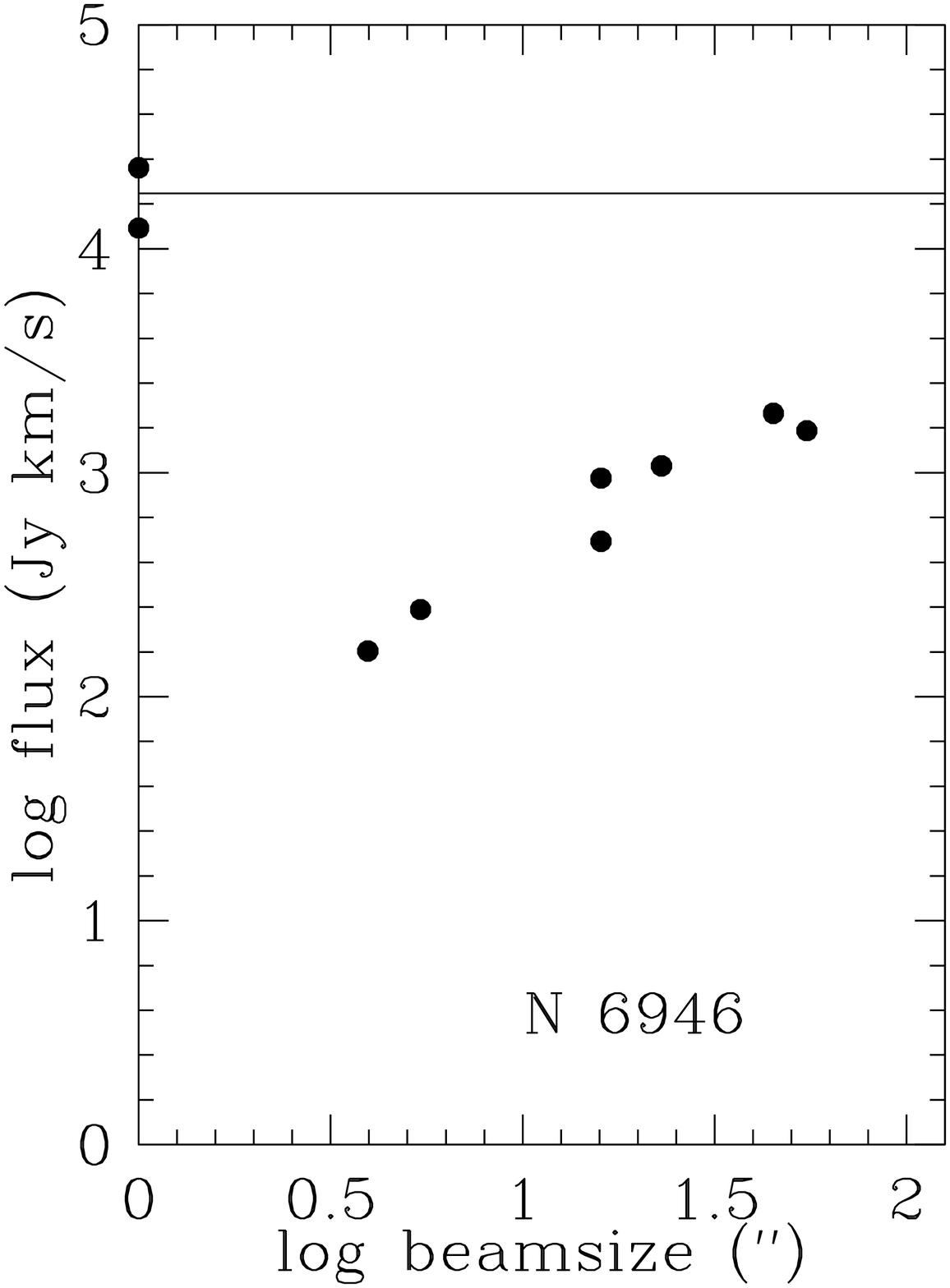}}}
   \hfill
 \resizebox{2.9cm}{!}{\rotatebox{0}{\includegraphics*{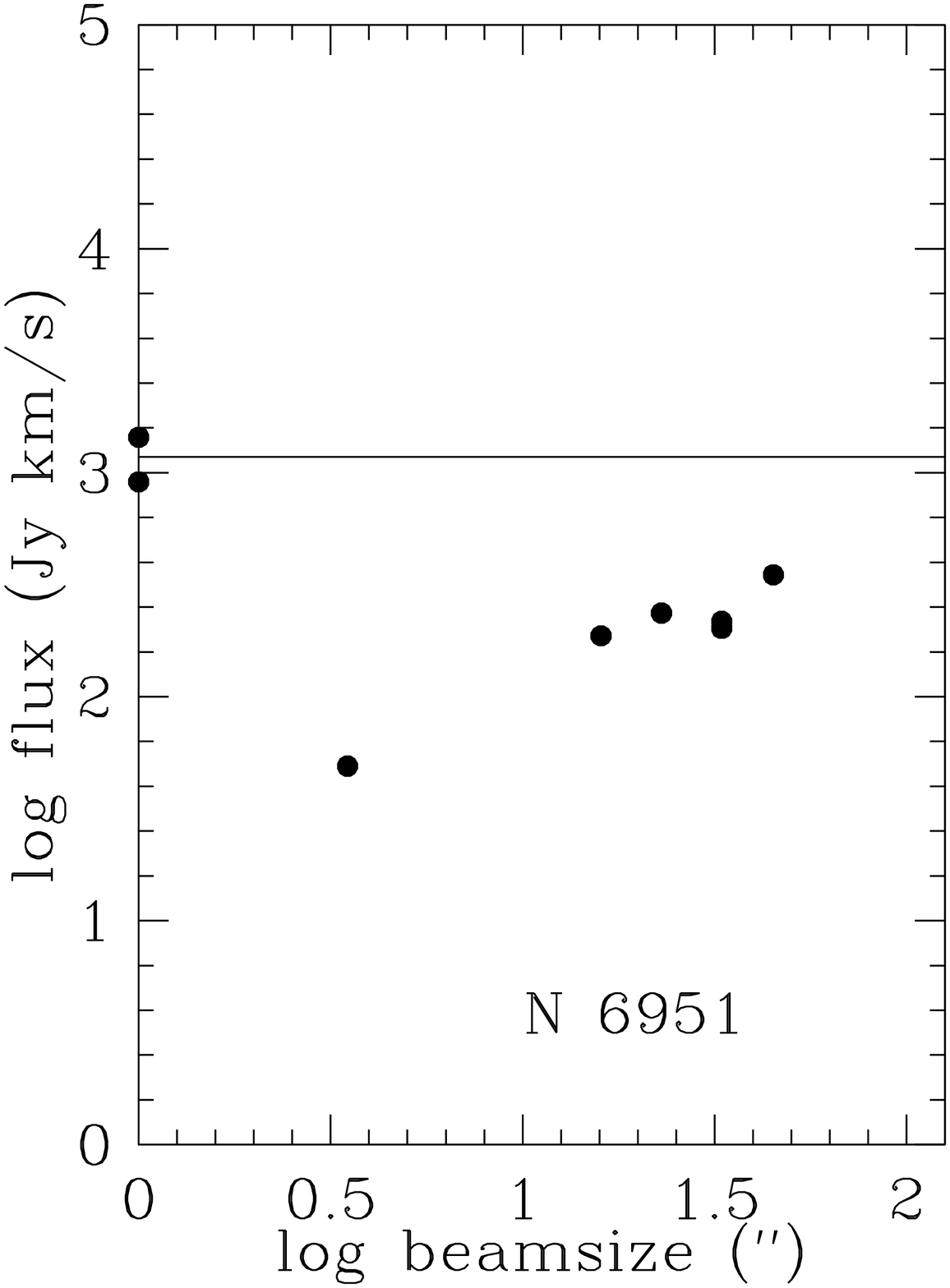}}}
 \caption{$J$=1-0 $\co$ multi-aperture photometry of galaxies observed
   with different telescopes. The points on the vertical axis refer to
   the integrated CO line flux of the entire galaxy.  In each panel,
   their average is marked by a horizontal line.  References to the
   measurements used in these diagrams and in the photometry analysis
   are given in Appendix A.
}
\label{cophot}
\end{figure*}

% Table 6 Spatial distribution
\begin{table*}
\caption[]{\label{size}Spatial distribution of CO emission} 
\begin{center}
{\small % 
\begin{tabular}{l|rccc||l|rccc||l|rccc} 
\noalign{\smallskip}     
\hline
\noalign{\smallskip}
 &\multicolumn{2}{c}{Whole galaxy$^a$}&\multicolumn{2}{c}{Central peak$^b$}&     &\multicolumn{2}{c}{Whole galaxy$^a$}&\multicolumn{2}{c}{Central peake$^b$}&     &\multicolumn{2}{c}{Whole galaxy$^a$}&\multicolumn{2}{c}{Central peak$^b$}\\
NGC&$\alpha$&$d_{\rm CO}$ &$\Omega_{\rm CO}$& $R_{\rm CO}$&NGC&$\alpha$ &$d_{\rm CO}$  &$\Omega_{\rm CO}$& $R_{\rm CO}$& NGC&$\alpha$&$d_{\rm CO}$ &$\Omega_{\rm CO}$& $R_{\rm CO}$  \\
IC   &     & $'$ $(\%)$& nsr  & kpc & IC   &     &$'$ $(\%)$& nsr   &  kpc & IC  &      &$'$ $(\%)$& nsr     & kpc  \\
 1)  & (2) & (3)       &  (4) & (5) & (1)  & (2) & (3)      &  (4)  &  (5) & (1) & (2)  & (3) &  (4)  &  (5) \\
\noalign{\smallskip}    
\hline                          
\noalign{\smallskip}    
 253 & 0.6 & 8.0 (43) & 16    & 0.3 & 3079 & 0.7 & 1.5 (19) &  5    & 0.5 & 4631 & 1.3  & 2.8 (18) & min      & ... \\
 278 & 1.9 & 1.0 (46) & min   & ... & 3175 & 0.9 & ...      &  8    & 0.9 & 4647 & 1.3  & 2.5 (85) & 13       & ... \\      
 470 & ... & ...      & 13$^1$& 1.1 & 3227 & 0.6 & 3.2 (60) &  4    & 0.6 & 4666 & 1    & 2.8 (60  & min      & ... \\      
 520 & 1   & 1.3 (29) & 11$^2$& 0.9 & 3256 & ... & ...      & 12    & 2.0 & 4736 & 1.1  & 4.5 (40) &  5.3     & 0.5 \\     
 613 & ... &  ...     & ...   & 1.1 & 3310 & 0.9 & 1.9 (63) & unr   & ... & 4826 & 1.2  & 1.4 (15) &  7.4     & 0.2 \\   
 628 & 2   & 3.4 (33) & min   & ... & 3504 & 0.6 & 2.0 (74) &7$^7$  & 1.1 & 4945 &...   & ...      &14.4$^{14}$& 0.4 \\ 
 660 & 0.6 & 6.5 (71) & 12    & 0.5 & 3556 & 0.6 & 5.2 (60) & ...   & ... & 5033 & 1.1  & 2.8 (27) &12.4$^{15}$& 0.8 \\
 891 & 0.2 &  ...     & 19    & 0.8 & 3593 & 0.8 & 1.5 (26) &5$^8$  & 0.5 & 5055 & 1.3  & 3.9 (31) & 4.2$^{10,15}$& 0.3\\
 908 & 1.2 & 2.1 (35) & ...   & ... & 3627 & 1.1 & 4.7 (51) &  8    & 0.4 & 5135 & 0.4  & ...      & 4.2$^{16}$& 2.0 \\
 972 & 0.4 & ...      & 18    & 2.3 & 3628 & 1.1 & 2.9 (20) &  7    & 0.5 & 5194 & ...  & ...      & min      & ...  \\  
Maf2 & ... & ...      & 14    & 0.2 & 3690 & 0.7 & 0.9 (32) & ...   & ... & 5236 & 1.4  & 3.7 (29) &  6.0     & 0.2 \\    
1055 & 1.3 & 2.3 (30) & min   & ... & 4030 & 0.8 & 2.7 (62) & ...   & ... & Circ & ...  & ...      &  ...     & 0.3 \\     
1068 & 1.2 & 5.0 (70) & 23    & 1.0 & 4038 & 1.4 & 1.2 (23) &4$^9$& 0.7 & 4444 & ...  & ...     &  ...     & 1.2 \\ 
1084 & 0.8 & 3.4 (100)& ...   & ... & 4039 & 1.2 & 1.1 (36) &5$^9$  & 0.6 & 5713 & 1.6  & 1.1 (40) &  9.8    & 1.8 \\ 
1097 & 0.5 & 2.3 (30) & 21$^3$& 0.9 & 4051 & 0.8 & 3.3 (62) &   6   & 0.7  & 5775 & 0.6 & 2.1 (50) &  7.4    & 1.5 \\  
1365 & 0.8 & ...      & 17$^4$& 1.2 & 4102 & 0.2 & ...      &11$^{10}$& 0.8 & 6000 & ... & 0.8 (40) & unr      & ... \\  
I342 & 1.3 & 8.7 (41) & 13    & 0.2 & 4254 & 1.4 & 2.8 (52) &6$^{11}$& 2.1  & 6240 & 0   & ...      & unr      & ... \\ 
1482 & 1.6 & 0.8 (30) & ...   & ... & 4258 & 0.7 & 3.3 (17) &5$^{12}$& 0.2  & 6764 & 0.8 & ...      &4.7$^{17}$ & 1.1 \\ 
1614 & 0.4 & 0.9 (71) & unr   & ... & 4293 & 0.7 & 0.8 (10) & ...   & ...  & 6946 & 1.0 & 6.5 (57) &  5.3    & 0.2 \\  
1672 & ... & ...      & ...   & 2.3 & 4303 & 1.0 & 2.0 (41) &  7    & 0.5  & 6951 & 0.8 & 3.1 (80) &6.7$^{10}$& 0.9 \\
1808 & 1.4 & ...      & 11    & 0.3 & 4321 & 1.1 & 4.0 (55) &  6    & 0.4  & 7469 & ... & 0.9 (60) & 4.7     & 2.6  \\ 
2146 & 1.1 & 1.2 (20) & 14    & 1.1 & 4388 & ... & ...      &  6    & 1.0  & 7541 & 1.2 & 1.2 (34) & ...     & 9.9  \\ 
2273 & 0.7 & 1.1 (35) & 11    & 1.0 & 4414 & 1.4 & 1.6 (44) & min   & ...  & 7674 & ... & ...      & 7.4     & 4.5  \\   
2559 & 0.9 & 1.3 (30) & 11    & 1.2 & 4457 & 0.5 & ...      & ...   & ...  & 7714 & ... & 0.9 (50) & ...     & ... \\  
2623 & 1.0 & 0.7 (32) &  5    & 2.3 & 4527 & 1.0 & 1.8 (30) &5$^{13}$& 0.6  & 7771 & 0   &  ...     & unr     & ...\\ 
2903 & 0.8 & 2.6 (21) &  5    & 0.2 & 4536 & 0.1 & ...      &4$^{10}$& 1.0  &      &     &          &         &    \\   
3034 & 1.1 & 1.8 (17) &23$^6$ & 0.7 & 4565 & 1.5 & ...      & min   & ...  &      &     &          &         &     \\
\noalign{\smallskip}     
\hline
\end{tabular}
} % 
\end{center}
Notes: $^a$ For an explanation of the columns, see Section 4.1. $^b$: For an explanation of the columns, see Section 4.2. $^c$: See Section 5.\\
References: 1. Rampazzo $\etal$ (2006); 2. Yun $\&$ Hibbard (2001);
3. Gerin $\etal$ (1988); 4. Sandquist, Aa, (1999); 6. Seaquist $\&$
Clark (2001); 7. Kuno $\etal$ (2000); 8. Garc\'ia-Burillo $\etal$
(2000); 9. Wilson $\etal$ (2000); Zhu $\etal$ (2003); Schulz $\etal$
(2007); 10. Kuno $\etal$ (2007); 11. Sofue $\etal$ (2003); 12. Cox
$\&$ Downes (1996); 13. Shibatsuka $\etal$ (2003); 14. Ott $\etal$
(2001) 15. Helfer $\etal$ (2003); 16. Regan $\etal$ (1999); 17. Eckart
$\etal$ (1991); 
\end{table*}

\section{CO maps and radial extent}

\subsection{Global CO flux and central fraction}

The literature provides $J$=1-0$\co$ observations at various
resolutions for about 50 sample galaxies that are accessible from the
northern hemisphere, and CO fluxes of entire galaxies have been
published by Stark $\etal$ (1987), Sage (1993), Young $\etal$ (1995),
and Chung $\etal$ (2009). These are summarized in Appendix A, and
examples of the multi-aperture photometry diagrams that can be
constructed from them are shown in Fig.\,\ref{cophot}.  In this
section CO intensities are expressed as line fluxes (Jy km/s) per beam
in order to emphasize their increase as larger areas are covered.

%Figure 8 Exponent Alpha and Fraction in 22" beam
\begin{figure}
  \resizebox{4.4cm}{!}{\rotatebox{0}{\includegraphics*{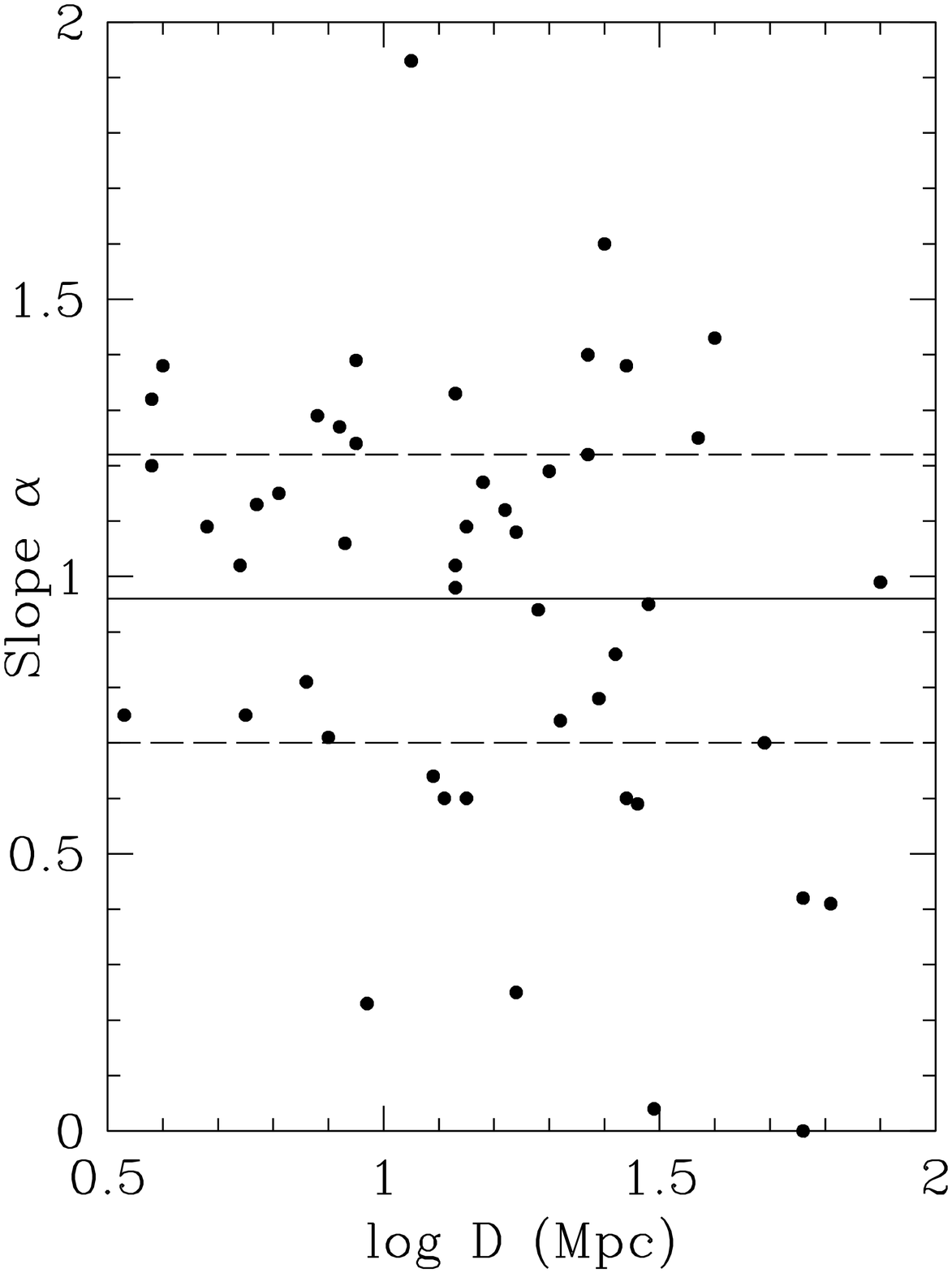}}}
  \resizebox{4.4cm}{!}{\rotatebox{0}{\includegraphics*{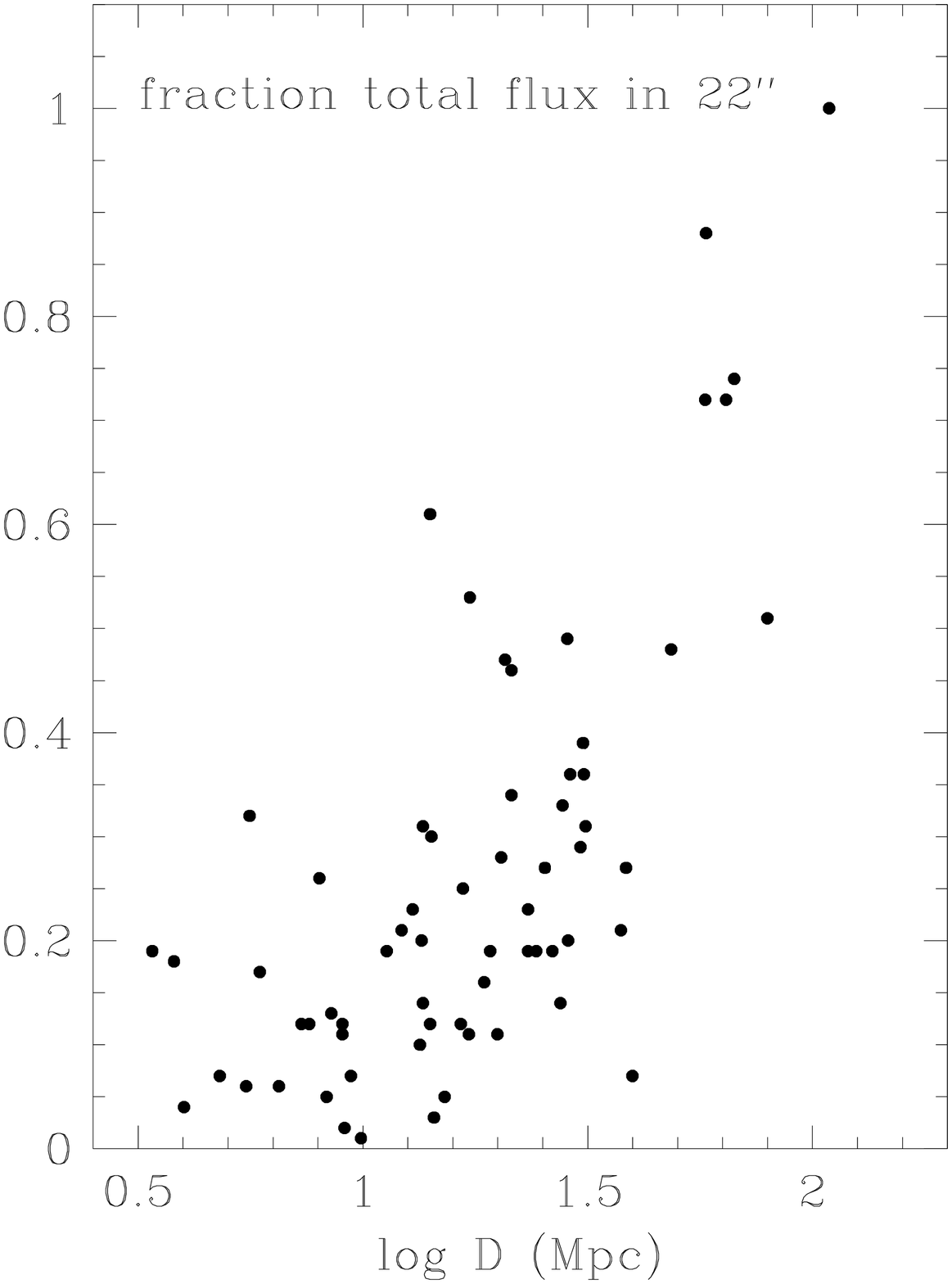}}}
  \caption{Left: Slope $\alpha$ derived from $J$=1-0 $\co$
    multi-aperture photometry as a function of galaxy
    distance. Completely unresolved galaxies have $\alpha$ = 0, and fully
    resolved galaxies have a constant CO surface brightness with
    $\alpha$ = 2. The solid line marks the mean value of the sample,
    the two dashed lines mark half-widths of the distribution. Right:
    Fraction of the total $J$=1-0 CO flux of the sample galaxies
    contained within a beam of FWHM $22"$ as a function of galaxy
    distance.  }
\label{frac22}
\end{figure}

%Figure 9 CO distribution histograms
\begin{figure}
  \resizebox{2.8cm}{!}{\rotatebox{0}{\includegraphics*{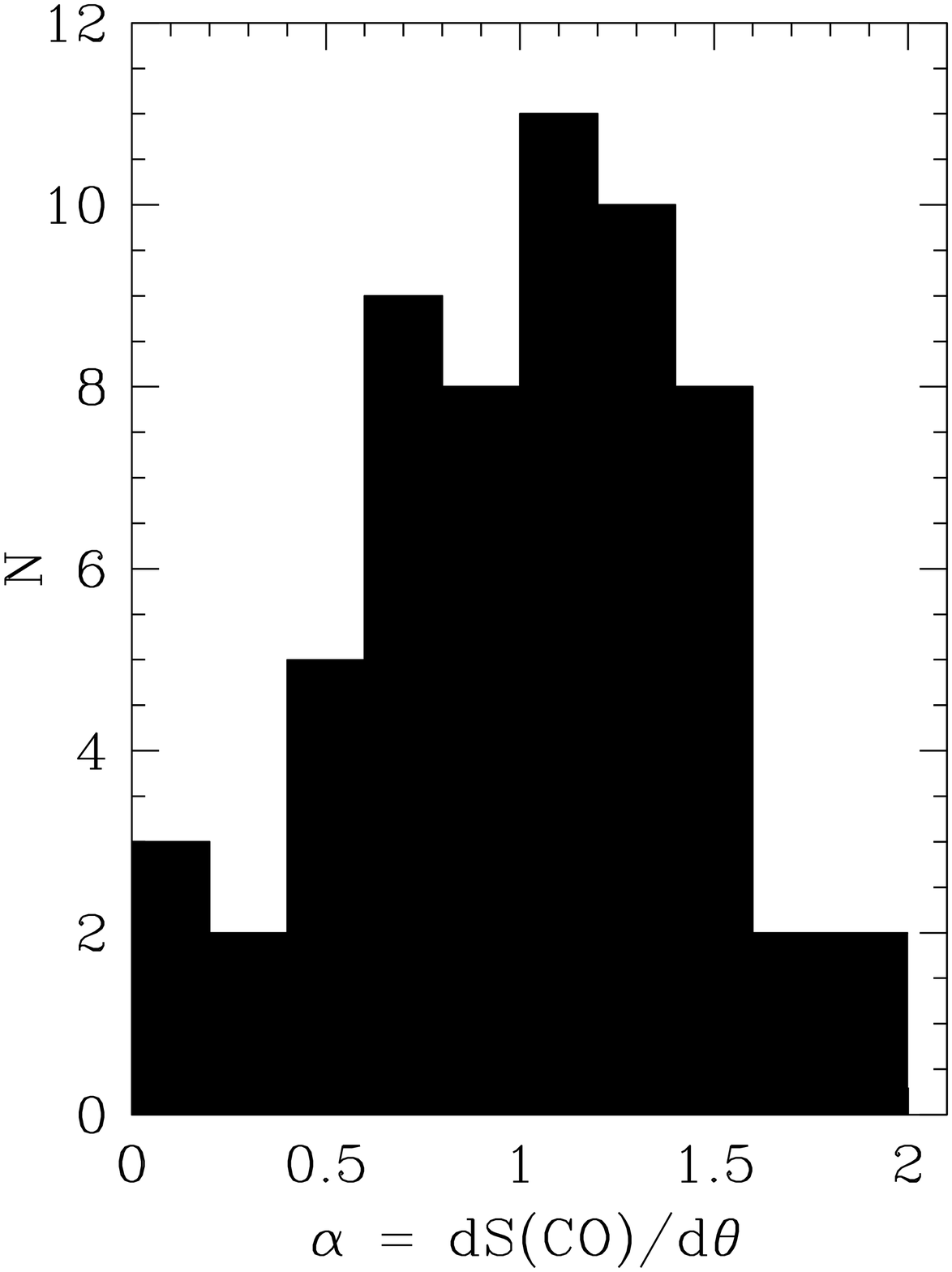}}}
  \resizebox{2.8cm}{!}{\rotatebox{0}{\includegraphics*{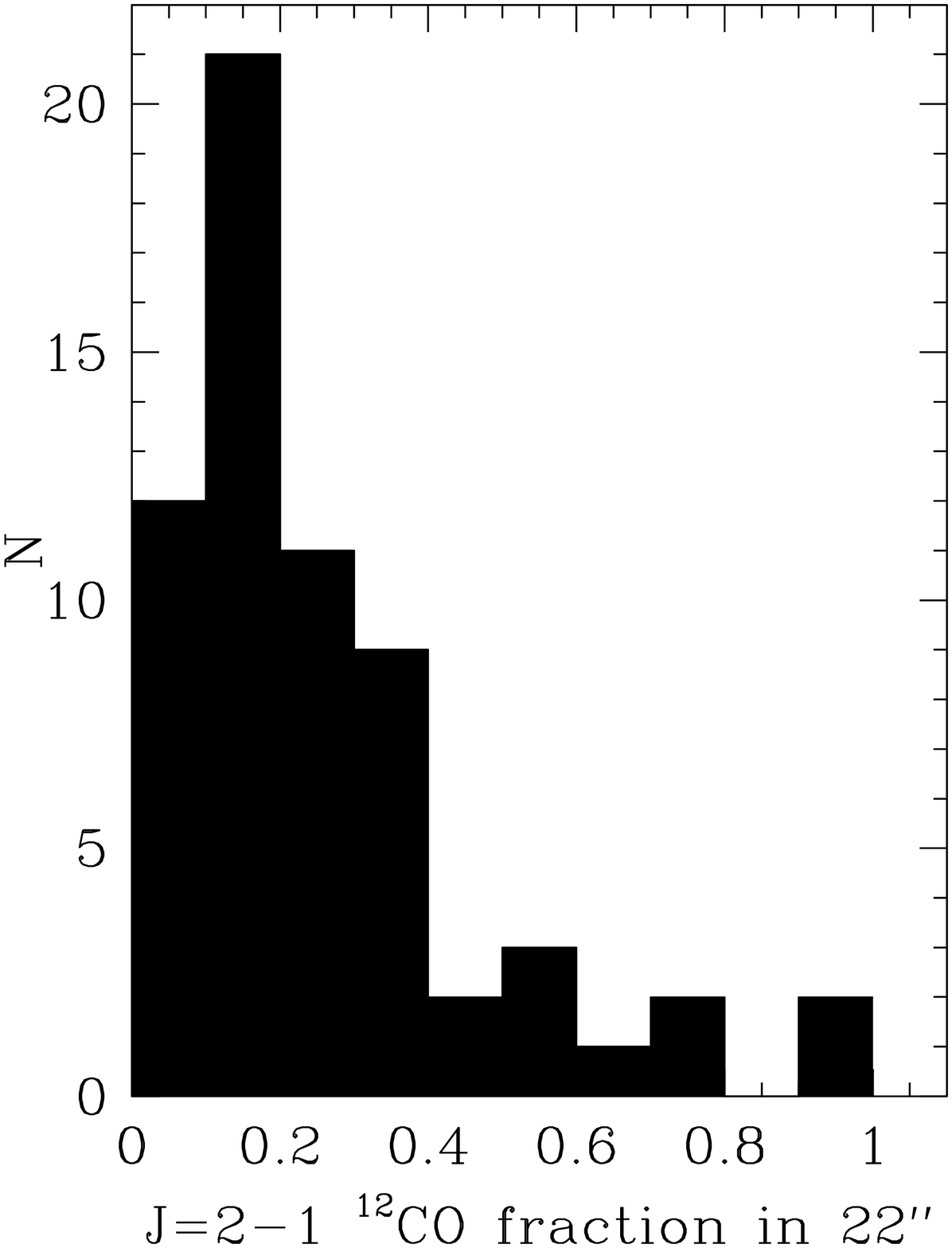}}}
  \resizebox{2.8cm}{!}{\rotatebox{0}{\includegraphics*{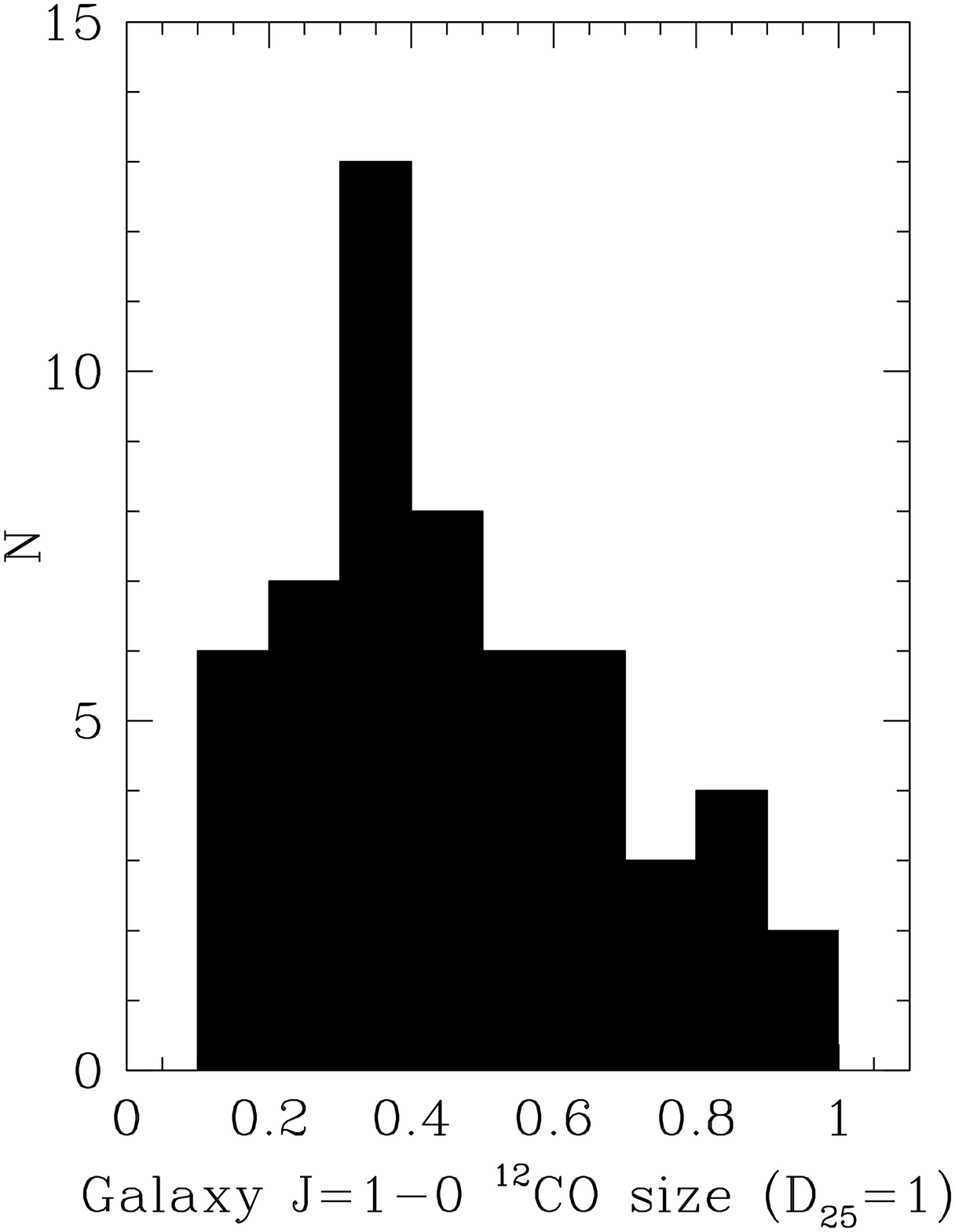}}}
  \caption{Distributions of the sample galaxies as a function of
    (left) slope $\alpha$, marking the change in measured $J$=1-0 CO
    flux as a function of increasing observing beam size, (center) the
    fraction $f_{22}$ of the extrapolated total galaxy CO flux
    detected in a $22"$ beam, and (right) the extrapolated galaxy CO
    size as a fraction of the optical size ($D_{25}$) (see text).  }
\label{co10bin}
\end{figure}

% Figure 10 Linear radii
\begin{figure}
\begin{minipage}[c]{3.9cm}
  \resizebox{3.9cm}{!}{\rotatebox{0}{\includegraphics*{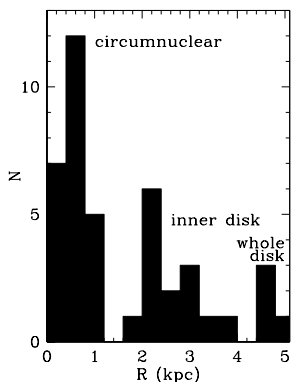}}}
\end{minipage}
\begin{minipage}[h]{3.9cm}
  \bigskip
  \bigskip
\caption{Histogram of the intrinsic (beam-deconvolved) radii
    of the central concentrations in galaxy CO maps. Three
    characteristic radii are distinguished (see text).}
\label{sizebin}
\end{minipage}
\end{figure}

%Figure 11 Sizes
\begin{figure}
\begin{minipage}[]{8.2cm}  
\resizebox{6.3cm}{!}{\rotatebox{0}{\includegraphics*{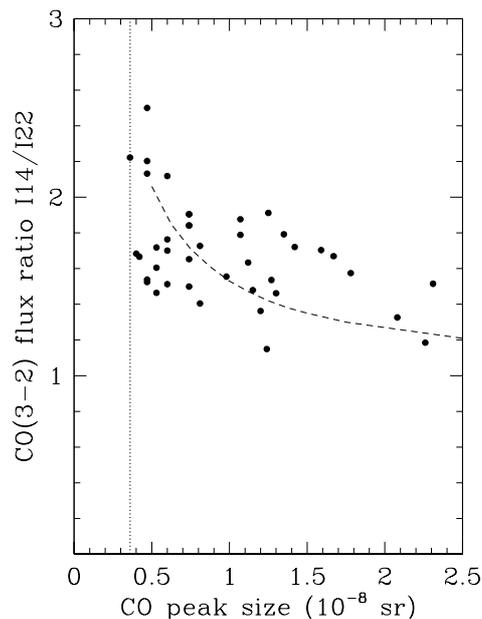}}}
\end{minipage}
\caption{Intensity  ratio of $J$=3-2 $\co$ emission in beams of
  $22"$ and $14"$ as a function of the effective surface area of the
  central CO  concentration taken from Table\,\ref{size}. Very
  extended  emission has a ratio of unity, and fully unresolved
  (point-like) sources  have a ratio of 2.25. The vertical line corresponds
  to the surface  area of a $14"$ beam. The dashed curve indicates the
  relation  expected for circular Gaussian sources without contamination
  by more extended emission.}
\label{sizecomp}
\end{figure}

%Figure 12 Beam ratios
\begin{figure}
\begin{minipage}[]{8.8cm}
  \resizebox{2.8cm}{!}{\rotatebox{0}{\includegraphics*{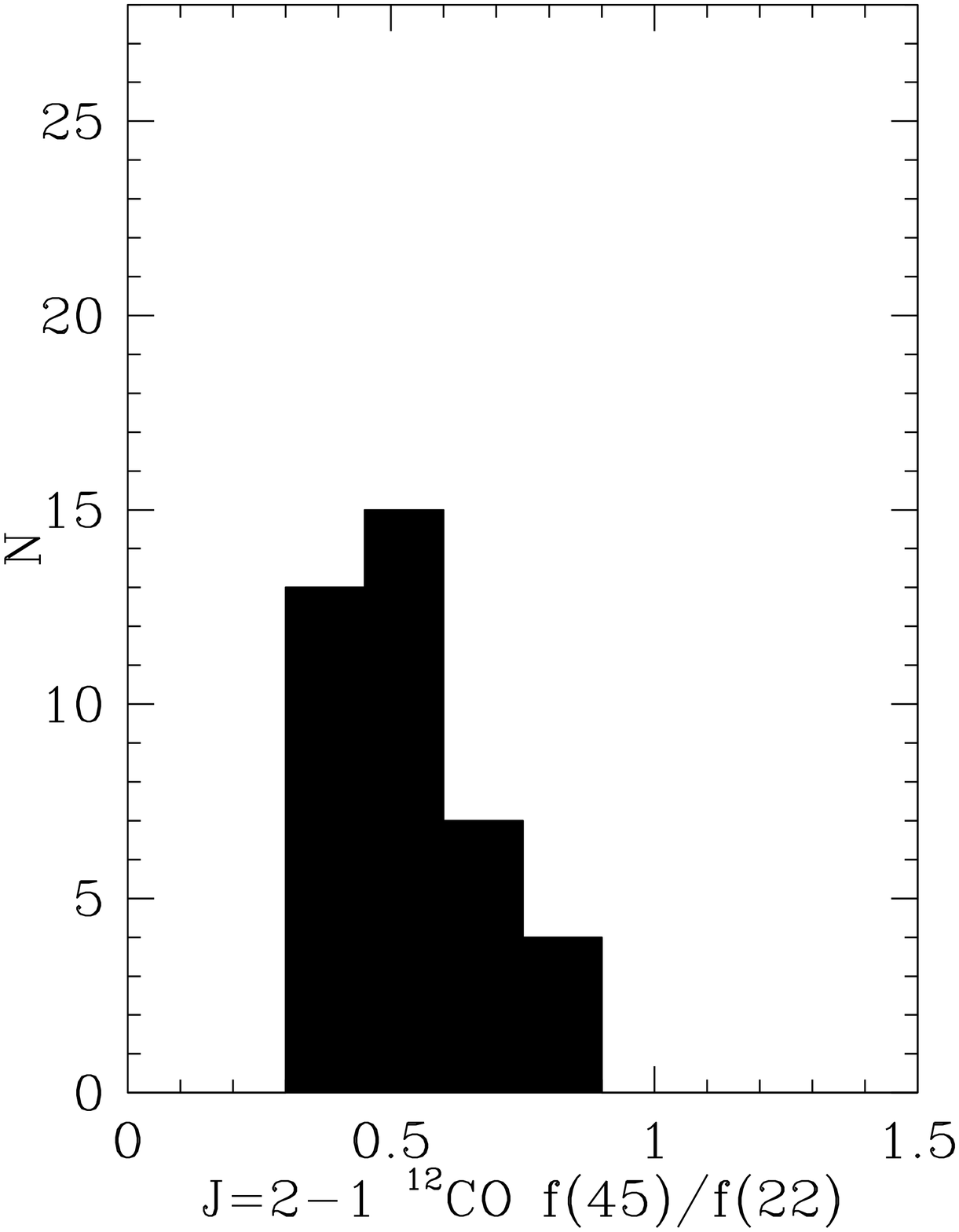}}}
  \resizebox{2.8cm}{!}{\rotatebox{0}{\includegraphics*{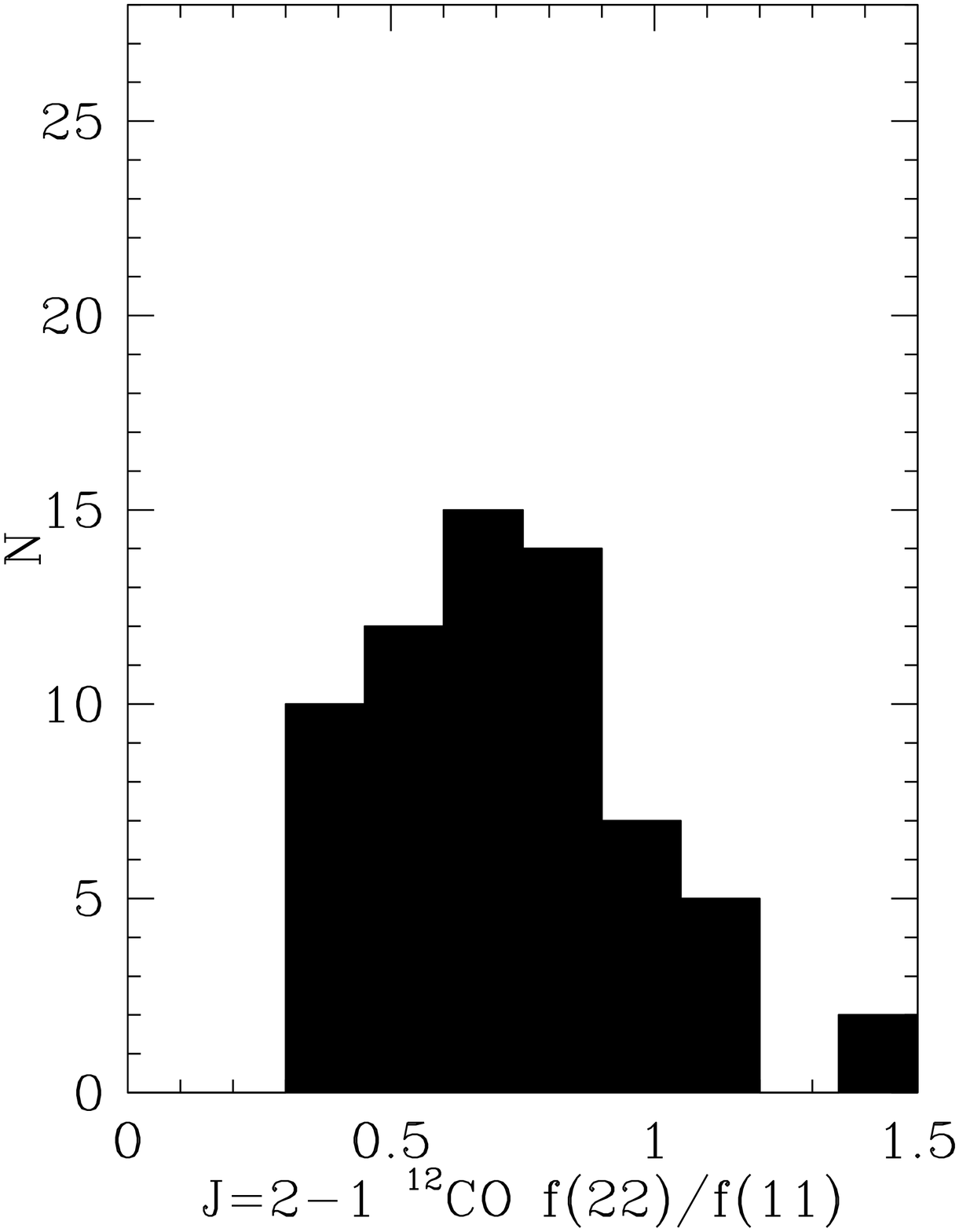}}}
  \resizebox{2.8cm}{!}{\rotatebox{0}{\includegraphics*{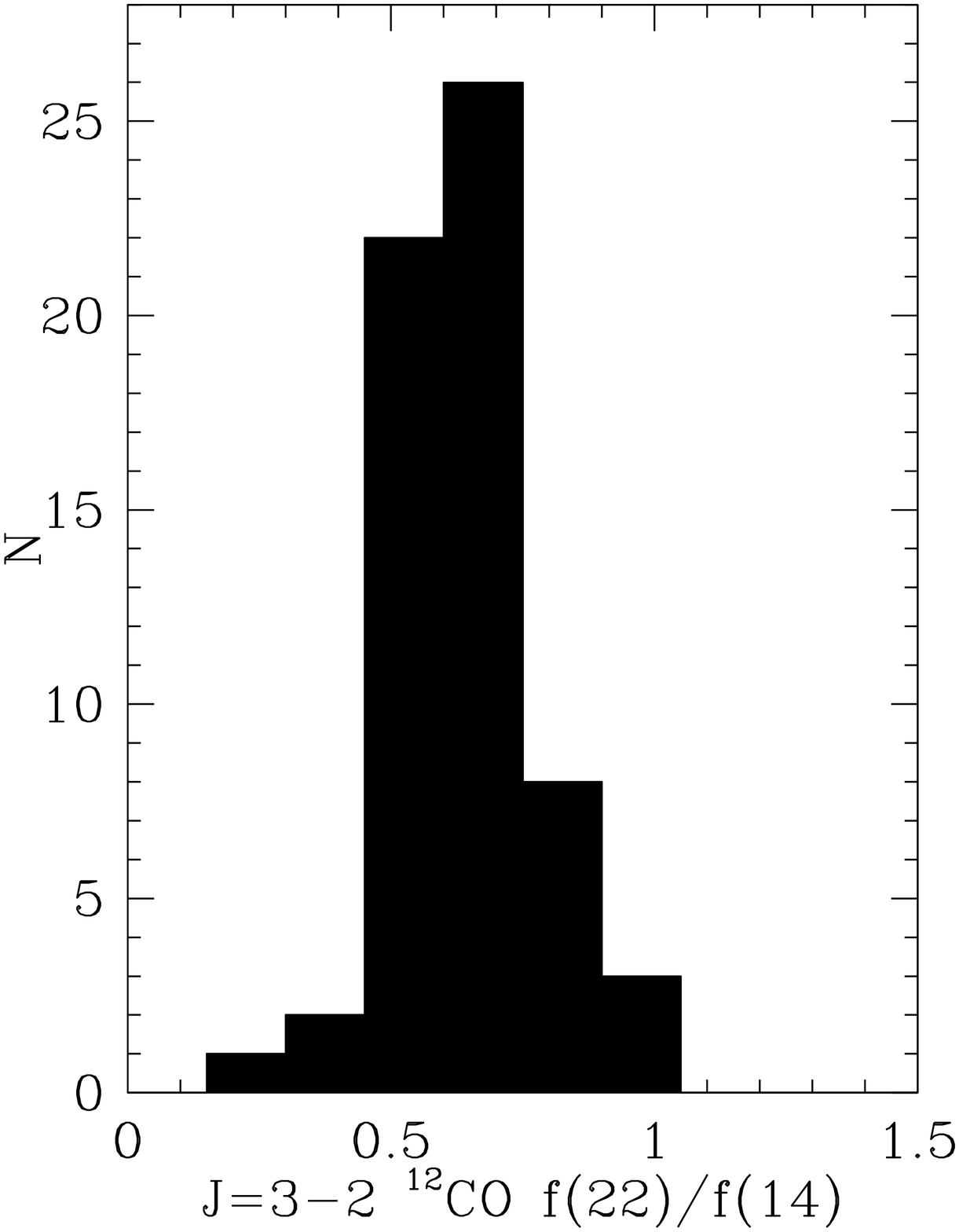}}}
\end{minipage}
\caption{Left: Histogram of $J$=2-1 CO intensity ratios in
  beams of $45"$ and $22"$. Center: Same for $J$=2-1 in $22"$ and $11"$.
  Right: Same for $J$=3-2 CO in $22"$ and $14"$ beams.}
\label{sizehist}
\end{figure}

We determined slopes $\alpha_{CO}$ (defined by
$F\propto\theta^{\,\alpha}$) describing the increase of flux $F$ with
beam-width $\theta$ \footnote{When we express CO intensities in
  temperature ($\kkms$) instead of flux units (Jy $\kms$), we have
  $\alpha'$ = $\alpha$-2.}. In extended sources much larger than the
sampling beams, the measured flux increases with the beam surface area
so that $\alpha$ = 2.  Point-like sources much smaller than the
sampling beams have identical fluxes in all beams so that
$\alpha_{CO}$ = 0. The observed CO emission does not represent either
extreme, as Figs.\,\ref{cophot} and \ref{frac22} illustrate.  The
average slope is close to unity, $\alpha_{CO}$ = $0.96\pm0.06$, with a
standard deviation of 0.42 (see Col. 2 in Table\,\ref{size} and
Fig.\,\ref{co10bin}) and is independent of galaxy distance. Assuming
that this sample is representative, we conclude that CO fluxes of
gas-rich spiral galaxies can be extrapolated from one beam to another
with a modest uncertainty of about $30\%$ by taking the linear beam
width ratio.

The galaxy CO extent ($d_{\rm CO}$) equals the angular size at which
the extrapolated CO flux in Fig.\,\ref{cophot}, for instance, reaches the total
CO flux taken from the literature. From internal consistency, we find
that the average error in the total fluxes is $26\%$ (cf. Appendix A),
which dominates the error in the global size. We list the
extrapolated global sizes in Col. 3 of Table\,\ref{size}, both
as an angular size in arcminutes and as a fraction of the optical
galaxy size $D_{25}$ (taken from Col. 5 in Table 1).  The
distribution in Fig.\,\ref{co10bin} is distinctly peaked at
$0.35\,D_{25}$ but the average value is slightly higher at
$(0.44\pm0.03)\,D_{25}$. As the average $\hi$ disk radius is 1.7
$D_{25}$ (cf. van der Kruit $\&$ Freeman, 2011), the extent of CO-emitting
gas is typically only 25$\%$ that of $\hi$: in late-type
galaxies, the molecular gas is much more concentrated than the atomic
gas.

\subsection{Inner galaxy CO concentrations}

For more than half of the sample galaxies, small maps of the central
CO emission at the relatively high resolution of $14"$ are provided by
JCMT $J$=3-2 $\co$ observations. These include the 24 galaxies shown
in Fig.\,\ref{galmap}, the 16 galaxies published in earlier papers
(Israel $\etal$ 1995, 2006; Israel $\&$ Baas 1999, 2001, 2003; Israel,
2009a, b), and some 20 more by the authors identified in the notes to
Table\,\ref{size}. The resolution of these maps is sufficient to
separate the emission of a central compact source from the extended
disk emission discussed in the previous section.  We determined both
the projected solid angle subtended by the central compact sources and
their radial extent (FWHM along the major axis).  We have at least
partial information for 73 galaxies in Table\,\ref{size}. Ten of
these do not have a central CO peak, but a central CO minimum instead
(e.g., NGC~628 in the upper left corner in Fig.\,\ref{galmap}). In 6
galaxies, the central CO peak is unresolved. Except for NGC~3310, all
are very distant galaxies, at distances of 60 Mpc or more. The
observed central peak solid angle ($\Omega_{\rm CO}$) of 52 galaxies
is listed in Col. 4 of Table\,\ref{size}. We corrected the central
peak FWHM radius ($R_{\rm CO}$) observed in 57 galaxies for finite
resolution by (Gaussian) deconvolution.  Column 5 lists the resulting
angular radii as well as the corresponding linear radii using the
distances from Table\,\ref{sample}. The distribution of the linear
radii is shown in Fig.\,\ref{sizebin}. As also shown in
Fig.\,\ref{galmap}, in most of the sample galaxies, a significant
amount of molecular gas is concentrated within a kiloparsec from the
nucleus (mean radius of 400 pc).  Another group of CO peak radii
$2\leq R_{\rm CO}\leq 4.5$ kpc represents galaxies with more extended
inner disk features such as `rings' (e.g., NGC~1068 and NGC~1097) or
bars (e.g., NGC~1365). All galaxies with a central CO minimum, absent
in the first group, are present in the second group with bright CO
emission.  Sakamoto $\etal$. (1999) obtained a similar result for 20
nearby spiral galaxies, many of which are also included in our
sample. Their average `local' scale length $r_{e}$ = 0.53 kpc and
average `global' scale length $R_{e}$ = 2.6 kpc closely correspond to
the first two peaks in Fig.\,\ref{sizebin}.  The occurrence of compact
circumnuclear molecular gas is probably more frequent than suggested
by Fig.\,\ref{sizebin} because galaxies with distances beyond 15-20
Mpc are imaged with relatively limited linear resolution, making it
hard to separate compact circumnuclear and extended inner disk
emission.

\subsection{CO size and beam-dependent intensity ratio}

In the absence of maps, beam-dependent intensity ratios are sometimes
used to estimate sizes. The map-derived solid angles in
Table\,\ref{size} can be used to determine the reliability of
(effective) source sizes recovered from the ratio of line intensities
in different apertures. In Fig.\,\ref{sizecomp} we show the ratio of
the $\co$(3-2) intensities in $14"$ and $22"$ beams
(Table\,\ref{jcmtdat3}) as a function of the measured solid angle
(Table\,\ref{size}). In each case, both intensities were derived
from the same map data-set. The observed points roughly follow the
dashed line that marks the expected relation for circularly symmetric
isolated compact peaks.  The observational scatter is increased by the
non-circularity of the peaks (points above the dashed line) and by the
presence of extended emission especially in case of barely resolved
peaks (points below the dashed line).

Figure\,\ref{sizecomp} suggests that the central peak diameters estimated
from homogeneous beam intensity ratios have errors of up to $\sim40\%$
that are mostly caused by unknown emission structure. In reality, the errors are
larger because this method is used precisely when no map is available.  In
this case, the combination of heterogeneous data results in additional
scatter.  In Fig.\,\ref{sizehist} we compare $22"$-to-$11"$ beam
intensity ratios from unrelated $J$=2-1 $\co$ JCMT and IRAM
measurements (center panel) with $45"$-to-$22"$ ($J$=2-1, left panel) and
$22$-to-$"/14"$ ($J$=3-2, right panel) beam intensity ratios extracted from
the same JCMT map data-sets. The dispersion of the ratios from
the heterogeneous data in the center panel is twice that of the
homogeneous ratios based on the same map data, which is especially clear for
the $J$=3-2 ratios in the rightmost panel.  The additional errors in
the heterogeneous intensity beam ratios increase the errors in the derived
source size to $70\%$ or more. As long as the actual morphology of the
emission remains unknown, more sophisticated treatments of the problem
(e.g., Yamashita $\etal$ 2017) do not significantly change these
uncertainties. With such uncertainties, the multi-aperture method is
only useful when no great accuracy is required.

\section{CO radiative transfer modeling}

The various transitions in our survey have been measured at different
resolutions, but a meaningful comparison requires intensities at the
same resolution. These are provided by the data measured directly
($J$=1-0, $J$=2-1) or indirectly ($J$=3-2, $J$=4-3) at a resolution of
$22"$.  Table\,\ref{normal} give all ratios in that aperture for all
galaxies with at least two measured line ratios. Galaxies with a
determination of the $\co$-to-$\thirco$ in the $J$=1-0 transition only
are separately listed in Table\,\ref{isotoponly}.  The $\co$ transition
ratios in Cols. 3 through 5 of Table\,\ref{normal} have typical errors
of $30\%$. The isotopologue ratios in Cols. 6 through 8 were
determined by fitting each $\thirco$ to its corresponding $\co$
profile rather than by a division of the $\co$ and $\thirco$
intensities in Tables\,\ref{sestdat} through \ref{jcmtdat3}. By
comparing the two methods, we find that the isotopologue ratios listed
here indeed have typical uncertainties close to those suggested in
Section 2.4.  In section 3.5 we noted that the $\co$-to-$\thirco$
isotopologue ratio is effectively independent of aperture in the
$J$=1-0 and $J$=2-1 transitions (cf. Appendix B), and we have assumed
that this is also true for the $J$=3-2 ratios measured in $14"$
apertures.  Complementary values taken from the literature are
identified by footnotes.

From the previous section, we determined that this normalized beam
covers between 3$\%$ and 11$\%$ of the total CO surface area of the
sample galaxies. When we restrict the sample to galaxies with
distances between 10 Mpc and 40 Mpc, we obtain the same result. The
fraction $f_{22}$ of all CO flux contained in an aperture of $22"$
(Col. 2 of Table\,\ref{normal}) is much higher, on average $26\%$. As
expected, Fig.\,\ref{frac22} shows that the individual values increase
with increasing distance $D$. The distribution of individual $f_{22}$
values is also shown in Fig.\,\ref{co10bin}.

We have modeled the data in Table\,\ref{normal} with the statistical
equilibrium radiative transfer code {\it RADEX} (Van der Tak $\etal$
2007).  It provides model line intensities as a function of three
input parameters per molecular gas phase: gas kinetic temperature
$T_{\rm k}$, molecular hydrogen density $n_{\h2}$ , and the CO column
density per unit velocity $N({\rm CO})$/d$V$. Each combination of
physical parameters uniquely determines a set of line intensities and
ratios. The opposite is not true because the same line ratio may
result from different combinations of physical input
parameters. Reverse tracing is therefore not a unique
process. Nevertheless, by comparing for each galaxy as many observed
line ratios as possible to extensive grids of precalculated model line
ratios, we may constrain and identify the physical parameters that
best describe the actual conditions.

The large linear beam sizes that apply to galaxy center observations
encompass molecular gas clouds at distinctly different temperatures
and densities, which require more than one model gas phase to produce
acceptable fits to the observations (see, e.g., Israel $\&$
Baas, 1999; Papadopoulos $\&$ Seaquist, 1999; Israel, 2009a, b). Good
model fits are easily obtained for data sets containing only $\co$
observations, but the high degree of degeneracy between $\h2$
temperature and density renders such excellent fits non-unique and not
very useful.  Not even long $\co$ ladders (such as those extending up
to $J$=13-12 obtained with {\it Herschel}-SPIRE) provide significant
constraints (e.g., see Meijerink $\etal$, 2013). Fortunately,
the degeneracy can be broken by measuring lines with low optical depth such
as $\thirco$ in addition to the mostly optically thick $\co$ lines at
the cost, however, of more physical parameters to be determined. Such
combinations of $\co$ with related species yield constraints that
although still not unique, are much tighter than those based on $\co$
alone.

% Table 7 Normalized ratios
\begin{table*}
\caption[]{\label{normal}Line intensity ratios normalized to $22"$ aperture$^a$} 
\begin{center}
{\small % 
\begin{tabular}{l|r|rrrrrr||l|r|rrrrrr} 
\noalign{\smallskip}     
\hline
  \noalign{\smallskip}
NGC &f$_{22}$&\multicolumn{3}{c}{Transition ratio}&\multicolumn{3}{c}{Isotopologue ratio}& NGC &f$_{22}$&\multicolumn{3}{c}{Transition ratio}&\multicolumn{3}{c}{Isotopologue ratio}\\
IC  &       &\multicolumn{3}{c}{$\co$(2-1) =1}   &\multicolumn{3}{c}{$\co/\thirco$}     & IC  &   &\multicolumn{3}{c}{$\co$(2-1) = 1}&\multicolumn{3}{c}{ $\co/\thirco$}\\
     &    &  1-0  & 3-2  & 4-3  & 1-0   & 2-1 & 3-2   &      &    &  1-0  & 3-2  & 4-3  & 1-0   & 2-1  & 3-2 \\
(1)  & (2)& (3)   & (4)  &  (5) &  (6)  & (7) & (8)   & (1)  & (2) & (3)  &  (4) &  (5) & (6)   & (7)  & (8) \\
\noalign{\smallskip}    
\hline                          
\noalign{\smallskip}    
 253 & 19 & 0.76  & 0.63 & 0.68 & 12.7  & 10.7 & 11.7 & 4030 & 19 & 1.20  & 0.5  & ...  &   6.6 &  9.9 & 10.1 \\ 
 278 & 19 & 1.07  & 0.71 & 0.43 &  9.0  &  8.1 &  8.0 & 4038 & 19 & 0.66  & 0.75 & ...  &  12.8 & 15.8 & 13.2 \\ 
 470 & .. & 1.22  & 0.79 & ...  & 15.0  & 15.3 & ...  & 4039 & 23 & 1.31  & 0.69 & ...  &  22.0 & 12.9 & 14.6 \\ 
 520 & 29 & 1.19  & 0.35 & ...  & 14.2  & 16.8 &  9.5 & 4051 & 23 & 1.71  & 1.35 & ...  &  18   &  21  & ...  \\ 
 613 & .. & 0.94  & 0.8  & ...  &  11.5 & 14.6 & 11.3 & 4102 & 53 & 0.83  & 0.5  & ...  &  12.5 & 12.6 & 12.7 \\ 
 628 & 19 & 1.2   & 0.6  & ...  &   6.3 &  9.7 & ...  & 4254 &  7 & 1.08  & 0.6  & ...  &   8.9 &  9.4 &  5.0 \\ 
 660 & 21 & 1.11  & 0.68 & 0.66 &  14.0 & 17.0 & 12.4 & 4258 & 26 & 1.04  &0.42  & ...  &15$^{1,2}$& ...& ...  \\ 
 891 &  7 & 2.2   & 0.32 & ...  &   7.8 & 10.2 & 10.9 & 4293 & 61 & 1.34  & 0.8  & ...  &  11.9 &  6.5 & 14.9 \\ 
 908 & 11 & 1.62  & 0.4  & ...  &   8.7 &  9.9 &  6   & 4303 & 14 & 1.29  & 0.54 & ...  &  18.1 & 12.2 & 16.4 \\
 972 & .. &.0.95  & 0.55 & ...  &  11.6 & 12.8 & 15.7 & 4321 & 12 & 1.47  & 0.63 & ...  &  10.0 & 10.7 & 10.8 \\ 
Maf2 & .. &.0.89  & 0.73 & 0.63 &   8.0 &  8.5 & 13.0 & 4414 & 12 & 1.37  & 0.51 & ...  &   7.5 &  8.5 &  7   \\ 
1055 & 10 & 1.42  & 0.4  & ...  &   7.3 &  8.9 & 11.4 & 4457 & 31 & 1.09  & 0.69 & ...  &  15   & 20   & 24   \\ 
1068 &  5 & 0.70  & 0.42 & 0.65 &  11.8 & 12.8 & 15.2 & 4527 & 20 & 0.85  & 0.4  & ...  & 13.3  & 17.6 & 10   \\ 
1084 & 16 & 0.98  & 0.5  & ...  &  13.2 &  8.8 & 10.0 & 4536 & 39 & 0.98  & 1.0  & ...  & 18.8  & 11.0 & 16.1 \\ 
1097 & 12 & 1.14  & 0.90 & ...  &  10.5 & 18.8 & ...  & 4631 & 12 & 1.28  & 0.67 & ...  &  15.1 & 15.3 & 6.9  \\ 
1365 & .. & 1.05  & 0.69 & ...  &  11.1 & 11.5 & 12.2 & 4666 & 14 & 1.39  & 0.68 & 0.29 &   9.7 & 10.3 & 14.5 \\ 
I342 &  3 & 0.93  & 0.70 & 0.64 &  10.2 &  7.4 & 10.8 & 4736 &  7 & 0.98  & 0.63 & ...  &   9.9 & 10.1 & 14.9 \\ 
1614 & 27 & 1.34  & 2.21 & ...  &  30.0 & ...  & ...  & 4826 & 18 & 0.89  & 0.49 & 0.65 &   8.3 &  7.2 &  8.4 \\ 
1792 & 72 & 0.94  & 0.5  & ...  &   7.0 &  7.5 &  8.0 & 4945$^3$  & 0.8   & 0.65 & ...  & 15.7  & 13.2 &  9.9 \\ 
1808 & .. & 0.95  & 1.03 & ...  &  16.5 & 12.6 & 17.1 & 5033 & 11 & 1.25  & 0.57 & ...  &   9.1 &  7.2 & 11.7 \\  
2146 & .. & 1.15  & 0.89 & ...  &  15.0 &  8.7 & 13.6 & 5055 &  5 & 1.28  & 0.5  & ...  &   7.5 &  8.3 &  8.4 \\
2273 & 25 & 1.01  & 0.65 & ...  &  10.5 & 12.2 & ...  & 5135 & .. & 1.64  & 1.33 & ...  & 23    & 11   & 22   \\ 
2415 & 49 & ...   & ...  & ...  &   ... & 11.2 & 20.0 & 5194 &  2 & 0.89  & 0.59 & 0.46 &   6.7 &  9.1 &  8.5 \\ 
2559 & 34 & 0.57  & ...  & ...  &   9.9 & 11.5 & 22.4 & 5236 &  4 & 0.90  & 0.59 & 0.44 &  13.6 &  8.6 & 10.0 \\  
2623 & 51 & 0.70  & 0.73 & ...  &   7   &  5   & ...  & Circ$^4$  & 1.2   & 0.5  & 0.3  & 16    & 10   & ...  \\
2903 & 12 & 1.34  & 0.91 & ...  &  11.2 &  8.6 & 12.5 & 5713 & 30 & 0.80  & 0.49 & ...  &  14.9 & 11.5 & 16.3 \\ 
3034 & 17 & 1.04  & 0.83 & 0.60 &  18.4 & 10.8 &  9.0 & 5775 & 36 & 1.39  & 0.58 & ...  &   9.1 &  9.4 & ...  \\ 
3044 & .. & 1.12  & 0.50 & ...  &  13.5 &  8.8 & ...  & 6000 & 36 & 0.98  & 0.7  & ...  &  13.7 & 10.1 & ...  \\ 
3079 & 47 & 1.18  & 0.5  & 0.70 &  15.8 & 14.6 & 7.9  & 6240 &100 & 1.00  & 1.14 & ...  &  29   & 40   & 26.5 \\ 
3175 & ...& 1.25  & 0.67 & ...  &  10.6 & 10.8 & 14.3 & 6764 & .. & 1.33  & 0.93 & ...  &  19   & 24   & ...  \\ 
3227 & 28 & 0.60  & 0.89 & ...  &  17.8 & 13.5 & 16.9 & 6946 &  6 & 0.95  & 0.49 & 0.46 &  13.7 & 14.3 & 10.6 \\ 
3310 & 19.& 0.90  & 1.38 & ...  &  12.4 & 12.0 & 15.2 & 6951 & 19 & 1.27  & 1.20 & ...  &  10.5 &  7.4 & 15.2 \\ 
3504 & 33 & 1.08  & 0.67 & ...  &  13.1 & 11.3 & 13.4 & 7331 &  3 & 2.4   & 0.46 & ...  &6$^{2,5}$&6.2&  5.7 \\
3593 & 32 & 1.51  & 0.4  & ...  &  12.4 &  9.1 & 13.8 & 7469 & 74 & 1.06  & 0.79 & 0.90 &  17   & 17   & 23   \\ 
3627 &  6 & 1.00  & 0.66 & ...  &  13.3 & 11.8 & 12.1 & 7541 & 21 & 0.47  & 0.4  & ...  &   8.2 & 11.5 & 10.3 \\ 
3628 & 13 & 1.25  & 0.70 & 0.55 &  12.0 & 11.9 &  9.1 & 7552 & ...& ...   & ...  & ...  &  10.8 & 9.1$^6$& ...\\ 
3690 & 50 & 1.07  & 0.62 & 0.73 &  23.2 & 20.0 & ...  & 7714 & 27 & 0.8   & 0.5  & ...  &   5.5 &  9.4 & ...  \\ 

  \noalign{\smallskip}                             
\hline                                           
\end{tabular}                                    
} %                                              
\end{center}
Note: $^a$: See Section 5.\\
References:  1. Cox $\&$ Downes (1996); 2. Krips $\etal$ (2010); 3. Dahlem $\etal$ (1993);
4. Hitschfeld $\etal$ (2008); 5. Vila-Vilaro $\etal$ (2015); 6. Aalto $\etal$ (1995).
\end{table*}

%Table 8
\begin{table}
\caption[]{\label{isotoponly}Galaxies with $J$=1-0 isotopologue ratio only} 
\begin{center}
{\small % 
\begin{tabular}{l|c||l|c||l|c} 
\noalign{\smallskip}     
\hline
  \noalign{\smallskip}
%NGC& $\frac{\co}{\thirco}$&NGC& $\frac{\co}{\thirco}$&NGC& $\frac{\co}{\thirco}$\\
NGC& $I_{\co}/I_{\thirco}$ &NGC& $I_{\co}/I_{\thirco}$ &NGC& $I_{\co}/I_{\thirco}$\\
(1)  & (2)  &(3)   &  (4) &  (5) & (6)  \\
\noalign{\smallskip}    
\hline                          
\noalign{\smallskip}    
1433 &  7.0 & 2369 & 14.9 & 4444 &  8.0 \\  
1448 & 13.0 & 2397 & 11.9 & 6221 & 12.0 \\
1482 & 13.9 & 3256 & 25   & 6300 & 20   \\
1559 &  5.8 & 3556 & 12.5 & 7552 & 10.8 \\
1566 & 16.0 & 3620 & 14.0 & 7590 & 12.0 \\
1672 & 10.6 & 3621 & 16.0 & 7771 & 13.9 \\
\noalign{\smallskip}     
\hline
\end{tabular}
} % 
\end{center}
\end{table}

We have modeled our data under the assumption that the emission is
dominated by two distinct model gas phases. This is an important and
necessary improvement over models assuming homogeneous single-phase
gas that tend to provide a poor fit to the data when overconstrained.
With only two gas phases, however, an ambiguity in temperature and
density remains. It would be more realistic to model the gas with a
smoothly changing temperature and density over a range of phases. This
is, however, unfeasible even with the present relatively extensive
data set because including more gas-phase components rapidly increases
the number of unconstrained free parameters, which renders the result
less rather than more realistic.

To simplify matters, we assumed that both gas phases have the
same fixed isotopologue abundance, and we considered abundances of 40 and
80, respectively.  In nearby galaxies such as NGC~253, NGC~4945, M~82,
IC~342, and NGC~1068 the lower value seems appropriate (Henkel et
al. 1994, 1998; Bayet $\etal$ 2004; Henkel $\etal$ 2014; Giannetti
$\etal$ 2014; Tang $\etal$ 2019). The higher value may be more
appropriate to (some of) the more distant very luminous galaxies such
as NGC 5135, NGC~6240, NGC~7469, and Mrk~231 (e.g., Henkel $\etal$
2014; Sliwa $\etal$ 2014; Tunnard $\etal$ 2015; see also Israel
$\etal$ 2015).
 
For any particular set of line ratios, the {\it RADEX} model-fit line
intensity, column density gradient, spatial density, and temperature in
the two phases do not vary independently. The beam-averaged CO
column density is sensitive only to the combined effect of these
variations, and its resulting dispersion of about 30$\%$ is much
lower than the uncertainty in each of the individual constituent model
parameters, as illustrated in Table\,\ref{examplefit}.

The fraction of gas-phase carbon contained in CO is a function of the
actual total carbon column densities $N_{\rm C}$. We determined for
each phase the fractional CO abundance [CO]/[C] as well as the total
beam-averaged carbon column densities N$_{\rm C}$ using the chemical
models presented by van Dishoeck $\&$ Black (1988) and updated by
Visser $\etal$ (2009). The detailed results of the two-phase modeling
are given in Table\,\ref{modelpar}, where we present for each galaxy
the model solution closest to the observations, regardless of the
other possible model ratios within the observational error.

These results were combined to derive the beam-averaged fractional CO
abundance [CO]/[C] and the beam-averaged total carbon column density
$N_{\rm C}$ (Cols. 2 and 3) in Table\,\ref{galmassx} for both of the
assumed isotopologue abunandances. The beam-averaged CO column density
$N_{\rm C}$ is the given by the producte:
$N_{\rm CO}\,=\,N_{\rm C}\,\times\, \frac{\rm [CO]}{\rm [C]}$.  Out of
72 galaxies, 64 ($90\%$) are successfully modeled with a
[$\co$]/[$\thirco$] abundance of 40, and 28 galaxies ($39\%$) even
require this abundance for successful modeling. Only 8 galaxies
($11\%$) need to be modeled with a high isotopological ratio of 80
instead. Half of the galaxy sample can be modeled with either ratio,
but in most cases, the lower ratio of 40 provides better fits. Four
galaxies (NGC 1614, NGC 4293, NGC 4527, and NGC 5236) have poor fits
at either abundance.

%Table 9 Physical parameters
\begin{table*}
\caption[]{\label{physpar}Physical properties of $22"$ central regions} 
\begin{center}
{\small % 
\begin{tabular}{l|rrrrr||l|rrrrr||l|rrrrr} 
\noalign{\smallskip}     
\hline
  \noalign{\smallskip}
  NGC  & \multicolumn{2}{c}{Carbon}&\multicolumn{2}{c}{Hydrogen}    &     &
  NGC  & \multicolumn{2}{c}{Carbon}&\multicolumn{2}{c}{Hydrogen}    &     &
  NGC  & \multicolumn{2}{c}{Carbon}&\multicolumn{2}{c}{Hydrogen}    &     \\
  IC   & $\frac{\rm [CO]}{\rm [C]}$&$N_{\rm C}$&$N_{\h2}$&$M_{\rm gas}$ & $X$ &
  IC   & $\frac{\rm [CO]}{\rm [C]}$&$N_{\rm C}$&$N_{\h2}$&$M_{\rm gas}$ & $X$ &
  IC   & $\frac{\rm [CO]}{\rm [C]}$&$N_{\rm C}$&$N_{\h2}$&$M_{\rm gas}$ & $X$\\
       &                           &e17       &  e21   & e7        &     &
       &                           &e17       &  e21   & e7        &     &
       &                           &e17       &  e21   & e7        & \\
       &$\%$                       &\multicolumn{2}{c}{$\cm2$}   &M$_{\rm \odot}$ &$X_{\rm \circ}$&
       &$\%$                       &\multicolumn{2}{c}{$\cm2$}   &M$_{\rm \odot}$ &$X_{\rm \circ}$&
       &$\%$                       &\multicolumn{2}{c}{$\cm2$}   &M$_{\rm \odot}$ &$X_{\rm \circ}$\\
(1)  & (2)  &(3)&  (4)&  (5)  & (6) & (1)  & (2)  &(3)&  (4)&  (5)  & (6) &  (1)  & (2)  &(3)&  (4)&  (5)  & (6)    \\
\noalign{\smallskip}    
\hline                          
\noalign{\smallskip}    
 253 & 30 & 210 &  21 &  8 & 0.10 & 2903 & 36 &  15 & 2.2 &  3 & 0.07 & 4457 & 59 &   9 & 1.1 & ...& 0.10\\   
 470 & 40 &   5 & 0.3 & 12 & 0.04 & 3034 & 39 &  83 & 8.1 & 14 & 0.06 & 4527 & 87 &  50 & ... & 18 & 0.23\\
 520 & 15 &  20 & 1.8 & 42 & 0.08 & 3044 & 09 &   3 & 0.2 & ...& 0.06 & 4536 & 28 &  10 & 0.6 & 18 & 0.05\\
 613 & 45 &  21 & 1.8 & ...& 0.13 & 3079 & 30 &  42 & 4.0 & 35 & 0.08 & 4631 & 22 &   4 & 0.2 & ...& 0.02\\
 628 & 60 &   8 & 0.7 & ...& 0.50 & 3175 & 34 &  12 & 1.0 &  4 & 0.11 & 4666 & 28 &  15 & 1.4 & ...& 0.09\\
 660 & 44 &  55 & 5.4 & 22 & 0.17 & 3227 & 28 &   5 & 0.3 &  4 & 0.02 & 4736 & 32 &   9 & 0.6 & 0.4& 0.07\\
 891 & 14 &  36 & 3.6 & 12 & 0.12 & 3310 & 34 &   2 & 0.2 &  2 & 0.10 & 4826 & 47 &  14 & 1.2 & 0.4& 0.07\\
 908 & 42 &  12 & 1.0 & ...& 0.17 & 3504 & 39 &   9 & 0.8 & 14 & 0.08 & 4945 & 35 & 125 &  15 & ...& 0.11\\
 972 & 30 &  13 & 0.8 & 23 & 0.06 & 3593 & 17 &  19 & 2.2 &  2 & 0.14 & 5033 & 21 &   9 & 0.6 &  4 & 0.06\\
Maf2 & 48 &  34 & 3.2 &  2 & 0.07 & 3627 & 48 &  10 & 1.1 &  1 & 0.08 & 5055 & 24 &  21 & 1.9 &  3 & 0.14\\
1055 & 23 &  26 & 1.7 & ...& 0.11 & 3628 & 21 &  80 & 7.7 & 11 & 0.19 & 5135 & 20 &   9 & 0.7 & 58 & 0.09\\
1068 & 34 &  36 & 3.5 & 40 & 0.10 & 3690 & 22 &  22 & 1.9 & ...& 0.15 & 5194 & 56 &  24 & 2.4 & ...& 0.25\\
1084 & 35 &   6 & 0.4 & ...& 0.06 & 4030 & 20 &  12 & 1.0 & ...& 0.12 & 5236 & 19 &  31 & 2.9 & 0.9& 0.08\\
1097 & 51 &  47 & 4.2 & 57 & 0.16 & 4038 & 42 &  15 & 1.1 & 15 & 0.11 & 5713 & 23 &  12 & 1.0 & 0.5& 0.10\\
1365 & 43 &  60 & 6.0 & 98 & 0.12 & 4039 & 14 &   8 & 0.6 &  9 & 0.06 & 5775 & 20 &   9 & 0.7 & 15 & 0.07\\
I342 & 37 &  21 & 2.0 &  2 & 0.06 & 4051 & 23 &  17 & 1.6 &  5 & 0.35 & 6000 & 23 &  15 & 1.2 & 26 & 0.08\\
1614 & 11 &  11 & 0.7 & 82 & 0.09 & 4102 & 30 &  15 & 1.3 & 11 & 0.09 & 6240 & 09 &   7 & 0.5 & ...& 0.04\\
1792 & 51 &   7 & 0.4 & ...& 0.08 & 4254 & 51 &   7 & 0.3 & 23 & 0.04 & 6764 & 21 &  11 & 0.9 & ...& 0.20\\
1808 & 32 &  16 & 1.4 &  7 & 0.05 & 4258 & 22 &   9 & 0.7 &  2 & 0.08 & 6946 & 38 &  43 & 3.9 &  2 & 0.08\\
2146 & 29 &  31 & 2.7 & 27 & 0.07 & 4293 & 24 &   4 & 0.4 & ...& 0.05 & 6951 & 22 &  15 & 1.4 &  2 & 0.24\\
2273 & 37 &   4 & 0.2 &  7 & 0.05 & 4303 & 18 &  10 & 0.8 &  4 & 0.08 & 7331 & 61 &  16 & 1.4 & ...& 0.45\\
2559 & 16 &  12 & 1.0 & 14 & 0.07 & 4321 & 29 &  34 & 3.3 & 14 & 0.20 & 7469 & 49 &   6 & 0.3 & ...& 0.03\\
2623 & 38 &   4 & 0.2 & 50 & 0.06 & 4414 & 23 &  14 & 1.2 & ...& 0.12 & 7541 & 83 &  19 & 1.5 & ...& 0.27\\
\noalign{\smallskip}     
\hline
\end{tabular}
} % 
\end{center}
Notes: $N_{\rm H}$ = 2000 $N_{\rm C}$ for the adopted gas-phase carbon abundance (Appendix C). $M_{\rm gas}$
= 1.35 $M_{\rm H}$ allowing for the presence of helium.
\end{table*}

\section{Gas-phase carbon budget}

\subsection{Carbon monoxide fraction}

\begin{figure*}
\begin{minipage}[c]{13.5cm}
  \resizebox{4.45cm}{!}{\rotatebox{0}{\includegraphics*{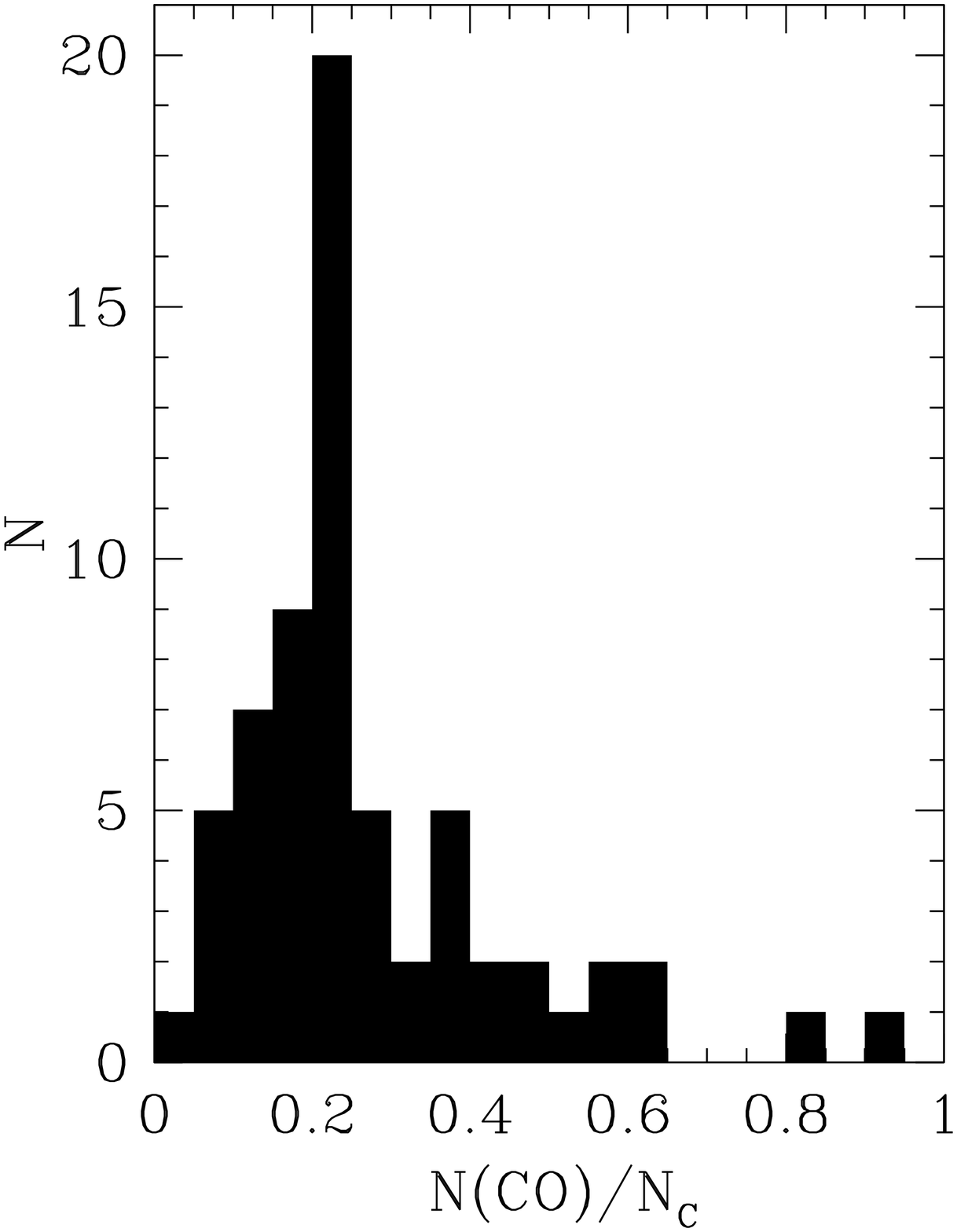}}}
  \resizebox{4.45cm}{!}{\rotatebox{0}{\includegraphics*{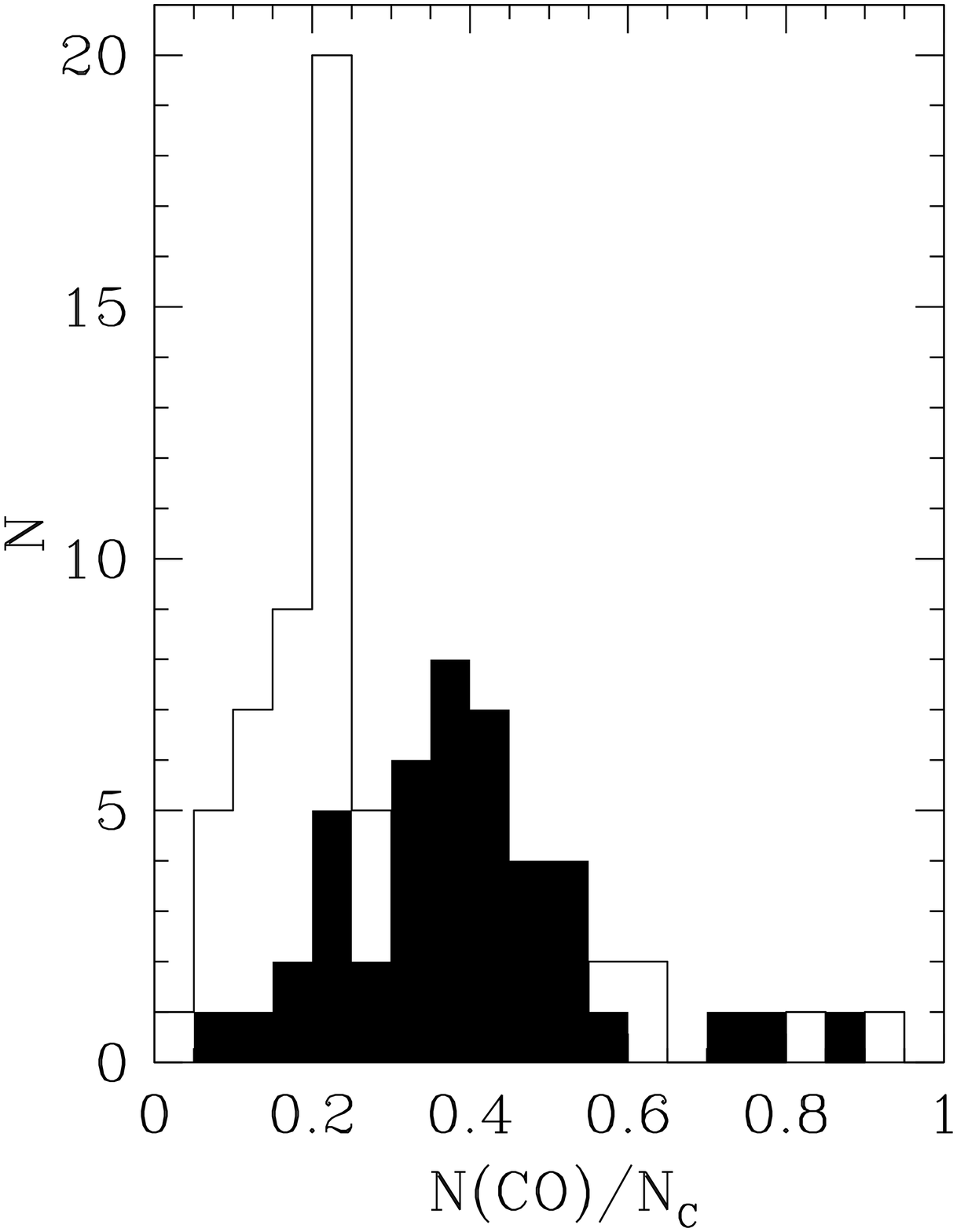}}}
  \resizebox{4.45cm}{!}{\rotatebox{0}{\includegraphics*{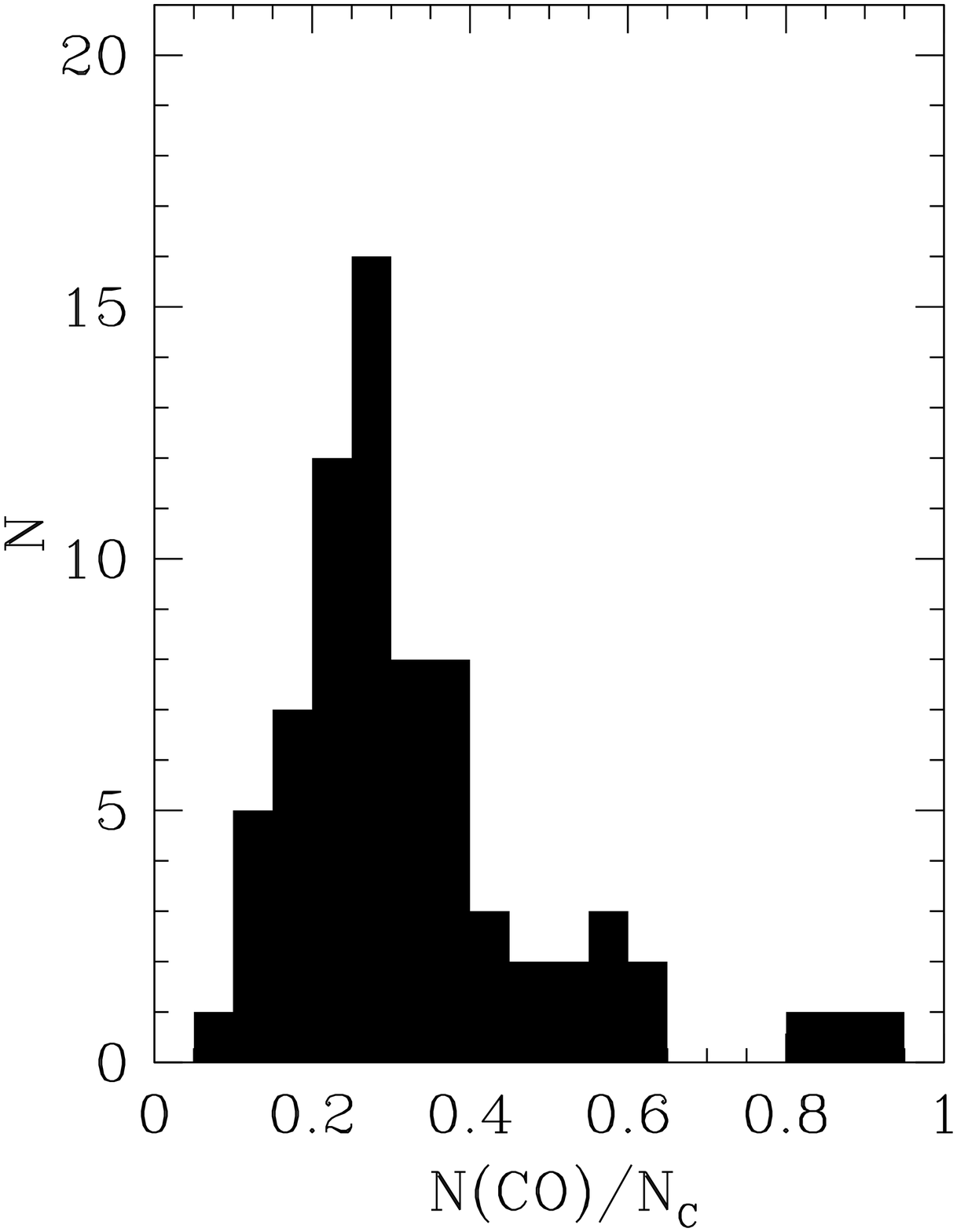}}}
\end{minipage}\hfill
\begin{minipage}[h]{4.45cm}
\bigskip
  \caption[]{Distribution of the fraction of all carbon contained in
    CO. Left: Results for an isotopological ratio of 40. Center:
    Results for an isotopological ratio of 80; for comparison, the
    distribution for the ratio of 40 is shown as well
    (unshaded). 
Right: Most probable distribution derived from both
    data sets (see text). Typically, one-third of the gas-phase carbon is
    in CO and two-thirds is in atomic or ionic form.}
\label{ccobin}
\end{minipage}
\end{figure*}

The distribution of the fractional CO abundances is shown in
Fig.\,\ref{ccobin} for the two isotopological abundances 40 (left) and
80 (center), with average values $f_{\rm CO}\,=\,0.28$ and
$f_{\rm CO}\,=\,0.38$ and standard deviations of $0.18$ and $0.17$,
respectively. We constructed the combined distribution (right) by
averaging the results for the galaxies that could be fit at either
ratio and by taking the single result for the galaxies that could
not. The combined distribution has an average value
$f_{\rm CO}\,=\,0.33$. The standard deviation $0.16$ exceeds the
uncertainty in the individual values and represents an intrinsic
spread of the ratios. The final adopted beam-averaged fractional CO
abundances and gas-phase total carbon column densities are summarized in
Table\,\ref{physpar} (Cols. 2 and 3).  

In Table\,\ref{ciiflux} we have selected the data for all galaxies for
which $\ci$ and $\cii$ measurements are also available. On average,
molecular carbon represents only one-third of all gas-phase carbon in
the observed galaxies. The remainder is atomic carbon either in
neutral (C$^{\circ}$) or in ionized form (C$^{+}$).  The $\ci$ and
$\cii$ line fluxes that are needed to further investigate this are
found in the literature.

%Table 10 CO, CI, and CII column densities
\begin{table*}
\caption[]{\label{ciiflux}Gas-phase carbon fractions}
\begin{center} 
{\small %
\begin{tabular}{l|rrrr|rrrr|rrrl}
\noalign{\smallskip}      
\hline
\noalign{\smallskip} 
Name  &f$_{[CII]}$&I[CII]&I(CI)$_{1-0}$ & I(CI)$_{2-1}$&N(C$^+$)&N(C$^{\circ}$)& N(CO) & N$_C$ &$\frac{\rm [CO]}{\rm [C]}$ & $\frac{\rm [C^{\circ}]}{\rm [C]}$&$\frac{\rm [C^+]}{\rm [C]}$&Sum \\
      & &\multicolumn{3}{c}{$\kkms$}&\multicolumn{4}{c}{$10^{17}\cm2$}&\multicolumn{4}{c}{}\\
 (1)  & (2)  & (3)   & (4)    & (5)    &  (6) &  (7)     &    (8)    & (9) & (10) & (11) & (12) & (13) \\
\noalign{\smallskip}
\hline                                                           
\noalign{\smallskip}
 253 & 0.86 & 569 & 256 & 154 & 59   &  70  &  66 & 215 & 0.30 & 0.33 & 0.28 & 0.91  \\
 891 & 0.58 &  53 &   8 & 4.8 &  6.6 &  4.0 & 5.0 &  36 & 0.14 & 0.11 & 0.18 & 0.43  \\
1068 & 0.74 & 111 &  65 &  52 & 12   &  19  &  13 &  36 & 0.33 & 0.57 & 0.28 & 1.18  \\
1097 & 0.72 &  52 &  21 &  15 &  5.4 &  13  &  25 &  47 & 0.51 & 0.28 & 0.12 & 0.91  \\
1365 & 0.73 & 128 &  64 &  51 & 13   &  56  &  26 &  60 & 0.43 & 0.93 & 0.22 & 1.58  \\
1482 & 0.90 &  88 &   6 & 4.8 &  9.2 &  2.6 & 2.0 &  18 & 0.11 & 0.15 & 0.53 & 0.79  \\
1614 & 0.90 &  43 &   7 &  7  &  4.5 &  1.3 & 1.7 &  11 & 0.14 & 0.12 & 0.41 & 0.67  \\
2146 & 0.89 & 431 &  24 &  29 & 45   &  5.8 & 9.0 &  31 & 0.28 & 0.20 & 1.57 & 2.05  \\
2623 & 0.75 &  31 &   3 &   3 &  3.2 &  1.0 & 1.5 &   4 & 0.38 & 0.25 & 0.76 & 1.39  \\
3034 & 0.60 & 892 & 123 & 185 & 93   &  27  &  33 &  83 & 0.39 & 0.33 & 1.12 & 1.74  \\
3079 & 0.8: &  86 &  28 & ... &  8.9 &  5.6 &  10 &  42 & 0.30 & 0.14 & 0.22 & 0.66  \\
3227 & 0.84 &  18 &  16 &  8  &  1.9 &  3.0 & 1.5 &   5 & 0.30 & 0.60 & 0.38 & 1.28  \\
3627 & 0.89 &  27 &  11 & 3.3 &  2.8 &  5.1 & 5.1 &  11 & 0.47 & 0.47 & 0.25 & 1.19  \\
3690 & 0.89 & 125 &  14 &  7  & 13   &  2.6 & 2.9 &  22 & 0.13 & 0.12 & 0.60 & 0.85  \\
4038 & 0.89 &  47 &   4 & 3.6 &  4.9 &  1.6 & 5.7 &  15 & 0.42 & 0.12 & 0.38 & 0.92  \\
4039 & 0.8: &  55 &   6 & 2.4 &  5.7 &  0.7 & 1.2 &   9 & 0.14 & 0.08 & 0.67 & 0.89  \\
4051 & 0.76 &   5 &  12 & 2.4 &  0.5 &  3.9 & 3.9 &  17 & 0.23 & 0.23 & 0.03 & 0.49  \\
4254 & 0.65 &  15 &   6 & 1.2 &  1.6 &  2.9 & 3.6 &   7 & 0.51 & 0.41 & 0.22 & 1.14  \\
4321 & 0.62 &  17 &   8 & 2.4 &  1.8 &  2.8 & 10  &  34 & 0.29 & 0.08 & 0.05 & 0.42  \\
4536 & 0.92 &  94 &   2 & 3.2 &  9.8 &  1.4 & 2.7 &  10 & 0.28 & 0.14 & 0.98 & 1.40  \\
4631 & 0.81 &  36 &   5 & 1.5 &  3.7 &  0.6 & 0.9 &   4 & 0.22 & 0.15 & 0.90 & 1.23  \\
4736 & 0.77 &  12 &   6 & 3.6 &  1.2 &  3.0 & 2.7 &   9 & 0.32 & 0.33 & 0.13 & 0.78  \\
4826 & 0.57 &  24 &  11 & 3.3 &  2.5 &  4.2 & 6.6 &  14 & 0.47 & 0.30 & 0.18 & 0.95  \\
4945 & 0.8: & 356 & 114 & 103 & 37   &  27  & 47  & 125 & 0.30 & 0.20 & 0.28 & 0.78  \\
5055 & 0.77 &  12 &   8 & 2.4 &  1.2 &  2.4 & 5.0 &  21 & 0.24 & 0.11 & 0.06 & 0.41  \\
5135 & 0.90 &  55 &  19 & 7.6 &  5.7 &  7.8 & 1.8 &   9 & 0.20 & 0.86 & 0.63 & 1.69  \\
5194 & 0.60 &  18 &   5 & 2.0 &  1.9 &  6.5 &  13 &  24 & 0.56 & 0.72 & 0.21 & 1.49  \\
5236 & 0.82 & 190 &  19 &  11 & 20   &  3.6 & 5.9 &  31 & 0.19 & 0.12 & 0.65 & 0.96  \\
5713 & 0.90 &  46 &   3 & 1.2 &  4.8 &  1.8 & 2.7 &  12 & 0.24 & 0.18 & 0.48 & 0.90  \\
6240 & 0.89 &  58 &  15 &  11 &  6.0 &  4.6 & 0.6 &   9 & 0.09 & 0.66 & 0.86 & 1.61  \\
6946 & 0.82 &  45 &  17 & 5.1 &  4.7 &  7.3 &  18 &  34 & 0.38 & 0.24 & 0.15 & 0.77  \\
7331 & 0.82 &  21 &   6 & 1.8 &  2.2 &  6.8 & 9.8 &  16 & 0.61 & 0.43 & 0.14 & 1.18  \\
7469 & 0.81 &  37 &  16 &  8  &  3.8 &  2.5 & 2.9 &   1 & 0.49 & 0.42 & 0.61 & 1.52  \\
\noalign{\smallskip}             
\hline
\end{tabular}
}% 
\end{center} 
Note: All data are reduced to the same $22"$ aperture. Column 1:
Galaxy name; Cols. 2, 3, and 6: fraction, intensity, and column density
of observed $\cii$ emission originating in neutral gas (see section
6.3); Cols. 4, 5, and 7: intensities and column density of the observed
$\ci$ emission (see section 6.2); Cols. 8 and 9: CO and total carbon
column densities (see Table\,\ref{modelpar}); Cols. 10, 11, and
12: relative contribution of CO, C$^{\circ}$, resp. C$^{+}$ to total
gas-phase carbon; Col. 13: sum of the three preceding columns; its
deviation from unity is a measure of their accuracy, especially that
of the least secure C$^{+}$ fraction (sections 6.3, 6.4).
\end{table*}

Figure 13 CO fraction of all C

\subsection{Neutral atomic carbon fraction} 

We took central $\ci$ line data from the compilations by Israel
$\etal$ (2015), Kamenetzky $\etal$ (2016), and Lu $\etal$ (2017). Most
of these were fluxes obtained with the SPIRE instrument onboard the
ESA {\it Herschel} Space Observatory \footnote {{\it Herschel} was an
  ESA space observatory with science instruments provided by
  European-led Principal Investigator consortia and with important
  participation by NASA.} in a $35"$ aperture. We expressed them as
integrated main-beam brightness temperatures in units of $\kkms$,
reduced to our `standard' beam by assuming identical $\ci$ and $\co$
distributions and filling factors and using the multi-aperture CO data
in Tables\,\ref{sestdat} and \ref{iramdat} to estimate the
$35"\rightarrow22"$ beam conversion factors (typically between 1.1 and
2.2). The resulting $\ci$ line intensities are given in
Table\,\ref{ciiflux}.

We cannot derive two-phase atomic carbon column densities in the same
way as the carbon monoxide column densities without additional
assumptions because only two $\ci$ transitions are available for
analysis. Fortunately, the $\ci$ intensities scale quite well with the
observed $\co$ intensities. We are uncertain of the cause, but it is
reasonable to expect that the $\ci$ emission either results from
photodissociation of the CO clouds in the beam or from material that is left
over in the formation of these CO clouds. In either case, neutral
carbon and carbon monoxide are closely related and associated with the
same $\h2$ gas.  We therefore used the $\h2$ densities, kinetic
temperatures, and relative filling factors from the CO analysis
(Table\,\ref{modelpar}) as {\it RADEX} input to determine model $\ci$
intensities. From these, we derived beam-averaged column densities of
$\ci$ in the same way as those of CO.  This procedure is less critical
for $\ci$ than for CO. With energy levels of 24 and 39 K and a
critical density of $10^{3}\,\cc$, the $\ci$ emission is thermalized
and close to being optically thin, roughly proportional to the
C$^{\circ}$ column, and only weakly dependent on temperature and
density (Schilke $\etal$ 1993, Stutzki $\etal$ 1997). We
calculated neutral carbon column densities separately for the $J$=1-0
and the $J$=2-1 transitions and for the two isotopologue abundances.
As expected, the average column densities are identical for the two
$\ci$ transitions, and use of the parameters of the $\co/\thirco$ = 40
case yields column densities lower than those for the $\co/\thirco$ =
80 case by a factor of 0.6 (standard deviation $35\%$), reflecting the
corresponding decrease in average optical depth. The final $\ci$
column densities in Table\,\ref{ciiflux} are the averages of the
independent determinations, as are the fractional [C$^{\circ}$]/[C]
abundances; their uncertainty is about $40\%$. The original SPIRE
measurements are quite accurate, therefore most of this uncertainty must be due
to assumptions in the analysis.

Our previous single gas-phase modeling of $\ci$ and CO intensities
of galaxy centers (Israel $\etal$ 2015) suggested a significantly
higher CO-to-C$^{\rm o}$ ratio.  However, the two studies measured
different quantities.  In the earlier study we used line ratios of the
mid-level $J$ transitions of $\co,$ representing the more highly
excited gas. We did not scale these results with line intensity, and
the derived column densities were not corrected for the
differentiating effects of beam filling factor and cloud velocity
width. They sampled purely local conditions rather than the global
conditions derived here.

The average gas-phase neutral atomic carbon fraction is 0.31.  When we
leave out the very high fractions derived for NGC~1365 (0.93) and
NGC~5135 (0.86), the average drops to 0.26. In general, the $\ci$
fraction is somewhat below the CO fraction. The derived $\ci$
fraction exceeds that of CO in less than $20\%$ of all cases

\subsection{Ionized atomic carbon fraction}

Ionized carbon $\cii$ line measurements useful for our purpose have
been carried out with the PACS instrument (Poglitsch $\etal$ 2010)
onboard the ESA $\it Herschel$ Space Observatory at a resolution of
$11"$ in square pixels of $9.4"\times9.4"$ size. We used the
compilations published by Fern\'andez-Ontiveros (2016), Croxall
$\etal$ (2017), D\'iaz-Santos $\etal$ (2017), and Herrera-Camus
$\etal$ (2018).  We interpolated intensities in the central nine PACS
pixels ($28.2"\times28.2"$) and in the single central pixel to those
expected in an intermediate $22"$ aperture. We also extrapolated
central PACS pixel intensities to those expected in a $22"$ aperture
assuming the emission to be point-like, with consistent results.

CO and $\ci$ emission can only originate in a neutral gas, but $\cii$
emission can also come from an ionized gas.  The fraction of the
$\cii$ emission from the neutral gas $f_{CII}$ relevant to our
analysis is estimated from the {\it Herschel} intensities of the [NII]
122$\mu$m (PACS) and $205\mu$m (SPIRE) lines in the usual way; for a
detailed description of this procedure and mapping results on many of
the sample galaxies, see Croxall $\etal$ (2017).  Typically, $~20\%$ of
the $\cii$ emission comes from ionized gas, and the neutral gas
fractions of interest to us range from 0.57 to 0.92 (mean 0.81,
average 0.75, with a standard deviation 0.21).  Table\,\ref{ciiflux}
provides normalized and corrected $\cii$ intensities $I[CII]$
available for all galaxies in which $\ci$ has also been measured. In
three cases, the actual fraction of $\cii$ emission arising from
neutral gas could not be determined; here we inserted the average
value, denoted by a colon.

The analysis of ionized carbon is more problematical than that of the
neutral carbon in the preceding section. Because the $\cii$ emission can be
more extended and associated with dense hydrogen gas that is not traced by
$\ci$ or CO emission, the CO parameters that we used to guide our
$\ci$ analysis are now of little use.  The 158$\mu$m $\cii$ line is
the only strong C$^+$ emission line in the far-infrared, and if it is
optically thin, the line-of-sight column density $N_{\rm C+}$ is
related to the line intensity $I_{\cii}$ (in $\kkms$) by Eq.\,(1)
from Pineda $\etal$ (2013):
$N_{\rm
  C+}\,=\,I_{\cii}\,\times\,(3.05\times10^{15}\,(1+0.5(1+2840/n)\,e^{91.2/T})$.
The temperatures and densities of the $\cii$-emitting gas cannot be
determined directly because there are three unknown parameters and only one
equation. The equation provides a lower limit
$N_{C+}\,=\,4.6\times10^{15}$ $I$(CII) to the column density in the
high-temperature, high-density limit, but no upper limit.  We
calculated C$^+$ column densities (Col. 6 of Table\,\ref{ciiflux})
for a more reasonable temperature $T_k\,=\,100$ K and density
$n(\h2)\,=\,3000\,\cc$, so that $N_{C+}\,=\,1.04\times10^{16}$
$I$(CII), doubling the high-$T$, high-$n$ limit. The corresponding
fractional abundances [C$^+$]/[C] are listed in Col. 12 of
Table\,\ref{ciiflux}.  They show a large spread; in particular, the
values for NGC~2146, M~82, and NGC~4536 are quite high, which might
indicate that their $\cii$ emission comes from gas that is denser and hotter
than we have assumed. The derived central C$^{+}$/C fractions and the
galaxy FIR luminosities are correlated, with considerable
scatter. Because we calculated ionized carbon column densities for
fixed temperatures and densities, this is loosely related to a
correlation between $\cii$ and FIR intensities.  No other
clear pattern seems to emerge from the data in Table\,\ref{ciiflux},
consistent with significant variation in the physical conditions of the ISM and
morphology even in galaxies that otherwise appear similar.

\subsection{Carbon budget in galaxy centers}

The combined column density of the three individual carbon gas-phase
components listed in Col. 13 in Table\,\ref{ciiflux} is generally
close to the total gas-phase column density derived independently of
the analysis of the $\co$ and $\thirco$ intensities sample
(Table\,\ref{modelpar}). The average ratio of the two values is
$1.06$, with a standard deviation of $0.45$.  The consistency of these
results underlines the validity of the analysis that produced them.

The relative amounts of the three gas components vary significantly
among the observed galaxies, but the average fractional CO,
C$^{\circ}$, and C$^+$ contributions to the total C are quite similar.
The respective contributions in the 33 galaxies are 0.32,
0.31, and 0.43, with standard deviations of 0.13, 0.23, and 0.36,
respectively. The three forms of carbon occur in comparable amounts in
the gas phase. Slightly less than one-third of all gas-phase carbon is
in molecular form, and somewhat less than half of all gas-phase carbon
is ionized.

As argued in section 6.1, the beam-averaged carbon monoxide column
density is well constrained. In section 6.2, we argued that the $\ci$
emission arises from the same molecular gas, so that the beam-averaged
neutral carbon column densities should likewise be robust. If the
actual CO-to-C and C$^{o}$-to-C ratios were constant across the sample, the
standard deviations would represent the measurement error in their
determination. More realistically, they serve as upper limits to the
uncertainty in the actual individual cases.  The ionized carbon column
densities are not so robust because they depend more strongly on
assumptions.  As we showed, there is a firm lower limit
corresponding to very hot ($T_{\rm k}>$250 K) and dense
(n$_{\h2}>10^{5}\,\cc$) gas, so that the column densities in
Table\,\ref{ciiflux} can be lowered by a factor of two at most. In
contrast, the $\cii$ line measurements do not define an upper
limit. For instance, column densities would be higher by a factor of
ten for modest temperatures $T_{\rm k}\approx50$ K and low densities
$n_{\h2}\approx300\,\cc$. This is very unlikely because it requires the
total carbon column densities $N_{\rm C}$ derived before to be
underestimated by factors of five, incompatible with the models
used. The corresponding very small CO fractions would also seem
incompatible with the high-metallicity environment of galaxy centers
as they are more characteristic of low-metallicity objects such as the
Magellanic Clouds (Requena-Torres $\etal$ 2016). As it is, modest
temperature decreases to 60 K, or equally modest density decreases to
1000 $\cc$, implying $70\%$ higher $\cii$ column densities, delineate
the limits of what is feasible in view of the various uncertainties
associated with Table\,\ref{ciiflux}.

%Figure 14 NH and MH versus D
\begin{figure}
  \resizebox{4.45cm}{!}{\rotatebox{0}{\includegraphics*{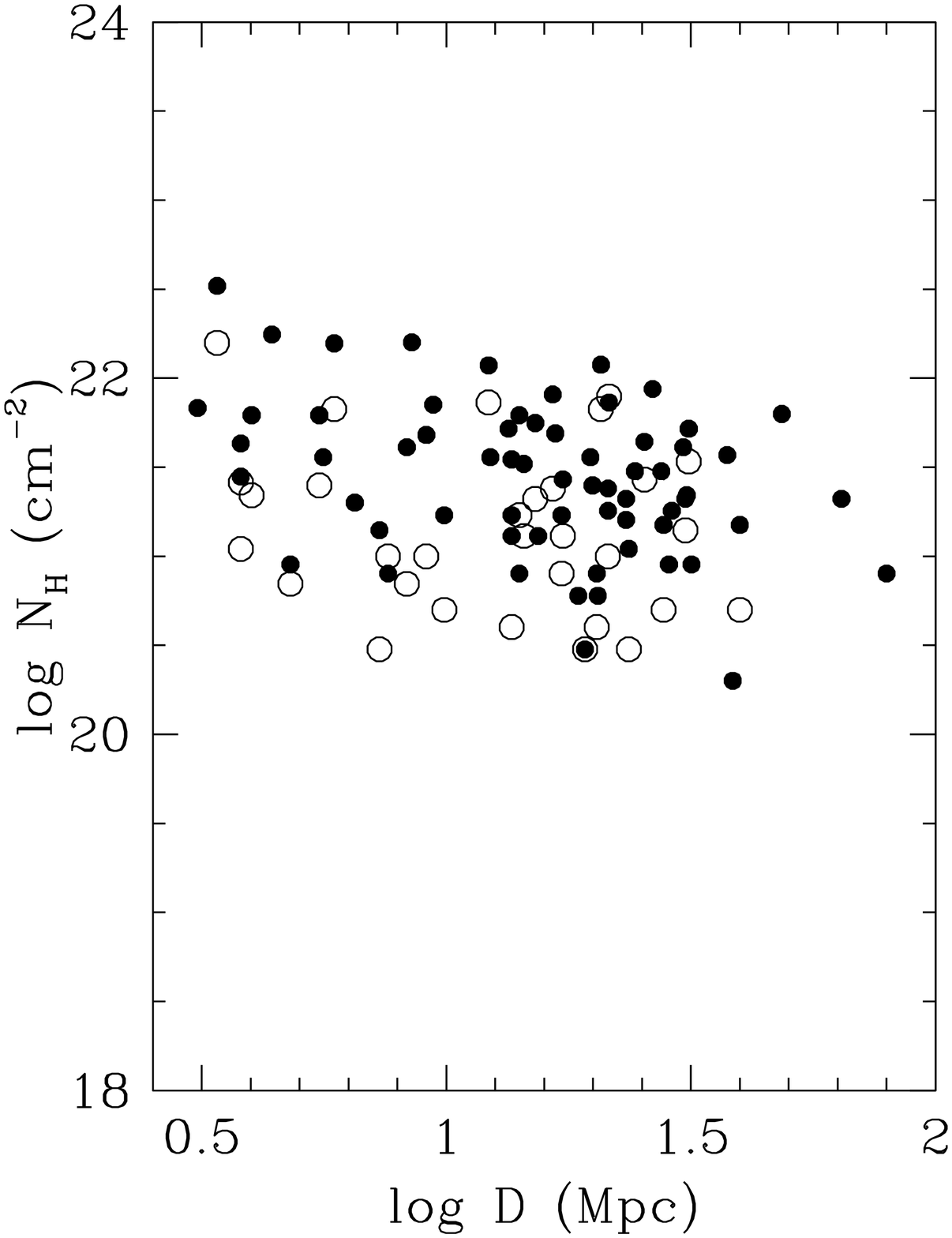}}}
  \resizebox{4.45cm}{!}{\rotatebox{0}{\includegraphics*{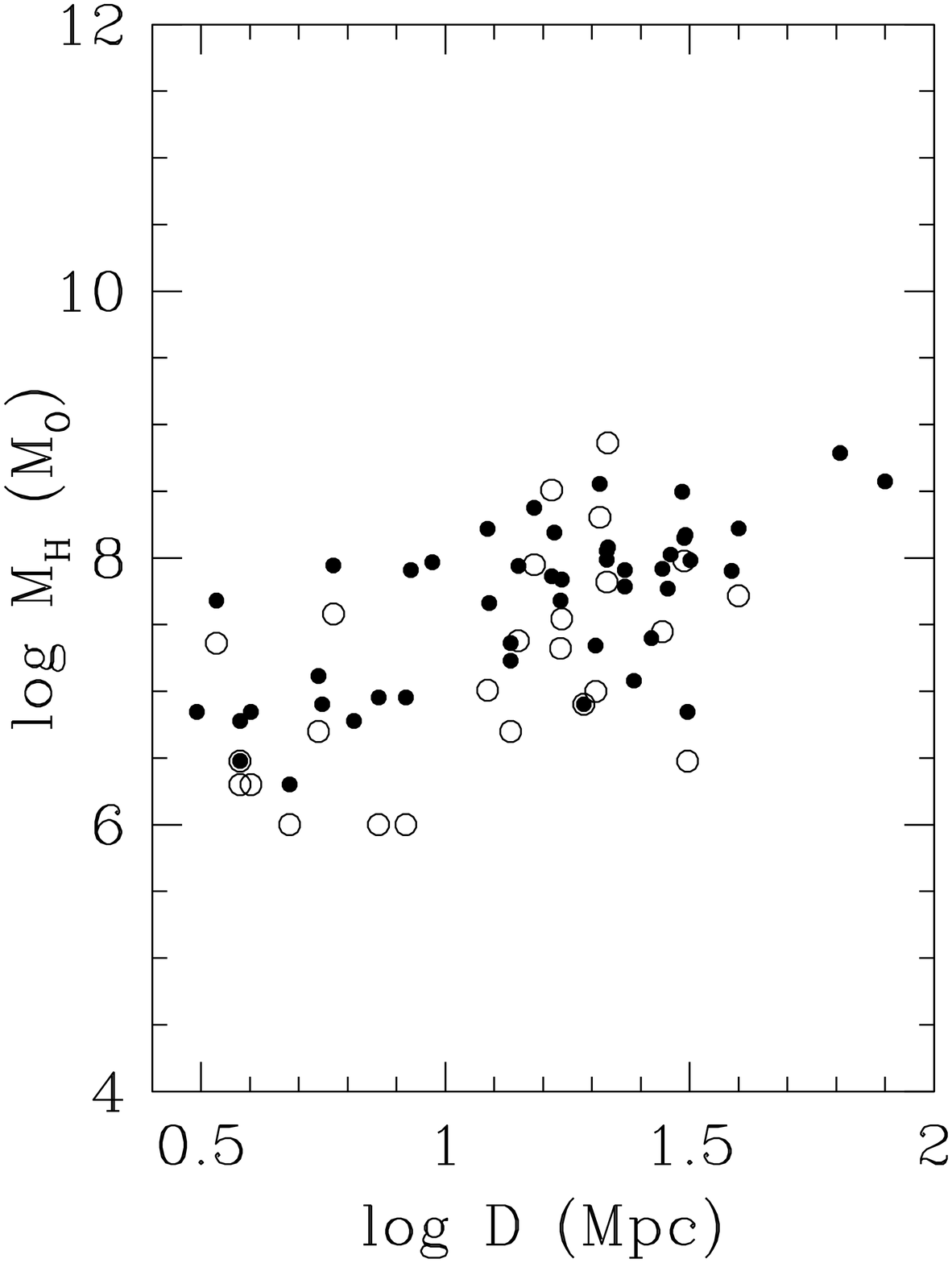}}}
  \caption{Calculated beam-averaged column densities N$_{\rm H}$
    (left) and central gas masses M$_{\rm H}$ (right) as a function of
    distance D. Filled circles: Values based on the preferred
    `nominal' carbon abundance. Open circles: Values bases on
    extrapolated maximum carbon abundances (see text). All values are
    based on an assumed isotopological ratio of 40. }
\label{galdis}
\end{figure}

%Figure 15 Distribution NH and MH
\begin{figure}
  \resizebox{4.45cm}{!}{\rotatebox{0}{\includegraphics*{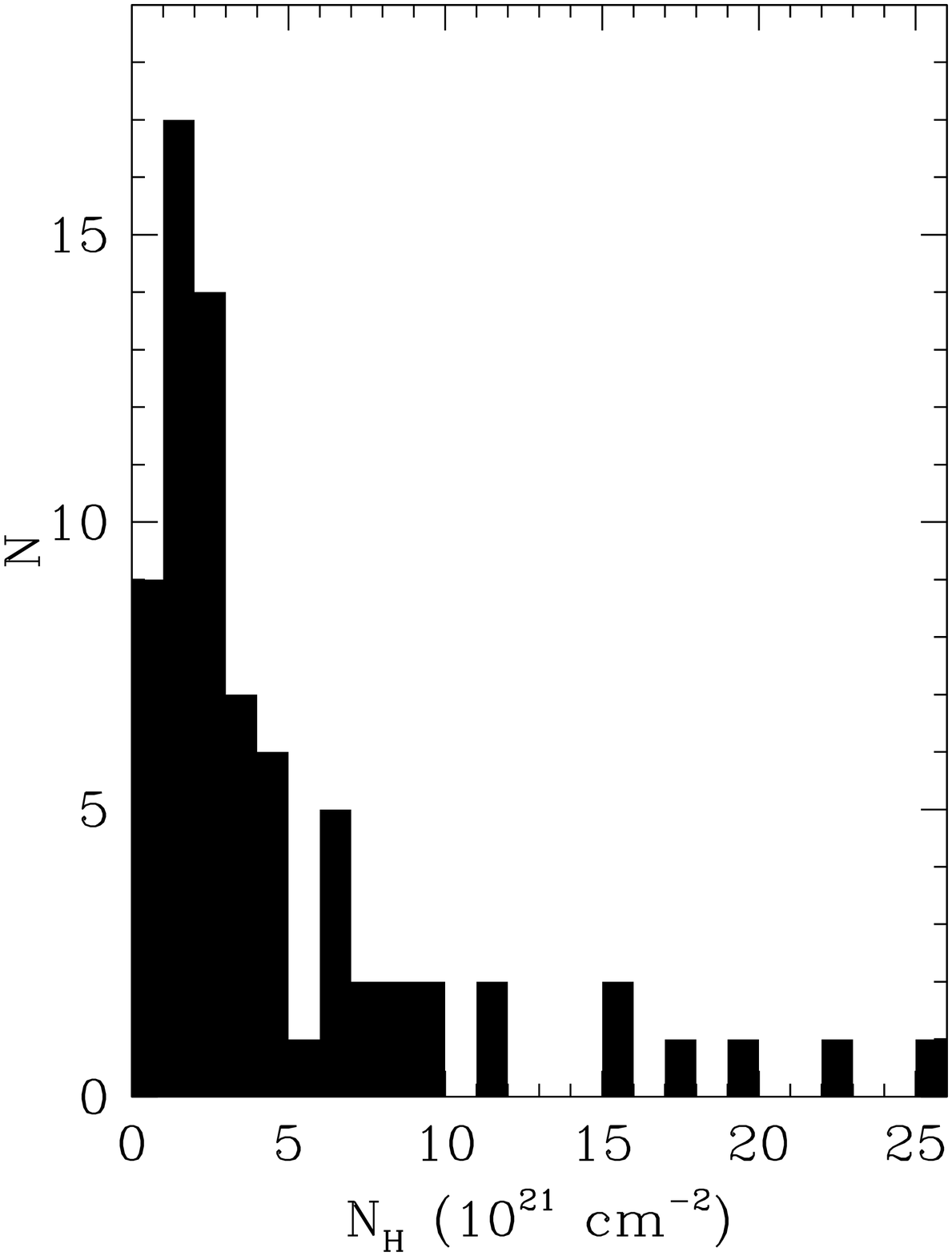}}}
  \resizebox{4.45cm}{!}{\rotatebox{0}{\includegraphics*{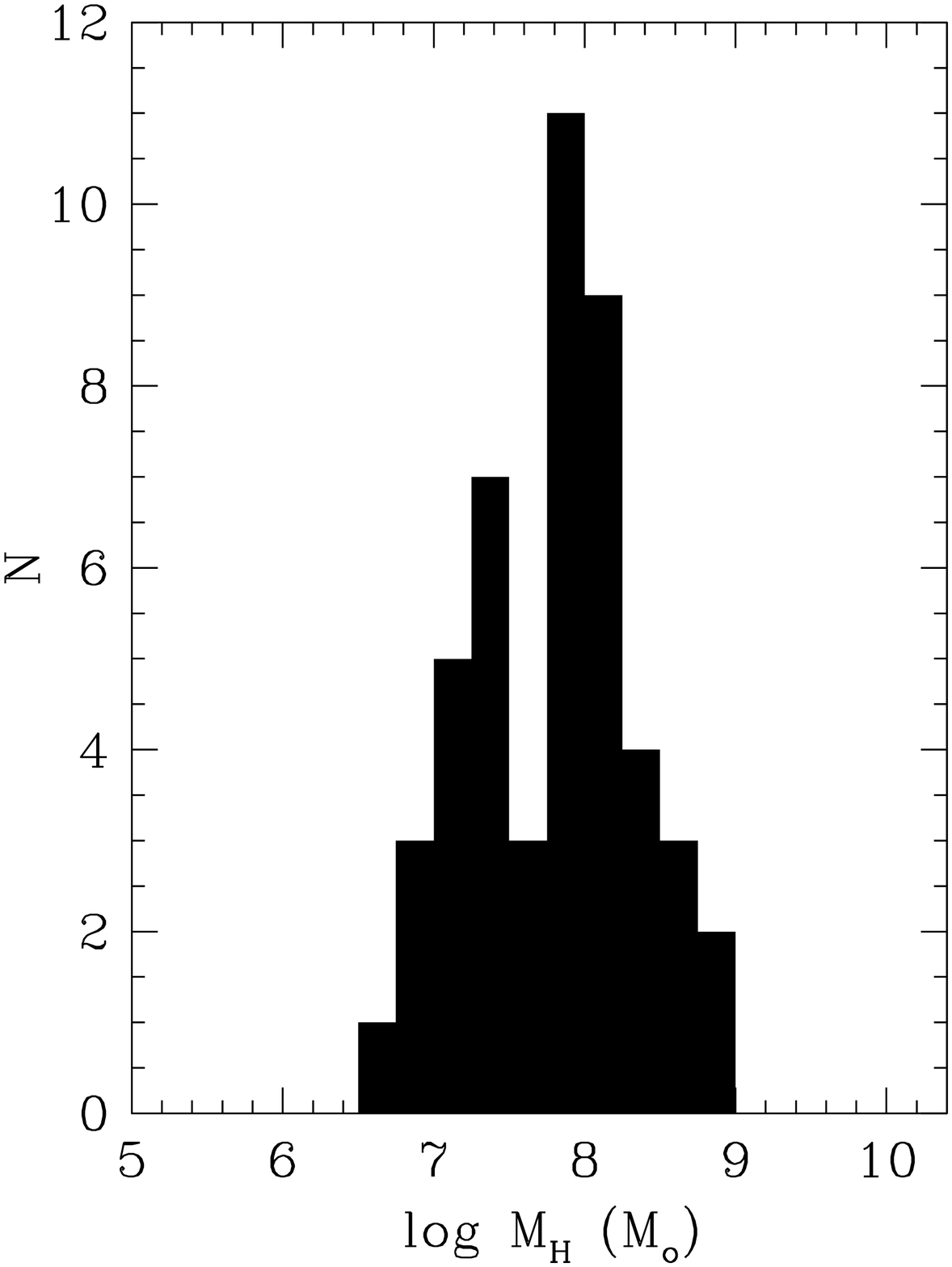}}}
  \caption{Distribution of the total hydrogen column densities (left)
    and total hydrogen masses (right). See text for sample definition
    and errors.}
\label{galhbin}
\end{figure}

\section{Hydrogen column density and mass}

Total hydrogen column densities $N_{\rm H}$ would follow directly from
the carbon column densities $N_{\rm C}$ if the gas phase
carbon-to-hydrogen abundance were directly known, which is not the
case. Instead, we must infer this abundance from our knowledge of (i)
the relative oxygen abundance [O]/[H], (ii) the relative carbon
abundance [C]/[O], and (iii), the fraction $\delta_C$ of all carbon
that is in the gas phase rather than locked up in dust grains. Based
on the detailed discussion in Appendix D, we adopt for all galaxy
centers a metallicity of twice that of the solar neigborhood,
identical [C] and [O] abundances and a carbon depletion factor
$\delta_{\rm C}\,=\,0.5\pm0.2$.  This yields a `nominal' (N) gas phase
ratio $N_{\rm H}/N_{\rm C}\,=\,(2\pm1)\times10^{3}$.  In
Table\,\ref{galmassx} we list the implied beam-averaged total hydrogen column
densities $N_{\rm H}$ for a range of assumptions, as well as the
derived molecular hydrogen column densities $N_{\h2}$ and overall
hydrogen gas masses $M_{\rm H}$. The adopted nominal values are
summarized in Table\,\ref{physpar}. The beam-averaged molecular
hydrogen column densities $N_{\h2}$ (Col. 4) are corrected for the
(small) contribition by $\hi$. The total gas masses $M_{\rm gas}$
(Col. 5) incorporate a $35\%$ contribution by helium. As discussed in
Appendix D, the uncertainty in individual values of $N_{\rm H}$ and
$M_{\rm H}$, hence also in $N_{\h2}$ and $M_{\rm gas}$, is a factor of
slightly more than two.

In Fig.\,\ref{galdis} we plot $N_{\rm H}$ and $M_{\rm H}$ as a
function of galaxy distance $D$ for both the nominal N and the
extrapolated E carbon abundance. The Galactic abundance case G is not
shown, but would be represented by points offset from the nominal
abundance points by +1.7 in the log.  The hydrogen gas column densities
peak in the center, causing beam-averaged column column densities to
decrease with galaxy distance as ever larger linear areas are covered
by the fixed $22"$ beam. At the same time, the encompassed hydrogen
mass increases with distance when the beam includes ever larger areas
of the galaxy disk.

Figure\,\ref{galhbin} shows the distributions of $N_{\rm H}$ and
$M_{\rm H}$. We merged the two isotopological abundance data sets and
averaged values where appropriate. The column densities have a
well-defined peak at about $N_{\rm H}=1.5\times10^{21}\,\cm2$, with a
tail to higher values primarily caused by long sight-lines through
highly tilted galaxies.  The mass distribution shows a wide range
of values from $10^{6}$ M$_{\odot}$ to $10^{9}$ M$_{\odot}$ with a
broad peak around a few times $10^{7}$ M$_{\odot}$ and a narrow peak
around $10^{8}$ M$_{\odot}$. Fig\,\ref{galdis} shows that with a few
exceptions, the higher masses are all found in galaxies at distances
of 10 Mpc or more and mostly refer to the `inner disks' in
Fig.\,\ref{sizebin}. The lower masses are found over a wider range of
distances, from 4 to 25 Mpc, and thus characterize both
'circum-nuclear disks' and low-mass 'inner disks'.

%Figure 16 Comparison of X(CO), X[CI], X[CII].
\begin{figure*}
\begin{minipage}[c]{13.5cm}
  \resizebox{4.45cm}{!}{\rotatebox{0}{\includegraphics*{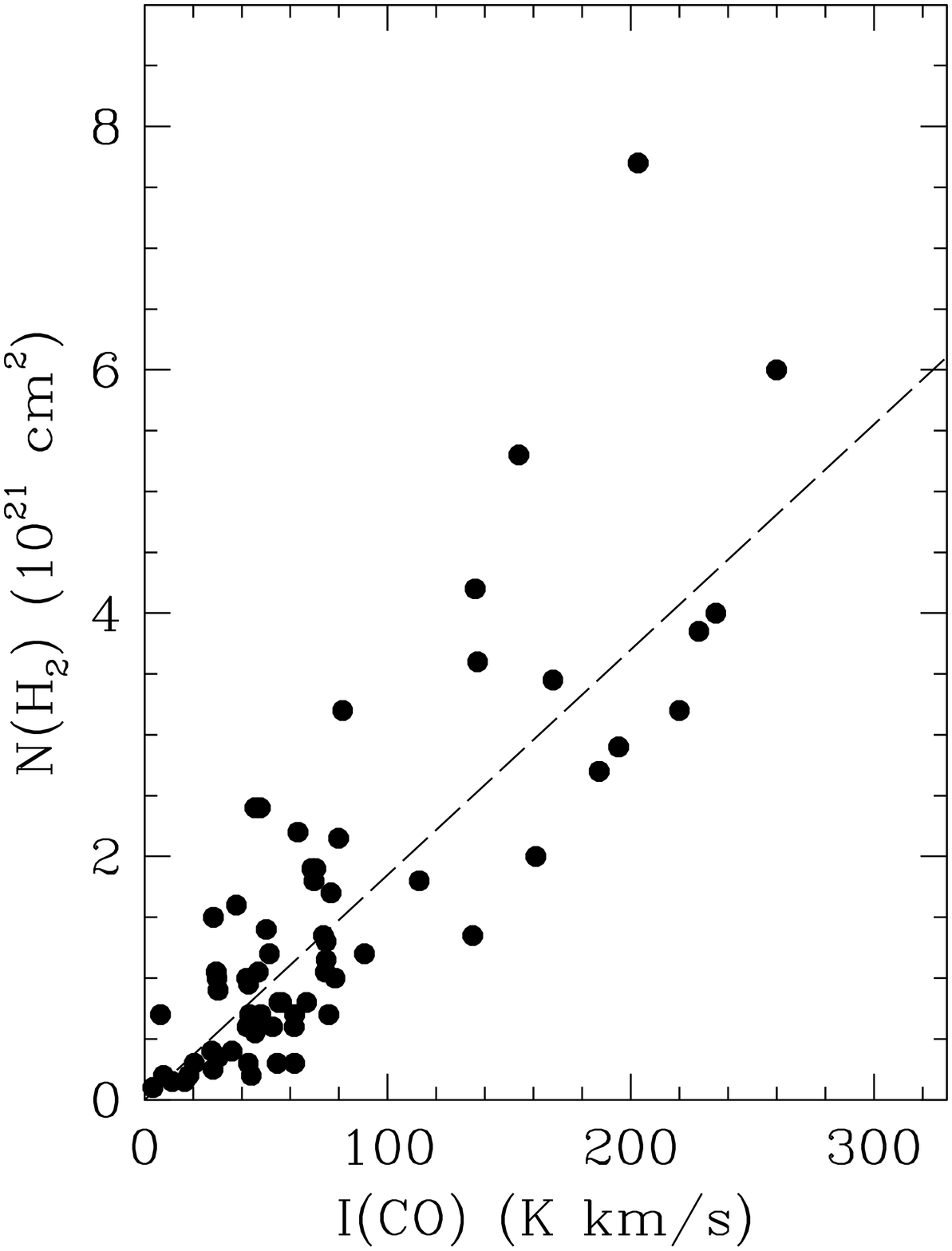}}}
  \resizebox{4.45cm}{!}{\rotatebox{0}{\includegraphics*{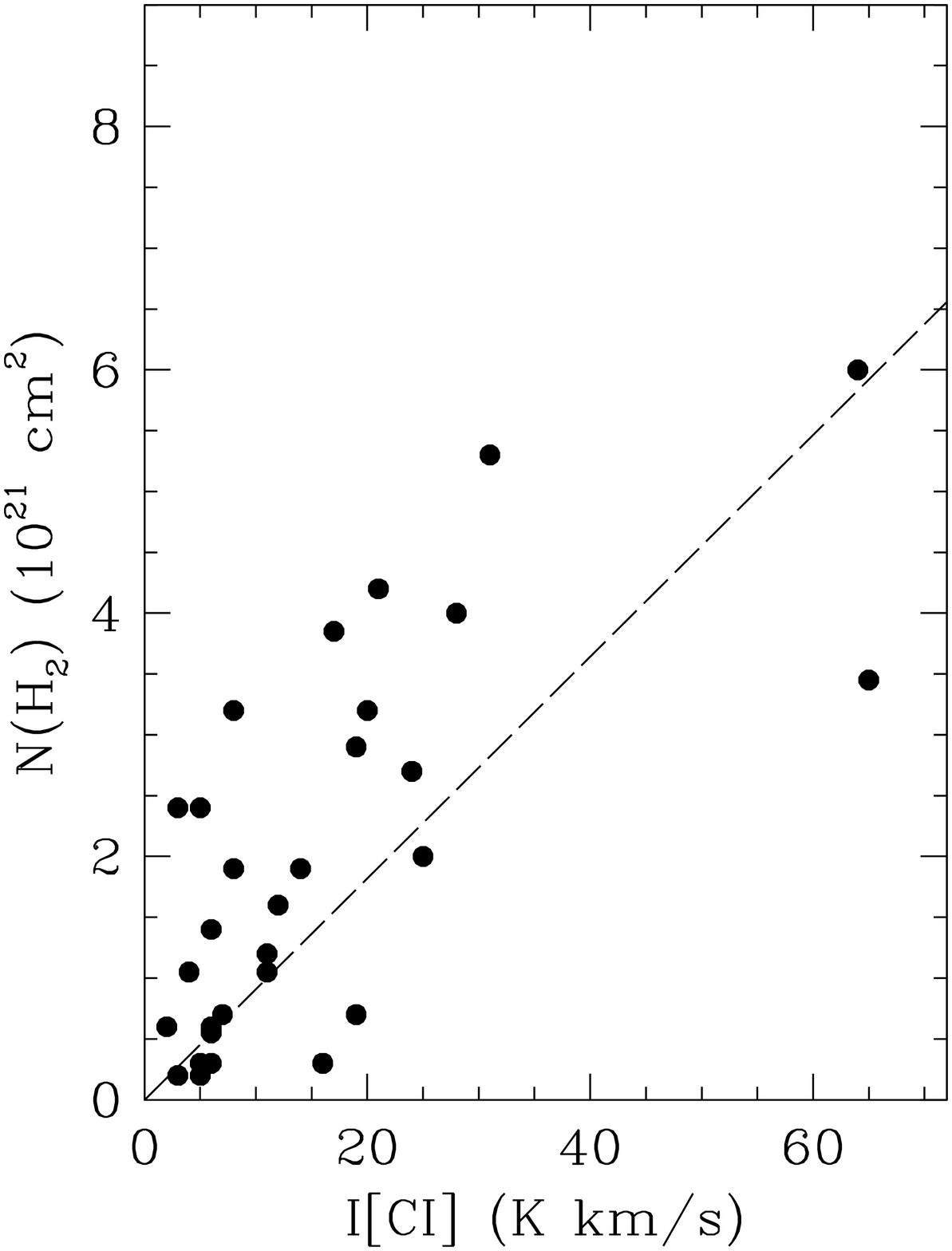}}}
  \resizebox{4.45cm}{!}{\rotatebox{0}{\includegraphics*{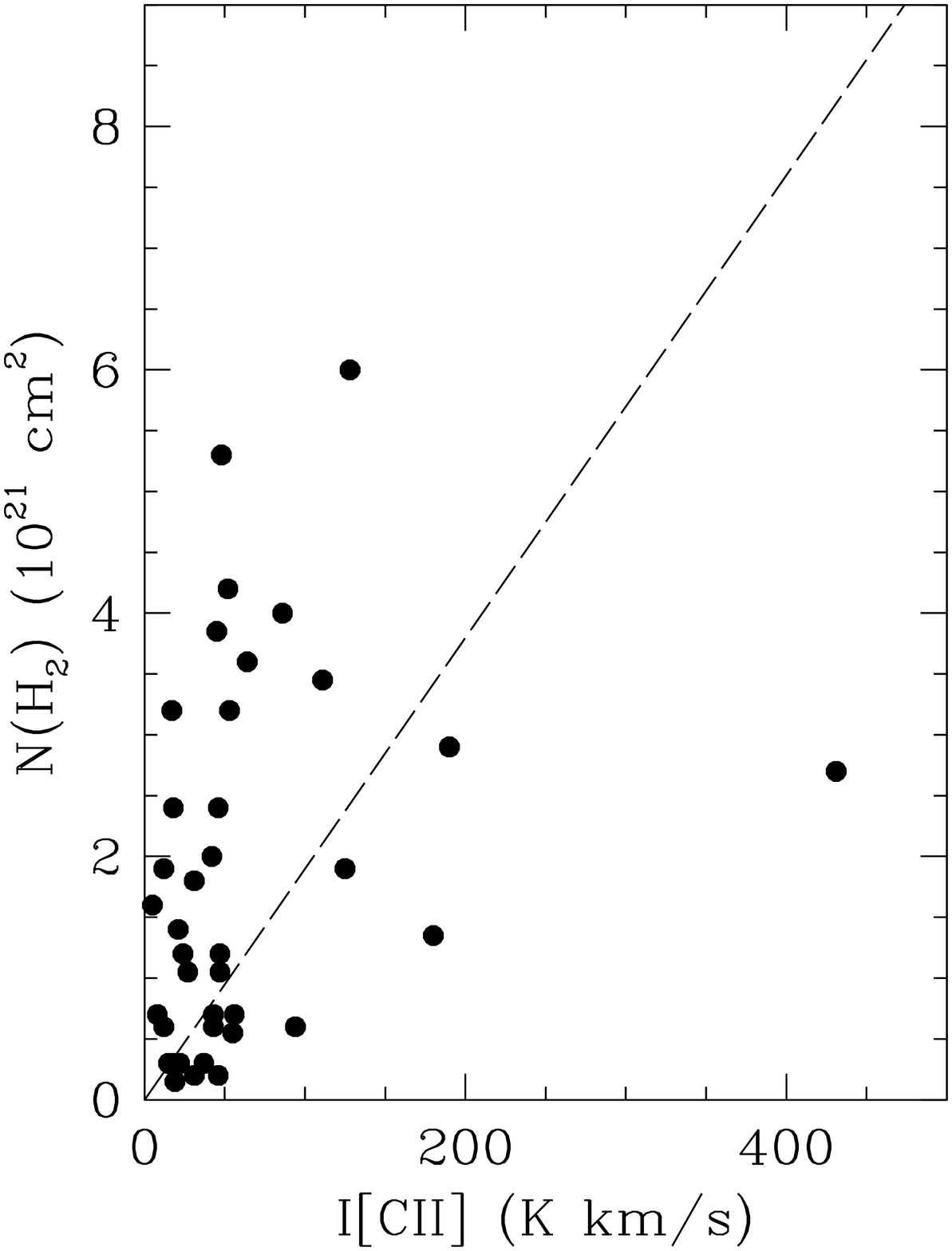}}}
\end{minipage}\hfill
\begin{minipage}[h]{4.45cm}
\medskip
\caption{Column densities $N_{\h2}$ as a function of (left) CO,
  (center) $\ci$, and (right) $\cii$ intensities. Points representing
  the nearby bright galaxies NGC~253, NGC~3034 (M82), and NGC~4945 are
  outside the box limits.  Dashed lines are linear regression fits to
  all data points, including these galaxies. The fits correspond to
  conversion factors $X$(CO) = $1.9\times10^{19}\,\cm2/\kkms$ and
  $X\ci\,=\,9.1\times10^{19}\,\cm2/\kkms$. There is no meaningful fit
  for $I(CII)$.  }
\label{xcomp}
\end{minipage}
\end{figure*}

% Figure 17 Distribution X.
\begin{figure*}
\begin{minipage}[c]{13.5cm}
  \resizebox{4.45cm}{!}{\rotatebox{0}{\includegraphics*{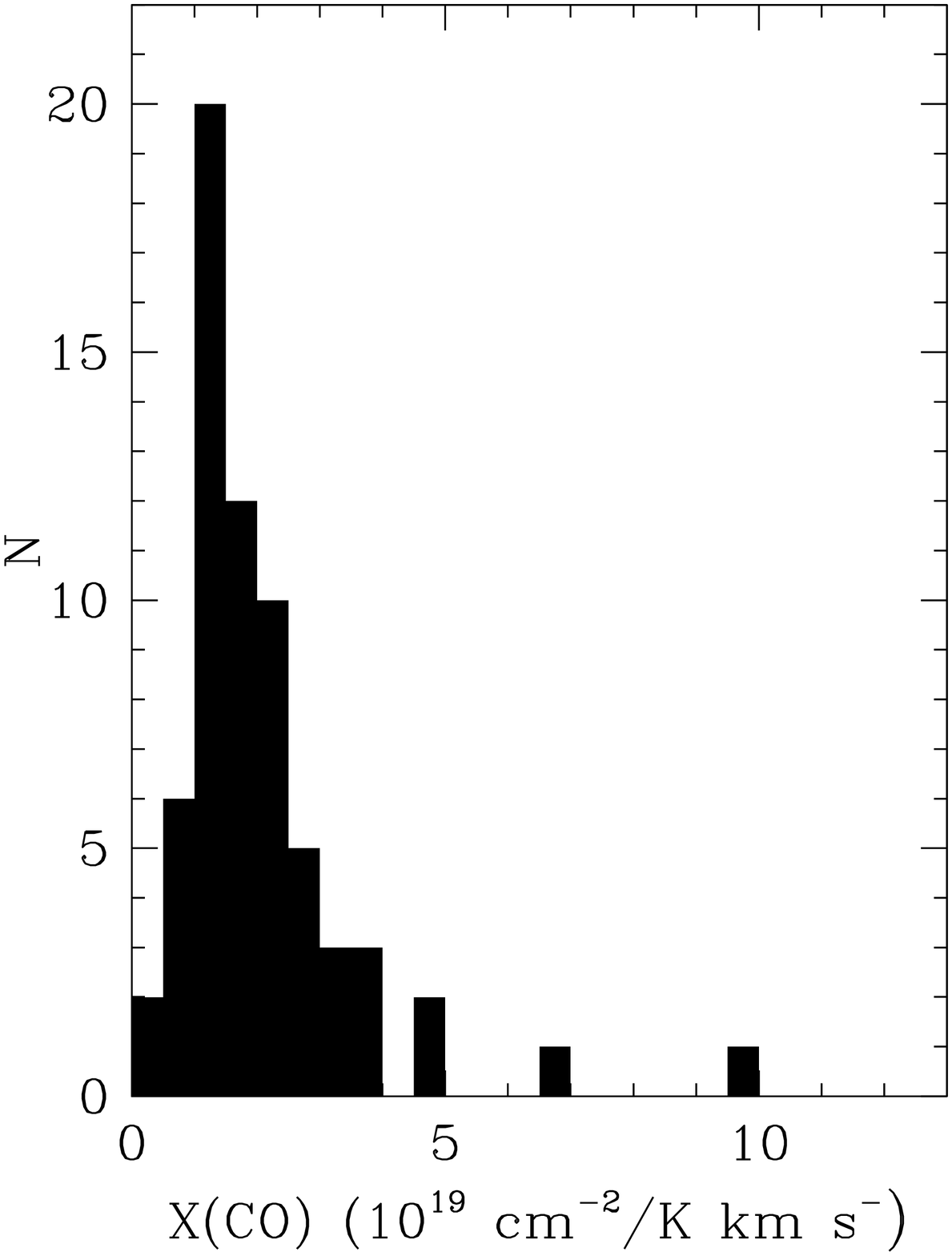}}}
  \resizebox{4.45cm}{!}{\rotatebox{0}{\includegraphics*{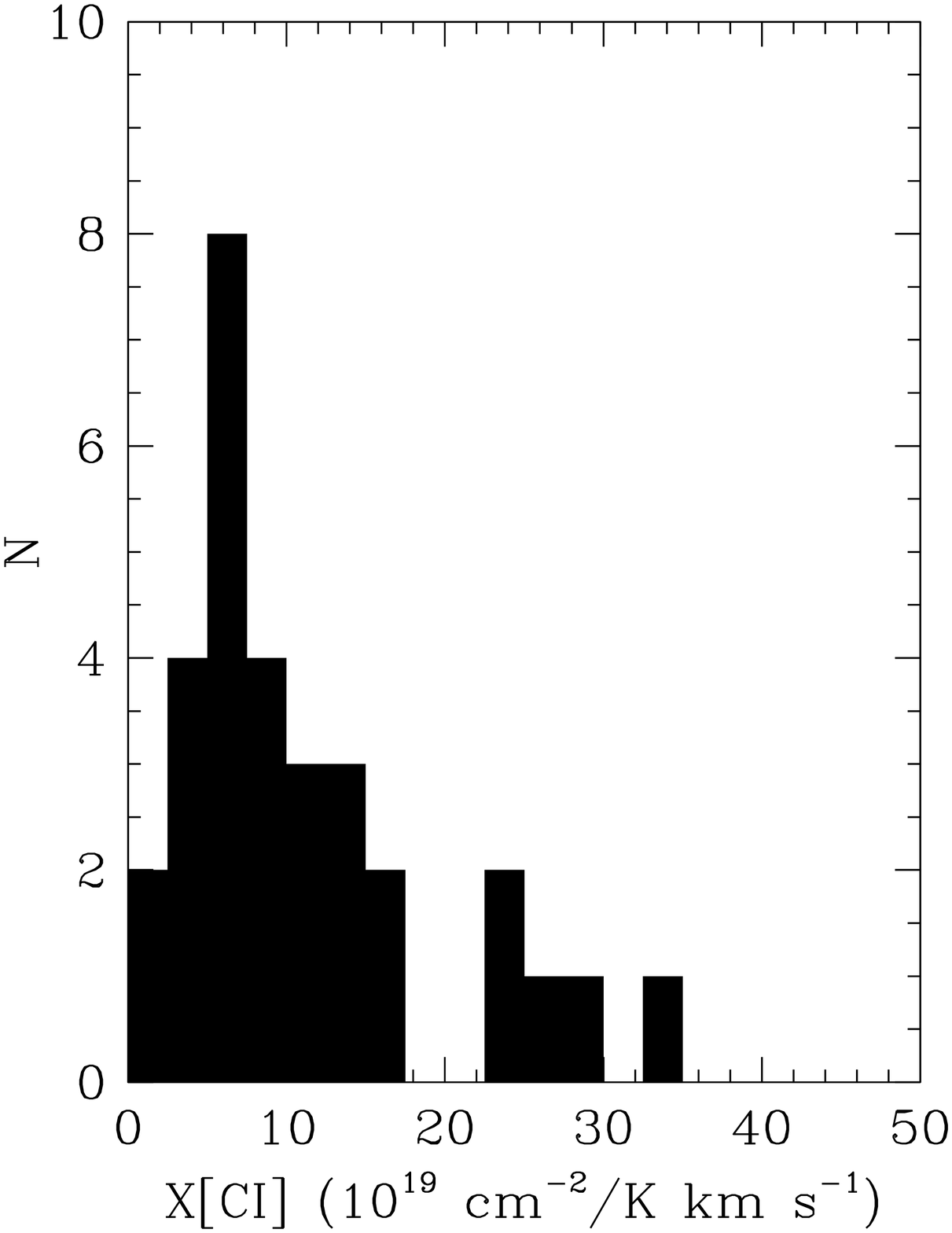}}}
  \resizebox{4.45cm}{!}{\rotatebox{0}{\includegraphics*{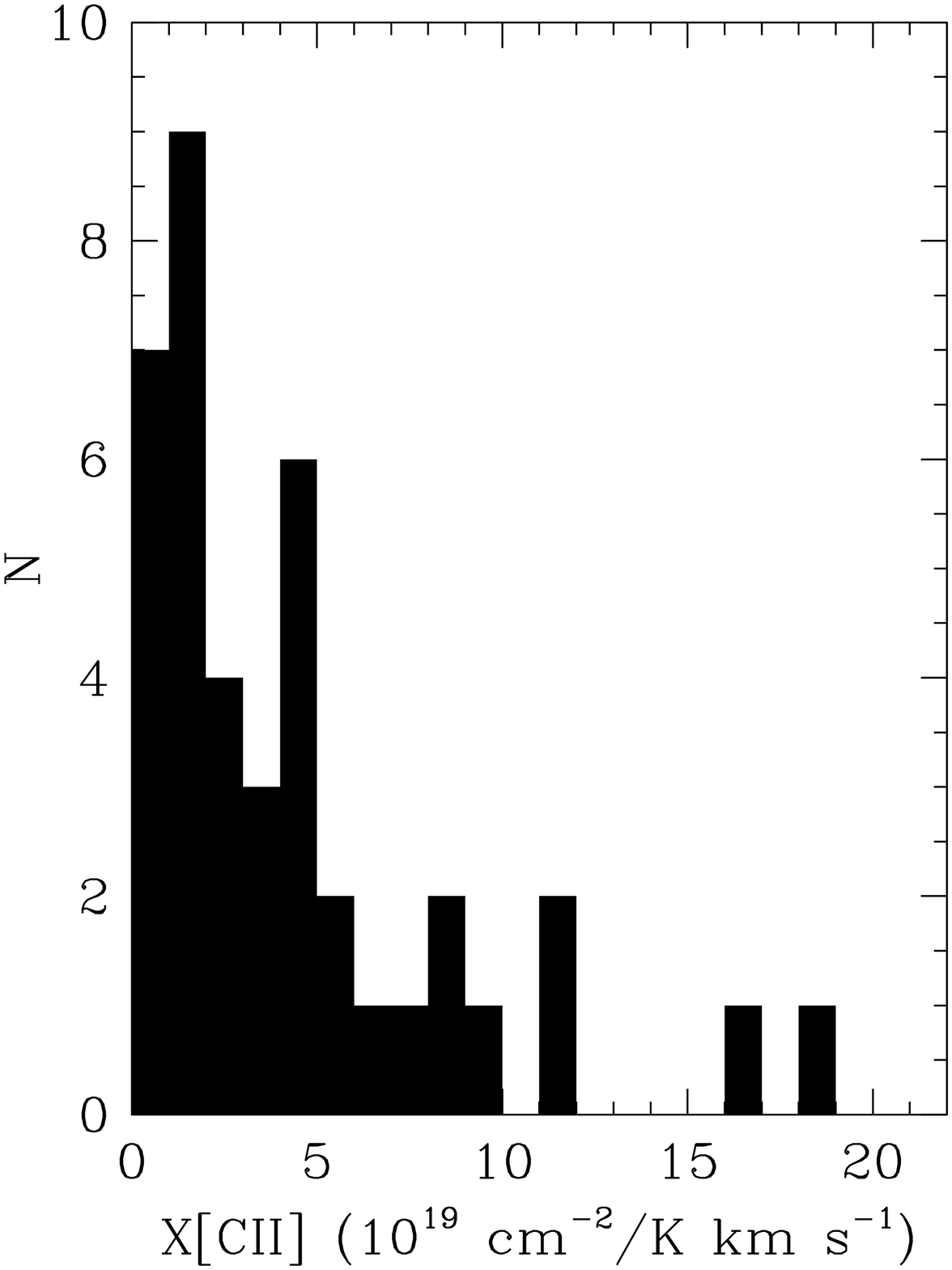}}}
\end{minipage}\hfill
\begin{minipage}[h]{4.45cm}
\bigskip
\caption[]{From left to right: Distributions of the CO-to$\h2$,
  [CI]-to$\h2$, and [CII]-to$\h2$ conversion factors $X$(CO), $X$[CI],
  and $X$[CII]. See text for sample definition and errors.}
\label{galh2bin}
\end{minipage}
\end{figure*}

\section{CO as a tracer of $\h2$}

\subsection{Conversion factors $X$(CO), $X$[CI], and $X$[CII]}

We are now in a position to consider to what extent CO, $\ci$, or
$\cii$ line intensyies trace $\h2$ molecular gas column densities.  In
Fig.\,\ref{xcomp} we plot the $N_{\h2}$ column densities from
Table\,\ref{physpar} as a function of the $J$=1-0 CO, $\ci$, and
$\cii$ line intensities from Tables\,\ref{sestdat} and
\ref{ciiflux}. In each panel, the dashed line marks the relation
between the two quantities with a slope corresponding to the $X$
factor. Because the same molecular hydrogen column densities are
plotted as a function of the observed line intensities, the results
are not subject to the uncertainties that plague the determination of
neutral and ionized carbon column densities discussed in Sections 6.2
and 6.3. The distribution of the individual $X$ values is shown in
Fig.\,\ref{galh2bin} for each line.  Inasmuch as the dispersion
exceeds the observational error, the dispersion around the mean $X$
value is an important quantity to judge the relative performance of
each of the three lines in predicting the molecular gas column
density.

The individual $I_{CO}$-to-$N(\h2)$ ratios occur in a fairly narrow range,
ten times below the commonly assumed Milky Way solar neighborhood
conversion factor $X_{\rm MW}\,=\,2.0\times10^{20}\,\cm2/\kkms$.  A
linear regression fit on $I_{CO}$ and $N(\h2)$ yields a high-quality
solution ($r^{2}$ = 0.78) corresponding to a conversion factor that is very
close to the average of the individual $I_{CO}$-to-$N(\h2)$ ratios: $X$(CO)
= $(1.9\pm0.2)\times10^{19}\,\cm2/\kkms$ . The average $X$-factor
applicable to galaxy centers is quite robust and well defined even
when individual CO-to-hydrogen conversion factors are still subject to uncertainties of a
factor of two.

A similar regression fit on the much lower intensities of the neutral
carbon line (Fig\,\ref{xcomp} central panel) yields a five times
higher value $X\ci\,=\,(9\pm2)\times10^{19}\,\cm2/\kkms$>. The
average of the individual values is almost twice as high, with
a large standard deviation. This is caused by the cluster of
low-intensity, low-column-density points in the lower left corner of
the central panel in Fig.\,\ref{xcomp}, and it suggests that $\ci$
intensities are useful but less reliable as $N(\h2)$ indicators than
CO.

The linear regression fit on all $\cii$ intensities gives a value
$(1.9\pm0.5)\times10^{19}\,\cm2/\kkms$ with very low significance
($r^{2}$ = 0.07). The average is more than twice as high,
$(4.4\pm0.8)\times10^{19}\,\cm2/\kkms$. These $\cii$
intensities include a contribution from ionized gas. If
corrections were included, the $N_{\h2}/I(CII)$ slopes would be
steeper by about $20\%$, but the relative distribution of points would
suffer little change.  In any case, from both Figs.\,\ref{galh2bin}
and \ref{xcomp} it follows that the present data do not define a
clear-cut single value for $X(\cii)$.

Following Wada $\&$ Tomisaka (2005), we also considered the $\co$
J=3-2 intensities as a tracer for $N(\h2)$.  Their three-dimensional,
non-LTE radiative transfer calculations for circum-nuclear molecular
gas disks predict that the J=3-2 line is more useful than the J=1-0
line as a tracer for $N(\h2)$ and they suggest a conversion factor of
$\sim2.7\times10^{19}\,\cm2/\kkms$ for this transition. Our linear
regression fit yields a value
$(2.1\pm0.4)\times10^{19}\,\cm2/\kkms$. There is considerable scatter
around this value and the average comes out higher, at
$(3.1\pm0.3)\times10^{19}\,\cm2/\kkms$.  We conclude that the
$X$(CO3-2) value calculated by Wada $\&$ Tomisaka (2005) is very close
to the actual value following from our work, but that $X$(CO1-0) is
still the better performer, contradictory to their expectations.

In a previous paper, Israel $\etal$ (2015) discussed the suggestion by
Papadopoulos $\etal$ (2004) and others that $\ci$ line emission might
provide a tracer of molecular hydrogen at least as good as CO
emission.  The above discussion, and indeed inspection of the CO, and
$\ci$ panels in Figs.\,\ref{galh2bin} and \ref{xcomp}, establishes that
the scatter is somewhat greater in the $\ci$ diagrams and that the
$X$[CI] distribution is less strongly peaked than the $X$(CO)
distribution. This result confirms and expands our earlier conclusion
that the $J$=1-0 $\co$ line should be preferred over the $\ci$ line as
a molecular gas tracer if both are available. If only the $\ci$ line
is accessible, as may be the case for redshifted objects, it should
be regarded as an acceptable substitute, provided an adequate
calibration can be established for the differing environmental
conditions.

The scatter in the $\cii$ diagrams is much greater than that in either
the $\co$ or $\ci$ counterparts, and our data do not establish a
convincing unique value of $X\cii$. It follows that $\cii$
intensities are not a useful tracer of extragalactic molecular gas
column densities, and given the minor contribution of $\hi$, are not a
useful tracer of total gas either.

\subsection{Low $X$(CO) and $\h2$ mass in  galaxy centers}

In the above, we have come to the conclusion that molecular hydrogen
column densities and masses in the centers of galaxies are an order of
magnitude lower than suggested by the `standard' conversion factor.
This conclusion depends to some extent on the correctness of the
carbon abundances assumed in the derivation.  If, for instance, these
were substantially lower, as in the Pilyugin $\etal$ (2014)
calibration already mentioned, the distribution of the individual
$X$ factors would still be similar to that depicted in
Fig.\,\ref{galh2bin}, but it would be shifted upward to a mean value
$X$(CO) = $5\times10^{19}\,\cm2/\kkms$.  Even then, the conversion
factors would still be over four times lower than the local Milky Way
disk factor $X_{\circ}$.

Our results fit the historical downward trend of the published values
of $X$ in galaxy centers as opposed to galaxy disks.Sandstrom $\etal$
(2013) studied the disks of 26 nearby galaxies, many in common with
our survey. They derived a more or less constant factor
$X\,=\,1.95\,\times\,10^{20}$ from $\co$(2-1), FIR, and $\hi$
emission, but suggested a lower central $X$. Follow-up $J$=1-0 $\co$ and
$\thirco$ observations of nine nearby galaxies led Cormier $\etal$
(2018) to a similar result, with a low average center
$X\,\sim\,0.15\,X_{\circ}$ which, fortuitously, happens to be close to
our result. Even lower $X$ factors of about $0.05\,X_{\circ}$ are now
discussed for the more extreme case of optically thin CO gas outflows
from the centers of nearby luminous galaxies (Alatalo $\etal$ 2011;
Sakamoto $\etal$ 2014; Oosterloo $\etal$ 2017)

As the low $X$ value of the ISM in the center of our own galaxy is thus
revealed to be characteristic of galaxy centers in general, it is
instructive to take a closer look at the well-studied Central
Molecular Zone (CMZ) in the Milky Way. The density of most of the gas
in the CMZ is not very high. High-density molecular tracers other than
CO are generally subthermally excited, implying densities of only
$\sim10^{4}\,\cc$ (Jones $\etal$ 2012). Observations of $J$=1-0 $\co$
and its optically thin isotope C$^{18}$O have shown very complex
molecular gas distributions characterized by moderate or low optical
depths (Dahmen $\etal$ 1998) on large scales. Observations of several
$\co$ and $\thirco$ transitions (but not including the $J$=1-0
transition) rule out excitation by a single component and instead
suggest the superposition of various warm gas phases (Requena-Torres
$\etal$ 2012). Requena-Torres and collaborators performed a two-component LVG analysis, analogous to the one
in this paper, that suggested a dominant phase with
$T_{\rm kin}\,\approx\, 200$ K and a density
$n(\h2)\,\sim3\times10^{4}\,\cc$ and a minor phase (20-30$\%$ by mass)
with a higher $T_{\rm kin}\,\approx$ 300-500 K,
$n(\h2)\,\sim2\times10^{5}\,\cc$. The lack of the low-$J$ transition
biases their analysis to higher densities and temperatures, whereas
our coverage of all lower transitions up to $J$=3-2 or $J$=4-3 samples
the lowest temperatures in phase 1, but may underestimate phase 2
temperatures.

The bulk of the CMZ molecular gas has temperatures
$T_{\rm kin}$ = 50 - 120 K, and the average gas temperature outside
the densest clouds is $65\pm10$ K (Ao $\etal$ 2013, Ginsburg $\etal$
2016). Both the temperature range and the average temperature are very
similar to those of our sample of galaxy centers
(cf. Table\,\ref{modelpar}), for which we find a mass-weighted mean
temperature $T_{\rm kin}\,=\,55\pm5$ K. The emission-weighted mean gas
density of the galaxy centers in our sample,
$n_{\h2}\,=\,(1.8\pm0.6)\times10^{4}\,\cc$, is likewise similar to the
values obtained for the CMZ.

Requena-Torres $\etal$ (2012) concluded that the gas sampled by them is
not organized by self-gravity, is unstable against tidal disruption,
and is transient in nature. Velocity dispersions of clouds in the
Galactic center region are five times higher than those of clouds in
the disk (Miyazaki $\&$ Tsuboi, 2000). This, the widespread presence
of shocked gas emitting in the $J$=2-1 SiO line (H\"uttemeister
$\etal$ 1998), and the high but variable gas temperatures throughout
the CMZ noted by Requena-Torres $\etal$ (2012) indicate that the
dense gas is dominated by turbulent heating.  Neither heating by UV
photons nor by cosmic rays can be important on global scales in the CMZ
(Ao $\etal$ 2013, Ginsburg $\etal$ 2016), as we have also concluded
in the individual cases of NGC~253 and NGC~3690 after detailed
modeling of molecular line data (Rosenberg $\etal$ 2014a, b).

The picture that emerges of the molecular gas in the CMZ, and by
implication also in the observed galaxy centers, is very different
from the picture of the star-forming molecular gas in the disk of the
Milky Way and other galaxies. In the CMZ, molecular gas appears to
occur mostly in extended diffuse clouds with modest optical depths in
CO but with relatively high surface filling factors.  This warm gas does
not only rotate rapidly around the nucleus, but is also
continuously stirred up and rather turbulent, even though the precise
mechanisms are not yet clear (see, e.g., the review by Mills, 2017).

That CO-to-$\h2$ conversion factors are much lower for molecular gas
in galaxy centers than in galaxy disks can be understood in terms of
different physical properties. For instance, Stacey $\etal$ (1991)
already suggested that higher excitation temperatures were responsible
for the drop in $X$(CO) implied by their $\cii$ measurements. Downes
$\&$ Solomon (1998) speculated that the central CO in luminous
galaxies is only moderately opaque and highly turbulent. They also
believed it to be subthermally excited, but higher $J$ CO intensities
show this to be an oversimplification.  More recently, models and
simulations have started to elucidate the effects of different
environments on empirical quantities such as $X$. In one example, Bell
$\etal$ (2006) used photon-dominated region (PDR) time-dependent chemical models to show that
increases in density, cosmic-ray ionization rate, metallicity, and
turbulent velocity all act to depress $X$-values. Bell et al. (2007)
explicitly noted that application of their models to M~51 and NGC~6946
yielded results that are very close to those obtained by us and ascribed the low
$X$ values primarily to high density and metallicity. In another
example, Narayanan $\etal$ (2011) combined smoothed particle hydrodynamics (SPH) simulations of galaxy
disks with physical and radiative transfer ISM models and the
CO-to-$\h2$ conversion factor in star-forming and merger galaxy
disks. Although their results do not directly apply to galaxy centers,
they do draw attention to the significant dependence of $X$ on kinetic
temperature and velocity dispersion.

The bulk of the gas in galaxy centers is, like that in the CMZ,
carbon rich, moderately dense, warm, and turbulent. With respect to
$\h2$ column density, the higher CO abundance and emissivity increase
the CO line intensity, whereas lower CO mean optical depth imply lower
$\h2$ column densities relative to CO intensity.  Thus, in galaxy
centers, the molecular gas radiates in CO with much enhanced
efficiency.  The gains thus made by galaxy center molecular gas with
respect to disk gas can be quantitatively estimated from the model
dependencies $X\,\propto\,T^{-1/2}$ and $X\,\propto\,\sigma^{-1/2}$
derived for the CO-to-$\h2$ conversion factor by Shetty $\etal$
(2011). The thrice higher carbon gas phase abundance leads to a three
times lower $X$. The elevated average kinetic temperature of 55 K
lowers $X$ by a further factor of two. Most of the central CO is of
modest optical depth, and taking our cue from the five times higher
CMZ velocity dispersions, another drop in $X$ by a factor of 2.2 is to
be expected.  Taking all together, we would expect
$X\,\approx\,1.5\times10^{19}\,\cm2/\kkms$ (or $0.075\,X_{\circ}),$ which
is very close to the average value
$X\,\approx\,1.9\times10^{19}\,(\cm2/\kkms)$ from Section 8.1. This
agreement shows that the conditions causing the low value of central
$X$ are reasonably well understood. It also shows that no single cause
prevails; all three contributing factors are of a similar magnitude,
and all three are needed. 

\section{Conclusions}

\begin{enumerate}
\item We determined intensities of galaxy centers in the lowest $J$
  transitions of $\co$ and $\thirco$. Out of a total of 126 galaxies
  112, 103, 88, and 24 were measured in the $J$=1-0, $J$=2-1, $J$=3-2,
  and $J$=4-3 transitions of $\co$, respectively, as well as 89, 71,
  and 61 in the $J$=1-0, $J$=2-1, and $J$=3-2 transitions of
  $\thirco$, respectively. In 15 galaxies, only the $J$=1-0 transition
  was measured and 30 galaxies lack $\thirco$
  measurements\footnote{The raw JCMT data are publicly available from
    the CADC JCMT Science Archive at
    https://www.cadc-ccda.hia-iha.nrc-cnrc.gc.ca/en/jcmt/ . Neither
    SEST nor IRAM data are archived, but the processed galaxy data
    from this paper can be downloaded in CLASS format from
    ftp://ftp.strw.leidenuniv.nl/pub/israel/data/galaxyfiles}.
\item Multi-aperture $J$=1-0 $\co$ fluxes from the survey and the
   literature show that CO luminosities increase roughly
  linearly with observing beam size. The extrapolated CO-emitting gas
  extends to $\sim 40\%$ of the optical size $D_{25}$ and $\sim 25\%$
  of the $\hi$ size. The molecular gas is thus much more concentrated
  than the atomic gas.
\item The $\co$ $J$=1-0 to $J$=4-3 transition ladder has relative 
  intensities 1.00 : 0.92 : 0.70 : 0.57. The mean isotopologue
  $\co$-to-$\thirco$ intensity ratios are $13.0\pm0.7$ ($J$=1-0),
  $11.6\pm0.6$ ($J$=2-1), $12.8\pm0.6$ ($J$=3-2), but individual values
  may be as low as 5 and as high as 25.
\item For more than 70 galaxies, physical parameters of a
  two-phase gas were determined with the use of non-LTE radiative
  transfer models ({\it RADEX}). On average, only one-third of the
  gas-phase carbon ($32\pm8\%$) is found to reside in CO. The full
  gas-phase carbon budget was determined for 45 galaxies, using
  literature data for neutral and ionized carbon line intensities. The
  intensities of the $\ci$ and $\co$ lines are closely related; on
  average, neutral carbon C$^{\circ}$ accounts for somewhat less of
  the gas-phase carbon ($30\pm4\%$).  Somewhat more than one-third of
  the gas-phase carbon is available for ionized carbon C$^{+}$. This
  condition is met if $\cii$ emission originates in a moderately dense
  and warm ($n\geq3000\,\cc$, $T\geq100$ K) gas.
\item Averaged over a $22"$ beam, mean total hydrogen column densities
  are $N_{\rm H}\,=\,(3.3\pm0.2)\times10^{21}\,\cm2$ and mean
  molecular hydrogen column densities are
  $N(\h2)\,=\,(1.5\pm0.2)\times10^{21}\,\cm2$. Total gas masses of
  central molecular zones up to one kiloparsec radius are typically a few
  times $10^{7}$ M$_{\odot}$, whereas the total molecular gas
  masses of the inner disk are typically $10^{8}$ M$_{\odot}$.
\item The observed $J$=1-0 CO intensities and the derived $N(\h2)$
  values yield CO-to$\h2$ conversion factors with a well-defined mean
  value $X(CO)\,=\,(1.9\pm0.2)\times10^{19}\,\cm2/\kkms$. This is a
  factor of ten below the `standard' solar neighborhood Milky Way
  factor $X_{\rm MW}$. The mean $\ci$-to$\h2$ conversion factor is
  $X[CI]\,=\,(9\pm2)\times10^{19}\,\cm2/\kkms$.  There is no
  meaningful conversion factor for $\cii$.
\item Use of a conversion factor based on $J$=3-2 $\co$ line
  intensities yields results that are better than those obtained with $\ci$,
  but not as good as those derived from the $J$=1-0 $\co$ line.
\item From comparisons with the well-studied CMZ in
  the Milky Way galaxy, it appears that the order-of-magnitude
  decrease of the CO-to-$\h2$ conversion factor in the central
  molecular zones of nearby other galaxies with respect to canonical
  galaxy disk conversion factors is caused in equal parts by the
  higher gas-phase carbon abundances in galaxy centers, elevated
  kinetic gas temperatures, and high molecular cloud velocity
  dispersions. 
\end{enumerate}

\section*{Acknowledgements}

Some of the SEST and IRAM observations and most of the JCMT
observations were made in service mode. With gratitude I acknowledge
my indebtedness to the many colleagues and facility staff, above all
the JCMT operators, who over many years helped to collect the large
and unique data base described in this paper. It is fitting to
commemorate the generous advice and assistance of Lars E.B. Johansson
(SEST) and Fred Baas (JCMT), both of whom sadly died before they could
see the results. I also thank Thijs van der Hulst for supplying his
unpublished SEST observations of NGC~613, NGC~1097, and NGC~1365 as
well as Rodrigo Herrera-Camus for sharing his machine-readable $\cii$
data in advance of publication. Critical comments by the referee,
Jonathan Braine, led to a substantial improvement of the paper. We
made extensive use of the JCMT archive part of the facilities of the
Canadian Astronomy Data Centre operated by the National Research
Council of Canada with the support of the Canadian Space Agency.

\begin{appendix}

\section{CO survey results and literature data}  
  
\subsection{$J$=1-0 fluxes versus observing beam size}

% Table A3 Sample
\begin{table}
\begin{center}
{\small %
\caption[]{\label{flux10comp}$J$=1-0 fluxes in different beams} 
\begin{tabular}{lr|rrr|l}
  \noalign{\smallskip}
  \hline
  \noalign{\smallskip}
NGC  &Tel$^a$& Beam  &  Flux$^b$       &          & Ref\\
     &       & ($"$  &  $Jy\kms$       &          &    \\ 
 (1) & (2)   & (3)   &  (4)            &  (5)     &(6)  \\
  \noalign{\smallskip}
  \hline
  \noalign{\smallskip}
 134 & SEST  & 45    &   432           &        & 18 \\
     & SEST  & 45    &   297           &        & 20 \\
     & SEST  & 45    &   323           &        & TP \\
     & ave   & 45    &   351$\pm$41    & (0.92) &    \\
 253 & BTL   & total & 44000           &        &  9 \\
     & FCRAO & total & 21060           &        & 15 \\
     & ave   & total & 32530$\pm$11470 &        &    \\
     & FCRAO & 45    & 10790           &        & 15 \\
     & FCRAO & 45    &  6767           &        & 61 \\
     & SEST  & 45    &  7130           &        & 64 \\
     & SEST  & 45    &  6067           &        & TP \\
     & ave   & 45    &  7689$\pm$1057  & (0.79) &    \\
     & IRAM  & 22    &  5295           &        & 61 \\
     & IRAM  & 22    &  4289           &        & 63 \\
     & IRAM  & 22    &  4841           &        & TP \\
     & ave   & 22    &  4808$\pm$291   & (1.01) &    \\
     & NOB   & 16    &   595           &        & 11 \\
     & NOB   & 16    &   756           &        & 62 \\
     & ave   & 16    &   675$\pm$81    &        &    \\
 278 & FCRAO & total &   480           &        & 15 \\
     & FCRAO & 45    &   328           &        & 15 \\
     & OSO   & 33    &   101           &        & 18 \\
     & IRAM  & 22    &    85           &        & 23 \\
     & IRAM  & 22    &    96           &        & TP \\
     & ave   & 22    &    91$\pm$6     & (0.93) &    \\
     & NOB   & 16    &    43           &        & 11 \\
     & NOB   & 16    &    36           &        & 13 \\
 520 & FCRAO & total &  1260           &        & 15 \\
     & NRAO  & total &  1260           &        &  1 \\
     & ave   & total &  1260$\pm$0     &        &    \\
     & FCRAO & 45    &   903           &        &  5 \\
     & FCRAO & 45    &   696           &        & 15 \\
     & SEST  & 45    &   531           &        & TP \\
     & ave   & 45    &   710$\pm$108   & (0.75) &    \\ 
     & OSO   & 33    &   344           &        & 20 \\
     & IRAM  & 22    &   211           &        & 24 \\
     & IRAM  & 22    &   278           &        & TP \\
     & ave   & 22    &   245$\pm$34    & (1.14) &    \\
     & NOB   & 16    &    94           &        & 11 \\
 613 & SEST  & 45    &   661           &        & 18 \\
     & SEST  & 45    &   448           &        & 25 \\
     & SEST  & 45    &   473           &        & TP \\
     & ave   & 45    &   527$\pm$67    & (0.90) &    \\
     & IRAM  & 22    &   328           &        & TP \\    
 628 & NRAO  & total &  2611           &        &  2 \\
     & FCRAO & total &  2160           &        & 15 \\
     & ave   & total &  2386$\pm$226   &        &    \\
     & NRAO  & 55    &   198           &        &  2 \\
     & NRAO  & 55    &   207           &        &  3 \\
     & ave   & 55    &   203$\pm$5     &        &    \\
     & FCRAO & 45    &    96           &        & 15 \\
     & IRAM  & 22    &    31           &        & TP \\
     & IRAM  & 22    &    19           &        & 23 \\
     & ave   & 22    &    25$\pm$6     & (0.76) &    \\
     & NOB   & 16    &    10           &        & 11 \\
     & NOB   & 16    &    15           &        & 13 \\
     & ave   & 16    &    13$\pm$3     &        &    \\
     & BIMA  & 6.2   &    10           &        & 58 \\
 660 & FCRAO & total &    2840         &        & 15 \\
     & BTL   & total &    2800         &        &  9 \\
     & ave   & total &    2820$\pm$20  &        &    \\
     & NRAO  & 55    &     497         &        &  4 \\
     & FCRAO & 45    &    1793         &        &  5 \\
     & FCRAO & 45    &     884         &        & 15 \\
     \noalign{\smallskip}
     \hline
     \noalign{\smallskip}
\end{tabular}
}% 
\end{center} 
\end{table}                                                        

\addtocounter{table}{-1}
\begin{table}
\begin{center}
\caption[]{continued} 
{\small %
\begin{tabular}{lr|rrr|l}
  \noalign{\smallskip}
  \hline
  \noalign{\smallskip}
NGC  & Tel   & Beam  &  Flux           &          & Ref\\
     &       & ($"$  &  $Jy\kms$       &          &    \\ 
 (1) & (2)   & (3)   &  (4)            &  (5)     &(6)  \\
  \noalign{\smallskip}
  \hline
  \noalign{\smallskip}
 660 & SEST  & 45    &     730         &        & TP \\
     & ave   & 45    &    1136$\pm$332 & (0.64) &    \\
     & OSO   & 33    &    1271         &        & 20 \\
     & IRAM  & 22    &     724         &        & TP \\
     & IRAM  & 22    &     465         &        & 23 \\
     & ave   & 22    &     595$\pm$130 & (1.22) &    \\
 695 & FCRAO & total &     220         &        & 15 \\
     & NRAO  & 55    &     210         &        &  7 \\
     & FCRAO & 45    &     200         &        & 15 \\
 891 & BTL   & total &   13000         &        &  9 \\
     & NRAO  & total &    7105         &        &  2 \\
     & FCRAO & total &    4570         &        & 15 \\
     & ave   & total &    8225$\pm$2497&        &    \\
     & NRAO  & 55    &    1355         &        &  2 \\
     & FCRAO & 45    &     607         &        & 15 \\
     & 0SO   & 33    &     542         &        & 18 \\
     & IRAM  & 22    &     644         &        & TP \\
     & IRAM  & 22    &     451         &        & 23 \\
     & ave   & 22    &     548$\pm$97  & (1.18) &    \\
 908 & FCRAO & total &    1330         &        & 15 \\
     & FCRAO & 45    &     436         &        & 15 \\
     & SEST  & 45    &     234         &        & 18 \\
     & SEST  & 45    &     452         &        & TP \\
     & IRAM  & 22    &     140         &        & TP \\
     & NOB   & 16    &      22         &        & 11 \\
     & ave   & 45    &     374$\pm$70  & (1.21) &    \\
 972 & FCRAO & total &     690         &        & 15 \\
     & FCRAO & 45    &     406         &        & 15 \\
     & IRAM  & 22    &     314         &        & TP \\
1055 & FCRAO & total &    2800         &        & 15 \\
     & FCRAO & 45    &     947         &        & 15 \\
     & SEST  & 45    &     544         &        & TP \\
     & ave   & 45    &     746$\pm$202 & (0.73) &    \\
     & OSO   & 33    &     260         &        & 18 \\
     & IRAM  & 23    &     361         &        & TP \\
     & IRAM  & 23    &     216         &        & 23 \\
     & ave   & 23    &     289$\pm$73  & (1.25) &    \\
1068 & BTL   & total &   19000         &        &  9 \\
     & FCRAO & total &   21060         &        & 15 \\
     & ave   & total &   20030$\pm$1030&        &    \\
     & NRAO  & 55    &    2555         &        &  6 \\
     & NRAO  & 55    &    2520         &        & 21 \\
     & FCRAO & 45    &    2306         &        & 15 \\
     & IRAM  & 22    &     790         &        & TP \\
     & IRAM  & 22    &    1025         &        & 23 \\
     & ave   & 22    &     908$\pm$118 & (0.87) &    \\
     & NOB   & 16    &     229         &        & 66 \\
     & BIMA  & 7.1   &     245         &        & 58 \\
     & BIMA  & 3.9   &      80         &        & 65 \\
1084 & FCRAO & total &     920         &        & 15 \\
     & FCRAO & 45    &     266         &        & 15 \\
     & SEST  & 45    &     220         &        & 18 \\
     & SEST  & 45    &     374         &        & TP \\
     & ave   & 45    &     287$\pm$46  & (1.30) &    \\
     & IRAM  & 22    &     143         &        & TP \\
     & IRAM  & 22    &     146         &        & 23 \\   
     & ave   & 23    &     145$\pm$2   & (0.99) &    \\
1097 & BTL   & total &    5400         &        &  9 \\
     & FCRAO & 45    &     158         &        & 15 \\
     & SEST  & 45    &     920         &        & TP \\
     & IRAM  & 22    &     639         &        & TP \\
1365 & NRAO  & 55    &    1680         &        &  6 \\
     & NRAO  & 55    &    1700         &        & 21 \\
     & NRAO  & 55    &    1670         &        & 69 \\
     \noalign{\smallskip}
     \hline
     \noalign{\smallskip}
\end{tabular}
}% 
\end{center} 
\end{table}                                                        

\addtocounter{table}{-1}
\begin{table}
\begin{center}
\caption[]{continued} 
{\small %
\begin{tabular}{lr|rrr|l}
  \noalign{\smallskip}
  \hline
  \noalign{\smallskip}
NGC  & Tel   & Beam  &  Flux           &        & Ref\\
     &       & ($"$  &  $Jy\kms$       &        &    \\ 
 (1) & (2)   & (3)   &  (4)            &  (5)   &(6)  \\
  \noalign{\smallskip}
  \hline
  \noalign{\smallskip}
1365 & ave   & 55    &    1683$\pm$9   &        &    \\
     & SEST  & 45    &    2035         &        & 18 \\
     & SEST  & 45    &    2117         &        & 68 \\
     & SEST  & 45    &    1928         &        & TP \\
     & ave   & 45    &    2026$\pm$55  & (0.95) &    \\
     & IRAM  & 22    &    1222         &        & TP \\
1433 & SEST  & 45    &     321         &        & 25 \\
     & SEST  & 45    &     276         &        & TP \\
     & ave   & 45    &     299$\pm$22  & (0.92) &    \\
1482 & FCRAO & total &     560         &        & 15 \\
     & NRAO  & 55    &     284         &        &  7 \\
     & FCRAO & 45    &     462         &        & 15 \\
     & SEST  & 45    &     645         &        & 18 \\
     & SEST  & 45    &     443         &        & 19 \\
     & SEST  & 45    &     442         &        & 22 \\
     & SEST  & 45    &     293         &        & TP \\
     & ave   & 45    &     460$\pm$102 & (0.64) &    \\
     & IRAM  & 22    &     151         &        & 22 \\
     & NOB   & 16    &      89         &        & 11 \\
1614 & FCRAO & total &     290         &        & 15 \\
     & NRAO  & 55    &     301         &        &  7 \\
     & FCRAO & 45    &     242         &        & 15 \\
     & SEST  & 45    &     163         &        & 17 \\
     & SEST  & 45    &     291         &        & 18 \\
     & SEST  & 45    &     227         &        & 20 \\
     & SEST  & 45    &     270         &        & TP \\
     & ave   & 45    &     239$\pm$22  & (1.13) &    \\
     & IRAM  & 22    &     203         &        & TP \\
     & IRAM  & 22    &      61         &        & 17 \\
     & IRAM  & 22    &     212         &        & 80 \\
     & ave   & 22    &     159$\pm$49  & (1.28) &    \\
1667 & NRAO  & 55    &     130         &        &  7 \\
     & NRAO  & 55    &     280         &        &  6 \\
     & NRAO  & 55    &     244         &        & 21 \\
     & ave   & 55    &     218$\pm$45  &        &    \\
1672 & SEST  & 45    &     855         &        & 25 \\
     & SEST  & 45    &     444         &        & TP \\
     & ave   & 45    &     650$\pm$205 & (0.68) &    \\ 
1808 & SEST  & 45    &    1512         &        & 20 \\
     & SEST  & 45    &    1604         &        & 26 \\
     & SEST  & 45    &    1739         &        & TP \\
     & ave   & 45    &    1618$\pm$66  & (1.07) &    \\ 
     & IRAM  & 22    &     635         &        & TP \\
2146 & FCRAO & total &    2840         &        & 15 \\
     & FCRAO & 45    &    1579         &        & 15 \\
     & OSO   & 33    &     824         &        & 18 \\
     & OSO   & 33    &    1042         &        & 20 \\
     & ave   & 33    &     933$\pm$109 &        &    \\
     & IRAM  & 22    &     879         &        & TP \\
     & IRAM  & 22    &     555         &        & 23 \\
     & ave   & 22    &   717$\pm$162   & (1.23) &    \\
2273 & FCRAO & total &     160         &        & 15 \\
     & FCRAO & 45    &     137         &        & 15 \\
     & NRAO  & 45    &     105         &        & 21 \\
     & ave   & 45    &     121$\pm$16  &        &    \\
     & IRAM  & 22    &      78         &        & TP \\
2369 & SEST  & 45    &     674         &        & 18 \\
     & SEST  & 45    &     702         &        & 20 \\
     & SEST  & 45    &     493         &        & TP \\
     & ave   & 45    &     623$\pm$80  & (0.79) &    \\
2397 & SEST  & 45    &     276         &        & 19 \\
     & SEST  & 45    &     269         &        & 22 \\
     & SEST  & 45    &     314         &        & TP \\
     & ave   & 45    &     286$\pm$17  & (1.10) &    \\
     \noalign{\smallskip}
     \hline
     \noalign{\smallskip}
\end{tabular}
}% 
\end{center} 
\end{table}                                                        

\addtocounter{table}{-1}
\begin{table}
\begin{center}
\caption[]{continued} 
{\small %
\begin{tabular}{lr|rrr|l}
  \noalign{\smallskip}
  \hline
  \noalign{\smallskip}
NGC  & Tel   & Beam  &  Flux           &        & Ref\\
     &       & ($"$  &  $Jy\kms$       &        &    \\ 
 (1) & (2)   & (3)   &  (4)            &  (5)   &(6)  \\
  \noalign{\smallskip}
  \hline
  \noalign{\smallskip}
2559 & FCRAO & total &    1100         &        & 15 \\
     & FCRAO & 45    &     743         &        & 15 \\ 
     & SEST  & 45    &     626         &        & 18 \\
     & SEST  & 45    &     607         &        & TP \\
     & ave   & 45    &     659$\pm$43  & (0.92) &    \\
     & IRAM  & 22    &     369         &        & TP \\
2623 & FCRAO & total &     170         &        & 15 \\   
     & NRAO  & 55    &     170         &        &  1 \\
     & NRAO  & 55    &     170         &        &  5 \\
     & NRAO  & 55    &     168         &        &  7 \\
     & ave   & 55    &     169$\pm$1   &        &    \\
     & FCRAO & 45    &     202         &        &  5 \\
     & FCRAO & 45    &     161         &        & 15 \\
     & ave   & 45    &     182$\pm$21  &        &    \\
     & IRAM  & 22    &      86         &        & TP \\
     & NOB   & 16    &      96         &        & 81 \\
2798 & FCRAO & total &     440         &        & 15 \\
     & FCRAO & 45    &     246         &        & 15 \\
     & OSO   & 33    &      75         &        & 18 \\
     & NOB   & 16    &      27         &        & 14 \\
2903 & BTL   & total &    4900         &        &  9 \\
     & NRAO  & total &    2783         &        &  2 \\
     & FCRAO & total &    2740         &        & 15 \\
     & ave   & total &    3474$\pm$712 &        &    \\
     & PMO   & 55    &    783          &        & 82 \\
     & NRAO  & 55    &    770          &        &  2 \\
     & ave   & 55    &    777$\pm$7    &        &    \\
     & FCRAO & 45    &    553          &        & 15 \\
     & IRAM  & 22    &    375          &        & TP \\
     & NOB   & 16    &    285          &        & 13 \\
     & BIMA  &  6    &    129          &        & 58 \\
2992 & NRAO  & 55    &    116          &        &  5 \\
     & NRAO  & 55    &    154          &        & 21 \\
     & ave   & 55    &    135$\pm$19   &        &    \\
     & SEST  & 45    &    155          &        & TP \\
     & NOB   & 16    &     25          &        & 13 \\
3034 & FCRAO & total &  18240          &        & 15 \\
     & PMO   & 55    &   7070          &        & 82 \\
     & FCRAO & 45    &   6670          &        & 15 \\
     & OSO   & 33    &   3116          &        & 71 \\
     & IRAM  & 22    &   3007          &        & 70 \\
     & IRAM  & 22    &   3196          &        & TP \\
     & ave   & 22    &   3102$\pm$95   & (1.03) &    \\
     & NOB   & 16    &   1001          &        & 11 \\
3079 & FCRAO & total &   2280          &        & 15 \\
     & NRAO  & 55    &   1509          &        & 21 \\
     & NRAO  & 55    &   1684          &        & 76 \\
     & ave   & 55    &   1597$\pm$88   &        &    \\
     & FCRAO & 45    &   1602          &        & 15 \\
     & SEST  & 45    &    845          &        & 20 \\
     & ave   & 45    &   1224$\pm$379  &        &    \\
     & OSO   & 33    &    870          &        & 18 \\
     & IRAM  & 22    &    996          &        & 23 \\
     & IRAM  & 22    &   1019          &        & 73 \\
     & IRAM  & 22    &   1105          &        & TP \\
     & IRAM  & 22    &   1153          &        & 80 \\
     & ave   & 23    &   1068$\pm$37   & (1.03) &    \\
     & OVRO  & 7     &    362          &        & 75 \\
     & NMA   & 1.7   &    125          &        & 74 \\
3175 & SEST  & 45    &    371          &        & 18 \\
     & SEST  & 45    &    357          &        & TP \\
     & ave   & 45    &    364$\pm$7    & (0.98) &    \\
     & IRAM  & 22    &    201          &        & TP \\
3227 & BTL   & total &    960          &        &  9 \\
     \noalign{\smallskip}
     \hline
     \noalign{\smallskip}
\end{tabular}
}% 
\end{center} 
\end{table}                                                        

\addtocounter{table}{-1}
\begin{table}
\begin{center}
\caption[]{continued} 
{\small %
\begin{tabular}{lr|rrr|l}
  \noalign{\smallskip}
  \hline
  \noalign{\smallskip}
NGC  & Tel   & Beam  &  Flux           &        & Ref\\
     &       & ($"$  &  $Jy\kms$       &        &    \\ 
 (1) & (2)   & (3)   &  (4)            &  (5)   & (6)\\
  \noalign{\smallskip}
  \hline
  \noalign{\smallskip}
3227 & NRAO  & 55    &    450          &        &  1 \\
     & NRAO  & 55    &    432          &        & 21 \\
     & IRAM  & 22    &    254          &        & 23 \\
     & IRAM  & 22    &    290          &        & TP \\
     & ave   & 22    &    272$\pm$18   & (1.07) &    \\
     & NOB   & 16    &     18          &        & 14 \\
3256 & SEST  & 45    &   1539          &        & 20 \\
     & SEST  & 45    &   1305          &        & 27 \\
     & SEST  & 45    &   1297          &        & TP \\
     & ave   & 45    &   1380$\pm$79   & (0.94) &    \\
3281 & NRAO  & 55    &  $<$84          &        &  6 \\
     & NRAO  & 55    &  $<$58          &        & 21 \\
     & SEST  & 45    &     26          &        & TP \\
3310 & FCRAO & total &    140          &        & 15 \\
     & NRAO  & 55    &     85          &        & 76 \\
     & FCRAO & 45    &     50          &        & 15 \\
     & IRAM  & 22    &     17          &        & 23 \\
     & IRAM  & 22    &     37          &        & TP \\
     & ave   & 22    &     27$\pm$10   & (1.37) &    \\
     & NOB   & 16    &     22          &        & 14 \\
3351 & FCRAO & total &    700          &        & 15 \\
     & FCRAO & 45    &    338          &        & 15 \\
     & OSO   & 33    &    233          &        & 18 \\
     & IRAM  & 22    &     80          &        & 23 \\
     & BIMA  & 6.3   &     31          &        & 58 \\
3504 & BTL   & total &   1500          &        &  9 \\
     & FCRAO & total &    410          &        & 15 \\
     & ave   & total &    807$\pm$426  &        &    \\
     & NRAO  & 55    &    511          &        &  4 \\
     & NRAO  & 55    &    193          &        &  5 \\
     & ave   & 55    &    352$\pm$159  &        &    \\
     & FCRAO & 45    &    317          &        & 15 \\
     & SEST  & 45    &    390          &        & 22 \\
     & ave   & 45    &    365$\pm$30   &        &    \\
     & IRAM  & 22    &    265          &        & TP \\
     & NOB   & 16    &    247          &        & 11 \\
3556 & FCRAO & total &    850          &        & 15 \\
     & FCRAO & 45    &    330          &        & 15 \\
     & OSO   & 33    &    149          &        & 18 \\
     & IRAM  & 22    &    253          &        & TP \\
     & NOB   & 16    &     38          &        & 11 \\
3593 & FCRAO & total &    910          &        & 15 \\
     & NRAO  & total &    634          &        & 2  \\
     & ave   & total &    772$\pm$138  &        &    \\
     & NRAO  & 55    &    567          &        & 2  \\
     & FCRAO & 45    &    431          &        & 15 \\
     & SEST  & 45    &    463          &        & TP \\
     & ave   & 45    &    447$\pm$16   & (1.04) &    \\
     & OSO   & 33    &    267          &        & 18 \\
     & IRAM  & 22    &    297          &        & TP \\
     & NOB   & 16    &    113          &        & 11 \\
3620 & SEST  & 45    &    902          &        & 18 \\
     & SEST  & 45    &    894          &        & TP \\
     & ave   & 45    &    898$\pm$4    & (1.000 &    \\
3627 & BTL   & total &   8300          &        &  9 \\
     & FCRAO & total &   4660          &        & 15 \\
     & ave   & total &   6480$\pm$1820 &        &    \\
     & PMO   & 55    &    542          &        & 82 \\
     & NRAO  & 55    &   1150          &        & 29 \\
     & FCRAO & 45    &    786          &        & 15 \\
     & SEST  & 45    &    516          &        & TP \\
     & PDBI  & 42    &    668          &        & 29 \\
     & ave   & 45    &    657$\pm$78   & (0.79) &    \\
     & PDBI  & 22    &    359          &        & 29 \\
     \noalign{\smallskip}
     \hline
     \noalign{\smallskip}
\end{tabular}
}% 
\end{center} 
\end{table}                                                        

\addtocounter{table}{-1}
\begin{table}
\begin{center}
\caption[]{continued} 
{\small %
\begin{tabular}{lr|rrr|l}
  \noalign{\smallskip}
  \hline
  \noalign{\smallskip}
NGC  & Tel   & Beam  &  Flux           &          & Ref\\
     &       & ($"$  &  $Jy\kms$       &          &    \\ 
 (1) & (2)   & (3)   &  (4)            &  (5)     &(6)  \\
  \noalign{\smallskip}
  \hline
  \noalign{\smallskip}
3627 & IRAM  & 22    &    343          &        & 29 \\
     & IRAM  & 22    &    423          &        & 23 \\
     & IRAM  & 22    &    378          &        & 30 \\
     & IRAM  & 22    &    350          &        & TP \\
     & ave   & 22    &    370$\pm$14   & (0.95) &    \\
     & NOB   & 16    &    168          &        & 28 \\
     & BIMA  & 6.5   &    129          &        & 58 \\
3628 & BTL   & total &  10000          &        &  9 \\
     & FCRAO & total &   3800          &        & 15 \\
     & ave   & total &   5867$\pm$2530 &        &    \\
     & PMO   & 55    &   1130          &        & 82 \\
     & NRAO  & 55    &    935          &        &  4 \\
     & ave   & 55    &   1033$\pm$98   &        &    \\
     & FCRAO & 45    &   1604          &        & 15 \\
     & SEST  & 45    &   1416          &        & TP \\
     & ave   & 45    &   1510$\pm$94   & (0.94) &    \\
     & IRAM  & 22    &    954          &        & TP \\
     & IRAM  & 22    &    611          &        & 23 \\
     & ave   & 22    &    783$\pm$171  & (1.22) &    \\
3690 & FCRAO & total &    610          &        & 15 \\
     & NRAO  & 55    &    610          &        &  1 \\
     & NRAO  & 55    &    140          &        &  5 \\
     & FCRAO & 45    &    449          &        & 15 \\
     & IRAM  & 22    &    264          &        & 24 \\
     & IRAM  & 22    &    323          &        & TP \\
     & ave   & 22    &    294$\pm$29   & (1.10) &    \\
     & NOB   & 16    &     29          &        & 11 \\
4030 & FCRAO & total &   1050          &        & 15 \\
     & FCRAO & 45    &    271          &        & 15 \\
     & SEST  & 45    &    437          &        & TP \\
     & ave   & 45    &    354$\pm$83   & (1.23) &    \\
     & IRAM  & 22    &    198          &        & TP \\
4038 & FCRAO & total &   1150          &        & 15 \\
     & NRAO  & 55    &    557          &        &  5 \\
     & FCRAO & 45    &    676          &        & 15 \\
     & SEST  & 45    &    567          &        & 20 \\
     & ave   & 45    &    622$\pm$55   &        &    \\
     & IRAM  & 22    &    220          &        & TP \\
     & NOB   & 16    &     67          &        & 11 \\
     & OVRO  & 3.9   &     91          &        & 31 \\
4039 & FCRAO & total &    920          &        & 15 \\
     & FCRAO & 45    &    685          &        & 15 \\
     & SEST  & 45    &    594          &        & 20 \\
     & ave   & 45    &    640$\pm$46   &        &    \\
     & IRAM  & 22    &    214          &        & TP \\
     & NOB   & 16    &    173          &        & 11 \\
     & OVRO  & 3.9   &     32          &        & 31 \\
4051 & BTL   & total &    790          &        &  9 \\
     & FCRAO & total &    740          &        & 15 \\
     & ave   & total &    765$\pm$25   &        &    \\
     & NRAO  & 55    &    259          &        & 21 \\
     & FCRAO & 45    &    218          &        & 15 \\
     & IRAM  & 22    &    178          &        & TP \\
     & BIMA  & 5.9   &     39          &        & 58 \\
4102 & FCRAO & total &    660          &        & 15 \\
     & NRAO  & 55    &    350          &        &  5 \\
     & FCRAO & 45    &    404          &        & 15 \\
     & OSO   & 33    &    339          &        & 18 \\
     & IRAM  & 22    &    351          &        & TP \\
     & NOB   & 16    &    277          &        & 13 \\
4254 & BTL   & total &   2611          &        &  8 \\
     & FCRAO & total &   3000          &        & 16 \\ 
     & FCRAO & total &   2830          &        & 10 \\
     & ave   & total &   2814$\pm$113  &        &    \\
     \noalign{\smallskip}
     \hline
     \noalign{\smallskip}
\end{tabular}
}% 
\end{center} 
\end{table}

\addtocounter{table}{-1}
\begin{table}
\begin{center}
\caption[]{continued} 
{\small %
\begin{tabular}{lr|rrr|l}
  \noalign{\smallskip}
  \hline
  \noalign{\smallskip}
NGC  & Tel   & Beam  &  Flux           &        & Ref\\
     &       & ($"$  &  $Jy\kms$       &        &    \\ 
 (1) & (2)   & (3)   &  (4)            &  (5)   &(6)  \\
  \noalign{\smallskip}
  \hline
   \noalign{\smallskip}
4254 & BTL   & 100   &   1547          &        &  8 \\
     & FCRAO & 45    &    445          &        & 16 \\
     & SEST  & 45    &    758          &        & TP \\
     & ave   & 45    &    602$\pm$157  & (1.26) &    \\
     & IRAM  & 22    &    201          &        & TP \\
     & NOB   & 16    &    113          &        & 11 \\
     & NMA   & 4.2   &     17          &        & 31 \\
     & NMA   & 2.6   &      8          &        & 12 \\
4258 & FCRAO & total &   1240          &        & 15 \\
     & NRAO  & 55    &    906          &        &  3 \\
     & FCRAO & 45    &    301          &        & 15 \\
     & SEST  & 45    &    356          &        & TP \\
     & ave   & 45    &    329$\pm$28   & (1.08) &    \\
     & IRAM  & 22    &    320          &        & 33 \\
     & NOB   & 16    &    221          &        & 34 \\
     & BIMA  & 5.7   &    103          &        & 58 \\
     & PDBI  & 3.9   &     50          &        & 32 \\
4293 & FCRAO & total &    279          &        & 16 \\
     & FCRAO & 45    &    265          &        & 16 \\
     & IRAM  & 22    &    169          &        & TP \\
     & NOB   & 16    &    161          &        & 11 \\    
4303 & BTL   & total &   1332          &        &  8 \\
     & FCRAO & total &   2280          &        & 16 \\
     & FCRAO & total &   1920          &        & 10 \\
     & ave   & total &   1844$\pm$339  &        &    \\
     & BTL   & 100   &   1148          &        &  8 \\
     & NRAO  & 55    &    238          &        &  4 \\
     & NRAO  & 55    &    588          &        & 76 \\
     & NMA   & 45    &    494          &        & 36 \\
     & FCRAO & 45    &    500          &        & 16 \\
     & ave   & 45    &    497$\pm$3    &        &    \\
     & IRAM  & 22    &    259          &        & TP \\
     & NOB   & 16    &    166          &        & 11 \\
     & NMA   & 13    &    193          &        & 35 \\
     & BIMA  & 6.4   &     65          &        & 58 \\
     & NMA   & 2.3   &     25          &        & 12 \\
4321 & BTL   & total &   4412          &        &  8 \\
     & FCRAO & total &   3340          &        & 16 \\
     & FCRAO & total &   2390          &        & 10 \\
     & ave   & total &   3381$\pm$584  &        &    \\
     & BTL   & 100   &   1423          &        &  8 \\
     & FCRAO & 45    &    630          &        & 16 \\
     & SEST  & 45    &    448          &        & TP \\
     & ave   & 45    &    539$\pm$9    & (0.83) &    \\
     & IRAM  & 22    &    392          &        & TP \\
     & IRAM  & 22    &    367          &        & 23 \\
     & IRAM  & 22    &    384          &        & 36 \\
     & ave   & 22    &    381$\pm$7    & (1.03) &    \\
     & NOB   & 16    &    151          &        & 11 \\
     & NOB   & 16    &     88          &        & 39 \\
     & ave   & 16    &    120$\pm$32   &        &    \\
     & BIMA  &  5.9  &     48          &        & 58 \\ 
     & ALMA  &  3.5  &     41          &        & 44 \\
     & NMA   &  2.5  &     27          &        & 38 \\
     & PDBI  &  1.6  &     11          &        & 37 \\
4385 & FCRAO & total &     70          &        & 15 \\
     & FCRAO & 45    &     54          &        & 15 \\
4388 & BTL   & total &    992          &        &  8 \\
     & FCRAO & total &    230          &        & 16 \\
     & ave   & total &    611$\pm$381  &        &    \\
     & BTL   & 100   &    227          &        &  8 \\
     & NRAO  & 55    &    233          &        & 21 \\
     & FCRAO & 45    &    113          &        & 16 \\
     & SEST  & 45    &    163          &        & TP \\
     \noalign{\smallskip}
     \hline
     \noalign{\smallskip}
\end{tabular}
}% 
\end{center} 
\end{table}                                                        

\addtocounter{table}{-1}
\begin{table}
\begin{center}
\caption[]{continued} 
{\small %
\begin{tabular}{lr|rrr|l}
  \noalign{\smallskip}
  \hline
  \noalign{\smallskip}
NGC  & Tel   & Beam  &  Flux           &        & Ref\\
     &       & ($"$  &  $Jy\kms$       &        &    \\ 
 (1) & (2)   & (3)   &  (4)            &  (5)   &(6)  \\
  \noalign{\smallskip}
  \hline
  \noalign{\smallskip}
4388 & ave   & 45    &    138$\pm$25   & (1.18) &    \\
4414 & FCRAO & total &    2740         &        & 15 \\
     & NRAO  & total &    1498         &        &  2 \\
     & ave   & total &    2119$\pm$621 &        &    \\
     & NRAO  & 55    &     973         &        &  2 \\
     & FCRAO & 45    &     729         &        & 15 \\
     & OSO   & 33    &     271         &        & 18 \\
     & IRAM  & 22    &     242         &        & TP \\
     & IRAM  & 22    &     268         &        & 23 \\
     & IRAM  & 22    &     249         &        & 40 \\
     & ave   & 22    &     253$\pm$10  & (0.96) &    \\  
     & NOB   & 16    &     190         &        & 11 \\
4457 & FCRAO & total &     490         &        & 15 \\
     & FCRAO & 45    &     197         &        & 15 \\
     & IRAM  & 22    &     139         &        & TP \\
4527 & BTL   & total &    2794         &        &  8 \\
     & FCRAO & total &    1800         &        & 15 \\
     & ave   & total &    2247$\pm$447 &        &    \\
     & BTL   & 100   &    2008         &        &  8 \\
     & FCRAO & 45    &    1260         &        & 10 \\
     & FCRAO & 45    &     662         &        & 15 \\
     & SEST  & 45    &     656         &        & TP \\
     & ave   & 45    &     859$\pm$200 & (0.76) &    \\
     & IRAM  & 22    &     414         &        & TP \\
     & NOB   & 16    &      62         &        & 11 \\
     & NMA   & 7.3   &     112         &        & 43 \\
     & NMA   & 4.6   &      95         &        & 42 \\
     & NMA   & 3.0   &      47         &        & 42 \\
4536 & FCRAO & total &     740         &        & 16 \\
     & NRAO  & 55    &     368         &        &  5 \\
     & FCRAO & 45    &     390         &        & 10 \\
     & FCRAO & 45    &     424         &        & 16 \\
     & SEST  & 45    &     280         &        & TP \\
     & ave   & 45    &     365$\pm$53  & (0.77) &    \\
     & OSO   & 33    &     288         &        & 18 \\
     & IRAM  & 22    &     290         &        & TP \\
     & NOB   & 16    &     290         &        & 11 \\
     & NMA   & 2.1   &      28         &        & 12 \\
4565 & BTL   & 100   &     545         &        & 47 \\
     & OSO   & 33    &     138         &        & 18 \\
     & IRAM  & 22    &      56         &        & 23 \\
     & NOB   & 16    &      35         &        & 46 \\ 
4593 & NRAO  & 55    &    $<$175       &        &  6 \\
     & NRAO  & 55    &      86         &        & 21 \\
     & SEST  & 45    &      32         &        & TP \\
     & IRAM  & 22    &      35         &        & TP \\
4631 & BTL   & total &    5200         &        &  9 \\
     & FCRAO & total &    1740         &        & 15 \\
     & PMO   & 55    &     546         &        & 82 \\
     & NRAO  & 55    &     329         &        &  4 \\
     & FCRAO & 45    &     549         &        & 15 \\
     & IRAM  & 22    &     227         &        & 19 \\
     & IRAM  & 22    &     206         &        & TP \\
     & ave   & 22    &     217$\pm$10  & (0.95) &    \\
4647 & BTL   & total &    2872         &        &  8 \\
     & FCRAO & total &     600         &        & 16 \\
     & ave   & total &    1736$\pm$1136&        &    \\
     & BTL   & 100   &     712         &        &  8 \\
     & FCRAO & 45    &     210         &        & 16 \\
     & NOB   & 16    &      62         &        & 11 \\
4666 & SEST  & total &    2780         &        & 48 \\
     & FCRAO & total &    2130         &        & 15 \\
     & ave   & total &    2455$\pm$325 &        &    \\
     & FCRAO & 45    &     696         &        & 15 \\
     \noalign{\smallskip}
     \hline
     \noalign{\smallskip}
\end{tabular}
}% 
\end{center} 
\end{table}                                                        

\addtocounter{table}{-1}
\begin{table}
\begin{center}
\caption[]{continued} 
{\small %
\begin{tabular}{lr|rrr|l}
  \noalign{\smallskip}
  \hline
  \noalign{\smallskip}
NGC  & Tel   & Beam  &  Flux           &        & Ref\\
     &       & ($"$  &  $Jy\kms$       &        &    \\ 
 (1) & (2)   & (3)   &  (4)            &  (5)   &(6)  \\
  \noalign{\smallskip}
  \hline
  \noalign{\smallskip}
4666 & SEST  & 45    &     580         &        & TP \\
     & ave   & 45    &     638$\pm$58  & (0.91) &    \\
     & IRAM  & 23    &     346         &        & TP \\
     & NOB   & 16    &     113         &        & 11 \\
     & OVRO  & 3.8   &      218        &        & 48 \\
4736 & BTL   & total &     3280        &        &  9 \\
     & FCRAO & total &     2560        &        & 15 \\
     & ave   & total &     2920$\pm$360&        &    \\
     & BTL   & 100   &      938        &        &  9 \\
     & PMO   & 55    &      252        &        & 82 \\
     & FCRAO & 45    &      405        &        & 15 \\
     & IRAM  & 22    &      198        &        & TP \\
     & IRAM  & 22    &      197        &        & 49 \\
     & ave   & 22    &      198$\pm$1  & (1.00) &    \\
     & NOB   & 16    &      146        &        & 11 \\
     & BIMA  & 5.9   &       45        &        & 50 \\
4826 & BTL   & total &     2600        &        &  9 \\
     & FCRAO & total &     2170        &        & 15 \\
     & ave   & total &     2385$\pm$215&        &    \\
     & NRAO  & 55    &      403        &        &  4 \\
     & FCRAO & 45    &     1033        &        & 15 \\
     & ave   & 45    &      718$\pm$315&        &    \\
     & OSO   & 33    &      456        &        & 20 \\
     & IRAM  & 22    &      425        &        & TP \\
     & IRAM  & 22    &      493        &        & 52 \\
     & ave   & 22    &      459$\pm$34 & (0.93) &    \\
     & NOB   & 16    &      336        &        & 11 \\
     & BIMA  &  6.2  &      103        &        & 58 \\
     & PDBI  &  2.6  &       34        &        & 51 \\
5033 & FCRAO & total &     1640        &        & 15 \\
     & NRAO  & 55    &      403        &        &  6 \\
     & NRAO  & 55    &      620        &        & 21 \\
     & ave   & 55    &      512$\pm$109&        &    \\
     & FCRAO & 45    &      302        &        & 15 \\
     & OSO   & 33    &      291        &        & 18 \\
     & OSO   & 33    &      248        &        & 20 \\
     & ave   & 33    &      270$\pm$22 &        &    \\
     & IRAM  & 22    &      248        &        & TP \\
     & IRAM  & 22    &       99        &        & 23 \\
     & ave   & 22    &      174$\pm$74 & (1.42) &    \\
     & NOB   & 16    &      142        &        & 11 \\
     & BIMA  & 5.7   &       41        &        & 58 \\
     & NMA   & 3.8   &       27        &        & 53 \\
5055 & NRAO  & total &     5719        &        &  2 \\
     & FCRAO & total &     5670        &        & 15 \\
     & ave   & total &     5695$\pm$25 &        &    \\
     & PMO   & 55    &      508        &        & 82 \\
     & NRAO  & 55    &      648        &        &  2 \\
     & NRAO  & 55    &     1184        &        &  3 \\
     & ave   & 55    &      780$\pm$206&        &    \\
     & FCRAO & 45    &      697        &        & 15 \\
     & OSO   & 33    &      382        &        & 18 \\
     & OSO   & 33    &      403        &        & 20 \\
     & ave   & 33    &      393$\pm$10 &        &    \\
     & IRAM  & 22    &      331        &        & TP \\
     & NOB   & 16    &      168        &        & 54 \\
     & BIMA  & 5.7   &       51        &        & 58 \\
5135 & NRAO  & 55    &      375        &        &  6 \\
     & NRAO  & 55    &      389        &        &  7 \\
     & NRAO  & 55    &      442        &        & 21 \\
     & ave   & 55    &      402$\pm$20 &        &    \\
     & SEST  & 45    &      340        &        & TP \\
     & IRAM  & 22    &      291        &        & TP \\
     &       &       &                 &        &    \\
     \noalign{\smallskip}
     \hline
     \noalign{\smallskip}
\end{tabular}
}% 
\end{center} 
\end{table}                                                        

\addtocounter{table}{-1}
\begin{table}
\begin{center}
\caption[]{continued} 
{\small %
\begin{tabular}{lr|rrr|l}
  \noalign{\smallskip}
  \hline
  \noalign{\smallskip}
NGC  & Tel   & Beam  &  Flux           &        & Ref\\
     &       & ($"$  &  $Jy\kms$       &        &    \\ 
 (1) & (2)   & (3)   &  (4)            &  (5)   &(6)  \\
  \noalign{\smallskip}
  \hline
  \noalign{\smallskip}
5194 & BTL   & total &    16000        &        &  9 \\
     & FCRAO & total &     9210        &        & 15 \\
     & ave   & total &   12605$\pm$3395&        &    \\
     & PMO   & 55    &     1050        &        & 82 \\
     & FCRAO & 45    &     1234        &        & 15 \\
     & IRAM  & 22    &      224        &        & TP \\
5236 & BTL   & total &    26000        &        &  9 \\
     & FCRAO & total &    16320        &        & 15 \\
     & ave   & total &   21160$\pm$4840&        &    \\
     & NRAO  & 55    &     1197        &        &  5 \\
     & FCRAO & 45    &     2050        &        & 15 \\
     & SEST  & 45    &     2400        &        & 79 \\
     & SEST  & 45    &     1486        &        & TP \\
     & ave   & 45    &     1979$\pm$266& (0.75) &    \\
     & IRAM  & 22    &      917        &        & TP \\
     & NOB   & 16    &       86        &        & 11 \\
5713 & FCRAO & total &      680        &        & 15 \\
     & NRAO  & 55    &     1043        &        & 76 \\
     & FCRAO & 45    &      254        &        & 15 \\
     & SEST  & 45    &      318        &        & TP \\
     & ave   & 45    &      286$\pm$32 & (1.11) &    \\
     & IRAM  & 23    &      213        &        & TP \\
     & NOB   & 16    &       77        &        & 11 \\
5775 & FCRAO & total &      630        &        & 15 \\
     & NRAO  & 55    &      382        &        & 76 \\
     & FCRAO & 45    &      314        &        & 15 \\
     & IRAM  & 22    &      225        &        & TP \\
6000 & FCRAO & total &     1000        &        & 15 \\
     & NRAO  & 45    &      904        &        & 15 \\
     & SEST  & 45    &      392        &        & 18 \\
     & SEST  & 45    &      406        &        & 19 \\
     & SEST  & 45    &      407        &        & 22 \\
     & SEST  & 45    &      425        &        & TP \\
     & ave   & 45    &      507$\pm$110& (0.84) &    \\
     & IRAM  & 22    &      361        &        & TP \\
     & NOB   & 16    &       67        &        & 11 \\
6090 & FCRAO & total &      200        &        & 15 \\
     & NRAO  & 55    &      151        &        &  7 \\
     & FCRAO & 45    &      145        &        & 15 \\
6215 & SEST  & 45    &      192        &        & 19 \\
     & SEST  & 45    &      216        &        & 20 \\
     & SEST  & 45    &      192        &        & 22 \\
     & SEST  & 45    &      206        &        & TP \\  
     & ave   & 45    &      202$\pm$6  & (1.02) &    \\
6221 & SEST  & 45    &      589        &        & 20 \\
     & SEST  & 45    &      582        &        & TP \\
     & ave   & 45    &      586$\pm$4  & (0.99) &    \\
6240 & FCRAO & total &      300        &        & 15 \\
     & NRAO  & 55    &      315        &        &  7 \\
     & FCRAO & 45    &      378        &        &  5 \\
     & FCRAO & 45    &      239        &        & 15 \\
     & SEST  & 45    &      331        &        & TP \\
     & ave   & 45    &      316$\pm$40 & (1.05) &    \\
     & IRAM  & 22    &      330        &        & TP \\
     & IRAM  & 22    &      277        &        & 24 \\
     & IRAM  & 22    &      345        &        & 80 \\
     & ave   & 22    &      317$\pm$21 & (1.04) &    \\
6764 & NRAO  & 55    &      249        &        &  5 \\
     & OSO   & 33    &       86        &        & 18 \\
     & IRAM  & 22    &      142        &        & TP \\
     & IRAM  & 22    &      110        &        & 55 \\
     & ave   & 22    &      126$\pm$16 & (1.13) &    \\
6810 & SEST  & 45    &      556        &        & TP \\
     & OSO   & 33    &      162        &        & 18 \\
     \noalign{\smallskip}
     \hline
     \noalign{\smallskip}
\end{tabular}
}% 
\end{center} 
\end{table}                                                        

\addtocounter{table}{-1}
\begin{table}
\begin{center}
\caption[]{continued} 
{\small %
\begin{tabular}{lr|rrr|l}
  \noalign{\smallskip}
  \hline
  \noalign{\smallskip}
NGC  & Tel   & Beam  &  Flux           &        & Ref\\
     &       & ($"$  &  $Jy\kms$       &        &    \\ 
 (1) & (2)   & (3)   &  (4)            &  (5)   &(6)  \\
  \noalign{\smallskip}
  \hline
  \noalign{\smallskip}
6810 & OSO   & 33    &      403        &        & 20 \\
     & ave   & 33    &      283$\pm$120&        &    \\
6946 & BTL   & All   &    23000        &        &  9 \\
     & FCRAO & All   &    12370        &        & 15 \\
     & ave   & All   &   17685$\pm$5315&        &    \\
     & NRAO  & 55    &    1540         &        & 77 \\
     & FCRAO & 45    &    1842         &        & 15 \\
     & IRAM  & 22    &    1072         &        & TP \\
     & NOB   & 16    &    946          &        & 11 \\
     & BIMA  & 5.4   &    245          &        & 58 \\
     & BIMA  & 4.0   &    160          &        & 78 \\
6951 & BTL   & total &   l910          &        &  9 \\
     & FCRAO & total &   1440          &        & 15 \\
     & ave   & total &   1675$\pm$235  &        &    \\
     & FCRAO & 45    &    350          &        & 15 \\
     & OSO   & 33    &    217          &        & 18 \\
     & OSO   & 33    &    202          &        & 20 \\
     & ave   & 33    &    210$\pm$8    &        &    \\
     & IRAM  & 22    &    236          &        & TP \\
     & NOB   & 16    &    187          &        & 11 \\
     & NMA   &  3.5  &     49          &        & 56 \\
7331 & BTL   & total &   5600          &        &  9 \\
     & FCRAO & total &   4160          &        & 15 \\
     & ave   & total &   4880$\pm$720  &        &    \\
     & FCRAO & 45    &    161          &        & 15 \\
     & OSO   & 33    &    250          &        & 18 \\
     & IRAM  & 22    &    179          &        & 23 \\
     & NOB   & 16    &     46          &        & 11 \\
7469 & FCRAO & total &    240          &        & 15 \\
     & NRAO  & 55    &    420          &        &  6 \\
     & NRAO  & 55    &    336          &        &  7 \\
     & NRAO  & 55    &    338          &        & 21 \\
     & ave   & 55    &    374$\pm$25   &        &    \\
     & FCRAO & 45    &    238          &        & 15 \\  
     & FCRAO & 45    &    403          &        &  5 \\
     & SEST  & 45    &    202          &        & TP \\
     & ave   & 45    &    281$\pm$62   & (0.72) &    \\
     & OSO   & 33    &    157          &        & 18 \\
     & IRAM  & 22    &    257          &        & TP \\
     & IRAM  & 22    &    296          &        & 80 \\
     & ave   & 22    &    277$\pm$20   & (0.93) &    \\
     & NOB   & 16    &     27          &        & 14 \\
7541 & FCRAO & total &    650          &        & 15 \\
     & NRAO  & 55    &    389          &        & 76 \\
     & FCRAO & 45    &    341          &        & 15 \\
     & SEST  & 45    &    401          &        & TP \\
     & ave   & 45    &    371$\pm$30   & (1.08) &    \\
     & OSO   & 33    &    626          &        & 18 \\
     & IRAM  & 22    &    134          &        & TP \\
     & NOB   & 16    &    118          &        & 11 \\
7552 & SEST  & 45    &    756          &        & 20 \\
     & SEST  & 45    &    792          &        & 57 \\
     & SEST  & 45    &    733          &        & TP \\
     & ave   & 45    &    760$\pm$17   & (0.96) &    \\
7582 & SEST  & 45    &    567          &        & 20 \\
     & SEST  & 45    &    750          &        & 57 \\
     & SEST  & 45    &    614          &        & TP \\
     & ave   & 45    &    702$\pm$52   & (0.87) &    \\
7590 & SEST  & 45    &    137          &        & 19 \\
     & SEST  & 45    &    138          &        & 22 \\
     & SEST  & 45    &    146          &        & TP \\
     & ave   & 45    &    140$\pm$3    & (1.04) &    \\
7674 & FCRAO & total &    350          &        & 15 \\ 
     & NRAO  & 55    &    140          &        &  7 \\
     \noalign{\smallskip}
     \hline
     \noalign{\smallskip}
\end{tabular}
}% 
\end{center} 
\end{table}                                                        

\addtocounter{table}{-1}
\begin{table}
\begin{center}
\caption[]{continued} 
{\small %
\begin{tabular}{lr|rrr|l}
  \noalign{\smallskip}
  \hline
  \noalign{\smallskip}
NGC  & Tel   & Beam  &  Flux           &          & Ref\\
     &       & ($"$  &  $Jy\kms$       &          &    \\ 
 (1) & (2)   & (3)   &  (4)            &  (5)     &(6)  \\
  \noalign{\smallskip}
  \hline
  \noalign{\smallskip}
7674 & FCRAO & 45    &    168          &        &  5 \\
     & FCRAO & 45    &    275          &        & 15 \\
     & ave   & 45    &    222$\pm$54   &        &    \\
7714 & FCRAO & total &    130          &        & 15 \\
     & NRAO  & 55    &    130          &        &  1 \\
     & NRAO  & 55    &    196          &        &  7 \\
     & ave   & 55    &    168$\pm$38   &        &    \\
     & FCRAO & 45    &    100          &        & 15 \\
     & SEST  & 45    &     42          &        & 17 \\
     & SEST  & 45    &     19          &        & TP \\
     & ave   & 45    &     54$\pm$24   & (0.35) &    \\
     & IRAM  & 22    &     16          &        & TP \\
     & IRAM  & 22    &     58          &        & 17 \\
     & ave   & 22    &     37$\pm$20   & (0.43) &    \\
     & NOB   & 16    &      7          &        & 14 \\
7771 & FCRAO & total &    540          &        & 15 \\
     & NRAO  & 55    &    420          &        &  7 \\
     & FCRAO & 45    &    381          &        & 15 \\
     & OSO   & 33    &    290          &        & 18 \\
     & IRAM  & 22    &    469          &        & TP \\
     & IRAM  & 22    &    462          &        & 24 \\
     & IRAM  & 22    &    489          &        & 80 \\
     & ave   & 22    &    473$\pm$8    & (0.99) &    \\
     & NOB   & 16    &    323          &        & 81 \\
I342 & BTL   & total & 100000          &        &  9 \\
     & FCRAO & total &  29220          &        & 15 \\
     & ave   & total &  64610$\pm$35390&        &    \\
     & FCRAO & 45    &   1512          &        & 15 \\
     & NOB   & 16    &    478          &        & 11 \\
     & BIMA  & 5.3   &     97          &        & 57 \\
Maf2 & FCRAO & 45    &    164          &        & 59 \\
     & IRAM  & 22    &    568          &        & 60 \\   
     \noalign{\smallskip}
     \hline
     \noalign{\smallskip}
\end{tabular}
}% 
\end{center} 
Notes:\\
$^a$ BTL: Bell Telephone Labs; SEST: Swedish-ESO Submillimeter
Telescope; OSO: Onsala Space Observatory; IRAM: Institut Radio
Astronomie Millim\'etrique - Pico Veleta);
NOB: Nobeyama; FCRAO: Five Colleges Radio Observatory;
NRAO: National Radio Astronomy Observatory; PMO: Purple Mountain
Observatory; CSO: Caltech Submillimeter Observatory; SMA: Smithsonian
Millimeter Array; OVRO: Owens Valley Radio Observatory; BIMA: Berkeley
Illinois Maryland Array. \\
$^b$ Conversion factors S$_{115GHz}$ --> T (see Papadopoulos $\etal$ 2012):
SEST S/T$_{A}^{*}$=27 Jy/K;
OSO S/T$_{A}^{*}$=31 Jy/K;
IRAM: S/T$_{A}^{*}$=6.3 Jy/K;
NRAO/ARO: S/T$_{R}^{*}$=35 Jy/K;
FCRAO/PMO: S/T$_{A}^{*}$=42 Jy/K;
NOB: S/T$_{mb}^{*}$=2.4 Jy/K; \\
\end{table}                                                        

\addtocounter{table}{-1}
\begin{table}
\caption[]{continued} 
{\small %
{\bf references to Table\,\ref{flux10comp}} 1.  Bushouse $\etal$ 1999;
2. Sage, 1993; Adler \& Liszt, 1989; Rickard, Turner, \& Palmer, 1985;
5. Sanders \& Mirabel, 1985; 6.  Heckman $\etal$ 1989; 7.  Sanders,
Scovile, \& Soifer, 1991; 8. Stark $\etal$ 1986; 9. Stark, Elmegreen,
\& Chance, 1987; 10. Chung $\etal$ 2009; 11. Komugi $\etal$ 2008;
12. Sofue $\etal$ 2003a; 14. Sofue $\etal$ 1994; 15. Young $\etal$
1995; 16. Kenney \& Young, 1988; 17. Chini $\etal$ 1992; 18. Elfhag
$\etal$ 1996; Chini, Kr\"ugel, \& Lemke, 1996; 20. Aalto $\etal$ 1995;
21. Maiolin $\etal$ 1997; 22. Albrecht, Kr\"ugel, \& Chini, 2007;
23. Braine $\etal$ 1993a; 24. Solomon, Downes, \& Radford, 1992; 25. Bajaja
$\etal$, 1995; 26. Dahlem, $\etal$ 1990; 27. Becker, \& Freudling,
1991; 28. Morukama-Matsui $\etal$ 2014; 29. Casasola $\etal$ 2011;
30. Reuter $\etal$ 1996; 31. Sofue $\etal$ 2003b; 32. Krause $\etal$
2007; 33. Cox \& Downes, 1996; 34. Sofue $\etal$ 1989; 35. Koda \&
Sofue, 2006; 36. Sempere \& Garc\'ia-Burillo, 1997;
37. Garc\'ia-Burillo $\etal$ 1998; 38. Sakamoto $\etal$ 1995;
39. Knapen $\etal$ 1993; 40. Braine, Combes, \& van Driel 1993;
41. Kawara $\etal$ 1990; 42. Shibatsuka $\etal$ 2003; 43. Sofue
$\etal$ 1993; 44. Vlahakis $\etal$ 2013; Golla \& Wielebinsky, 1994;
46. Sofue, \& Nakai, 1992; 47. Richmond \& Knapp, 1985; 48. Walter,
Dahlem, \& Lisenfeld, 2004; 49. Gerin, Casoli, \& Combes, 1991;
50. Wong \& Blitz, 2000; 51. Garci\'a-Burillo $\etal$ 2003;
52. Casoli, \& Gerin, 1993; 53. Kohno $\etal$ 2003 54. Kuno $\etal$
1997; 55. Eckart $\etal$ 1991; 56. Kohno, Kawabe, \& Vila-Vilar\'o,
1999; 57. Claussen \& Sahai, 1992; 58. Helfer $\etal$ 2003; 59. Mason,
\& Wilson, 2004; 60. Weliachew, Casoli, \& Combes, 1988; 61. Bayet
$\etal$ 2004; 62. Sorai $\etal$ 2000; 63. Harrison, Henkel, \&
Russell, 1999; 64. Houghton $\etal$ 1996; 65. Helfer \& Blitz, 1995;
66. Kaneko $\etal$ 1989 67. Scovile, Young, \& Lucy, 1983;
68. Sandqvist, J\"ors\"ater, \& Lindblad, 1995; 69. Sandqvist, Elfhag,
\& J\"ors\"ater, 1988; 70. Mao $\etal$ 1988; 71. Olofson, \& Rydbeck,
1984; 72. Young \& Scoville, 1985; 73. Braine $\etal$ 1997; 74. Koda
$\etal$ 2002; 75. Young, Clausen, \& Scoville, 1988; 76. Tinney
$\etal$ 1990; 77. Crosthwaite \& Turner, 2007; 78. Regan \& Vogel,
1995; 79. Lundgren $\etal$ 2004; 80. Costagliola $\etal$ 2011;
81. Yamashita $\etal$ 2017; 82. Tan $\etal$ 2011.
}
\end{table}

Table\,\ref{flux10comp} collects $J$=1-0 $\co$ data from this paper
and from the literature as a function of observing beam size. Column
2 identifies the telescope used. The beam size is listed in Col. 3;
'total' indicates the extrapolated line flux of the whole galaxy taken
from the reference cited. Column 4 gives the line flux in units of
Jy$\kms$. The factors required to convert flux density into temperature are
given in a footnote. In all cases where two or more fluxes (including
the new measurements from this paper) are available at the same
resolution, their average (marked `ave') is also given in Col. 4. In
Col. 5 we list the ratio of our new value to this
average. The IRAM fluxes presented in this paper are extracted from
profiles with generally better S/Ns and better
baselines than those taken from the literature.  For the same galaxy
and the same aperture, fluxes given in the literature can differ by as
much as a factor of two. Given this spread, we did not attempt to
identify discrepant fluxes {\bf or to eliminate} them from the compilation.

The average ratios of the fluxes for beam sizes of $45"$ and $22"$
measured by us to those collected from the literature are given in
Table\,\ref{102132stat}. They are close to unity and have modest
standard deviations. This shows that the results of the present survey and
those of previous work are consistent.  The data in
Table\,\ref{flux10comp} have been used to construct
Fig.\,\ref{cophot}.

\subsection{Comparison CO(2-1) and CO(3-2) survey results} 

%Table 2 Line Intensities
\begin{table}
\caption[]{\label{co2132comp}Galaxy center line intensities from JCMT, IRAM, and HHSMT} 
\begin{center} 
{\small % 
\begin{tabular}{l|rrr|rrrr} 
\noalign{\smallskip}     
\hline
\noalign{\smallskip} 
NGC  &\multicolumn{3}{c}{CO(2-1)}&\multicolumn{4}{c}{CO(3-2)}\\
IC   &TP$^a$&B93$^b$&Ratio&TP$^c$&D01$^c$&M10$^c$&Ratio\\
     &J$22"$&I$22"$&      &J$22"$&H$22"$& H$22"$&      \\
 (1) & (2)  &  (3) & (4)  & (5)  &  (6) & (7)   & (8)  \\ 
\noalign{\smallskip}     
\hline
\noalign{\smallskip} 
 253 & 1360 &  ... & ...  &  850 &  680 &  ...  & 1.25$^f$\\
 278 & 19.6 &  17  & 1.15 & 14.0 &   15 &  ...  & 0.93$^f$\\
 628 &  4.2 &  1.8 & 2.33 &  2.7 &  ... &  ...  & ...  \\
 660 &  149 & 112  & 1.33 & 94.0 &  ... &  71.1 & 1.32 \\
 891 & 61.6 &  86  & 0.72 &  ... &   71 &  17.3 & ...  \\
 972 & 70.5 &  ... & ...  & 39.0 &  ... &  37.0 & 1.05 \\
Maf2 &  247 &  ... & ...  &  179 &  139 & 157.1 & 1.14 \\
1055 & 54.4 &   58 & 0.94 &  ... &  ... &  19.6 & ...  \\
1068 &  239 &  240 & 1.00 &  101 &  ... & 116.0 & 0.87 \\
1084 & 30.7 &   21 & 1.46 &  ... &  ... &  14.9 & ...  \\
I342 &  173 &  ... & ...  &  121 &  148 &  ...  & 0.82$^f$\\
2146 &  164 &  138 & 1.19 &  137 &  172 &  66.9 & 0.80$^f$\\
2559 & 69.7 &  ... & ...  & 40.3 &  ... &  37.3 & 1.08 \\
2903 & 59.5 &  ... & ...  & 53.8 &  ... &  59.6 & 0.91 \\
3034 &  657 &  ... & ...  &  548 &  770 & 1056  & 0.71$^f$\\
3227 & 48.3 &   42 & 1.15 & 41.7 &  ... &  18.1 & 2.30 \\
3310 &  8.7 &  9.5 & 0.92 & 11.7 &  ... &   6.9 & 1.64 \\
3351 &  ... &  28  & ...  & 29.9 &  ... &  26.0 & 1.15 \\
3627 & 74.3 &   77 & 0.96 & 48.7 &  ... &  33.4 & 1.46 \\
3628 &  162 &  142 & 1.14 &  114 &  206 & 140.7 & 0.81 \\
3690 & 64.4 &  ... & ...  & 40.3 &  ... &  36.4 & 1.11 \\
3982 & 13.9 &  ... & ...  & 12.7 &  ... &   6.7 & 1.90 \\
4038 & 63.6 &  ... & ...  & 52.0 &  ... &  44.0 & 1.18 \\
4039 & 35   &  ... & ...  & 23.8 &  ... &  15.2 & 1.57 \\
4258 & 44.3 &  ... & ...  & 32.7 &  ... &  64.4 & 0.51 \\
4303 & 42.6 &  ... & ...  & 23.0 &  ... &  33.7 & 0.68 \\
4321 & 55.6 &   57 & 0.98 & 35.3 &  ... &  56.9 & 0.62 \\
4414 & 37.1 &   46 & 0.81 & 18.7 &  ... &  37.6 & 0.50 \\
4457 & 27.5 &  ... & ...  & 19.0 &  ... &  20.3 & 0.94 \\
4565 & 10.1 &    9 & 1.12 &  ... &  ... &   8.3 & ...  \\
4631 & 34.3 &  ... & ...  & 23.2 &   37 &  17.7 & 1.31 \\
4666 & 53.1 &  ... & ...  & 35.3 &  ... &  36.7 & 0.96 \\
4736 & 42.7 &  ... & ...  & 27.4 &  ... &  21.3 & 1.29 \\
4826 &  102 &  ... & ...  & 49.9 &  ... &  86.8 & 0.57 \\
5033 & 42.5 &   33 & 1.29 & 24.1 &  ... &  16.7 & 1.44 \\
5194 & 54.2 &  ... & ...  & 31.7 &   47 &  44.4 & 0.71 \\
5236 &  251 &  ... & ...  &  137 &  234 & 153.9 & 0.89 \\
6240 & 70.2 &  ... & ...  & 80.4 &  ... &  74.9 & 1.07 \\
6946 &  240 &  ... & ...  &  117 &  129 & 132.3 & 0.88 \\
7331 & 15.5 &   19 & 0.82 &  7.1 &  ... &   7.2 & 0.99 \\
7469 & 52.0 &  ... & ...  & 40.7 &  ... &  35.2 & 1.16 \\
\noalign{\smallskip}     
\hline
%Average&    &      & 1.13 &      &      &       & 1.05 \\
%SD   &      &      & 0.37 &      &      &       & 0.43 \\                 
%n    &      &      &   17 &      &      &       &   36 \\
%\noalign{\smallskip}     
%\hline
\end{tabular}
}
\end{center}
Notes: $^a$: This paper, measured JCMT intensities;
$^b$: Braine $\etal$ (1993a) convolved IRAM intensities; 
$^c$: This paper, convolved JCMT intensities;
$^d$: Dumke $\etal$ (2001) peak intensities in HHST maps;
$^e$: Mao $\etal$ (2010), measured HHST intensities;
$^f$: Ratio based on Dumke $\etal$ (2010) intensity
\end{table}                                                            

\begin{table}
\caption[]{\label{102132stat}Statistics of the line intensity comparisons$^a$} 
\begin{center} 
{\small % 
\begin{tabular}{l|rr|r|r} 
\noalign{\smallskip}     
\hline
\noalign{\smallskip} 
Line     & CO(1-0) & CO(1-0) & CO(2-1) & CO(3-2) \\
Telescope& SEST    & IRAM    & JCMT    & JCMT    \\
         & SEST    & IRAM    & IRAM    & HHSMT   \\
Beam size&S:S 45$"$&I:I $22"$&J:I $22"$&J:H $22"$\\
 (1)     & (2)     &  (3)    & (4)     & (5)     \\ 
\noalign{\smallskip}     
\hline
\noalign{\smallskip} 
Average  & 0.95    & 1.04    & 1.13    &  1.05   \\
SD       & 0.31    & 0.29    & 0.37    &  0.43   \\                 
n        &   43    &  27     &   17    &   36    \\
\noalign{\smallskip}     
\hline
\end{tabular}
}
\end{center}
Note: $^a$ Average: average ratio of value from this paper over
literature values in same beam; SD: standard deviation; n: number of
values in the average
\end{table}

The literature provides far fewer data for the $J$=2-1 and $J$=3-2 CO
transitions. Braine $\etal$ (1993a) used the IRAM telescope with early
receivers and backends to measure a large number of galaxies in the
$J$=2-1 transition, of which 17 are in common with our survey.  They
convolved small maps to match the $J$=2-1 intensities to the $J$=1-0
beam. These are compared to the JCMT $J$=2-1 measurements in
Table\,\ref{co2132comp}. Our results are somewhat higher on average,
but the dispersion (standard deviation) is significant. The other
large extragalactic IRAM $J$=2-1 survey by Albrecht $\etal$ (2007)
unfortunately has few objects in common with our survey, as is the
case for the JCMT $J$=3-2 survey by Yao $\etal$ (2003). Of more
interest are the $J$=3-2 CO surveys conducted with the 10m Heinrich
Hertz Submillimeter Telescope (HHSMT) with a beam size of $22"$
(Mauersberger $\etal$ 1999; Dumke $\etal$ 2001; Mao $\etal$ 2010),
which have several galaxies in common with our survey. Data from the
latter two surveys are also summarized in Table\,\ref{co2132comp}
together with JCMT intensities from maps convolved to $22"$. We
disregarded the Mauersberger $\etal$ (1999) results because they are
superseded by the Mao $\etal$ (2010) survey and suffer from serious
calibration and pointing issues. To a lesser extent, these also plague
the later two surveys, as inspection of Cols. 6 and 7 shows. This
issue is discussed in more detail by Mao $\etal$ (2010). It is
therefore not surprising that the ratios of the HHSMT to the JCMT
intensities vary either way by factors of up to 2.5. Notwithstanding
the relatively large dispersions, the average intensities are quite
consistent.

\section{Observed and literature isotopologue ratios}

% Table B1 Sample
\begin{table*}[h]
\begin{center}
{\small %
\caption[]{\label{isotopcomp}Detailed comparison of $J$=1-0 $\co$/$\thirco$ } 
\begin{tabular}{lr|rrrrrrrr|l}
  \noalign{\smallskip} \hline \noalign{\smallskip}
  NGC/IC&This&\multicolumn{8}{c}{Telescope$^a$ and beam-width (arcsec)}&Reference\\
        &Paper& BTL & NRAO & AROKP&FCRAO & PMO &SEST & OSO & IRAM & \\
        & 22 & 100 & 55 & 55 & 45 & 45 & 45 & 33 & 22 & \\
  (1)   &(2) & (3) & 4) &(5) &(6) &(7) &(8) &(9) &(10)& (11) \\
  \noalign{\smallskip}
  \hline
  \noalign{\smallskip}
  253 & 12.7&11.4&16.6& ...& 11.9 & ...& ...& ...& ... &4             \\
  278 & 9.0& ...& ... &11.3& ...  & ...& ...& ...& ... &28            \\
  520 &14.2& ...& ... & ...& ...  & ...& ...&11.5& ... &10            \\
  628 & 6.3& ...& 6.7 &10.8& ...  & ...& ...& ...& 9.9 &8,28,31       \\
  660 &14.0& ...& ... & ...& ...  & ...& ...&14.0&16.5 &6/10,26       \\
  891 & 7.8& ...& 8.4 & 9.9& ...  & ...& ...& ...& 8.5 &8,28,19       \\
 1068 &11.8& ...& ...&...&13.4/10.7&...& ...& ...& 14  &11,4,1        \\
 Maf2 & 8.6& ...& 9.0/8.2&...& ...& ...& ...& ...& ... &8/12          \\
 I342 &10.2&11.0&11.1/16.0&...&8.4/8.7&...&...&...&... &13,8/12,4/11  \\
 1365 &11.1& ...& 6.4 & ...& ...  & ...& ...& ...& 13  &23,5          \\
 1614 &30.0& ...& ... & ...& ...  & ...&$>$26&...& 29.4&10,26         \\
 1808 &16.5& ...& ... & ...& ...  & ...& 16.4&...& ... &6/10          \\
 2146 &15.0& ...& ... & ...& 14.9 & ...& ...&12.2& ... &4,6/10        \\
 2369 &14.9& ...& ... &... & ...  & ...& 15 & ...& ... &6             \\
 2903 &11.2& ...& 9.5 &11.3& 12.8 & ...& ...& ...& 12.3&8,28,2,31     \\
 3034 &18.4& ...&15.9 & ...& 27.3 &21.9& ...& ...&14.3 &8,4,2,25      \\
 3044 &13.5& ...& ... & 6.3& ...  & ...& ...& ...& ... &28            \\
 3079 &15.8& ...& ... & ...& 15.3 & ...& ...&10.9& 17.1&4,6/10,26     \\
 3227 &17.8& ...& ... & ...& ...  & ...& ...& ...& 17  &5             \\
 3256 &25  & ...& ... & ...& ...  & ...&35/26&...& ... &6/8/10/7      \\
 3556 &12.5& ...& ... &14.3&  8.7 & ...& ...& ...& 12.5&28,4,26       \\
 3593 &12.4& ...& ... & ...& 10.6 & ...& ...& ...& ... &4             \\
 3627 &13.3& ...& 10  & ...& 11.2 &10.4& ...& ...& 15.2&8,4,2,31      \\
 3628 &12.0& ...& 8.2 & ...&  9.2 &11.1& ...& ...& ... &8,4,2         \\
 4030 & 6.6& ...& ... & 7.7& ...  & ...& ...& ...& ... &28            \\
 4038 &12.8& ...& ... & ...& ...  & ...& 16 & ...& 11  &6,16          \\
 4039 &22.0& ...& ... & ...& ...  & ...& 19 & ...& ... &6             \\
 4051 &18  & ...& ... & ...& ...  & ...&...& ...&13/16.7&5/29         \\
 4254 & 8.9& ...& ... &11.5& ...  & ...& ...& ...& 8.4 &31            \\
 4321 & 9.3& ...& ... &... &  7.6 & ...& ...& ...& 10.9&11,31         \\
 4414 & 7.5& ...& 6.3 &... & ...  & ...& ...& ...&  9  &18            \\
 4527 &13.3& ...& ... &11.3&  6.1 & ...& ...& ...& ... &4,28          \\
 4631 &15.1&... & ... &... & 16.3 &10.5& ...& ...& 11.5&4,2,19        \\
 4666 & 9.7& ...& ... &10.5& ...  & ...& ...& ...& ... &28            \\
 4736 & 9.9& ...& 8.3 & ...& ...  & 7.2& ...& ...& ... &8,2           \\
 4826 & 8.3& ...& 5.3 & ...& ...  & ...& ...& 5.1& ... &8,6/10        \\
 5033 & 9.1& ...& ... & ...& ...  & ...& ...& 9.3&8/8.4&6/10,5/29     \\
 5055 & 6.2& ...& ... & 7.2&  5.7 & 7.7& ...& 6.2&  7.2&28,4,2,6/10,31\\
 5135 &23  & ...& ... & ...& ...  & ...& ...& ...& 26  &5             \\
 5194 & 6.7&15.9& 6.5 & ...&5.4/10.4&8.5&...& ...&...  &13,12,4/11,2  \\
 5236 &13.7&15.8&13.1 & ...& 11.0 & ...& ...& ...& ... &13,12,11      \\
 5775 & 9.1& ...& ... &13.0& ...  & ...& ...& ...& ... &28            \\
 6221 & 12 & ...& ... & ...& ...  & ...& 18 & ...& ... &6             \\
 6240 & 29 & ...& ... & ...& ...  & ...& 44 & ...&28.8/45&8,26/1      \\
 6764 & 19 & ...& ... & ...& ...  & ...& ...& ...& 11/17&5/20         \\
 6946 &13.7& ...&11.8/11.1&...&15.0/17.0&...&...&...&15.2&12/8,11/4,31\\
 7469 & 17 & ...& ... & ...& ...  & ...& ...& ...& 20.8 &26           \\
 7541 & 8.2& ...& ... &15.1& ...  & ...& ...& ...& ...  &28           \\
 7552 &10.8& ...& ... & ...& ...  & ...&14.2& ...& ...  &6            \\
 7771 &13.9& ...& ... & ...& ...  & ...& ...& ...& 13.6 &26           \\
     \noalign{\smallskip}
     \hline
     \noalign{\smallskip}
\end{tabular}
}% 
\end{center} 
Notes: $^a$ BTL: Bell Telephone Labs; NRAO: Kitt Peak millimeter
telescope operated by National Radio Astronomy Observatory; AROKP:
Kitt Peak millimeter telescope operated by Arizona Radio Observatory;
FCRAO: Five Colleges Radio Astronomy Observatory; PMO: Delingha
Telescope operated by Purple Mountain Observatory; SEST: Swedish-ESO
Submillimeter Telescope; OSO: Onsala Space Observatory; IRAM:
Institut Radio Astronomie Millim\'etrique - Pico Veleta \\
\end{table*}

Large-scale extragalactic $\thirco$ hence $\co$/$\thirco$ isotopologue
surveys are lacking in all $J$ transitions. There are, however, a
significant number of $J$=1-0 $\thirco$ small-scale surveys and
individual measurements from which isotopologue ratios can be
constructed. In Table\,\ref{isotopcomp} we collect these ratios
for the $J$=1-0 transition as could be found in the literature for the
galaxies in the present sample.  For Cols. 3 through 10, we
determined the ratio of each value to the corresponding value in
Col. 2 (i.e., the ratio determined by the measurements in this
paper). At the bottom of the table we present for each column the
average of these ratios as well as its standard deviation. In all
columns the average is close to unity, implying that for resolutions
between $22"$ and $55"$, the isotopologue ratio in the $J$=1-0
transition does not vary significantly with beam width. In
Table\,\ref{isotop21comp} we summarize the much more sparse data for
the $J$=2-1 transition, as well as the ratios of the value determined
in this paper to the literature value, their average, and their
standard deviation. With one exception ($J$=1-0 AROKP), the standard
deviations are all between 0.22 and 0.31.
  
These standard deviations represent the combined error of the
isotopologue ratio derived from the literature and our value. In both
transitions, our data are derived from simultaneous dual-frequency
measurements, which compared to the literature data have well-defined
baselines and relatively high S/Ns so that their
contribution to the listed standard deviations is substantially below
that of the literature values.

% Table B2 Sample
\begin{table}[b]
\begin{center}
{\small %
\caption[]{\label{isotop21comp}Literature comparison of $J$=2-1 $\co$/$\thirco$ } 
\begin{tabular}{lr|rr|rl}
  \noalign{\smallskip}
  \hline
  \noalign{\smallskip}
NGC  & This &  Lit.& Ratio& Telescope$^{a}$        & Ref       \\
     & Paper& Value&      &                       &            \\
 (1) & (2)  & (3)  &   4) &  (5)                  &(6)  \\
  \noalign{\smallskip}
  \hline
  \noalign{\smallskip}
 660 & 17.0 & 21.8 & 0.78 & O     & 6,10      \\  
1068 & 12.8 & 10   & 1.28 & J     & 1         \\
1365 & 11.5 & 10/13& 0.98 & J/SMA & 5/22      \\
1808 & 12.6 & 15.5 & 0.81 & S     & 6,10      \\
2146 &  8.7 &  9.9 & 0.88 & O     & 6,10      \\
3034 & 10.8 & 11.9 & 0.91 & J     & 24        \\ 
3034 & 14.3 & 11.3 & 1.29 & I     & 25         \\
3227 & 13.5 & 25   & 0.54 & J     & 5         \\
3256 & 16   & 10/21/23/17 & 0.90 & S/SMA & 6,10,8/27 \\  
4039 & 12.9 & 12.0 & 13/13/10 & C/J/I & 3/15/16   \\ 
4038 & 15.8 & 19.8 & 27/16/16 & C/J/I & 3/15/16   \\
4051 & 21   & 14   & 1.50 & J     & 5         \\
4414 &  8.5 &  9   & 0.94 & I     & 18         \\
4736 & 10.1 & 13.6 & 0.74 & C     & 3          \\
4826 &  5.1 &  5.3 & 0.96 & O     & 6,10      \\
5033 &  8.3 &  6   & 1.38 & J     & 5          \\
5135 &  9.1 & 13   & 0.70 & J     & 5         \\
6240 & 40   & 53   & 0.75 & J     & 1          \\
6764 & 24   & 15   & 1.60 & I     & 20        \\         
7331 &  6.2 &  6   & 1.03 & I     & 29        \\ 
7469 & 17   & 15   & 1.13 & J     & 5          \\
  \noalign{\smallskip}
  \hline
  \noalign{\smallskip}
\end{tabular}
}% 
\end{center} 
Notes: $^a$ S: Swedish-ESO Submillimeter Telescope (SEST); O: Onsala
Space Observatory (OSO); I: Institut Radio Astronomie Millim\'etrique
- Pico Veleta (IRAM); J: James Clerk Maxwell Telescope (JCMT);
C: Caltech Submillimeter Telescope (CSO) SMA: Smithsonian Millimeter Array\\
\end{table}

\addtocounter{table}{-1}
\begin{table}
  \caption[]{continued} {\small %
    {\bf References to Tables\,\ref{isotopcomp} and
      \ref{isotop21comp}} 1.  Papadopoulos $\etal$ 2012; 2. Tan
    $\etal$ 2011; 3.  Glenn $\&$ Hunter 2001; 4.  Paglione $\etal$
    2001; 5.  Papadopoulos $\etal$ 1998; 6.  Aalto $\etal$ 1995; 7.
    Garay $\etal$ 1993; 8.  Casoli $\etal$ 1992; 10. Aalto $\etal$
    1991; 11. Young $\&$ Sanders 1986; 12. Rickard $\&$ Blitz 1985;
    13. Encrenaz $\etal$ 1979; 15. Zhu $\etal$ 2003; 16. Schulz
    $\etal$ 2007; 18. Braine, Combes \& van Driel, 1993b; 19. Golla
    $\&$ Wielebinsky, 1994 20. Eckart $\etal$ 1991; 21. Sakamoto
    $\etal$ 1997; 22. Sakamoto $\etal$ 2007; 23. Sandqvist $\etal$
    1988; 24. Petitpas $\&$ Wilson 2000; 25. Mao $\etal$ 2000;
    26. Costagliola $\etal$ 2011; 27. Sakamoto $\etal$ 2014;
    28. Vila-Vilaro $\etal$ 2015; 29. Krips $\etal$ 2010; 30. Muraoka
    $\etal$ 2016; 31.  Cormier $\etal$ 2018.  }
\end{table}

% Table B3 Statistics
\begin{table*}[h]
\begin{center}
{\small %
\caption[]{\label{isotopcompstat}$J$=1-0 $\co$/$\thirco$ comparison statistics} 
\begin{tabular}{l|rrrrrrrr|r}
  \noalign{\smallskip}
  \hline
  \noalign{\smallskip}
          &\multicolumn{8}{l}{$J$=1-0}                             &$J$=2-1\\
Telescope$^a$& BTL & NRAO & AROKP&FCRAO & PMO &SEST & OSO & IRAM & Div \\
Beam ($"$)   & 100 & 55   & 55   & 45   & 45  & 45  & 33  & 22   & --- \\
  (1)        &(2)  & (3) & 4)   &(5)   &(6)   &(7)  &(8)  &(9)   &(10)  \\
  \noalign{\smallskip}
  \hline
  \noalign{\smallskip}
Average$^b$& 0.83&1.13  & 0.92 & 1.11 &1.09 &0.86 &1.22 & 1.00 & 0.99\\
SD        & 0.27 & 0.27 & 0.39 & 0.31 &0.28 &0.18 &0.25 & 0.22 & 0.28\\
n         &    4 &   19 &   14 & 22   & 7   &   9 &   7 & 30   & 21  \\
  \noalign{\smallskip}
  \hline
  \noalign{\smallskip}
\end{tabular}
}% 
\end{center}
Notes: $^a$ BTL: Bell Telephone Laboraory (USA); NRAO: National Radio
Astronomy Observatory (USA); AROKP: Arizona Radio Observatory Kitt
Peak (USA); FCRAO: Five Colleges Radio Astronomy Observatory (USA);
PMO: Purple Mountain Observatory (PRC) SEST: Swedish-ESO Submillimeter
Telescope (Chile); OSO: Onsala Space Observatory (Sweden); IRAM:
Institut Radio Astronomie Millim\'etrique (Spain) $^b$ Average of
relative isotopologue ratios, i.e., values from Table 3 in this paper
derived from literature values; SD: Standard deviation; n:
number of values in average\\

\end{table*}

\section{CO radiative transfer modeling}

%Table x example fit
\begin{table*}
\caption[]{\label{examplefit} Two-phase fitting: example}
\begin{center} 
{\small %
\begin{tabular}{l|rrr|r|rrr|rrrrr|rrc}
\noalign{\smallskip}      
\hline
\noalign{\smallskip} 
f1 & n1 & N1/dV & T1 & f2 & n2 & N2/dV & T2 & $\thirco$10& $\thirco$21 & $\thirco$32 & $\co$1 & $\co$3 &\multicolumn{2}{c}{Mean}& N(CO)\\ 
   &    &       &    &    &    &       &    &            &             &             &        &        &T$_{mb}$&N/dV  & 10$^{17}\cm2$\\
(1)& (2)& (3)   & (4)& (5)& (6)&  (7)  & (8)& (9)        & (10)        &    (11)     & (12)   & (13)   &  (14) & (15)     & (16)        \\
\noalign{\smallskip}
\hline                                                           
\noalign{\smallskip}
\multicolumn{16}{l}{NGC 2903 $I_{\rm CO(2-1)}$ = 59.5 $\kkms$}\\
0.25 & 1e4 & 1e17 &  30 & 0.75 & 1e2 & 1e17 & 100 & 10.3 &  8.9 & 12.7 & 1.28 & 0.90 & 12.09 & 1.00 & 4.9 \\ 
0.30 & 1e4 & 1e17 &  30 & 0.70 & 1e2 & 1e17 & 100 & 10.4 &  8.5 & 11.5 & 1.25 & 0.89 & 12.90 & 1.00 & 4.6 \\ 
0.30 & 3e3 & 1e17 &  60 & 0.70 & 1e2 & 1e17 & 100 & 12.2 &  9.3 & 13.3 & 1.23 & 0.84 & 15.95 & 1.00 & 3.7 \\ 
0.70 & 1e3 & 6e16 & 150 & 0.30 & 1e3 & 3e17 & 150 & 11.8 &  8.6 & 11.3 & 1.28 & 0.78 & 32.30 & 1.32 & 2.4 \\ 
0.75 & 1e3 & 1e17 & 150 & 0.25 & 1e3 & 3e17 & 100 & 10.9 &  8.3 & 11.3 & 1.29 & 0.78 & 34.50 & 1.50 & 2.6 \\                     
0.80 & 1e3 & 1e17 & 150 & 0.20 & 1e3 & 3e17 & 150 & 12.0 &  8.5 & 11.3 & 1.30 & 0.78 & 35.15 & 1.40 & 2.4 \\ 
0.85 & 1e3 & 1e17 & 150 & 0.15 & 5e2 & 6e17 & 150 & 10.6 &  8.6 & 11.7 & 1.28 & 0.78 & 34.23 & 1.75 & 3.0 \\ 
{\bf 0.90} & 1e3 & 1e17 & 150 & 0.10 & 1e3 & 6e17 & 100 & 11.3 &  8.5 & 11.3 & 1.29 & 0.78 & 33.88 & 1.50 & 2.6 \\
0.90 & 1e3 & 1e17 & 150 & 0.10 & 5e2 & 1e18 & 150 & 10.6 &  8.5 & 11.5 & 1.28 & 0.78 & 34.40 & 1.90 & 3.3 \\ 
obs  &     &      &     &      &     &      &     & 11.2 &  8.6 & 12.5 & 1.34 & 0.91 &       &      &     \\
\noalign{\smallskip}
\hline                                                           
\noalign{\smallskip}
\multicolumn{16}{l}{NGC 4736 $I_{\rm CO(2-1)}$ = 42.7 $\kkms$}\\
0.25 & 3e3 & 3e17 &  10 & 0.75 & 3e3 & 3e16 &  60 &  9.6 & 10.0 & 14.2 & 0.97 & 0.70 & 18.0 & 0.98 &  2.2 \\
0.25 & 5e2 & 3e17 &  60 & 0.75 & 3e3 & 3e16 &  60 & 10.4 & 11.1 & 14.6 & 1.01 & 0.72 & 23.4 & 0.98 &  1.8 \\
0.30 & 3e3 & 1e16 & 150 & 0.70 & 3e3 & 6e16 &  20 & 10.2 &  9.5 & 15.4 & 1.03 & 0.68 & 13.2 & 0.45 &  1.5 \\
0.35 & 3e3 & 6e15 & 150 & 0.65 & 3e3 & 6e16 &  20 &  9.6 &  9.2 & 15.2 & 1.01 & 0.68 & 11.7 & 0.44 &  1.6 \\
{\bf 0.40} & 1e5 & 1e16 &  20 & 0.60 & 3e3 & 6e16 &  20 & 10.1 &  9.5 & 15.1 & 1.00 & 0.72 & 11.9 & 0.40 &  1.4 \\
0.40 & 3e3 & 6e15 & 100 & 0.60 & 3e3 & 6e16 &  20 & 10.1 &  9.5 & 15.4 & 1.03 & 0.65 & 11.4 & 0.38 &  1.4 \\
0.45 & 1e5 & 6e15 &  20 & 0.55 & 3e3 & 6e16 &  20 &  9.9 &  9.5 & 15.8 & 0.99 & 0.71 & 10.6 & 0.36 &  1.5 \\
0.45 & 3e3 & 1e17 &  10 & 0.55 & 3e3 & 3e16 &  60 &  9.5 & 10.2 & 15.4 & 1.00 & 0.68 & 16.8 & 0.62 &  1.6 \\
obs  &     &      &     &      &     &      &     &  9.9 & 10.1 & 14.9 & 0.98 & 0.63 &      &      &      \\
\noalign{\smallskip}
\hline                                                           
\noalign{\smallskip}
\multicolumn{16}{l}{NGC 5033 $I_{\rm CO(2-1)}$ = 42.5 $\kkms$}\\
0.10 & 1e3 & 1e17 & 100 & 0.90 & 3e3 & 6e16 &  20 &  9.0 &  7.4 & 12.2 & 1.26 & 0.64 & 14.1 & 0.64 &  2.7 \\
0.10 & 1e3 & 1e17 & 150 & 0.90 & 3e3 & 6e16 &  20 &  9.4 &  7.5 & 12.2 & 1.26 & 0.65 & 14.4 & 0.64 &  2.7 \\
0.10 & 3e3 & 6e16 &  30 & 0.90 & 3e3 & 6e16 &  20 &  8.8 &  7.0 & 11.6 & 1.23 & 0.63 & 13.1 & 0.60 &  2.0 \\
0.20 & 3e3 & 6e16 &  30 & 0.80 & 3e3 & 6e16 &  20 &  8.9 &  7.1 & 11.5 & 1.22 & 0.64 & 13.7 & 0.60 &  1.9 \\
{\bf 0.20} & 3e3 & 6e16 &  30 & 0.80 & 3e3 & 6e16 &  20 &  9.1 &  7.1 & 11.4 & 1.22 & 0.64 & 13.7 & 0.60 &  1.9 \\
{\bf 0.30} & 3e3 & 6e16 &  30 & 0.70 & 3e3 & 6e16 &  20 &  9.2 &  7.1 & 11.4 & 1.22 & 0.65 & 14.9 & 0.60 &  1.7 \\
0.30 & 5e2 & 1e17 &  60 & 0.70 & 1e5 & 3e16 &  10 &  8.7 &  7.4 & 11.7 & 1.29 & 0.65 &  7.4 & 0.51 &  2.9 \\
0.30 & 3e3 & 1e17 &  20 & 0.70 & 5e2 & 6e15 & 150 &  9.2 &  7.7 & 11.9 & 1.30 & 0.60 &  7.0 & 0.34 &  2.1 \\
0.30 & 5e2 & 1e17 &  30 & 0.70 & 3e3 & 6e16 &  30 &  9.4 &  7.7 & 11.4 & 1.22 & 0.65 & 16.4 & 0.72 &  1.9 \\
0.30 & 1e3 & 1e17 &  30 & 0.70 & 3e3 & 6e16 &  30 &  9.2 &  7.4 & 11.3 & 1.21 & 0.67 & 17.5 & 0.72 &  1.8 \\
0.35 & 3e3 & 1e17 &  30 & 0.65 & 1e3 & 6e16 &  30 &  8.9 &  7.6 & 11.3 & 1.28 & 0.64 & 15.4 & 0.74 &  2.0 \\
0.35 & 3e3 & 6e16 &  30 & 0.65 & 3e3 & 6e16 &  20 &  9.3 &  7.1 & 11.3 & 1.21 & 0.65 & 14.6 & 0.60 &  1.8 \\
0.35 & 3e3 & 6e16 &  30 & 0.65 & 1e4 & 3e16 &  10 &  9.6 &  7.0 & 11.3 & 1.22 & 0.63 &  9.8 & 0.42 &  1.8 \\
0.35 & 1e3 & 1e17 &  30 & 0.65 & 3e3 & 6e16 &  30 &  9.0 &  7.4 & 11.4 & 1.21 & 0.67 & 17.3 & 0.77 &  1.9 \\
0.40 & 3e3 & 6e16 &  30 & 0.60 & 3e3 & 3e16 &  10 &  9.6 &  7.6 & 11.3 & 1.28 & 0.60 &  9.9 & 0.42 &  1.8 \\
0.40 & 3e3 & 1e17 &  20 & 0.60 & 1e3 & 1e16 &  60 &  8.9 &  7.5 & 11.8 & 1.27 & 0.61 & 15.4 & 0.46 &  1.3 \\
0.40 & 1e3 & 1e17 &  30 & 0.60 & 3e3 & 6e16 &  30 &  8.8 &  7.5 & 11.5 & 1.22 & 0.66 & 19.3 & 0.76 &  1.7 \\
0.45 & 1e3 & 3e16 & 100 & 0.55 & 3e3 & 1e17 &  20 &  8.9 &  7.4 & 12.1 & 1.26 & 0.66 & 15.9 & 0.69 &  1.8 \\
0.45 & 3e3 & 1e17 &  20 & 0.55 & 1e3 & 1e16 & 100 &  8.7 &  7.5 & 12.0 & 1.21 & 0.63 & 10.7 & 0.51 &  2.0 \\
0.45 & 3e3 & 1e17 &  20 & 0.55 & 1e3 & 6e16 &  60 &  8.7 &  7.4 & 12.1 & 1.28 & 0.66 & 16.4 & 0.78 &  2.0 \\
0.45 & 1e3 & 3e16 &  60 & 0.55 & 3e3 & 1e17 &  20 &  8.8 &  7.4 & 11.8 & 1.28 & 0.63 & 13.4 & 0.69 &  2.2 \\
obs  &     &      &     &      &     &      &     &  9.1 &  7.2 & 11.7 & 1.25 & 0.57 &      &      &      \\
\noalign{\smallskip}
\hline                                                           
\end{tabular}
}
\end{center}
Note: The solution used in Table\,\ref{modelpar} is marked in bold in the first column.
\end{table*}

Our modeling assumes that the measured intensities are described by
two distinct model gas phases. As noted in Section 5, this is an
oversimplification, but two phases is the most that is allowed by the
present data without introducing major additional assumptions.
We preferred to fit the more diagnostic and more accurate
$\co$-to-$\thirco$ ratios, and we assumed that both phases have the same
isotopological abundance. The modeling was accomplished by searching a
grid of line intensity ratios resulting from the superposition of two
model gas clouds with kinetic temperature T$_{\rm k}$ between 10 K and
150 K, densities n$_{\h2}$ between $10^{2} \cc$ and $10^{5} \cc$, and
CO velocity gradients $N(CO)$/d$V$ between
$6 \times 10^{15} \cm2/\kms$ and $1 \times 10^{18} \cm2/\kms$ for
ratios matching the observed set, with the relative weights of the two
as a free parameter.  This parameter space contains the full range of
physical conditions, from translucent gas to dense clouds, that the
$\co$ and $\thirco$ transitions included in this paper are expected to
distinguish. Because we consider only the lower $J$ $\co$ and $\thirco$
transitions, our models lack sensitivity to molecular gas at the very
high densities and temperatures that are sampled by the higher CO
transitions or by molecular species such as HCN or HCO$^{+}$.
Measurements of $\co$ and $\thirco$ at higher $J$ would provide more
information on very high-pressure molecular gas but would not add much
to the present model results on the cooler and less dense
lower-pressure gas of which the bulk of the molecular ISM consists.

%Table 4  RADEX Model Results
\begin{table*}
\caption[]{\label{modelpar}{\it RADEX} model results}
\begin{center}
{\small % 
\begin{tabular}{lr|rrrrrrc|rrrrrrc}
\noalign{\smallskip}      
\hline
\noalign{\smallskip} 
Name   &Size &\multicolumn{7}{c}{[$\co$]/[$\thirco$]=40}       & \multicolumn{6}{c}{[$\co$]/[$\thirco$]=80} \\
&$d_{22}$&$n_1$&$T_1$&$n_2$&$T_2$&$f_{12}$&$N(CO)$&$\frac{\rm [CO]}{\rm [C]}$
        &$n_1$&$T_1$&$n_2$&$T_2$&$f_{12}$&$N(CO)$&$\frac{\rm [CO]}{\rm [C]}$\\
       &     &     &    &     &    &       &$10^{17}$&    &  &    &     &    &       & $10^{17}$ &     \\
       & kpc &$\cc$& K  &$\cc$&K   &       &$\cm2$  &    &$\cc$&K  &$\cc$&K   &       & $\cm2$ &      \\ 
(1)    & (2) & (3) & (4)& (5) & (6)& (7)   & (8)  &(9  ) &(10)&(11)&(12) &(13)& (14)  & (15) & (16) \\
\noalign{\smallskip}      
\hline 
\noalign{\smallskip} 
 253   & 0.4 & 3e3 & 60 & 3e3 & 150& 15:85 &  38  & 0.23 & 1e4 & 20 & 3e3 &  60& 15:85 &  94  & 0.37 \\
 278   & 1.2 & 1e5 & 20 & 1e2 & 150& 50:50 &  2.3 & 0.39 & ... & ...& ... & ...&...    & ...  &...   \\
 470:  & 3.4 & 1e5 & 20 & 5e2 & 150& 45:55 &  0.7 & 0.13 & 5e2 & 50 & 1e3 & 150& 40:60 &  1.8 & 0.45 \\
 520   & 3.2 & 5e2 & 30 & 1e4 & 20 & 30:70 &  3.0 & 0.15 & ... & ...& ... & ...&...    & ...  &...   \\ 
 613   & 2.1 & 1e5 & 60 & 5e2 & 20 & 20:80 &  3.2 & 0.18 & 1e3 & 30 & 1e4 & 20 & 10:90 &  16  & 0.71 \\
 628:  & 1.1 & 3e3 & 20 & 1e2 & 30 & 50:50 &  4.8 & 0.60 & ... & ...& ... & ...&...    & ...  &...   \\
 660   & 1.3 & 1e5 & 60 & 1e2 & 80 & 10:90 &  9.4 & 0.16 & 1e5 & 100& 1e2 & 100& 15:85 &  37  & 0.72 \\
 891   & 1.0 & 1e5 & 20 & 5e2 & 20 & 10:90 &  5.0 & 0.14 & ... & ...& ... & ...&...    & ...  &...   \\
 908   & 2.1 & 1e5 & 30 & 1e2 & 20 & 15:85 &  5.0 & 0.42 & ... & ...& ... & ...&...    & ...  &...   \\
 972   & 2.4 & 1e3 & 60 & 1e4 & 20 & 30:70 &  2.1 & 0.23 & 3e3 & 20 & 3e3 & 30 & 10:90 &  6.3 & 0.37 \\
Maf2   & 0.3 & 1e3 & 100& 1e4 & 30 & 10:90 &  16  & 0.48 & ... & ...& ... & ...&...    & ...  &...   \\
1055  & 1.4 & 1e5 & 20 & 5e2 & 20 & 10:90 &  6.0 & 0.23 & ... & ...& ... & ...&...    & ...  &...   \\
1068  & 1.6 & 3e3 & 30 & 3e3 & 150& 10:90 &  6.2 & 0.22 & 1e5 & 10 & 3e3 & 150& 45:55 &  19  & 0.45 \\
1084  & 2.0 & 1e4 & 20 & 3e3 & 60 & 90:10 &  0.9 & 0.30 & 1e5 & 10 & 3e3 & 30 & 25:75 &  3.1 & 0.39 \\
1097: & 1.8 & 1e5 & 60 & 1e2 & 60 & 15:85 &  9.4 & 0.26 & 1e5 & 60 & 1e2 & 60 & 30:70 &  44  & 0.76 \\
1365  & 2.3 & 1e2 & 100& 1e5 & 20 & 40:60 &  22  & 0.37 & 1e4 & 10 & 1e4 & 20 & 35:65 &  29  & 0.49 \\
342   & 0.4 & 1e3 & 20 & 3e3 & 100& 30:70 &  7.8 & 0.37 & ... & ...& ... & ...&...    & ...  &...   \\
1482::& 2.7 & 1e5 & 10 & 1e2 & 30 & 20:80 &  1.3 & 0.06 & 3e3 & 20 & 5e2 & 20 & 10:90 &  2.0 & 0.15 \\
1614: & 6.8 & 3e3 & 150& 5e2 & 10 & 10:90 &  1.5 & 0.14 & ... & ...& ... & ...&...    & ...  &...   \\
1792  & 1.6 & 1e4 & 10 & 1e4 & 20 & 15:85 &  3.6 & 0.51 & ... & ...& ... & ...&...    & ...  &...   \\
1808  & 1.3 & 1e5 & 60 & 1e3 & 150& 45:55 &  1.8 & 0.10 & 3e3 & 25 & 3e3 & 125& 50:50 &  7.6 & 0.54 \\
2146  & 1.8 & 1e3 & 150& 1e5 & 10 & 40:60 &  5.3 & 0.21 & 3e3 & 60 & 1e2 & 100& 40:60 &  13  & 0.36 \\
2273: & 3.0 & 1e3 & 30 & 1e3 & 150& 10:90 &  0.8 & 0.19 & 1e3 & 60 & 1e5 & 10 & 15:85 &  1.7 & 0.55 \\
2559  & 2.3 & 1e4 & 60 & 1e4 & 10 & 45:55 &  1.9 & 0.16 & ... & ...& ... & ...&...    & ...  &...   \\
2623: & 8.4 & 1e4 & 30 & 3e3 & 100& 15:85 &  1.5 & 0.38 & ... & ...& ... & ...&...    & ...  &...   \\
2903  & 0.8 & 1e3 & 100& 1e3 & 150& 10:90 &  2.7 & 0.38 & 1e3 & 150& 1e2 & 150& 10:90 &  7.3 & 0.33 \\
3034  & 0.6 & 1e5 & 60 & 1e3 & 150& 10:90 &  20  & 0.25 & 3e3 & 150& 3e3 & 30 & 15:85 &  46  & 0.53 \\
3044: & 2.2 & 1e5 & 10 & 1e3 & 60 & 25:75 &  0.4 & 0.13 & 1e5 & 10 & 3e3 & 30 & 15:85 &  0.2 & 0.05 \\
3079  & 2.2 & 1e5 & 100& 1e2 & 30 & 10:90 &  8.3 & 0.14 & 1e3 & 60 & 3e3 & 20 & 10:90 &  11  & 0.45 \\
3175  & 1.5 & 1e4 & 20 & 5e2 & 150& 15:85 &  1.6 & 0.23 & 3e3 & 100& 5e2 & 10 & 15:85 &  7.7 & 0.45 \\
3227  & 2.2 & 3e3 & 60 & 1e4 & 60 & 20:80 &  0.8 & 0.21 & 1e4 & 50 & 1e5 & 10 & 15:85 &  2.1 & 0.35 \\
3310  & 2.0 & 1e5 & 100& 1e3 & 100& 20:80 &  0.5 & 0.27 & 3e3 & 30 & 1e4 & 30 & 20:80 &  0.4 & 0.41 \\
3504  & 3.0 & 1e2 & 60 & 3e3 & 60 & 40:60 &  3.5 & 0.44 & 3e3 & 30 & 3e3 & 30 & 25:75 &  3.4 & 0.34 \\
3593  & 0.6 & 3e3 & 20 & 5e2 & 100& 15:85 &  2.2 & 0.12 & 1e5 & 10 & 5e2 & 20 & 50:50 &  4.2 & 0.21 \\
3627  & 0.7 & 1e5 & 20 & 5e2 & 30 & 45:55 &  3.7 & 0.37 & 1e5 & 10 & 1e3 & 60 & 40:70 &  6.5 & 0.59 \\
3628  & 0.9 & 1e5 & 30 & 1e2 & 30 & 20:80 &  17  & 0.21 & ... & ...& ... & ...&...    & ...  &...   \\
3690: & 5.2 & 1e5 & 30 & 1e2 & 150& 15:85 &  1.6 & 0.05 & 1e2 & 50 & 3e3 & 100& 50:50 &  4.2 & 0.38 \\
4030  & 2.8 & 1e4 & 20 & 3e3 & 10 & 30:70 &  2.4 & 0.20 & ... & ...& ... & ...&...    & ...  &...   \\
4038  & 2.5 & 1e4 & 150& 1e2 & 100& 30:70 &  5.5 & 0.50 & 3e3 & 20 & 1e4 & 20 & 15:85 &  5.9 & 0.33 \\
4039  & 2.5 & 3e3 & 60 & 5e2 & 100& 25:75 &  0.7 & 0.09 & 3e3 & 60 & 5e2 & 100& 15:85 &  1.7 & 0.19 \\
4051: & 1.4 &  ...& ...& ... & ...& ...   & ...  & ...  & 1e2 & 50 & 5e2 & 20 & 10:90 &  3.9 & 0.23 \\
4102  & 1.8 & 1e5 & 10 & 3e3 & 60 & 10:90 &  2.9 & 0.22 & 1e4 & 20 & 3e3 & 60 & 10:90 &  6.5 & 0.38 \\
4254  & 4.2 & 1e5 & 10 & 1e4 & 20 & 15:85 &  3.6 & 0.51 & ... & ...& ... & ...&...    & ...  &...   \\
4258::& 0.9 &  ...& ...& ... & ...& ...   & ...  & ...  & 1e5 & 10 & 3e3 & 30 & 25:75 &  2.0 & 0.22 \\
4293  & 1.5 & 1e3 & 150& 1e3 & 60 & 25:75 &  1.0 & 0.24 & ... & ...& ... & ...&...    & ...  &...   \\
4303  & 1.5 & 3e3 & 10 & 3e3 & 30 & 50:50 &  1.1 & 0.12 & 1e5 & 10 & 3e3 & 20 & 30:70 &  2.5 & 0.23 \\
4321  & 1.5 & 1e4 & 30 & 1e2 & 60 & 10:90 &  6.5 & 0.21 & 1e5 & 10 & 1e2 & 150& 20:80 &  14  & 0.37 \\
4414::& 1.0 & 1e5 & 30 & 5e2 & 20 & 15:85 &  3.2 & 0.23 & ... & ...& ... & ...&...    & ...  &...   \\
4457  & 1.4 &  ...& ...& ... & ...&  ...  & ...  & ...  & 1e4 & 60 & 1e2 & 100& 20:80 &  5.3 & 0.59 \\
4527  & 1.4 &  ...& ...&  ...& ...& ...   & ...  & ...  & 1e4 & 10 & 1e4 & 20 & 10:90 &  44  & 0.87 \\
4536  & 3.3 & 1e2 & 100& 3e3 & 100& 45:55 &  2.2 & 0.20 & 3e3 & 100& 1e5 & 10 & 40:60 &  3.2 & 0.35 \\
4631  & 0.8 & 1e3 & 60 & 1e4 & 20 & 25:75 &  0.9 & 0.22 & ... & ...& ... & ...&...    & ...  &...   \\
4666  & 2.9 & 5e2 & 100& 5e2 & 100& 10:90 &  3.3 & 0.22 & 3e3 & 30 & 5e2 & 30 & 25:75 &  5.4 & 0.34 \\
4736  & 0.5 & 1e5 & 20 & 3e3 & 20 & 40:60 &  1.4 & 0.36 & 3e3 & 20 & 3e3 & 30 & 30:70 &  3.9 & 0.28 \\
4826  & 0.4 & 1e5 & 10 & 3e3 & 100& 35:65 &  6.5 & 0.47 & ... & ...& ... & ...&...    & ...  &...   \\
4945  & 0.5 & 3e3 & 30 & 1e4 & 20 & 20:80 &  22  & 0.20 & 3e3 & 60 & 1e4 & 20 & 20:80 &  71  & 0.50 \\
5033  & 1.8 & 3e3 & 30 & 3e3 & 20 & 25:75 &  1.9 & 0.21 & ... & ...& ... & ...&...    & ...  &...   \\
5055  & 0.9 & 1e5 & 20 & 5e2 & 20 & 25:75 &  5.0 & 0.24 & ... & ...& ... & ...&...    & ...  &...   \\
5135  & 6.1 &  ...& ...& ... & ...& ...   & ...  & ...  & 3e3 & 30 & 1e5 & 10 & 25:75 &  1.8 & 0.20 \\
5194  & 1.0 & 1e5 & 10 & 3e3 & 20 & 15:85 &  13  & 0.56 & ... & ...& ... & ...&...    & ...  &...   \\
5236  & 0.4 & 1e4 & 20 & 3e3 & 60 & 40:60 &  5.9 & 0.19 & ... & ...& ... & ...&...    & ...  &...   \\
5713  & 3.3 & 3e3 & 10 & 1e4 & 10 & 30:70 &  2.4 & 0.18 & 1e4 & 30 & 1e5 & 10 & 35:65 &  2.9 & 0.29 \\
\noalign{\smallskip}                                                                   
\hline                                                                                 
\end{tabular}
}%                                                                                     
\end{center}
\end{table*}                                                                           
\addtocounter{table}{-1}
\begin{table*}
\caption[]{continued} 
\begin{center} 
{\small %
\begin{tabular}{lr|rrrrrrr|rrrrrrr}
\noalign{\smallskip} 
\hline
\noalign{\smallskip} 
(1)    & (2) & (3) & (4)& (5) & (6)& (7)   & (8)  &(9)   &(10) &(11)&(12) &(13)&(14)   &(15)  & (16)\\
\noalign{\smallskip}      
\hline
\noalign{\smallskip} 
5775: & 3.1 & 1e3 & 100& 1e3 & 20 & 20:80 & 0.20 &  1.8 & ... &... & ... & ...&...    &...   & ...  \\
6000: & 3.3 & 3e3 & 60 & 3e3 & 30 & 50:50 & 0.17 &  1.9 & 3e3 & 100& 1e5 &10  & 25:75 & 0.29 &  4.6 \\
6240  &11.6 & ... &... & ... &... &  ...  &...   &  ... & 1e4 & 20 & 1e2 &125 & 15:85 & 0.09 &  0.6 \\
6746: & 4.1 & ... &... & ... &... &  ...  &...   &  ... & 1e4 & 30 & 1e2 &100 & 10:90 & 0.21 &  2.3 \\
6946  & 0.6 & 1e3 & 60 & 1e4 & 20 & 20:80 & 0.22 &  6.8 & 1e5 & 10 & 1e4 & 20 & 20:80 & 0.53 &  29  \\
6951  & 2.6 & 1e4 & 20 & 1e2 & 100& 35:65 & 0.22 &  3.3 & ... &... & ... & ...&...    &...   & ...  \\
7331  & 1.5 & 1e5 & 10 & 1e2 & 150& 15:85 & 0.61 &  9.8 & ... &... & ... & ...&...    &...   & ...  \\
7469  & 7.1 & ... &... & ... &... &...    &...   &  ... & 5e2 & 50 & 3e3 &100 & 35:65 & 0.49 &  2.9 \\
7541  & 4.0 & 1e5 & 10 & 1e4 & 30 & 10:90 & 0.83 &  16  & ... &... & ... & ...&...    &...   & ... \\
7714: & 4.1 & 3e3 & 10 & 3e3 & 60 & 45:55 & 0.22 &  0.2 & ... &... & ... & ...&...    &...   & ... \\
\noalign{\smallskip}             
\hline
\end{tabular}
}% 
\end{center} 
Note: 
Column 1: Name; a single colon denotes that one of three isotopologue
ratios is lacking, a double colon denotes that two are lacking;
Col. 2: projected linear diameter (in kpc) of a $22"$ aperture at
galaxy distance from Table\,\ref{sample};
Cols. 3 through 9 contain two-phase modeling results assuming an
intrinsic $\co$/$\thirco$ abundance of 40;
Cols. 10 through 16 assume an abundance of 80 (see text).
Columns 3 and 10: $n_{1}$ is the $\h2$ volume density of the first gas phase;
Cols. 4 and 11: $T_{1}$ is the kinetic gas temperature $T_{k}$ of the first gas phase;
Cols. 5 and 12: $n_{2}$ is the $\h2$ volume density of the second gas phase;
Cols. 6 and 13: $T_{2}$ is the kinetic gas temperature $T_{k}$ of the second gas phase;
Cols. 7 and 14: $f_{12}$ is the relative contributions of gas phases 1 and 2 to
the observed velocity-integrated $J$=2-1 $\co$ emission;
Cols. 8 and 15: $\frac{\rm [CO]}{\rm [C]}\,=\,N(CO)/N_{\rm C}$ is
the fraction of all gas-phase carbon in CO.
Columns 9 and 16: $N_{\rm C}$ is the beam-averaged column density of all
gaseous carbon in units of $10^{17}\,\cm2$.
\end{table*}                                         

Residual degeneracies occur in the majority of galaxies modeled and
the observed ratios can usually be fit with a range of comparable
solutions. Thus, the model solutions obtained are not unique but
instead only constrain values to limited regions of parameter space.
This is illustrated in Table\,\ref{examplefit} where we show the
search results for a few galaxies. Although each of the solutions is
acceptable, individual fit parameters sometimes vary
considerably. However, these variations are not independent and the
final beam-averaged column densities resulting from the combination of
the two phases normalized by the observed CO(2-1) line intensity is
much less variable. The highest and lowest value differ by a
factor of two or less, and dispersions are typically 30\%\ or
less.

In Table\,\ref{modelpar} we list the model solution that was
closest to the observations, even when other model ratios were only
marginally different and within the observational error; for examples,
again see Table\,\ref{examplefit}. We rejected solutions in which
the denser gas component is also hotter than the more tenuous
component because we consider the large pressure imbalances implied
by such solutions physically implausible, certainly on the observed kiloparsec
scales.  Table\,\ref{modelpar} gives the model $\h2$ gas
volume densities $n_{\h2}$ and kinetic temperatures $T_{\rm k}$ for
each of the two phases as well as their relative contributions
$f_1$:$f_2$ = $f_{12}$ to the observed $J$=2-1 $\co$
intensity. Because of the residual degeneracy in the two-phase
modeling, kinetic temperatures and gas densities are uncertain by
factors of two and three, respectively.  Moreover, because our analysis is
limited to the lower three $\co$ and $\thirco$ transitions (only
14 galaxies were also observed in $J$=4-3 $\co$), we cannot
meaningfully distinguish temperatures above 100 K or densities above
$10^{4}$ $\cc$, even though our analysis may formally do so. The
results in Table\,\ref{modelpar} must therefore be taken as
representative rather than individually accurate.

Notwithstanding this reservation, successful fits at either abundance
generally suggest modest kinetic temperatures well below 100 K,
typically between 20 K and 30 K for the densest phase, usually with
densities above 3000 $\cc$. The less dense phase exhibits greater
variety in temperature and density. In most galaxies, two-thirds or
more of the $J$=2-1$\co$ emission come from a single gas phase, often
at relatively low kinetic temperatures of 10-30 K.  Overall, no more
than one-third of the detected $J$=2-1 $\co$ emission is contributed by
high-density ($n\,>\,3000\,\cc$) gas. About one-third of the observed
galaxies can be modeled with only a difference in density between the
two phases, and another one-third with only a difference in temperature.
The variation in molecular gas suggested by our modeling even in
apparently similar galaxies will at least partly reflect the rapidly
changing state of the molecular gas in galactic centers as a function
of inflow and outflow rates, as well star formation and AGN activity.
Each of these processes may significantly influence the balance of
central gas phases in any particular galaxy on timescales much
shorter than the Hubble time.

\section{Hydrogen amount}

\subsection{H column density and mass}

%Table 5 H2 column densities
\begin{table*}
\caption[]{Physical parameters of molecular gas in galaxy centers}
\begin{center}
{\small % 
\begin{tabular}{l|cc|rr|ccccc|c|cc|l}
\hline
\noalign{\smallskip} 
  NGC& $\frac{\rm [CO]}{\rm [C]}$ 
     &$N_{\rm C}$
     &\multicolumn{2}{c}{Abundance}  
     &\multicolumn{3}{c}{N$_{\rm H}$}
     & N$_{\rm HI}$
     & N$_{\h2}$
     & M$_{\rm H}$
     &\multicolumn{2}{c}{X=N$_{\h2}$/I$_{CO}$}&\\
     &&($10^{17}\,\cm2$)&O&C
     &\multicolumn{5}{c}{($10^{21}\,\cm2$)}
     &($10^7$ M$_{\odot}$)
     &\multicolumn{2}{c}{($10^{19}\,\cm2/\kkms$)}
     &Ref\\
     &40-80&40-80&&
     &E$^{40}$&G$^{40}$&N$^{40}$-N$^{80}$
     &
     &N$^{40}_h$-N$^{80}_h$
     &N$^{40}$-N$^{80}$
     &N$^{40}$-N$^{80}$&N$^{40}_h$-N$^{80}_h$
     &\\ 
(1)  & (2)     & (3)       & (4)  & (5)  & (6) & (7)  & (8)      & (9) & (10)    & (11)   & (12)    &(13)     &(14) \\
\noalign{\smallskip}
\hline              
\noalign{\smallskip}
 253 & 0.23-0.37 & 165-254 & 9.17 & 9.32 &  16 & 104  &   33-51  & 0.7 &  16-25  &   5-7  & 1.6-2.5 & 1.6-2.4 & A,B,G,1,2\\
 278 & 0.39 ...  &   6 ... & 9.30 & 9.54 & 0.3 &  3.5 &  1.1 ... & 1.1 &    ...  &   ...  & 2.7 ... &     ... & F,3,8 \\
 470 & 0.45 0.35 &   5-4   & ...  & ...  & ... &  3.0 &  0.9-0.8 & ... & 0.3-0.2 &   9-8  & 1.7-1.5 & 0.9-0.6 & ... \\
 520 & 0.15 ...  &  20 ... & ...  & ...  & ... &  13  &  4.1 ... & abs & 1.8 ... & 31 ... & 1.8 ... & 1.6 ... & 4 \\
 613 & 0.18-0.71 &  18-23  & ...  & ...  & ... &  12  &  3.5-4.6 & ... & 1.5-2.1 &   ...  & 2.5-3.3 & 2.2-3.0 & ... \\
 628 & 0.60 ...  &   8 ... & 9.30 & 9.54 & 0.6 &  5.0 &  1.7 ... & 0.3 & 0.7 ... &   ...  &  11 ... & 9.9 ... & A,B,G,5  \\
 660 & 0.16-0.72 &  59-51  & 9.10 & 9.21 & 7.3 &  37  &   12-10  & abs & 5.7-4.9 &  17-15 & 3.8-3.3 & 3.7-3.2 & F,6 \\
 891 & 0.14 ...  &  36 ... & ...  & ...  & ... &  23  &  7.1 ... & 6.4 & 3.6 ... &  9 ... & 2.6 ... & 2.4 ... & 7 \\
 908 & 0.42 ...  &  12 ... & ...  & ...  & ... &  7.8 &  2.5 ... & ... & 1.0 ... &   ...  & 4.1 ... & 3.4 ... & ... \\
 972 & 0.23-0.37 &   9-17  & 9.11 & 9.23 & 1.0 &  5.6 &  1.8-3.4 & 1.7 & 0.7-0.9 &  11-22 & 1.3-2.5 & 1.0-1.3 & F,8 \\
Maf2 & 0.48 ...  &  34 ... & ...  & ...  & ... &  22  &  6.8 ... & 1.1 & 3.2 ... &  1 ... & 1.6 ... & 1.4 ... & 9 \\
1055 & 0.23 ...  &  26 ... & 9.16 & 9.30 & ..  &  16  &  5.2 ... & ... & 1.7 ... &   ...  & 2.5 ... & 2.2 ... & G \\
1068 & 0.22-0.45 &  28-43  & 9.24 & 9.43 & 2.1 &  18  &  3.9-8.6 & 0.3 & 2.7-4.2 &  24-36 & 1.6-2.6 & 1.6-2.5 & B,G,10 \\
1084 & 0.30-0.39 &   3-8   & ...  & ...  & ... &  1.9 &  0.6-1.6 & ... & 0.1-0.6 &   ...  & 1.0-2.7 & 0.3-2.0 & ... \\
1097 & 0.26-0.76 &  36-58  & 9.27 & 9.48 & 2.4 &  23  &  7.3-12  & 1.0 & 3.1-5.3 &  32-52 & 2.7-4.3 & 2.3-3.9 & E,G,11 \\
1365 & 0.37-0.49 &  60-59  & 9.08 & 9.18 & 7.9 &  38  &   12-12  & 0.1 & 6.0-5.9 &  73-72 & 2.3-2.3 & 2.3-2.3 & A,B,G,12,13 \\
I342 & 0.37 ...  &  21 ... & 9.35 & 9.61 & 1.1 &  14  &  4.3 ... & 0.3 & 2.0 ... &  1 ... & 1.3 ... & 1.2 ... & A,G,14 \\
1614 & 0.06-0.15 &  11 ... & ...  & ...  & ... &  6.6 &  2.1 ... & abs & 0.7 ... & 61 ... & 2.4 ... & 1.7 ... & 15 \\
1792 & 0.51 ...  &   7 ... & ...  & ...  & ... &  4.2 &  1.3 ... & 0.7 & 0.4 ... &   ...  & 2.3 ... & 1.5 ... & 16 \\
1808 & 0.10-0.54 &  18-14  & ...  & ...  & ... &  11  &  3.6-2.8 & abs & 1.5-1.2 &   5-4  & 1.3-1.0 & 1.1-0.8 & 17\\
2146 & 0.21-0.36 &  25-37  & ...  & ...  & ... &  16  &  4.9-7.4 & 1.0 & 2.2-3.2 &  16-23 & 1.2-2.0 & 1.2-1.7 & 18 \\
2273 & 0.19-0.55 &   4-3   & ...  & ...  & ... &  2.7 &  0.9-0.5 & 0.4 & 0.2-0.1 &   6-4  & 2.5-1.6 & 1.4-0.4 & 8 \\
2559 & 0.16 ...  &  12 ... & ...  & ...  & ... &  7.7 &  2.4 ... & ... & 1.0 ... & 10 ... & 1.6 ... & 1.3 ... & ... \\
2623 & 0.38 ...  &   4 ... & ...  & ...  & ... &  2.7 &  0.8 ... & abs & 0.2 ... & 37 ... & 2.3 ... & 1.1 ... & 15 \\
2903 & 0.38-0.33 &   7-22  & 9.36 & 9.63 & 0.4 &  4.6 &  1.4-4.4 & 0.8 & 2.5-1.8 &   1-2  & 0.9-2.8 & 0.6-2.2 & A,B,D,G,19 \\
3034 & 0.25-0.53 &  79-87  & 9.19 & 9.37 & 6.7 &  50  &   16-17  & abs & 7.7-8.5 &   9-11 & 1.2-1.3 & 1.1-1.2 & E,20 \\
3044 & 0.13-0.05 &   3-4   & ...  & ...  & ... &  1.9 &  0.6-0.8 & ... & 0.1-0.2 &   ...  & 2.7-3.5 & 0.7-1.5 & ... \\
3079 & 0.14-0.45 &  59-25  & 9.12 & 9.25 & 6.7 &  38  &   12-5.0 & abs & 5.7-2.3 &  36-15 & 2.5-1.1 & 2.4-0.9 & F,21 \\
3175 & 0.23-0.45 &   7-17  & ...  & ...  & ... &  4.2 &  1.3-3.4 & ... & 0.4-1.5 &   2-4  & 1.5-4.0 & 1.0-3.4 & ... \\
3227 & 0.21-0.35 &   4-6   & 9.18 & 9.34 & 0.4 &  2.4 &  0.8-1.2 & abs & 0.2-0.4 &   2-3  & 0.6-1.0 & 0.3-0.6 & G,22 \\
3310 & 0.27-0.41 &   2-1   & 8.96 & 8.98 & 0.3 &  0.9 &  0.3-0.3 & 0.8 & 0.2-0.1 &   1-1  & 4.5-2.6 & 2.6-1.3 & F,G,23,24 \\
3504 & 0.44-0.34 &   8-10  & 9.26 & 9.47 & 0.5 &  4.8 &  1.5-2.0 & 0.1 & 0.7-0.9 &   8-11 & 1.4-1.8 & 1.3-1.7 & F,G,25  \\
3593 & 0.12-0.21 &  18-20  & ...  & ...  & ... &  11  &  3.6-4.0 & ... & 2.6-1.8 &   1-1  & 2.9-3.2 & 2.6-2.8 & ... \\
3627 & 0.37-0.59 &  10-11  & ...  & ...  & ... &  6.4 &  2.0-3.0 & 0.1 & 0.8-1.3 & 0.7-0.7& 1.4-2.0 & 1.2-1.8 & 26  \\
3628 & 0.21 ...  &  80 ... & ...  & ...  &...  &  50  &   16 ... & abs & 7.7 ... &  8 ... & 3.9 ... & 3.8 ... & 27 \\
3690 & 0.05-0.38 &  32-11  & ...  & ...  & ... &  20  &  6.3-2.2 & abs & 2.9-0.9 &   ...  & 4.6-1.6 & 4.5-1.4 & 28 \\
4030 & 0.20 ...  &  12 ... & ...  & ...  & ... &  7.8 &  2.5 ... & ... & 1.0 ... &   ...  & 2.9 ... & 2.4 ... & ...\\
4038 & 0.50-0.33 &  11-18  & ...  & ...  & ... &  6.6 &  1.5-3.6 & 0.8 & 0.7-1.4 &   8-14 & 2.3-3.8 & 1.4-3.0 & 29 \\
4039 & 0.09-0.19 &   8-9   & ...  & ...  &...  &  5.0 &  1.6-1.8 & 0.8 & 0.6-0.5 &   6-7  & 1.7-2.0 & 1.2-1.1 & 29 \\
4051 & ...  0.23 & ... 17  & ...  & ...  & ... &  ... &  ... 3.4 & 0.3 & ... 1.6 &  ... 4 & ... 7.7 & ... 7.0 & 30 \\
4102 & 0.22-0.38 &  13-17  & 9.16 & 9.30 & 1.3 &  8.5 &  2.7-3.4 & abs & 1.1-1.5 &   7-9  & 1.8-2.3 & 1.5-2.0 & F,31 \\
4254 & 0.51 ...  &   7 ... & 9.28 & 9.50 & 0.5 &  4.7 &  1.5 ... & 0.8 & 0.3 ... & 17 ... & 1.7 ... & 0.8 ... & A-E,G,32,33 \\
4258 & ...  0.22 & ... 9   & 9.13 & 9.27 & ... &  ... &  ... 1.9 & 2.4 & ... 0.7 &  ... 1 & ... 2.3 & ... 1.6 & B,G,34 \\
4293 & 0.24 ...  &   4 ... & ...  & ...  & ... &  2.6 &  0.8 ... & abs & 0.4 ... &    ... & 1.1 ... & 1.0 ... & 35 \\
4303 & 0.12-0.23 &   9-11  & 9.35 & 9.62 & 0.4 &  5.5 &  1.7-2.2 & 0.4 & 0.7-0.9 &   2-3  & 1.6-2.0 & 1.3-1.7 & A-D,G,32,33 \\
4321 & 0.21-0.37 &  31-37  & 9.31 & 9.56 & 1.7 &  20  &  6.2-7.4 & 0.4 & 2.9-3.5 &   9-10 & 3.8-4.5 & 3.6-4.3 & A-C,E,G,36 \\
4414 & 0.23 ...  &  14 ... & ...  & ...  & ... &  2.1 &  0.7-... & 2.0 & 1.2 ... &   ...  & 2.7 ... & 2.3 ... & 37 \\
4457 & ...  0.59 & ... 9   & ...  & ...  & ... &  11  &  3.5-1.8 & abs & 1.5-0.6 &    ... & 3.2-1.6 & 2.7 1.2 & 35 \\
4527 & ...  0.87 & ... 50  & ...  & ...  & ... &  ... &  ...  10 & ... &    .... & ... 13 & ... 4.9 & ... 4.6 & ... \\
4536 & 0.20-0.35 &  11-9   & 9.07 & 9.17 & 1.4 &  6.6 &  2.1-1.7 & abs & 0.7-0.5 &  14-12 & 1.7-1.4 & 1.1-0.8 & E,32 \\
4631 & 0.22 ...  &   4 ... & 8.92 & 8.92 & 1.0 &  2.6 &  0.8 ... & 4.6 & 0.2 ... &   ...  & 0.9 ... & 0.4 ... & E,G,38 \\
4666 & 0.22-0.34 &  15-16  & ...  & ...  & ... &  9.4 &  3.0-3.2 & ... & 1.3-1.4 &   ...  & 2.6-2.2 & 1.7-1.9 & ... \\
4736 & 0.36-0.28 &   4-14  & 9.03 & 9.10 & 0.7 &  2.8 &  0.9-2.0 & 0.3 & 0.3-0.9 & 0.2-0.3& 1.0-2.4 & 0.7-2.0 & B,D,G,39 \\
4826 & 0.47  ... &  14 ... & 8.98 & 9.04 & 2.6 &  8.9 &  2.8 ... & 1.1 & 1.2 ... & 0.3 ...& 1.6 ... & 1.3 ... & E,40\\
4945 & 0.20-0.50 & 108-141 & ...  & ...  & ... &  65  &   36-22  & abs &  18-11  &   .... & 2.4-1.5 & 2.8-1.5 & 41 \\
5033 & 0.21  ... &   9 ... & 9.18 & 9.34 & 0.8 &  5.4 &  1.7 ... & 1.1 & 0.6 ... &  3 ... & 1.6 ... & 1.2 ... & B,G,19  \\
5055 & 0.24  ... &  21 ... & 9.45 & 9.78 & 0.7 &  13  &  4.2 ... & 1.3 & 1.9 ... &  2 ... & 3.0 ... & 2.7 ... & A,B,G,19 \\ 
5135 & ...  0.20 & ... 9   & ...  & ...  & ... &  ... &  ... 1.8 & ... & ... 0.7 & ... 43 & ... 4.6 & ... 1.8 & ... \\
5194 & 0.56 ...  &  24 ... & 9.39 & 9.70 & 1.0 &  15  &  4.8 ... & 0.1 & 2.4 ... &   ...  & 5.0 ... & 4.9 ... & A,B,D,G 42 \\
5236 & 0.19 ...  &  31 ... & 9.24 & 9.45 & 2.2 &  20  &  6.2 ... & 0.3 & 2.9 ... & 0.7 ...& 1.6 ... & 1.5 ... & A,B,G,43\\
5713 & 0.18-0.29 &  14-10  & 9.07 & 9.17 & 1.9 &  8.6 &  2.9-2.0 & ... & 1.2-0.8 &0.4- 0.3& 3.2-1.8 & 2.7-1.4 & E \\
5775 & 0.20 ...  &   9 ... & ...  & ...  & ... &  5.7 &  1.8 ... & 7.5 & 0.7 ... & 11 ... & 1.9 ... & 1.4 ... & 44 \\
\noalign{\smallskip}
\hline
\end{tabular}
}%
\end{center}
\label{galmassx}
\end{table*}

\addtocounter{table}{-1}
\begin{table*}
\caption[]{continued} 
\begin{center} 
{\small %
\begin{tabular}{l|cc|rr|ccccc|c|cc|l}
\noalign{\smallskip}  
\hline
\noalign{\smallskip} 
(1)     & (2) & (3)  & (4)  & (5) & (6)  & (7)  & (8)  &(9) &(10)& (11)&(12) &(13) &(14) \\
\noalign{\smallskip}      
\hline
\noalign{\smallskip} 
6000 & 0.17-0.29 &  11-16  & ...  & ...  & ... &  6.9 &  2.2-3.2 & ... & 15-22 & 0.9-1.4 & 1.5-2.1 & 1.2-1.8 & ...\\
6240 & ...  0.09 & ... 7   & ...  & ...  & ... &  ... &  ... 1.5 & abs & ... 1 &     ... & ... 1.1 & ... 0.7 & 45 \\
6746 & ...  0.21 & ... 11  & ...  & ...  & ... &  ... &  ... 2.2 & ... &   ... & ... 0.9 & ... 4.9 & ... 3.9 & ...\\
6946 & 0.22-0.53 &  31-54  & 9.22 & 9.40 & 2.5 &  20  &  6.2-11  & 0.8 &  1-2  & 2.7-5.0 & 1.4-2.4 & 1.2-2.2 & A,B,G,46 \\ 
6951 & 0.22 ...  &  15 ... & ...  & ...  & ... &  9.5 &  3.0 ... & 0.2 &  1 ...& 1.4 ... & 3.0 ... & 2.7 ... & 47 \\ 
7331 & 0.61 ...  &  16 ... & 9.22 & 9.40 & 1.3 &  10  &  3.3 ... & 1.2 &    ...& 1.4 ... & 21  ... & 9   ... & B,G,48 \\ 
7469 & ...  0.49 & ... 6   & ...  & ...  & ... &  ... &  ... 1.3 & abs &    ...& ... 0.3 & ... 2.5 & ... 0.6 & 15 \\ 
7541 & 0.83 ...  &  19 ... & ...  & ...  & ... &  12  &  3.7 ... & 0.7 &    ...& 1.5 ... & 6.6 ... & 5.4 ... & 49 \\ 
7714 & 0.22 ...  &   1 ... & ...  & ...  & ... &  0.7 &  0.2 ... & 1.3 &    ...& ... ... & 1.1 ... & ... ... & 50 \\ 
\noalign{\smallskip}             
\hline
\end{tabular}
}% 
\end{center} 
Notes: \\
Abundances in Cols. 4 and 5 are 12+log[O]/[H] and 12+log[C]/[H],
respectively.  In Cols. 6 through 8, total hydrogen column densities
were derived assuming a carbon dust grain depletion factor of 0.5.  In
Cols. 10 through 13, values marked with subscript h are derived with
correction for the $\hi$ contribution in Col. 9 for galaxies with a
tilt b/a $\geq$ 0.6; a contribution N($\hi$)=$0.5\times10^{21}$ (see
section 7.2) was assumed for all other galaxies including those
lacking $\hi$ data. Mass listed is total hydrogen mass; to obtain
total gas mass, multiply by 1.35. I(CO) is taken from
Table\,\ref{sestdat}.  E(xtrapolated) is the lower limit assuming
the extrapolated carbon abundance from Col. 5; G(alactic) is the
upper limit assuming solar neighborhood carbon abundances; N(ominal)
is the most representative intermediate carbon abundance discussed in
the text. The small galaxies NGC~3310, NGC~4631, and NGC~4826 have
much lower abundances than the other galaxies; for these we set N
equal to G. Superscripts $^{40}$ and $^{80}$ denote results for
isotopological abundances
$[\co]/[\thirco]=40$ and $[\co]/[\thirco]=80$, respectively.\\
Entries in Col. 17 are as follows:\\
Reference to $\hi$ data: 1. Lucero $\etal$ (2015); 2. Puche $\etal$
(1991); 3. Knapen $\etal$ (2004); 4. Stanford (1990); 5. Shostak
$\etal$ (1984) 6. van Driel $\etal$ (1995); 7. Swaters $\etal$ (1997);
8. WHISP database, hhtps://www. astro.rug.nl/$~$whisp/ 9. Hurt $\etal$
(1996); 10. Brinks $\etal$ (1997); 11. Ondrechen $\etal$ (1989b);
12. Ondrechen $\etal$ (1989a) 13. J\"ors\"ater \& van Moorsel (1995)
14. Crosthwaite $\etal$ (2000); 15. Hibbard $\etal$ (20001);
16. Dahlem (1992) 17. Saikia $\etal$ (1990); 18. Taramopoulos $\etal$
(2001) 19. Wevers $\etal$ (1986); 20. Yun $\etal$ (1993); 21. Irwin \&
Seaquist (1991) 22. Mundell $\etal$ (1995); 23. Mulder $\etal$ (1995);
24. Kregel $\etal$ (2001); 25. van Moorsel G.A. (1983); 26. Zhang
$\etal$ (1993); 27. Wilding $\etal$ (1993); 28. Stanford \& Wood
(1989); 29. van der Hulst (1979)' 30. Liszt \& Dickey (1995);
31. Verheijen \& Sancisi (2001); 32. Warmels (1988); 33. Cayatte
$\etal$ (1990); 34. van Albada (1980); 35. Chung $\etal$ (2009)
36. Knapen $\etal$ (1993) 37. Braine, Combes \& van Driel (1993b);
38. Rand (1994); 39. Mulder \& van Driel (1993); 40. Braun $\etal$
(1994); 41. Ott $\etal$ (2001); 42. Tilanus \& Allen (1991);
43. Tilanus \& Allen (1993); 44. Irwin (1994); 45. Baan $\etal$
(2007); 46. Tacconi \& Young (1986); 47. Haan $\etal$ (2008);
48. Bosma (1981); 49. Chengalur $\etal$ (1994); 50. Smith \& Wallin
(1992).\\ References to abundance data: A: Vila-Costas $\&$ Edmunds,
1992; B: Zaritzky $\etal$ C. Skillman $\etal$, 1996; D. Moustakas $\&$
Kennicutt, 2006; E. Moustakas $\etal$, 2010; F. Robertson $\etal$,
2013; G. Pilyugin $\etal$, 2014.
\end{table*}            

Total hydrogen column densities $N_{\rm H}$ would follow directly from
the carbon column densities $N_{\rm C}$ if the gas phase
carbon-to-hydrogen abundance were directly known, which is not the
case. Instead, we must infer this abundance from our knowledge of (i)
the relative oxygen abundance [O]/[H], (ii) the relative carbon
abundance [C]/[O], and (iii), the fraction $\delta_C$ of all carbon
that is in the gas phase rather than locked up in dust grains.

Optical spectroscopy of disk $\hii$ regions has yielded oxygen
  abundances that are expressed as metallicities 12+log(O/H) for various
galaxies. The results for 34 galaxies from our sample are summarized in
Col. 4 of Table\,\ref{galmassx}. They are taken from the
compilations of extragalactic HII region abundances by Vila-Costas
$\&$ Edmunds (1992) and later publications based on similar methods
(references given in the table) \footnote{Some studies (cf. Pilyugin
  $\etal$ 2012, 2014) derive abundances that are systematically lower by a
  factor $\sim$2.5. See the discussion by Peimbert $\etal$
  2017}. Unfortunately, individual measurements of HII regions in
galaxy centers may suffer errors of up to factors of two. Frequently, no
suitable HII regions occur in the very center of a galaxy. In these
cases, central metallicities are deduced from a linear extrapolation
of metallicity gradients to zero radius, but the non-negligible
dispersion of individual abundances causes relatively large errors in
the slopes of the metallicity gradient. In addition, there is some evidence
that these gradients flatten at small radii so that linear
extrapolations to zero radius overestimate the central metallicity.
The data in Table\,\ref{galmassx} define a mean zero-radius abundance
12+log(O/H)=9.2$\pm$0.2, which is three times the solar neighborhood
metallicity.  However, in view of the uncertainties involved, we
consider an intermediate metallicity twice solar to be more
reasonable. This implies an elemental ratio [O]/[H] = 10$^{-3}$, which
we consider uncertain by a factor of one and a half.

The relation of carbon to oxygen, the carbon abundance [C]/[O], has
been investigated at solar metallicity and below by various authors
(e.g., Kobulnicky $\&$ Skillman, 1998; Garnett $\etal$ 1999, 2004;
Esteban $\etal$ 2014, Berg $\etal$ 2016). From solar metallicity
downward, [C]/[O] drops linearly proportional to [O]/[H]
to a metallicity of about a tenth solar and then flattens.  The few
available data suggest equal [C]/[H] and [O]/[H] abundances at
metallicities just above solar. It is unlikely that the trend observed
at subsolar abundances can be extrapolated much farther as this would
quickly lead to unrealistically high carbon fractions.  Accordingly,
we assume equal carbon and oxygen abundances at supersolar
metallicities, so that [C]/[H] $\approx\,10^{-3}$.

A final source of uncertainty is the carbon-depletion factor
  $\delta_{\rm C}$.  In galaxy disks, as much as two-thirds of all carbon
may be tied up in dust particles, rendering it unavailable for the
gas phase (see Jenkins, 2009). However, turbulence and shocks may
cause substantial dust grain erosion in galaxy centers, leading to a
higher carbon gas fraction. It is thus reasonable to adopt a depletion
factor $\delta_{\rm C}\,=\,0.5\pm0.2$.  Taken together, these three
considerationsC suggest that the best result is obtained with an
intermediate `nominal' (N) phase ratio
$N_{\rm H}/N_{\rm C}\,=\,(2\pm1)\times10^{3}$.

In Table\,\ref{galmassx} we present the beam-averaged total hydrogen
column densities $N_{\rm H}$ based on this nominal gas-phase carbon
ratio (columns headed 'N'), which is taken to be the same for all galaxies
unless otherwise noted. We present results for isotopologue ratios of
40 and 80, respectively, denoted by superscript. For comparison, we
also list column densities ('E') assuming the individual extrapolated
high gas-phase carbon abundance from Col. 5 to apply, as well as
column densities ('G') assuming for all galaxies the same low solar
neighborhood gas-phase carbon abundance.  The 'E' and 'G' columns
represent the extreme lower and upper limits.

Only the extrapolated case E total hydrogen column densities
$N_{\rm H}$ in Table\,\ref{galmassx} are based on individually
determined carbon abundances; they have a dispersion of 0.09,
corresponding to a factor 1.25. The case N as well as case G values
are based on a fixed value for the whole sample. We assume that these
are characterized by the same dispersion as case E.  The mean values of
the beam-averaged case N column densities for isotopological
abundances of 40 and 80 differ by a factor of 1.6, thus introducing an
error of $27\%$ in the combined data set. Taking into account the
uncertainty in the beam-averaged carbon column densities themselves
(see Section 6.1), we find a combined uncertainty in each derived set
of hydrogen column densities slightly over a factor of two.

The final uncertainty lies, however, in our choice of the assumed
metallicity-dependent carbon abundance.  We chose a nominal value (N)
for $N_{\rm H}$ that is on average a factor of four higher than the
extrapolated value (E) and is by definition a factor of three lower
than the Galactic value (G) in Col. 3. This is a
collective rather than an individual uncertainty, however.

\subsection{$\h2$ column density}

The column density of molecular hydrogen N($\h2)$ follows from that of
total hydrogen N$_{\rm H}$ after subtraction of the neutral hydrogen
contribution N($\hi$). In Col. 9 of Table\,\ref{galmassx} we list the
$\hi$ column densities from the literature at resolutions similar to
those of the normalized CO beam used in this paper; almost all were
originally obtained with either the Westerbork Synthesis Radio
Telescope (NL) or the Very Large Array (USA). Most of the $\hi$ column
densities are relatively low. The few high values originate in
strongly tilted galaxies where the long lines of sight include gas at
large radii. Maps show that the distributions of CO and $\hi$ in
galaxies are anticorrelated: CO usually peaks in the center where
$\hi$ maps frequently exhibit clear central holes. For galaxies with
$b/a\,\geq\,0.6,$ we took the actual N($\hi$) values from Col. 9, and
for tilted galaxies with $b/a\,<\,0.6$ as well as those where no $\hi$
data were found, we set N($\hi$) to $0.5\times10^{21}\,\cm2$, the
average for the galaxies with $b/a\,\geq\,0.6$.

Beam-averaged molecular hydrogen column densities assuming nominal
abundances are listed in Table\,\ref{galmassx} for the two
isotopological abundances. The values in Cols. 10 and 13 are
corrected for the contribution of $\hi$. We did not separately list
uncorrected column densities as these are just the values in Col. 8
divided by two. For most galaxies, the $\hi$ contribution is minor,
typically $15\%$. In a recent study, Gerin $\&$ Liszt (2017) reached
an almost identical conclusion for the inner Milky Way ($R\,<$ 1.5
kpc) using a completely different line of reasoning.

For carbon abundances as low as those of the Solar Neighbourhood (case
G), the $\hi$ contribution is in fact negligible (typically
$\leq 5\%$).  However, if the carbon abundances were as high as
suggested by the full extrapolation of the abundance gradients (case E),
the total hydrogen column densities would be reduced to the levels of
$\hi$. This would leave no room for molecular hydrogen, implying
improbably high CO emissivities as well as improbably low gas-to-dust
ratios. The high case E carbon abundances are therefore ruled out
unless almost all carbon is in dust and very little in the gas
phase. This is not expected in (dynamically) active environments such
as galaxy centers.

The errors in the derived $N(\h2)$ values are almost identical to
those discussed in the preceding section: about a factor of 2.5 for
for the nominal (N) case and slightly less for the low abundance (G)
case. This level of uncertainty is similar to the systematic uncertainty
represented by the difference between the two cases.

Funally, we derived the CO-to-$\h2$ conversion factors
$X\,=\,N(\h2)/I_{\rm CO(1-0)}$ for the nominal carbon abundance with
and without $\hi$ subtraction for each of the two isotopologue
abundances considered. The real but relatively small effects of $\hi$
correction and isotopologue abundance are illustrated by Cols. 12 and
13 in Table\,\ref{galmassx}.

\end{appendix}

\end{document}